\documentclass[preprint,12pt]{elsarticle}
\usepackage{amssymb,siunitx}
\usepackage{color}
\usepackage{graphicx}
\usepackage{ulem}
\usepackage{amsmath}
\usepackage{hyperref}

\DeclareUnicodeCharacter{2500}{\textendash}

\newcommand{\apjl}{The Astrophysical Journal Letters}
\newcommand{\apjs}{The Astrophysical Journal Supplement}

\newcommand{\aap}{Astronomy \& Astrophysics}

\begin{document}
\begin{frontmatter}
\title{Baryon Cycles in the Biggest Galaxies}
\author[1]{Megan Donahue\corref{corr}}
\ead{donahu42@msu.edu}

\author[1]{G. Mark Voit}
\ead{voit@msu.edu}
\cortext[corr]{Corresponding author.}
\address[1]{Michigan State University, Physics \& Astronomy Department, East Lansing, Michigan USA}

\begin{abstract}
The universe's biggest galaxies have both vast atmospheres and supermassive central black holes.  This article reviews how those two components of a large galaxy couple and regulate the galaxy's star formation rate.  Models of interactions between a supermassive black hole and the large-scale atmosphere suggest that the energy released as cold gas clouds accrete onto the black hole suspends the atmosphere in a state that is marginally stable to formation of cold clouds.  A growing body of observational evidence indicates that many massive galaxies, ranging from the huge central galaxies of galaxy clusters down to our own Milky Way, are close to that marginal state.  The gas supply for star formation within a galaxy in such a marginal state is closely tied to the central velocity dispersion ($\sigma_v$) of its stars. We therefore explore the consequences of a model in which energy released during black-hole accretion shuts down star formation when $\sigma_v$ exceeds a critical value determined by the galaxy's supernova heating rate.
\end{abstract}

\end{frontmatter}



\tableofcontents

\section{Introduction \label{sec:Intro}}

At the center of every massive galaxy is a supermassive black hole deeply connected to the star-formation history of the galaxy hosting it.  Such a behemoth black hole has a mass of $10^6$--$10^{10} \, M_\odot$, far exceeding that of a star, or even of most clusters of stars. It reveals itself most prominently when it accretes gas from its surroundings.  The gravitational potential energy released during accretion powers either copious emission of radiation across the entire electromagnetic spectrum or strong bipolar outflows, and sometimes both, emanating from a relatively compact region centered on the black hole.  During those periods of high activity, the region around the black hole is known as an active galactic nucleus (AGN).  

Astronomers now consider AGNs essential to a complete understanding of galaxy evolution because of two important discoveries made roughly two decades ago. First, the mass of a galaxy's central supermassive black hole correlates with galactic properties measured beyond where the black hole's gravity dominates the galaxy's dynamics. Those properties include the central velocity dispersion of the galaxy's stars and the total stellar mass of its bulge \citep{1998AJ....115.2285M, 2000ApJ...539L...9F,2000ApJ...539L..13G}.  Second, the AGNs in the universe's biggest galaxies drive outflows with kinetic power that approximately matches the radiative energy losses from gas surrounding the galaxy \cite{Churazov+02,Birzan+04}, a scenario originally envisioned by Binney and Tabor \cite{TaborBinney1993MNRAS.263..323T,BinneyTabor_1995MNRAS.276..663B}.


If numerical simulations of cosmological galaxy formation were able to reproduce the observed properties of large galaxies without including energy input from AGNs, then these correlations could be dismissed as merely circumstantial.  However, numerical simulations of large galaxies that do not include the effects of feedback fail to match observations of real ones \cite{Balogh+2001,Borgani+2004,Benson2010PhR...495...33B}.  Also, a galaxy's central black hole mass does not simply correlate with its stellar mass. Its connection with the stars must be more intimate, because the central black hole masses observed among galaxies of similar stellar mass \textit{anticorrelate} with the star formation rates of those galaxies \citep{Terrazas_2016ApJ...830L..12T,Terrazas_2017ApJ...844..170T}.  The relationships found between supermassive black holes and their host galaxies therefore appear to have a causal origin resulting from what has come to be known as \textit{AGN feedback} \cite[e.g., ][]{SilkRees1998AA...331L...1S,Haehnelt+1998,SomervilleHopkins+2008,Cattaneo_2009Natur.460..213C,Fabian12,CrainEAGLE+2015}. 

The term ``feedback" is not used here as a sound engineer might use it.  To a sound engineer, feedback describes a self-amplifying loop, or ``positive feedback." In the context of galaxy evolution, ``feedback" usually means ``negative feedback" that regulates galaxy growth by limiting star formation.  An AGN can limit star formation within a galaxy by ejecting potentially star-forming gas clouds from a galaxy. Or an AGN can prevent the hotter gas surrounding the galaxy from cooling and increasing the amount of cold star-forming gas inside the galaxy.  These feedback mechanisms can be self-regulating if AGN feedback events are triggered by accumulations of potentially star-forming gas that accretes onto the supermassive black hole, powering the AGN.

This review summarizes the evidence connecting supermassive black holes with the evolution of galaxies through the black hole's influence on a galaxy's most diffuse gaseous component, its \textit{circumgalactic medium} (CGM).  Star formation in a galaxy depends on its gas supply, which enters a galaxy through the CGM.  Feedback processes then return much of that gas to the CGM, and some of it eventually recycles from the CGM back through the galaxy.  In low-mass galaxies, supernova explosions can provide enough energy to drive this ``baryon cycle."  But AGNs are essential components of baryon cycles in the biggest galaxies.  

\subsection{Galaxy Formation: A Primer \label{sec:GalForm}}

Many readers of this article will already be familiar with the basic features of galaxy formation.  Some might not be. Here we provide a brief summary of galaxy-formation ideas that the rest of the article will presume are common knowledge.  

Galaxy formation started with the Big Bang, the moment when the universe began to expand from a hot, dense state \cite{Voit_2005RvMP...77..207V}.  The current expansion rate is determined by measuring the distances to galaxies well outside of our own Local Group and comparing them with the speeds at which those galaxies are receding from us, as measured through the redshifts of their spectra \cite{Hubble1929,Lemaitre1927}. Initially, the universe's expansion slowed with time, at the rate expected for a universe filled with a blend of matter and radiation, but the expansion has been accelerating for the last several billion years, presumably because a mysterious form of ``dark energy" has come to dominate its global dynamics \cite{1998AJ....116.1009R,1999ApJ...517..565P}.

Measurements of the current expansion rate, as well as its early deceleration and later acceleration, provide fairly precise determinations of the universe's matter and energy contents \cite{2016ApJ...826...56R}. It currently consists of about 30\% matter and 70\% dark energy, with a negligible radiation energy density, but those proportions change with time. Energy density in the form of non-relativistic matter declines as $a^{-3}$, where $a$ is the universe's scale factor, and the energy densities of radiation and relativistic matter decline as $a^{-4}$. The dark energy density apparently remains nearly constant.  That constancy is why dark energy eventually dominates the universe's dynamics as the universe expands.

Observations of the remnant radiation of the Big Bang, also known as the cosmic microwave background (CMB), corroborate all of those inferences from the universe's expansion rate and provide additional information about the universe's structure \cite{2016A&A...594A..13P,2013ApJS..208...19H}.  Analyses of the extremely subtle spatial fluctuations in the CMB show that the fraction of matter in baryonic\footnote{For astronomical purposes, the term ``baryonic" matter represents not only matter comprised of baryons (protons, neutrons) but also the associated leptons (electrons).} form is only about one-sixth of the total amount of matter.  The rest, presumed to consist of as yet undiscovered particles that interact only through gravity and perhaps also the weak force, is called ``dark matter."  

Galaxies and clusters of galaxies grew from the subtle differences in matter density \cite{Lemaitre1933,Tolman1934,Lifshitz1946} also responsible for the CMB fluctuations \cite{SunyaevZeldovich1970b}.  In regions of the early universe that were denser than average, deceleration of the universe's expansion was slightly greater than average.  The density contrasts of those overdense regions, compared to the universe as a whole, grew slowly at first and later hastened as gravity locally halted the expansion and caused those regions to collapse upon themselves.\footnote{Dark energy is relevant only over volumes very large compared to the sizes of these gravitationally bound regions.} As clumps of matter fell back toward the center of the local potential well, stochastic gravitational forces between them deflected their trajectories, causing them to settle into gravitationally bound but randomly oriented orbits.  

This ``virialization" of the infalling matter resulted in a gravitationally bound object known as a ``halo" with a gravitational potential energy roughly twice the magnitude of the total kinetic energy of the moving parts \cite{Poincare1911,Eddington1916}.  In fact, the first evidence for dark matter came from application of this virialization concept to galaxy clusters, which showed that the orbital speeds of those galaxies implied the presence of vastly more matter in the cluster than could be ascribed to stars in the cluster's galaxies  \cite{Zwicky1933,Zwicky1937}.  Non-baryonic dark matter cannot lose orbital energy by radiating photons, but a halo's baryonic matter is able to lose orbital energy through inelastic collisions that do radiate photons.  The baryonic matter therefore settles toward the halo's center, where it forms stars and produces an observable galaxy.  

Galaxies were first noticed as smudges of light called ``nebulae."  Distance measurements with standard-candle techniques \cite{Leavitt1912} later revealed them to be analogs of our own Milky Way ranging from millions to billions of light-years away \cite{Hubble1925}.  Motions of stars and gas clouds in galaxies indicate that dark matter provides most of the gravity binding them to the galaxy. A typical large galaxy has more than 100 billion stars, but the total amount of matter, as first inferred from orbits of gas clouds in disk galaxies \cite{Rubin+1980,RogstadShostak1972} and later confirmed through gravitational lensing \cite[e.g.,][]{Bolton+2008} exceeds the stellar mass inferred from the total amount of starlight.\footnote{For more background, consult \textit{The History of Dark Matter} by \citet{HistoryOfDarkMatter2018RvMP}.}  In fact, the mass represented by stars is only a small fraction of the baryonic mass that should have accompanied the dark matter during its collapse to form a halo.  The majority of a halo's baryonic matter is therefore presumed to be in the form of diffuse gas surrounding the central galaxy \cite{2016A&A...594A..13P,2017ARA&A..55..389T,Bregman_2018ApJ...862....3B,Macquart+2020}.

The diffuse and gaseous portion of this baryonic matter around most galaxies is very hard to detect.  Therefore, its location and properties are difficult to definitively determine.  However, the most massive galaxies in the universe---the Brightest Cluster Galaxies (BCGs) that inhabit the centers of galaxy clusters---have diffuse gaseous atmospheres that emit easily observable X-rays \cite{Sarazin_1986RvMP...58....1S}. 
Observations of both X-rays and starlight from galaxy clusters confirm that their proportions of stellar mass, baryonic mass, and dark matter are consistent with the universe's overall proportions, as inferred from CMB studies \cite{2012ApJ...754..119M,2013ApJ...767..116M,2014MNRAS.443.1973V,2015MNRAS.449..685H,2016ApJ...821..116U}.

Many questions remain about the details of this broad-brush picture. Despite the apparently simple origins of galaxies from gravitational growth of small-amplitude density fluctuations in the early universe \cite{PressSchechter1974}, the appearances and histories of galaxies can vastly differ. Gravitational collapse alone would give rise to ``self-similar" structures, in the sense that the more massive things would look similar to scaled-up versions of less massive things. But astronomers knew even in the early 20th century that galaxies come in many varieties. So there is clearly more to galaxy formation than simple gravitational collapse followed by radiative cooling of baryonic gas that goes on to form stars.

In the consensus $\Lambda$CDM model of cosmological structure formation, gravitational clumping of cold dark matter (CDM) makes halos. Eventually, dark energy with properties resembling Einstein's cosmological constant ($\Lambda$) accelerates the halos apart. This model has been hugely successful in reproducing the large-scale distribution of matter \citep[e.g.,][]{BOSS2017MNRAS.470.2617A, DESclusters_2020PhRvD.102b3509A, DEScosmo_2019MNRAS.483.4866A}. But some puzzles remain when the $\Lambda$CDM model is applied to galaxy evolution. In the $\Lambda$CDM model, galaxy formation is a hierarchical process.  Small halos form first, and larger halos grow through mergers of smaller ones.  One might therefore expect larger and more massive galaxies to appear younger than smaller ones.  

However, galaxy surveys show the opposite trend  \cite{Cowie+1996AJ....112..839C,Behroozi+2013ApJ...770...57B}.
Massive elliptical galaxies formed most of their stars very early in the history of the universe, while lower-mass disk galaxies formed their stars later and are still making them \citep{Behroozi+2013ApJ...762L..31B, Behroozi+2013ApJ...770...57B}. Something about massive galaxies causes them to start forming stars earlier and to stop sooner than their lower mass cousins.  Also, the stellar populations of massive elliptical galaxies at the centers of high-mass halos have star-formation histories that are not consistent with simple hierarchical assembly through mergers of lower-mass galaxies like those we observe today
\cite{DeLuciaBlaizot2007,Guo_2011MNRAS.413..101G}.  Our best guess as to why star formation in massive galaxies ceases earlier is that additional physics, beyond the physics of gravitational infall and cosmological shock heating, alters the baryon cycle that feeds star formation.

\subsection{The Baryon Cycle: A Preview \label{sec:BaryonCycle}}

This review adopts a particular point of view about how AGN feedback couples with the baryon cycle in the universe's biggest galaxies.  Motivated by evidence that has been accumulating for the past couple of decades, we consider the connections between feedback output from the supermassive black hole in a massive halo's central galaxy and the prevailing conditions in its CGM.  Throughout most of the article, we will treat a massive halo as though it were evolving in isolation, without considering the effects of mergers with similarly massive halos.  We will therefore be focusing on how the asymptotic state of a relaxed galactic atmosphere shapes the baryon cycle that determines how a halo's central galaxy evolves.  But as the article draws to a close, we will return to the role of mergers and satellite galaxies and how they affect AGN feedback.

\begin{figure}
\centering
\includegraphics[scale=0.6]{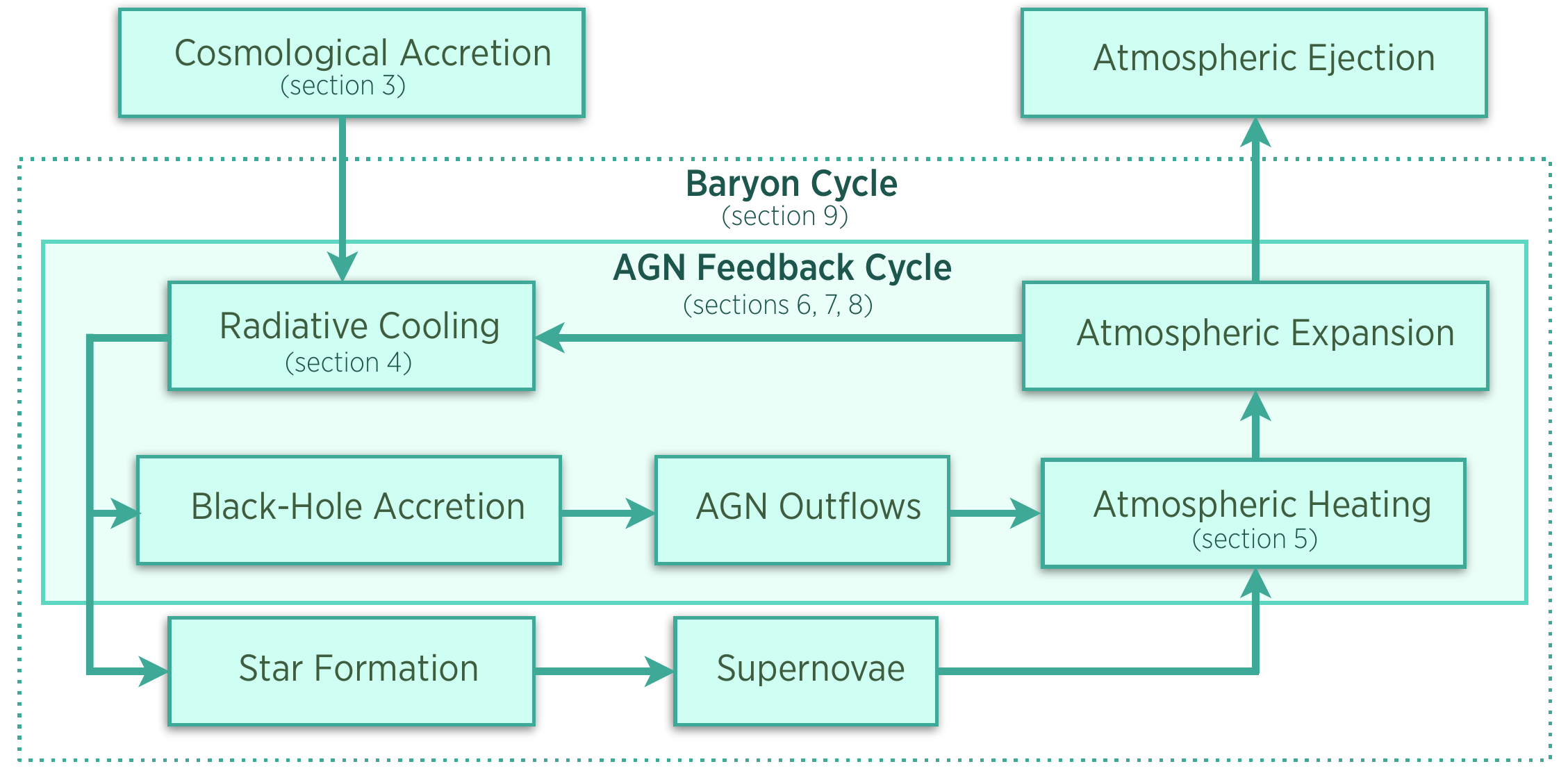}
\caption{Schematic diagram of the galactic baryon cycle, with references to sections of this article. \label{figure:BaryonCycle}}
\end{figure}

Figure~\ref{figure:BaryonCycle} provides a graphical representation of the key ideas and how they are related. The article itself proceeds as follows: \\

\begin{itemize}

    \item Section 2 describes the observational evidence for atmospheres around galaxies and introduces the basic physical properties of galactic atmospheres.
    
    \item Section 3 outlines how those atmospheres first accumulated through cosmological accretion.  That process drives shocks that raise the specific entropy of accreting gas. It would have resulted in atmospheres with self-similar structure, if the shocked gas had been unable to cool.
    
    \item Section 4 explains how radiative cooling broke self-similarity by allowing the atmospheric gas to shed some of the thermal energy it gained through accretion.  Cooling lowers the specific entropy of galactic atmospheres and allows galaxy formation to happen.
    
    \item Section 5 summarizes the observations indicating that energetic feedback emanating from a halo's central galaxy intervenes to limit cooling and condensation of its CGM, thereby limiting the central galaxy's ability to form stars. 
    
    \item Section 6 focuses on the central galaxies of galaxy clusters---the biggest galaxies in the universe---because that is where observations of AGN feedback and its effects on the CGM are most complete.  Multiwavelength observations of central cluster galaxies suggest that AGN feedback produces ``weather" in the CGM that regulates a central galaxy's ability to form stars.
    
    \item Section 7 discusses a set of physical models that provide a theoretical framework for how the regulation mechanism works.  The ``weather" appears to depend on how susceptible the CGM is to producing a ``rain" of cold clouds (sometimes called \textit{precipitation}).  Some of those clouds then fall toward the galaxy's center and can fuel feedback from the central black hole (through a process sometimes called \textit{chaotic cold accretion}).
    
    \item Section 8 presents observational evidence supporting the hypothesis that precipitation is a key part of the self-regulating feedback loop in massive galaxies and possibly in lower-mass galaxies as well.  It also points out some observationally testable predictions that models incorporating precipitation make about the state of the CGM.
    
    \item Section 9 explores the implications of these ideas for how both AGN feedback and stellar feedback regulate star formation in galaxies. In galaxies like the Milky Way, feedback stimulated by precipitation can link a galaxy's star formation rate to the depth of its central potential well, allowing galaxies with deeper central potentials to form stars more rapidly.  However, growth of a galaxy's central potential well is ultimately self-limiting because of how it focuses precipitation onto the central black hole, forging a tight connection between AGN feedback and circumgalactic pressure that suppresses further star formation.
    
\end{itemize}

\section{Properties of Galactic Atmospheres \label{sec:AmbientCGM}}

This review treats all of the gas gravitationally bound to a galaxy's halo as an \textit{atmosphere} because many of the concepts that apply to the atmospheres of planets and stars also apply to the gas in and around galaxies.  The most important of those concepts is hydrostatic equilibrium, the balance between gas pressure and gravity that prevents an atmosphere from collapsing.  Galactic atmospheres are never perfectly static, but hydrostatic equilibrium is often a useful approximation because of how it links a galaxy's atmospheric temperature ($T$) to the depth of its halo's potential well and the atmosphere's pressure ($P$) and density ($\rho$) to its total gas mass.  While significant departures from hydrostatic equilibrium can happen around massive galaxies and are quite likely around lower mass galaxies, especially during mergers, much can be learned by treating those departures as perturbations of the hydrostatic state that a galactic atmosphere would settle into, if left undisturbed.

A galaxy's atmosphere consists of all of the gas gravitationally bound to its halo. According to this definition, a galaxy's atmosphere includes both its interstellar medium (ISM) and its CGM.  Much more is known about the ISM component of a galactic atmosphere, because it is comparatively easy to observe in many different bands of the electromagnetic spectrum.  Many books and review articles already provide excellent introductions to the ISM in all kinds of galaxies \cite{Spitzer1998,Draine2011,OsterbrockFerland2006, 2001RvMP...73.1031F,2005ARA&A..43..337C}, so this section will say little about the ISM, focusing instead on the properties of the CGM.

The section begins with a discussion of a galactic atmosphere's boundaries and then introduces what observations reveal about the characteristics of the CGM, along with the physical concepts used to interpret those observations. It concludes by outlining the principles that govern a galactic atmosphere's structure.

\subsection{Atmospheric Boundaries}
\label{sec:boundaries}

One reason to consider all of a galaxy's gravitationally bound gas to be a unified atmosphere is that there is no consensus on where the ISM ends and the CGM begins.  In our view, this distinction is artificial because a galaxy's atmosphere is all of a piece.  Even in a disk galaxy, the ISM sits beneath the CGM and is compressed by its weight.  The ISM receives gas that cools out of the CGM, enriches it with new elements as stars form and explode, and returns much of the ISM gas to the CGM through hot, highly enriched outflows collectively driven by the galaxy's supernovae \cite{2004RvMP...76..125M,Putman_2012ARA&A..50..491P}.  Star formation in the ISM is the most obvious stage of the baryon cycle but would consume most of the ISM gas within a couple of gigayears (1~Gyr = $10^9$ years) if supply from the CGM were interrupted \cite[e.g., ][]{Krumholz+2012ApJ...745...69K}.  The CGM's properties therefore limit how rapidly the baryon cycle of a galaxy can operate.

There is also no consensus on the outer boundary of the CGM.  Beyond it lies the universe's most diffuse baryonic component, the intergalactic medium (IGM).  As observations of the CGM have accumulated, much of what used to be called the IGM is now considered by many to belong to the CGM. Indeed, the Warm-Hot Intergalactic Medium (WHIM), defined to be intergalactic gas between $10^5$~K and $10^7$~K, is not so easy to distinguish from the CGM, as traced by O~VI absorption lines \cite{McQuinnWerk_2018ApJ...852...33M,Werk_2014ApJ...792....8W,Rudie+2019ApJ...885...61R}.  

One commonly applied CGM demarcation line is the halo's virial radius ($r_{\rm vir}$), which likewise has no consensus definition.  This article will use a definition for the virial radius specified in terms of how the mean matter density within it compares with the cosmological critical density
\begin{equation}
    \rho_{\rm cr}(z) \equiv \frac {3 H^2(z)} {8\pi G}
\end{equation}
where $G$ is the gravitational constant and $H(z)$ is the Hubble expansion parameter corresponding to cosmological redshift $z$.\footnote{The cosmological redshift $z$ of a distant galaxy's light is related to the universe's scale factor $a$ at the time that light was emitted via $a = 1 / (1+z)$. At the present time ($z=0$), the total matter density $\rho_{\rm M}$ amounts to a fraction $\Omega_{\rm M} \equiv \rho_{\rm M}/\rho_{\rm cr}(0) \approx 0.3$ of the critical density. If dark energy with a constant energy density makes the universe's total mass-energy density equal to $\rho_{\rm cr}$, then $H(z) = H_0 E(z) = H_0 \sqrt{\Omega_{\rm M} (1+z)^3 + (1- \Omega_{\rm M})}$.}  
The radius encompassing a matter density $\Delta$ times the critical density is therefore
\begin{equation}
    r_\Delta 
       \equiv \left( \frac {3 M_\Delta} 
                           {4 \pi \Delta \rho_{\rm cr}} \right)^{1/3}
       =  \left( \frac {2 G M_\Delta} 
                       {\Delta H^2} \right)^{1/3}
       =  \left( \frac {2} {\Delta} \right)^{1/2}
                 \frac {v_{\rm c}(r_\Delta)} {H}
    \label{eq:r_Delta}
\end{equation}
where $M_{\Delta}$ is the mass contained within $r_\Delta$ and $v_{\rm c}(r_\Delta) = (G M_\Delta / r_\Delta)^{1/2}$ is the halo's circular velocity at $r_\Delta$.  Numerical simulations show that orbits of dark matter particles in the outer parts of a cosmological halo gradually shift from mostly infalling to mostly isotropic in the neighborhood of $\Delta = 200$ \cite{ColeLacey1996MNRAS.281..716C,BryanNorman1998ApJ...495...80B}, and so our working definition for the virial radius will be $r_{200 {\rm c}}$.  The letter ``c" in the subscript is a reminder that $\Delta$ is defined with respect to $\rho_{\rm cr}$.\footnote{A precise definition of the halo's outer radius matters when you want to state the mass associated with a halo to within a factor of two. Be aware that elsewhere in the literature the outer radius is sometimes defined with respect to the mean background matter density instead of the critical density.}

\begin{figure}[!t]
\centering
\includegraphics[width=5.4in,trim=0.1in 0.0in -0.1in 0.0in,clip]{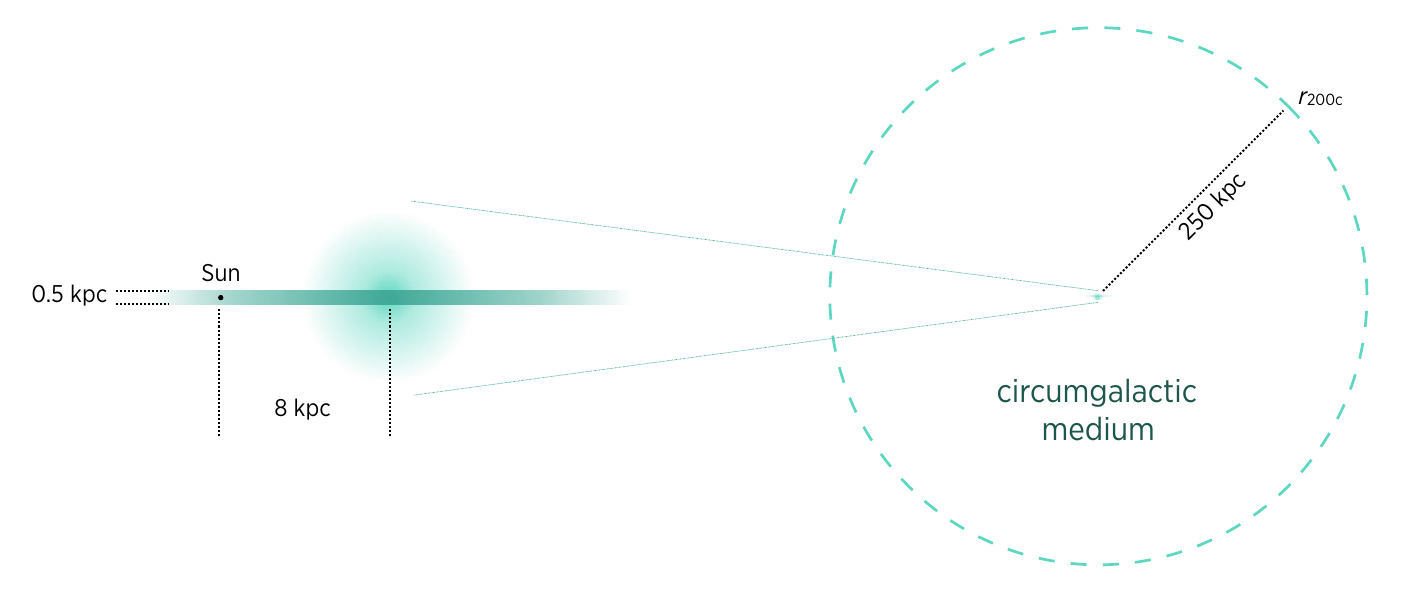}
\caption{Schematic diagram showing how the extent of the Milky Way's circumgalactic medium compares with its visible regions.  Distances shown are approximate, and the galaxy's virial radius is comparable to $r_{200 {\rm c}}$. \label{Figure:MilkyWaySchematic}}
\end{figure}

Most of the atmosphere belonging to a galaxy centered in a halo of mass of $M_{200 {\rm c}} = M_{12} \times 10^{12} \, M_\odot$ therefore lies within  
\begin{equation}
    r_{200 {\rm c}} \approx (210 \, {\rm kpc}) \, M_{12}^{1/3} \, E^{-2/3}(z)
\end{equation}
where $E(z) \equiv H(z)/H_0$ and $H_0 \equiv H(z=0)$.
At that radius, the circular velocity is $v_{\rm c} (r_{200 {\rm c}}) \approx (144 \, {\rm km \, s^{-1}}) \, M_{12}^{1/3} \, E^{1/3}(z) $.  We will consider all of the gas inside of $r_{200 {\rm c}}$ to be part of the galaxy's atmosphere, regardless of whether it is gravitationally bound.  In a disk galaxy, the angular momentum of the inner atmosphere (i.e.~the ISM) allows it to settle into a relatively thin layer (of thickness $\lesssim 1 \, {\rm kpc}$) executing approximately circular orbits around the center.  Atmospheric gas outside of that layer (i.e.~the CGM, see Figure~\ref{Figure:MilkyWaySchematic}) is either pressure supported or executes motion with a significant radial component.  The interface between the ISM and CGM is not well-defined in a disk galaxy\footnote{Arguably, the most physically meaningful distinction is to separate the two components according to angular momentum, considering the ISM to consist of gas that is primarily rotationally supported and the CGM to consist of gas supported primarily by pressure or turbulence.} and is nearly meaningless in an elliptical galaxy.  Likewise, the interface at $r_{200 {\rm c}}$ is also indistinct.  At least some of the gas outside of $r_{200 {\rm c}}$ was previously within that arbitrary boundary and may have been gently pushed beyond it by thermally-driven expansion of the CGM \cite{Voit_pNFW_2019ApJ...880..139V}.  

\subsection{Discovery of the Milky Way's CGM}

Evidence that the Milky Way itself has an extended hot atmosphere first arrived in 1956 \citep{Spitzer1956}, when Lyman Spitzer became aware of Guido M\"unch's absorption-line observations of stars above the Milky Way's disk \citep{MunchZirin_1961ApJ...133...11M}, which went unpublished for another 5 years.\footnote{Guido M\"unch passed away at age 99 on April 29, 2020, during the writing of this review.}  Those observations provided information about the temperature ($T \sim 100$~K) and particle density ($n \sim 10 \,{\rm cm^{-3}}$) of intervening cold gas clouds above the galactic disk.  Spitzer reasoned that a hotter and much more diffuse ambient medium was necessary to confine those clouds. He then inferred the temperature and density of the hot medium by assuming the hot medium was hydrostatic within the Milky Way's potential well at a pressure $nkT \sim 10^{-13} \, {\rm erg \, cm^{-3}}$.  His resulting estimates of ambient CGM properties above the Sun's location in the galactic disk differ little from the current best estimates, more than 6 decades later 
\cite{HenleyShelton_2013ApJ...773...92H,Fang2013ApJ...762...20F,MillerBregman_2013ApJ...770..118M,MillerBregman_2015ApJ...800...14M}.

\subsubsection{Hydrostatic Equilibrium}
\label{sec:HSE}

Hydrostatic gas with pressure $P$ and mass density $\rho$ confined within a gravitational potential well obeys the equation
    \begin{equation}
        \nabla P  =  - \rho \nabla \phi
        \label{eq:HSE}
    \end{equation}
where $\phi$ is the gravitational potential.  To see how this equation links gravity and temperature, divide both sides by $P$ and then multiply by the vector ${\bf r}$ between the bottom of the potential well and the location of interest, giving
    \begin{equation}
        {\bf r} \cdot \nabla \ln P   
            = - 2 \frac {T_\phi} {T}
            \; \; .
        \label{eq:HSE_ln}
    \end{equation}
In this version of the hydrostatic equilibrium equation, the gravitational temperature
    \begin{equation}
        T_\phi \equiv \frac {\mu m_p} {2k} 
            \left( {\bf r} \cdot \nabla \phi \right)
                \label{eq:Tphi}
    \end{equation}
(where $\mu m_p$ is the mean mass per particle) reflects the depth of the potential well, and the ratio $T_\phi/T$ determines the logarithmic slope of the pressure gradient.  In an extended hydrostatic atmosphere that is gravitationally confined and approximately spherical, the dimensionless quantity ${\bf r} \cdot \nabla \ln P \approx d \ln P / d \ln r$ must be of order unity,  implying $T \sim T_\phi$.  This relationship leads to a characteristic CGM temperature
\begin{equation}
    T_\phi \approx 1.5 \times 10^6 \, {\rm K}
        \left( \frac {v_{\rm c}} {200 \, \, {\rm km \, s^{-1}}} \right)^2
\end{equation}
where $v_{\rm c}$ is the local circular velocity at radius $r$, for a galaxy like the Milky Way.  That was Spitzer's reasoning, and modern X-ray spectroscopy of O~VII and O~VIII emission lines from gas in the Milky Way's halo does indeed indicate an ambient CGM temperature $T \approx 2 \times 10^6$~K \cite{MillerBregman_2013ApJ...770..118M,MillerBregman_2015ApJ...800...14M}.

\subsubsection{Cooling Time}

Spitzer also had the information necessary to estimate the time scale on which such an atmosphere could lose its current thermal energy content through radiative cooling, a time scale we now call the \textit{cooling time}.  Cosmic gas in collisional ionization equilibrium at $\sim 10^6$~K is almost completely ionized, making $u = 3P/2$ a good approximation for the thermal energy density $u$.  Also, the photons it generates through two-body inelastic collisions easily escape the optically thin galactic atmosphere without re-thermalizing.  The resulting rate of radiative energy loss is usually expressed in terms of a cooling function $\Lambda(T,Z,n)$, defined so that $du/dt = - n_e n_i \Lambda (T,Z,n)$ in a medium of electron number density $n_e$, ion number density $n_i$, and particle number density $n$.\footnote{The precise definition of $\Lambda(T,Z,n)$ depends on the pair of particle densities used to define it \cite{CLOUDY2013RMxAA..49..137F}. Here we have chosen $n_e n_i$, but sometimes the proton density $n_p$ or hydrogen density $n_{\rm H}$ is used instead of $n_i$, necessitating 10\%--20\% changes in the numerical value of $\Lambda$. In ambient CGM gas, $\Lambda$ does not depend strongly on density until $n$ is low enough for photoionization to significantly alter the abundances of key ions that would otherwise be determined by collisional ionization equilibrium \cite{sd93}. In the present-day universe, photoionization does not significantly affect radiative cooling of CGM gas with $T \gtrsim 10^{5.5}$~K and $n \gtrsim 10^{-4.5} \, {\rm cm^{-3}}$ \citep{Wiersma_2009MNRAS.393...99W}.} Abundances of elements other than H and He are expressed here in terms of $Z$, defined to be their collective fractional contribution to the total gas mass. 
\begin{figure}[!t]
    \centering
    \includegraphics[width=5.4in]{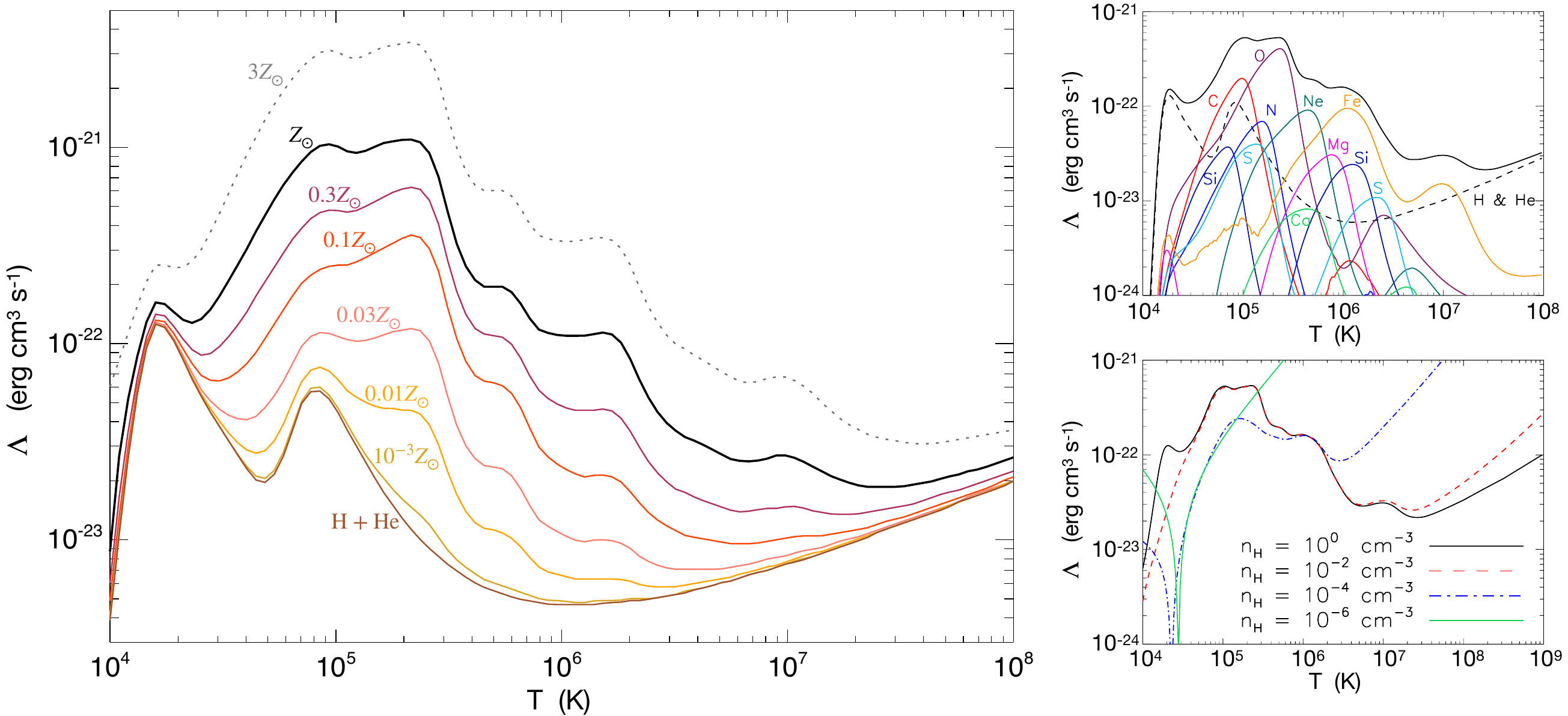}
    \caption{Dependences of the radiative cooling function $\Lambda$ on temperature $T$, heavy-element abundance $Z$, and hydrogen density $n_{\rm H}$. \textit{Left:} Cooling functions from \cite{sd93} for plasmas in collisional ionization equilibrium with differing heavy-element abundances, as labeled. \textit{Top right:} Illustration of how the contributions of various elements produce the temperature-dependent structure of $\Lambda$ for a plasma with solar abundances (from \citep{Wiersma_2009MNRAS.393...99W}).  \textit{Bottom right:} Illustration of how photoionization changes $\Lambda$ in low-density galactic atmospheres  (from \citep{Wiersma_2009MNRAS.393...99W}).  This example is for a plasma exposed to the cosmic ionizing UV background at $z = 3$.  As gas density drops, photoionization reduces radiative cooling of gas at $\sim 10^5$~K because heavy elements in that gas are more highly ionized.  Meanwhile, inverse Compton scattering increases $\Lambda$ at high temperatures (see \citep{Wiersma_2009MNRAS.393...99W} for details.)  
    \label{Figure:CoolingCurve}}
\end{figure}
Pure radiative cooling therefore causes cumulative radiative losses from CGM gas to become comparable to its thermal energy content on a timescale 
    \begin{equation}
        t_{\rm cool} \equiv 
            \frac {3}{2} \frac {P} {n_e n_i \Lambda (T,Z,n)}
            \; \; .
            \label{eq:tcool}
    \end{equation}

Radiative cooling at very high temperatures ($T > 2 \times 10^7$~K) comes predominantly from free-free collisions that produce bremsstrahlung radiation, but other mechanisms dominate at lower temperatures.  Spitzer was a pioneer of plasma physics who recognized that collisionally excited line emission would be the most important coolant in gas at $T \sim 10^6$~K, if elements other than H and He were present in solar proportions.  Accounting for those elements, particularly the emission lines from Fe ions, allowed him to estimate a cooling time of $\sim 1$~Gyr for the Milky Way's ambient CGM gas. 

\subsubsection{Specific Entropy}

Compression during radiative cooling generally causes changes in temperature. The sense of those changes may be counterintuitive because the temperature changes depend on the atmosphere's global configuration.  For example, nuclear fusion in stars becomes possible because the temperature of a self-gravitating interstellar gas cloud \textit{increases} as it radiates away its internal thermal energy.  Consequently, we learn more about how radiative cooling affects CGM gas by paying closer attention to changes in specific entropy than to changes in temperature.

According to the first law of thermodynamics, changes in the thermal energy density $u$ of a uniform system with a constant number of particles are governed by
\begin{equation}
    d \left( \frac {u} {n} \right) \: = \:  
       kT \, ds - P \, d \left( \frac {1} {n} \right)
\end{equation}
where the specific entropy per particle, $s$, is expressed as a dimensionless quantity.  In a gas with $u = 3P/2$, this equation can be rearranged to give
\begin{equation}
    ds \: = \: \frac {3} {2} \, d \ln T - d \ln n 
       \: = \: \frac {3} {2} \, d \ln K
\end{equation}
where the entropy index (or adiabat) $K$ is defined so that $K \propto T n^{-2/3}$.  This result is independent of the normalization of $K$, so one is free to choose a definition for $K$ suiting the observations used to measure it. In X-ray studies of galactic atmospheres, the most widely used definition is $K \equiv kT n_e^{-2/3}$ because spatially-resolved X-ray spectroscopy reveals $kT$ and its gradients, and X-ray surface brightness profiles tell us how electron density $n_e$ depends on radius.  That is why the quantity $kT n_e^{-2/3}$ is sometimes loosely called the ``entropy" of a particular layer in a galactic atmosphere.\footnote{Mirroring the looseness of the literature, this review paper usually refers to $K$ as either the ``entropy" or the ``specific entropy" of atmospheric gas, when in fact the atmosphere's specific entropy is the natural logarithm of $K^{3/2}$ plus a constant determined by the composition of the gas, as specified by the Sackur-Tetrode equation \cite{2013AnP...525A..32G}.}

Given this definition, pure radiative cooling causes $K$ to change at the rate 
\begin{equation}
    \frac {d \ln K} {dt} \: = \: \frac {2} {3} \frac {ds} {dt} 
        \: =  \: - \frac {2} {3} \frac {n_e n_i \Lambda} {P} 
        \: = \: - \frac {1} {t_{\rm cool}}
        \; \; .
        \label{eq:tcool_K}
\end{equation}
Notice that defining $t_{\rm cool}$ with a prefactor of 3/2, as in equation (\ref{eq:tcool}), has resulted in a  timescale for entropy change that is simply equal to $t_{\rm cool}$, regardless of how the gas temperature is changing. That outcome helps to justify an apparent inconsistency in equation (\ref{eq:tcool}), which sets $t_{\rm cool}$ equal to the time required for gas to radiate away all of its current thermal energy while somehow remaining in a state of both constant density and constant temperature.\footnote{A common alternative in the literature replaces the prefactor 3/2 in equation (\ref{eq:tcool}) with 5/2 to give the cooling time in a state of constant pressure and constant temperature.  However, that expression is not literally self-consistent either, because there is no reason for both the pressure and the temperature of CGM gas to remain constant as it cools.}

\subsubsection{Implications for the Baryon Cycle}

Spitzer's 1956 paper astutely interpreted his finding that $t_{\rm cool} \sim 1$~Gyr at $r \sim 10$~kpc in the atmosphere above the Milky Way's disk.  By that time, measurements of the universe's expansion had established its age to be greater than 1~Gyr by about an order of magnitude.  Spitzer correctly reasoned that the Milky Way's atmospheric conditions could not be primordial, because the atmosphere or ``corona," as he called it, had already had plenty of time to cool and change.  In his words, 
    \begin{quote}
    ``The problem of origin of such a corona reduces primarily to the problem of heating."
    \end{quote}
He concluded that somehow heating of the atmosphere has to compensate for radiative cooling, in approximate time-averaged balance.
    
Taking that logic one step further, Spitzer envisaged a baryon cycle for the Milky Way broadly similar to our current understanding of it.  He imagined that matter ejected from massive stars in the disk might move outward into the corona, heating it by dissipating kinetic energy into the ambient gas.  In turn, thermal instability of the corona might sustain star formation in the galactic disk by supplying it with cold clouds.  

Spitzer recognized that the gas mass of the atmospheric layer at $\sim 10$~kpc was only $\sim 10$\% of the disk's gas mass, which was problematic for his nascent baryon-cycle scenario, and so he speculated that this gas reservoir was continually replenished by some other source.  Today we know that a disk galaxy's ``corona" extends much farther than Spitzer imagined, with a total gas mass at least as great as the total stellar mass of the galaxy immersed in it \cite{MillerBregman_2015ApJ...800...14M}.  That was not at all obvious in 1956, but a few years later X-ray telescopes began to blaze a pathway to that realization, starting with observations of galaxy clusters and progressing to smaller halos as X-ray observations became more sensitive. 

\subsection{Atmospheres of Galaxy Clusters} 
\label{sec:GalaxyClusters}

Few astronomers prior to the dawn of X-ray astronomy in the mid-1960s expected galaxy clusters to be luminous X-ray sources.\footnote{Theorists occasionally pondered the possibility of a significant gaseous component  \cite{Limber1959ApJ...130..414L,vanAlbada1960BAN....15..165V} but did not take the leap of equating the gas temperature to the equivalent of the galaxy velocity dispersion ($kT \sim  \mu m_p \sigma_v^2$), maybe due to reluctance to assume clusters of galaxies were held together by huge sums of invisible (dark) matter.} Clusters were notable as large collections of galaxies, mostly elliptical galaxies with old stellar populations, but astronomers were fairly certain that the vast majority of stars are not very impressive X-ray sources. Nevertheless, the earliest X-ray telescopes, essentially Geiger counters launched on small rockets, found that clusters of galaxies were among the brightest X-ray sources in the sky.  

Pointings of rocket-borne telescopes toward M87, the central galaxy of the Virgo Cluster, provided the first detections of a hot cluster atmosphere in the mid-1960s \citep{1966Sci...152...66B,Bradt_1967ApJ...150L.199B}.  The first orbiting satellite dedicated to X-ray astronomy, \textit{Uhuru}, could achieve an angular resolution $\sim 30^\prime$ using a scanning collimator. \textit{Uhuru} observations confirmed in 1971 that the hot atmosphere of the Coma Cluster \citep{1971ApJ...167L..81G} is extended, and revealed $\sim20$ nearby X-ray clusters \citep{BahcallN1974}. With those observations, various workers
 \cite{Field1972ARA&A..10..227F,1973ApJ...184L.105L,1975ApL....16..141L, Gull1975MNRAS.173..585G,CavaliereFuscoFemiano1976A&A....49..137C} were able to infer the approximate density and hydrostatic temperature of a gas bound by the cluster and estimated that $\sim1/8 - 1/15$ of the mass of the clusters Coma, Perseus, and Virgo is in the form of hot, X-ray emitting baryons. 
 The 1978 launch of the \textit{Einstein} Observatory \cite{1979ApJ...230..540G} delivered the first true X-ray imaging instruments to Earth orbit, with angular resolution as fine as $\sim 3^{\prime\prime}$, enabling the first detailed studies of the hot atmospheres of galaxies and clusters  \citep{Sarazin_1988}.

X-ray observations of galaxy clusters with \textit{Uhuru} and \textit{Einstein}  showed that cluster atmospheres have X-ray luminosities $L_{\rm X} \sim 10^{43-45}$ erg s$^{-1}$ and temperatures $kT \sim 2$--10~keV and are usually centered on the brightest cluster galaxy. (See references in Sarazin's classic review \cite{Sarazin_1988}.) Sufficiently long observations with recent and current X-ray telescopes (e.g., XMM-\textit{Newton}, \textit{Chandra}, \textit{Suzaku}) now detect X-ray emission from gas beyond a massive cluster's virial radius \cite{2009MNRAS.395..657G,2019SSRv..215....7W}.  According to the definition we have adopted in this review article, all of that gas belongs to the galactic atmosphere of the BCG, but \S \ref{sec:Heating} will present observations showing that feedback from the central galaxy influences only the innermost $\sim 10$\% of that atmosphere.



\subsubsection{Radial Profiles}
\label{sec:RadialProfiles}

\begin{figure}[!t]
\centering
\includegraphics[width=5.4in,trim=0.1in 0.0in -0.1in 0.0in,clip]{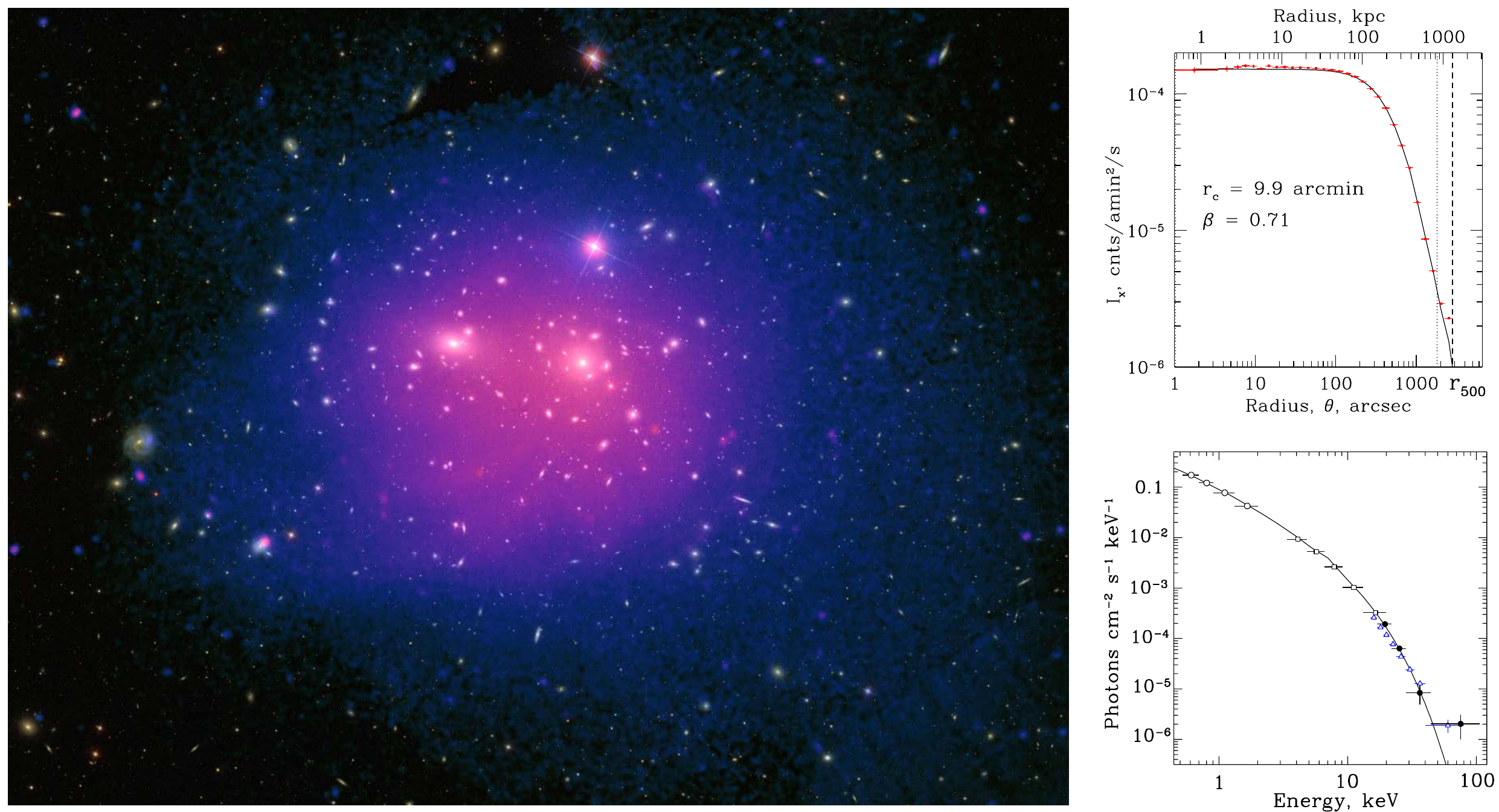}
\caption{Observations of the Coma Cluster, a massive galaxy cluster in Coma Berenices. \textit{Left:} X-ray image of the cluster's hot atmosphere (purple and blue), from XMM-\textit{Newton}, overlaid on a visible-light image from the Sloan Digital Sky Survey (ESA and \cite{Sanders+2020A&A...633A..42S}). \textit{Top right:} X-ray surface brightness profile of the atmosphere (red points) along with a line showing the best fit of the parametric model in equation (\ref{eq:beta_model})  \cite{Lyskova2019MNRAS.485.2922L}. \textit{Bottom right:} Broadband X-ray spectrum of the atmosphere from the ROSAT (open circles), RXTE (open squares), and INTEGRAL (filled circles) satellites \cite{Lutovinov+2008ApJ...687..968L}. 
\label{Figure:Coma}}
\end{figure}

Modern X-ray telescopes allow a hot atmosphere's radial profiles of density and temperature to be derived from a single X-ray observation.  The data can be represented as a three-dimensional cube of detection events, each recording the energy of the triggering photon and the two-dimensional position on the sky from which it came.  Aggregating all the events coming from a particular spot on the sky results in an X-ray spectrum produced by all the hot gas projected along that line of sight (see Figure \ref{Figure:Coma}).  

Interpreting that spectrum requires some modeling.
The X-ray spectrum from an atmospheric layer of hot intracluster gas generally has a continuum component of thermal bremsstrahlung with a high-energy exponential cutoff indicating the gas temperature.  On top of that continuum are emission lines with strengths depending on both the gas temperature and elemental abundances. Emission from the Fe K line is often prominent and constrains the iron abundance of that gas layer.  The Fe/H ratio is typically $\sim 0.3$ times the solar value, except within $\sim 100$~kpc of the cluster center, where iron abundances can reach solar values 
\cite{Rosati+2002ARA&A..40..539R,DeGrandiMolendi2001ApJ...551..153D,KravtsovBorgani2012ARA&A..50..353K}.  


In general, the atmospheric layers along a given line of sight differ in density, temperature, and sometimes abundance.  The X-ray emitting atmosphere must therefore be modeled to obtain the atmosphere's radial profiles. Those models usually rely on supplemental assumptions such as spherical symmetry and sometimes on parametric functional forms for the density profile and gravitational potential. In astronomical X-ray data analysis, unlike optical data analysis, the model is folded through telescope detector response and scattering models and then compared to the observed distribution of X-ray events as a function of energy and detector location. Some binning is usually required for computational expediency. Direct deprojection procedures that subtract emission contributed by outer layers from lines of sight encompassed within them have been commonly used because a cluster X-ray source is conveniently strongly peaked in the center, and its continuum emission is optically thin, while the hot gas is virtually free of obscuring dust particles \cite{Sarazin_1988}.  However, high-resolution deprojection requires a high-quality observation with enough photon events to constrain radial changes in gas temperature. So increasingly, X-ray astronomers use forward modeling to derive a hot atmosphere's radial profiles \cite{sp10, 2013ApJ...767..116M}. 

Fitting the data with a projected parametric model is simpler but less accurate. The most commonly used density-profile model 
\begin{equation}
  n(r) = n_0 \left[ 1 + \left( \frac{r}{r_{\rm core}} \right)^2 \right]^{-3\beta/2}
  \label{eq:beta_model}
\end{equation}
is known as a beta model \cite{CavaliereFuscoFemiano1976A&A....49..137C}.  Its functional form is meant to resemble the radial number-density distribution of the cluster's galaxies, but the central density $n_0$, core radius $r_{\rm core}$, and slope parameter $\beta$ are free parameters.  Fits of the beta model to galaxy-cluster observations generally give $\beta \approx 2/3$ and $r_{\rm core} \sim 100$~kpc (see Figure \ref{Figure:Coma}) and show that the distribution of gas density at large radii ($r \gg r_{\rm core}$) is similar to the averaged radial distribution of both stellar mass and total mass.

\subsubsection{Measuring Cluster Mass and Baryon Fraction} 
\label{sec:ClusterMass}

An estimate for a cluster's distribution of total mass with radius can be derived from its deprojected gas density and temperature profiles, assuming hydrostatic equilibrium.  In a spherically symmetric potential, combining equations (\ref{eq:HSE_ln}) and (\ref{eq:Tphi}) with the equation of state gives
\begin{equation}
    M(r) = \frac {kTr} {G \mu m_p} 
           \left( - \frac {d \ln n} {d \ln r}
                  -  \frac {d \ln T} {d \ln r} \right) 
                  \; \; ,
\end{equation}
where $M(r)$ is the total mass enclosed within radius $r$.  Comparing cluster mass estimates obtained through this method with those obtained through gravitational lensing measurements based on the distorted images of background galaxies show that the two methods usually agree to within $\sim 20$\%. Since the gravitational lensing mass estimate is not affected by the equilibrium state of the object responsible for lensing (i.e. the galaxy cluster), this agreement confirms that a cluster's atmosphere is usually not far from hydrostatic equilibrium \cite{2013ApJ...767..116M,2014MNRAS.439....2V,Donahue+2014ApJ...794..136D,2015MNRAS.450.3633S,2015MNRAS.449..685H,2016MNRAS.457.1522A}.  A typical galaxy cluster's total mass turns out to be $M_{200 {\rm c}} \sim 10^{14-15} \, M_\odot$.

The fraction of a cluster's mass in the form of baryons can be obtained from the same set of X-ray observations by integrating the gas density profile and adding it to the cluster's total stellar mass.  Doing so shows that the baryonic mass fraction of a galaxy cluster is similar to the universal baryonic mass fraction derived from CMB observations.  Massive galaxy clusters are therefore very nearly ``closed boxes" that retain a large majority of the baryons originally associated with the halo's dark matter
\cite[e.g., ][]{Gunn1977,Davis+1985,1993Natur.366..429W,Sun+09}.

\subsubsection{Central Cooling Time}
\label{sec:CentralCoolingTime}

Measurements of a cluster's gas density also inform estimates of atmospheric cooling time as a function of radius. Early X-ray measurements showed that  $t_{\rm cool} \lesssim 10$~Gyr for gas in the cores of many clusters of galaxies \cite{Fabian1994}. More recently, \textit{Chandra} and XMM observations have shown that the central regions ($r \lesssim 10$~kpc) of nearly half of all galaxy clusters have $t_{\rm cool} \lesssim 1$~Gyr \cite{McNamaraNulsen2012NJPh...14e5023M,Cavagnolo+08,Cavagnolo+09,DunnFabian2008}, raising the the same issue that Lyman Spitzer raised about the Milky Way in 1956.  How can gas persist in a state that is able to radiate away all of its thermal energy during a time period much less than the universe's age?  

Sections \ref{sec:Cooling}, \ref{sec:Heating}, \ref{sec:Weather}, and \ref{sec:Balance} will have a lot more to say about the central cooling times of galaxy clusters, the energy sources that compensate for cooling, and the feedback loop that keeps them in long-term balance. Before getting to that, we will whet the reader's appetite with an observational clue: the presence of atmospheric cavities in galaxy clusters.

\subsubsection{X-ray Cavities}

The \textit{Chandra} X-ray Observatory 
became one of the most important instruments for studying baryon cycles in big galaxies not long after its launch in 1999.  Its mirrors are the highest quality mirrors ever made for X-ray astronomy, capable of focusing high-energy photons with sub-arcsecond angular resolution \cite{2019A&G....60f6.19W}.  The angular resolution of \textit{Chandra} will probably remain unparalleled among X-ray telescopes for at least another decade.  Not even the European Space Agency's \textit{Athena} mission, slated for launch no earlier than 2031, will have angular resolution as fine as \textit{Chandra's}.\footnote{It is sobering to the authors to contemplate we might not live to see a telescope with X-ray vision that surpasses \textit{Chandra's}. That might happen with the \textit{Lynx} mission (\url{http://www.lynxobservatory.com.}), proposed to the 2020 US Decadal Survey for Astrophysics.}

\textit{Chandra's} sharp X-ray vision enabled it to detect large regions of depressed X-ray surface brightness in and around the central galaxies of galaxy clusters, revealing the presence of huge, evacuated cavities in their hot atmospheres \cite{2007ApJ...659.1153W,2000ApJ...534L.135M,2001ApJ...562L.149M,2002ApJ...569L..79H,2005ApJ...635..894F,2000MNRAS.318L..65F,2005MNRAS.360L..20F,2005ApJ...625..748C,2009ApJ...697.1481C,2004ApJ...606..185C,2001ApJ...558L..15B,2004ApJ...612..817B}.  Hints of such cavities were present in earlier X-ray observations of NGC~1275, the central galaxy of the Perseus Cluster \cite{Boehringer+1993MNRAS.264L..25B}.  \textit{Chandra} showed not only that X-ray cavities were commonplace but also that they are exclusive to clusters with central cooling times of $\lesssim 2$~Gyr \cite{Cavagnolo+08,DunnFabian2008}.

\begin{figure}[!t]
\centering
\includegraphics[width=5.3in]{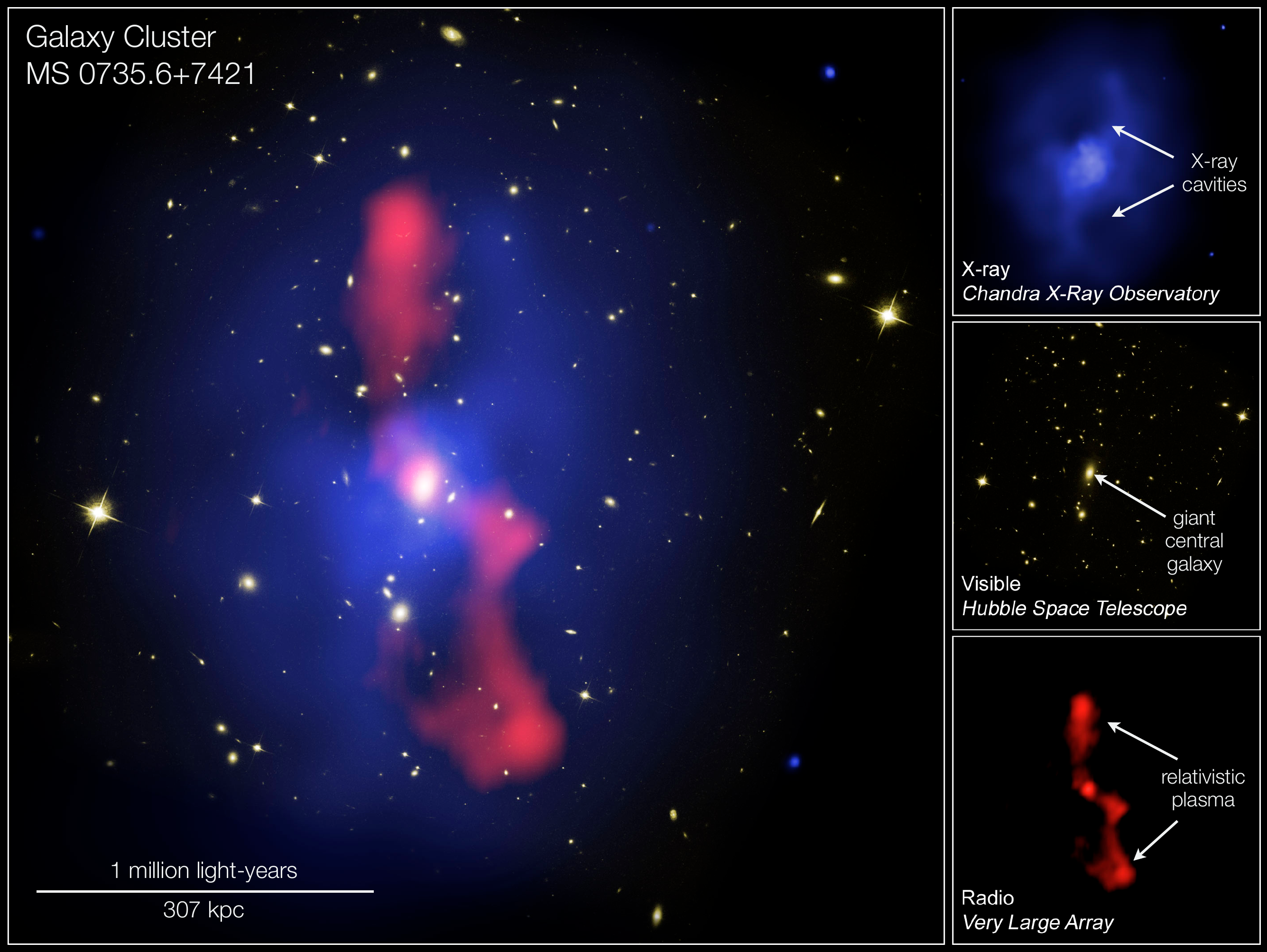}
\caption{Enormous X-ray cavities inflated by relativistic plasma in the galaxy cluster MS 0735.6+7421 (X-ray data from \citet{McNamara_2005Natur.433...45M,Vantyghem+2014}; image credit NASA/CXC/Univ. of Waterloo; Optical: NASA/STScI; Radio: NRAO/VLA) \label{Figure:MS0735}}
\end{figure}

Radio observations of synchrotron emission coinciding with those evacuated regions demonstrate that they are filled with relativistic electrons, linking the X-ray cavities with outflows of relativistic plasma from the central galaxy's AGN (see Figure \ref{Figure:MS0735}).  The amount of energy required to excavate the cavities can be estimated by multiplying cavity volume by the surrounding gas pressure.  Cavity energies of $\sim 10^{59} \, {\rm erg}$, equivalent to the kinetic energy of $\sim 10^8$ supernovae, are not uncommon.  The most extreme examples reach $\sim 10^{61} \, {\rm erg}$, equivalent to the kinetic energy of $\sim 10^{10}$ supernovae \citep{McNamara_2005Natur.433...45M,Nulsen_Hydra_2005ApJ...628..629N}.  So clearly, the central AGN is putting out plenty of energy capable of offsetting the atmosphere's radiative losses.  Later in this article, we will explore how that feedback energy couples with the atmosphere and regulates the AGN's power output (\S \ref{sec:BH_Heating}, \S \ref{sec:Balance}, \S \ref{sec:Implications}).

\subsection{Atmospheres of Galaxy Groups}
\label{sec:GalaxyGroups}

\begin{figure}[!t]
\centering
\includegraphics[width=5.4in,trim=0.1in 0.0in -0.1in 0.0in,clip]{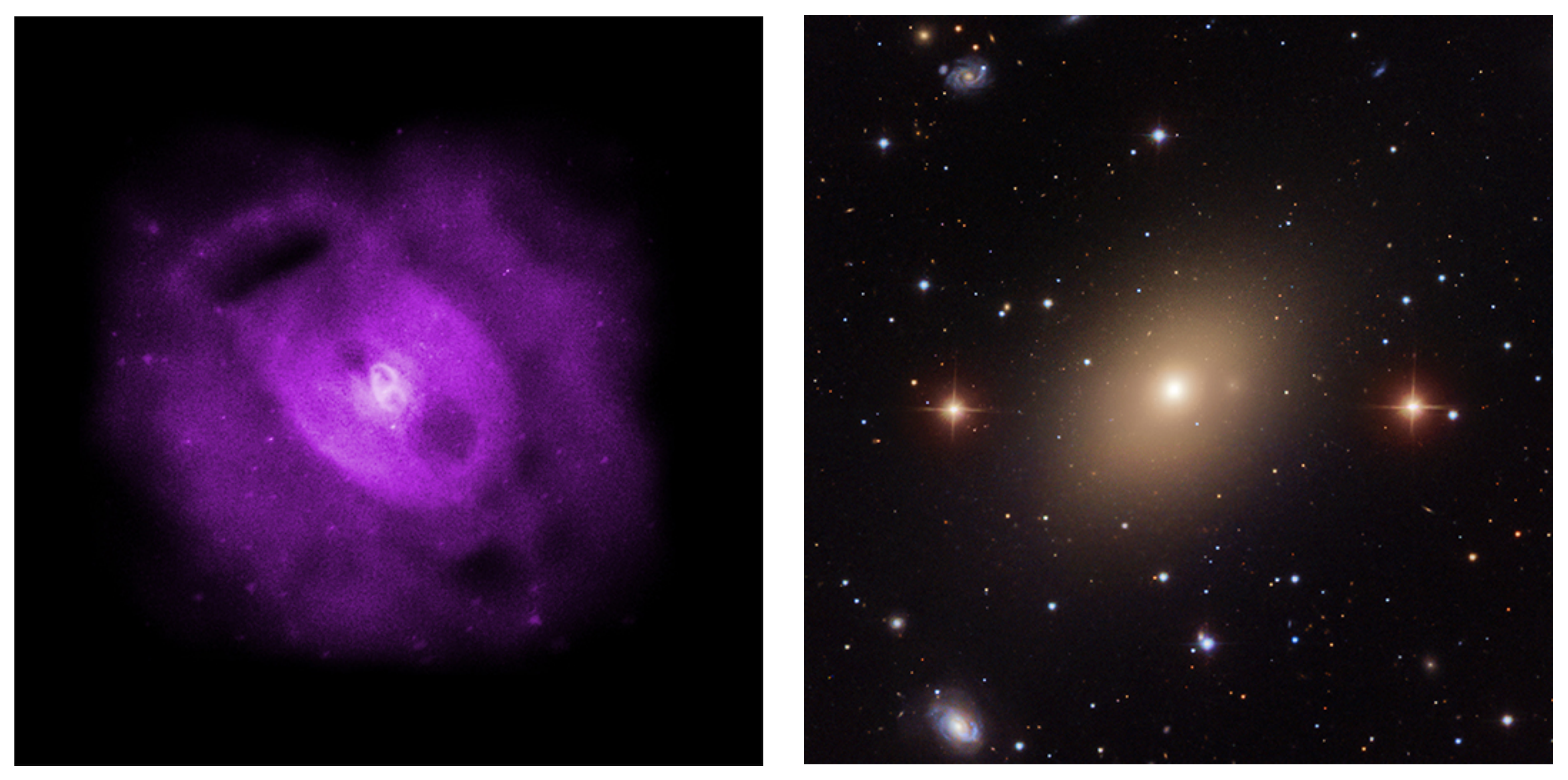}
\caption{X-ray (left) and optical (right) images of the massive elliptical galaxy NGC~5813, which occupies the center of a galaxy group about 105 million light-years from Earth (from \cite{Randall+2015ApJ...805..112R} and Chandra X-ray Observatory/NASA).  The \textit{Chandra} X-ray image clearly shows several X-ray cavities (dark patches within the purple X-ray image) resulting from multiple AGN outbursts, while the optical image appears relatively undisturbed. \label{Figure:NGC5813}}
\end{figure}

Galaxy groups are the smaller cousins of galaxy clusters, residing in halos roughly an order of magnitude less massive ($\sim 10^{13-14} \, M_\odot$).  They are also extended X-ray sources (see Figure \ref{Figure:NGC5813}), with X-ray luminosities of $L_{\rm X} \sim 10^{41-43} \, {\rm erg \, s^{-1}}$ and ambient temperatures of $kT \sim 0.5$--$2 \, {\rm keV}$ \citep[e.g.,][]{Sun+09}.  However, X-ray observations of galaxy groups are more difficult, partly because they are less luminous than galaxy clusters, but also because their atmospheric gas is less dense, especially toward the center, which suppresses a group's X-ray surface brightness.  

Two features of the X-ray properties of galaxy groups show that the central gas deficiency depends on halo mass, with lower mass halos having lower central gas density.  First, the observed relationship between $L_{\rm X}$ and $kT$ shows a stronger dependence on temperature than the simplest models predict (see \S~\ref{sec:Lx-T}).  Second, the overall ratio of gas mass to total mass appears to be smaller in groups.  Both properties indicate that feedback energy input near the center inflates the atmosphere and pushes a significant fraction of a galaxy group's baryons to greater altitudes \cite{Sun+09}.

\subsubsection{X-ray Luminosity--Temperature Relation}
\label{sec:Lx-T}

\begin{figure}[!t]
\centering
\includegraphics[width=5.4in,trim=0.1in 0.0in -0.1in 0.0in,clip]{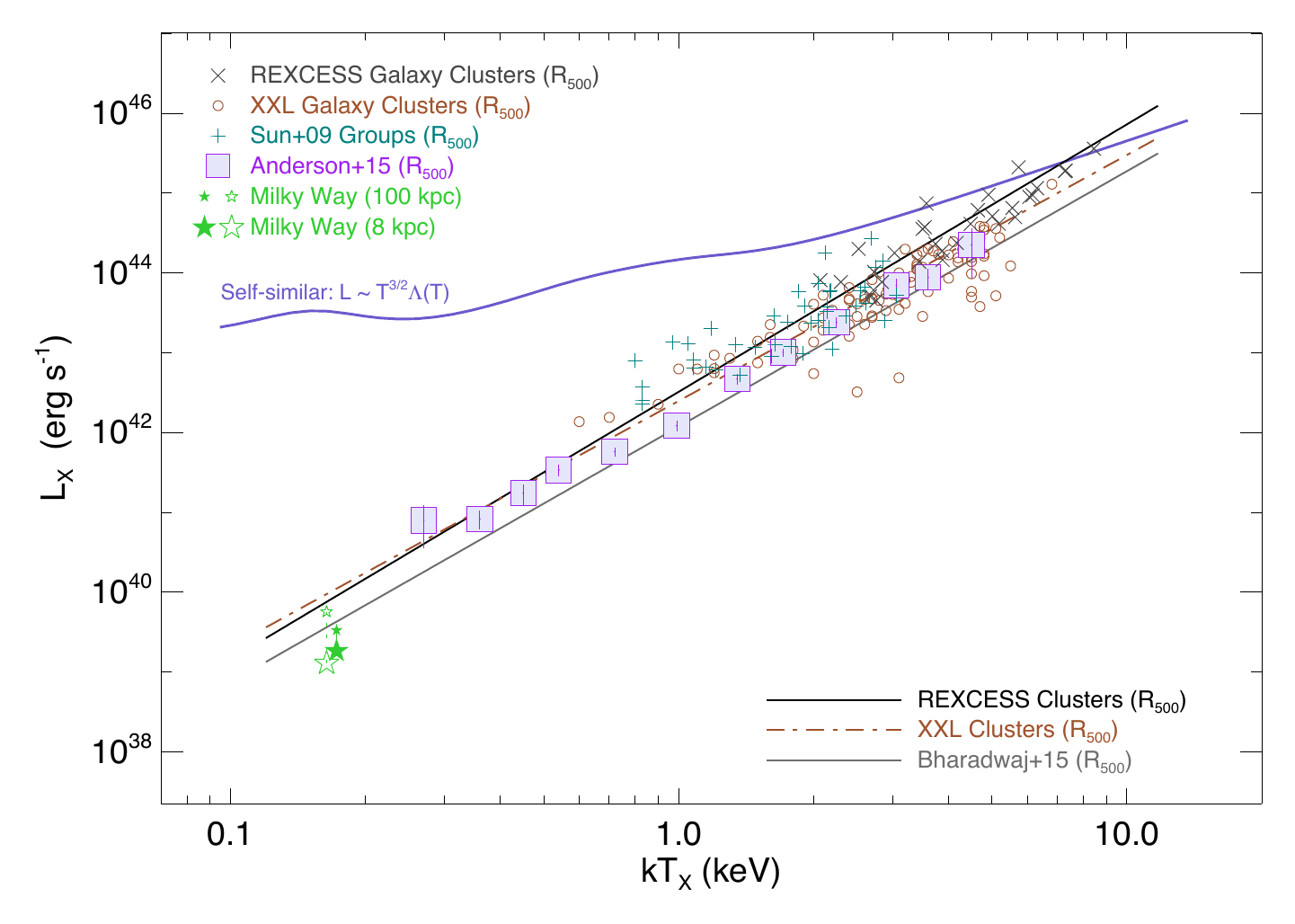}
\caption{Observed relationships between X-ray luminosity and gas temperature for galaxy groups, galaxy clusters, and the Milky Way, spanning halo masses from $10^{12} \, M_\odot$ to $10^{15} \, M_\odot$. A solid purple line shows the $L_{\rm X}$--$T$ relation that a set of atmospheres with self-similar structure would produce.  (Adapted from \citep{Voit2018_LX-T-R}, where citations and discussions of the data sources can be found). 
\label{Figure:LX_T}}
\end{figure}

One of the earliest findings of X-ray galaxy cluster surveys was a strong correlation between X-ray luminosity and gas temperature \citep{Mushotzky_1984PhST....7..157M}. Many subsequent surveys have shown that the X-ray luminosity from within $r_{500 {\rm c}}$ is approximately proportional to $T^3$ \cite{1991MNRAS.252..414E,David+1993ApJ...412..479D,
1997ApJ...482L..13M,1997ApJ...482L..13M,2011MNRAS.412.2391A,1999MNRAS.305..631A,2014Ap&SS.349..415B,2002MNRAS.336..409B,2008AN....329..135B,2007A&A...472..739B,2007ApJS..172..561B,2007A&A...466..805C,2014ApJ...794...48C,2011A&A...535A.105E,2004A&A...417...13E,1999ApJ...519L..51F,2012MNRAS.424.2086H,2005ApJ...633..781K,2010MNRAS.406.1773M,2016MNRAS.463.3582M,Maughan_2012_LX-T,2006MNRAS.365..509M,2020A&A...636A..15M,2004A&A...428..757O,2006ApJ...640..673O,Pratt_2009_REXCESS_LX-T,2011A&A...535A...4R,2013A&A...558A..75T,1999ApJ...524...22W,2007A&A...467..437Z,2016MNRAS.463..820Z}.\footnote{These studies generally use the radius $r_{500 {\rm c}}$ inside which the enclosed mean mass density is $500 \rho_{\rm cr}$ instead of $r_{200 {\rm c}}$ because measuring X-ray surface brightness is difficult beyond $r_{500 {\rm c}}$.}  However, the observed $L_{\rm x}$--$T$ relationship is not the one expected for a set of similarly-structured (``self-similar") atmospheres (Figure \ref{Figure:LX_T}).

If galaxy clusters and groups all had the same atmospheric structure as a function of $r/r_{500 {\rm c}}$, then the expected $L_X$--$T$ relation derived from the integral
\begin{equation}
    L_{\rm X}(r_{500 {\rm c}}) \; = \;  
        \int_0^{r_{500 {\rm c}}} n_e(r) n_i(r) \Lambda [T(r)] 
                        \cdot 4 \pi r^2 \, dr 
    \; \; ,
\end{equation}
would reduce to $L_X \propto  M_{500 {\rm c}} \Lambda(\bar{T}) H^2(z)$, 
where $\bar{T} \propto T_\phi$ is an appropriately averaged gas temperature.  The relationship between atmospheric temperature and halo mass, derived from  $T_\phi \propto M_{500 {\rm c}}/r_{500 {\rm c}}$ and $M_{500\rm c} \propto r_{500{\rm c}}^3 H^2(z)$, is $T \propto M_{500 {\rm c}}^{2/3} H^{2/3}(z)$.  Furthermore, radiative cooling of gas with $kT \gtrsim 2 \, {\rm keV}$ is dominated by thermal bremsstrahlung, for which $\Lambda \propto T^{1/2}$.  Combining these scaling relations then gives the prediction $L_{\rm X} \propto T^2 H(z)$ for galaxy clusters ($kT \gtrsim 2 \, {\rm keV}$), and the relation predicted for galaxy groups has a weaker temperature dependence because $\Lambda(T)$ is nearly independent of temperature in the 0.5--2 keV temperature range of galaxy groups.  

Figure~\ref{Figure:LX_T} shows how poorly the self-similar assumption fares. The observed discrepancy between the predictions of self-similar models and observations of the $L_X$--$T$ relation implies that the volume-averaged value of $n^2$ in the atmospheres of galaxy groups is smaller than in galaxy clusters. This trend indicates that non-gravitational feedback processes cause the properties of hot gas in groups of galaxies to systematically differ from those in the cores of massive galaxy clusters.

\subsubsection{Mass-Dependent Gas Fraction}
\label{sec:group_baryons}

\begin{figure}[!t]
\centering
\includegraphics[width=5.3in]{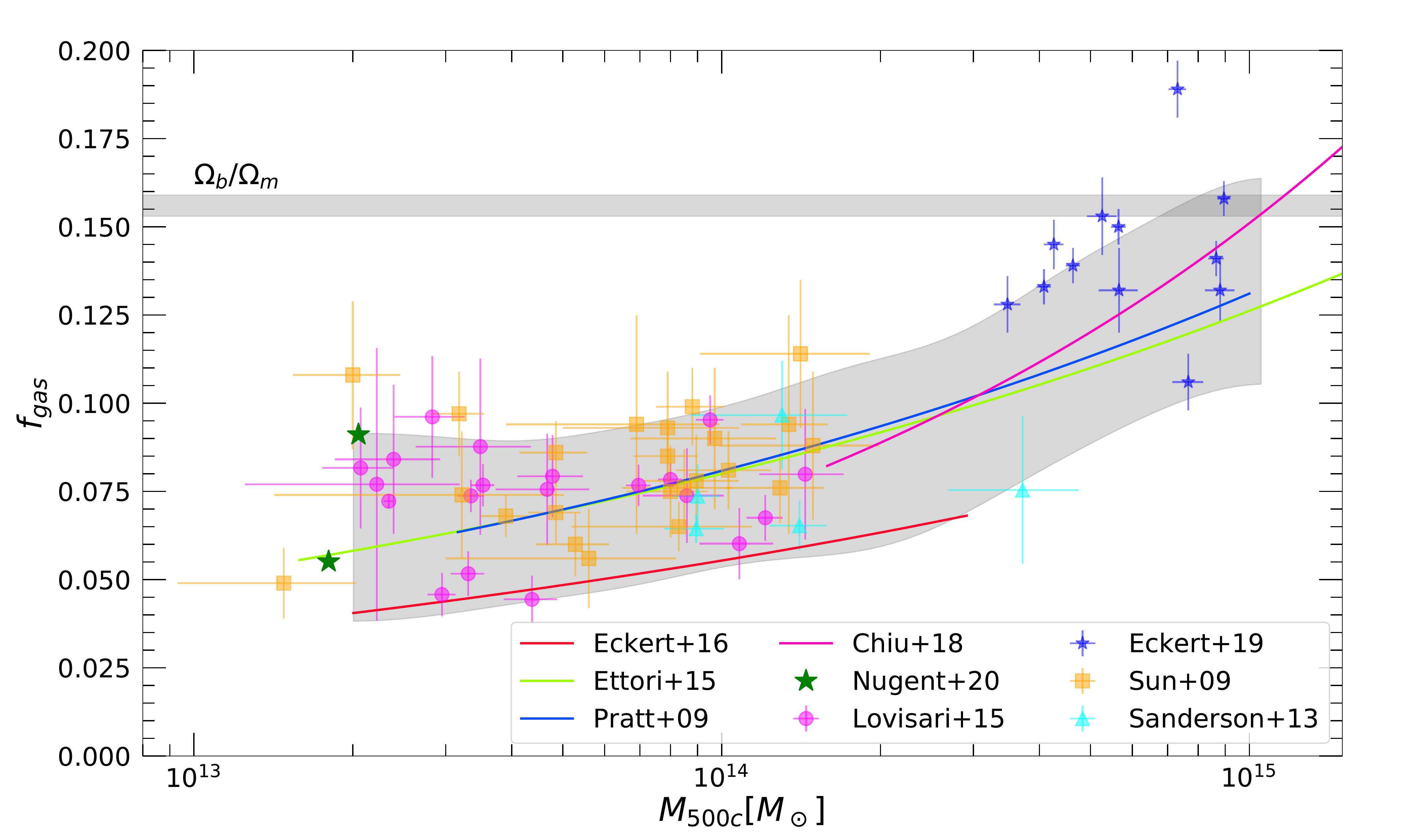}
\caption{Dependence on halo mass ($M_{500 {\rm c}}$) of the mass fraction $f_{\rm gas}$ of observable atmospheric gas  (from \citet{Eckert_2021Univ....7..142E}). A gray band labeled $\Omega_b / \Omega_m$ shows the cosmic baryon fraction inferred from CMB observations.  Dark blue stars (from \citep{Eckert_2019A&A...621A..40E}) show that $f_{\rm gas}$ is close to the cosmic mean in massive galaxy clusters.  Many different X-ray analyses have shown that $f_{\rm gas}$ is approximately half the cosmic mean in galaxy groups (see \citep{Eckert_2021Univ....7..142E} for citations and more details).
\label{Figure:fgas_M500c}}
\end{figure}

Deprojection of X-ray surface brightness shows that the volume-averaged value of gas density is indeed smaller in groups, out to at least $r_{500 {\rm c}}$, beyond which the X-ray signal of a group atmosphere can no longer be distinguished from the X-ray background \citep{Mulchaey_2000ARA&A..38..289M,2003ApJS..145...39M,Sun+09,2017MNRAS.472.1482O}.  This finding is typically stated in terms of the fraction $f_{\rm gas} = M_{\rm gas} / M_{\rm total}$ of a group's total mass that is in the form of detectable gas. Observations show a trend similar to $f_{\rm gas} \propto M_{500 {\rm c}}^{0.3}$ inside of the radius $r_{2500 {\rm c}}$ containing a mean mass density $2500 \rho_{\rm cr}$ \citep{Sun+09}. The dependence on halo mass is not as strong inside of the larger radius $r_{500 {\rm c}}$, indicating that group atmospheres become more like cluster atmospheres at larger radii.  However, Figure~\ref{Figure:fgas_M500c} shows that $f_{\rm gas}$ out to $r_{500c}$ in groups is only about half the value observed in galaxy clusters.  Stars cannot make up the difference, because the fraction of total group mass in the form of stars is only 2--3\% \citep{Eckert_2021Univ....7..142E}.

\subsubsection{Scattering of CMB Photons}
\label{sec:SZE}

X-ray observations show that galaxy groups are deficient in baryonic gas but do not show where the missing gas might be 
\cite{2010ApJ...719..119D,2013ApJ...778...14G,1994ApJ...421L..63H,2006ApJ...652..917L,2012NJPh...14d5004S}.  Much of it has probably been pushed beyond the virial radius, where it can still be detected through electron scattering of CMB photons
\cite{2018MNRAS.478.3072C,2016MNRAS.455..258C,2006ApJ...652..917L,2016MNRAS.463.4533V}.  Along lines of sight through hot gaseous halos like those of galaxy groups and clusters, electron scattering produces a distortion of the CMB spectrum known as the Sunyaev-Zel'dovich Effect (SZE), named after \citet{SunyaevZeldovich1972} who applied the CMB-scattering physics worked out by \citet{Weymann1965}. 

The magnitude of an atmosphere's thermal SZE signal is proportional to the probability that a CMB photon passing through it will scatter off an electron, making the signal proportional to the atmosphere's electron column density.  It is also proportional to the mean fractional energy change ($kT/m_e c^2$) of the scattered photon.  The overall thermal SZE signal ($Y_{\rm SZ}$) from a cluster or group, integrated over its footprint in the plane of the sky, is therefore proportional to the product of gas mass and temperature, which yields the prediction $Y_{\rm SZ} \propto M_{500 {\rm c}}^{5/3}$ for hot atmospheres with similar structure and a baryonic gas mass fraction that is independent of halo mass.

Thermal SZE observations of individual clusters became feasible in the late 1990s \cite{1999PhR...310...97B,2002ARA&A..40..643C,2001ApJ...552....2G,2004ApJ...617..829B} and they show the expected scaling with mass and X-ray properties \cite{2016JCAP...08..013B,2013A&A...550A.131P,2013A&A...550A.129P,2011A&A...536A..10P,2011A&A...536A..12P,2019AstL...45..403L}.  Placing similar constraints on thermal SZE scaling among lower-mass halos still requires stacking of many smaller objects binned according to halo mass.  Stacks of CMB observations acquired with the \textit{Planck} telescope \citep{Planck_LRGstacks_2013A&A...557A..52P,Greco_2015ApJ...808..151G} are consistent with a $Y_{\rm SZ} \propto M_{500 {\rm c}}^{5/3}$ scaling that extends from the masses of galaxy clusters down through the masses of galaxy groups.

That finding may seem to conflict with the mass-dependent gas deficits implied by X-ray observations.  However, the \textit{Planck} signal from all but the closest groups comes from a region of the sky several times larger than $r_{500}$ and therefore includes gas beyond the virial radius.  The most plausible hypothesis for reconciling the X-ray evidence for an $f_{\rm gas}$ deficit within $r_{500 {\rm c}}$ with the lack of evidence for a gas deficit in the SZE observations is to assume that an excess of hot gas outside of $r_{500}$ compensates for the deficit within $r_{500}$.
Testing that hypothesis using SZE techniques requires either sensitive SZE observations with greater angular resolution than \textit{Planck} (e.g., with Advanced ACT-Pol, SPT-3G, Mustang2, or ALMA  \cite{2017AN....338..305P,2018A&A...612A..39R,2020ApJ...891...90R}) or $\textit{Planck}$ observations of nearby galaxy groups \citep[e.g.,][]{Pratt_SZ_2021arXiv210501123P}.

\subsection{Atmospheres of Massive Galaxies}
\label{sec:MassiveGalaxies}

Individual massive galaxies reside in halos of mass $\sim 10^{12-13} \, M_\odot$, two orders of magnitude less massive than the halos of galaxy clusters.  Elliptical galaxies dominate the high end of this mass range.  The Milky Way and other massive disk galaxies are mostly toward the low-mass end.  Isolated massive galaxies have atmospheres with X-ray luminosities $\sim 10^{39-42} \, {\rm erg \, s^{-1}}$ and temperatures $\sim 0.2$--0.5~keV, generally making them much harder to observe with X-ray telescopes than groups or clusters of galaxies.  However, recent observations using a variety of techniques spanning the electromagnetic spectrum are providing important clues.

\subsubsection{X-ray Observational Challenges}

The atmospheres around isolated galaxies are extremely difficult to image with current X-ray telescopes. Such galaxies have very low X-ray surface brightness, and their low-energy X-ray photons are absorbed by cooler gas in our own Galaxy.  The column density of neutral hydrogen gas that we must look through to see outside of our own Galactic disk is $N_{\rm HI} \sim 10^{20} \, {\rm cm^{-2}}$, meaning that the X-ray opacity of the associated heavier elements in the Milky Way's gas blocks extragalactic photons with energies $\lesssim 0.2$--0.5 keV along many lines of sight.  A third issue is that many individual compact sources, such as a galaxy's X-ray binaries, contribute X-ray photons that can be mistaken for atmospheric emission, if those point sources are not masked or if their contribution to the galaxy's X-ray spectrum cannot be accurately modeled.

Early X-ray studies of the most luminous atmospheres around individual elliptical and lenticular galaxies were possible with \textit{Einstein} because the luminosities of those atmospheres ($\sim 10^{41-42} \, {\rm erg \, s^{-1}}$) were considerably greater than the collective luminosity of the galaxy's point sources \citep{Sarazin_1988,Fabbiano1989}.  Direct detections of lower-luminosity atmospheres required \textit{Chandra's} much greater spatial resolution, which allows  precise masking of a galaxy's brightest point sources.  That capability has enabled direct detections of hot atmospheres in and around many lower-mass elliptical galaxies and also a handful of disk galaxies \citep[e.g.][]{2004ApJ...610..213T,2011ApJ...730...84L,2013MNRAS.428.2085L,2010PNAS..107.7168W,Juranova2020_arXiv200801161J,2020ApJ...897...63D,2019ApJ...885...38J,Kim_atlas_2019ApJS..241...36K,2020A&A...637A..12K}

At the lowest luminosity levels, X-ray emitting galactic atmospheres can be detected only through stacking of many such observations of similar galaxies.  The most extensive efforts to date show that the relation between X-ray luminosity and halo mass ($L_{\rm x} \propto M_{500 {\rm c}}^{1.9}$) continuously extends all the way down to halos of mass $M_{500 {\rm c}} \approx 10^{12.5} \, M_\odot$, which have $L_{\rm X} \sim 10^{40} \, {\rm erg \, s^{-1}}$ \citep{Anderson_2015MNRAS.449.3806A}.  In stacked halos of even lower mass, atmospheric X-ray emission becomes indistinguishable from the collective luminosity of stellar X-ray point sources.

\subsubsection{X-ray Coronae}
\label{sec:Coronae}


Another feature of some individual elliptical galaxies in galaxy groups and clusters is the presence of a bright, compact ``X-ray corona" extending only 2--4 kpc from the galaxy's center, at a temperature distinctly different from that of the surrounding gas. This type of corona is not what Lyman Spitzer had in mind for the Milky Way, as it is much more compact and has an even shorter cooling time, often as short as $t_{\rm cool} \lesssim 10^8 \, {\rm yr}$ near the center \cite{2014MNRAS.439.1182S,Sun09,Sun_2007ApJ...657..197S,2005ApJ...619..169S}.  The pronounced difference between a corona's temperature and the surrounding atmospheric temperature suggests that the corona's characteristics are determined by the properties of its galaxy, while the characteristics of gas outside the corona reflect the properties of the larger halo.

\begin{figure}[!t]
\centering
\includegraphics[width=5.3in]{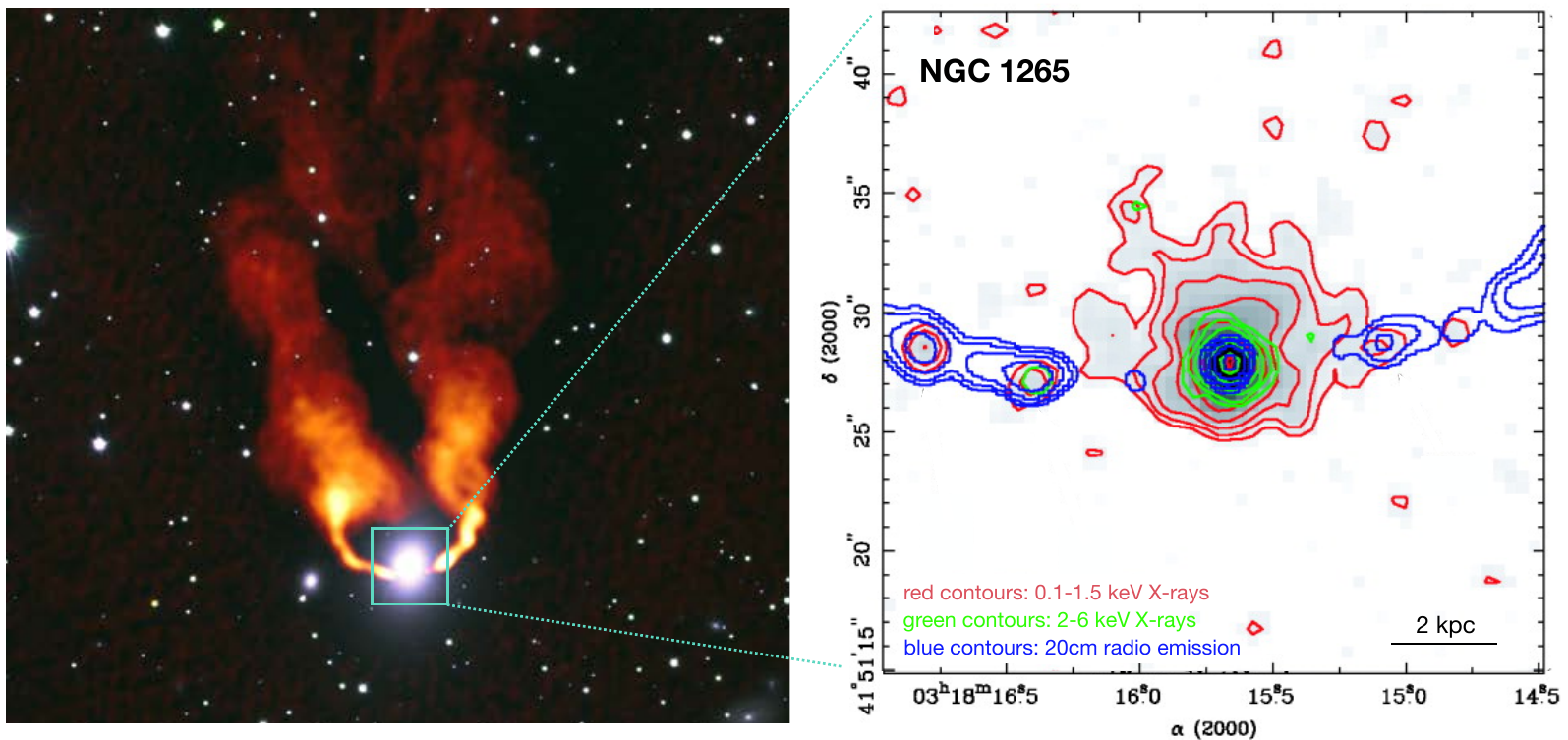}
\caption{The X-ray corona of NGC 1265, a bright radio galaxy in the Perseus Cluster. \textit{Left:} A VLA image of the galaxy's bipolar radio jets (orange), overlaid on a visible-light image showing the galaxy's starlight (from \citet{Gendron-Marsolais_2020MNRAS.499.5791G},  NRAO/AUI/NSF and Sloan Digital Sky Survey.).  The jets are swept back as the galaxy moves through the cluster's atmosphere.  \textit{Right:} Structure of the X-ray corona (from \citet{Sun_2005ApJ...633..165S}).  Red contours show 0.5--1.5~keV X-ray emission from the corona's gas, which extends only $\sim 2$~kpc from the galaxy's center.  Green contours show 2--6~keV X-ray emission associated with the AGN. Blue contours show the radio-emitting jets, which appear to drill through the corona without significantly disrupting it.
\label{Figure:NCG1265_corona}}
\end{figure}

Radio observations reveal that some of these coronae contain AGNs with enough power output to disrupt the corona within a relatively short time period \cite{Sun_2007ApJ...657..197S}.  As in the cores of galaxy clusters, a short central cooling time appears strongly associated with feedback from the central AGN, but the feedback seems less well tuned to the X-ray cooling rate of the galaxy's atmosphere \citep{Sun09}.  Morphological observations of the radio-emitting outflows show that some of them drill narrow channels through the X-ray coronae and deposit most of their energy more than 10~kpc from the galaxy's center, allowing the corona's gas to continue fueling the central engine without major disruption (see Figure \ref{Figure:NCG1265_corona}). We will discuss the nature of X-ray coronae further in \S \ref{sec:BH_Heating}.

\subsubsection{Multiphase Atmospheric Gas}
\label{sec:UV_COS}

During the past decade, observers using the Hubble Space Telescope's Cosmic Origins Spectrograph (COS)
\cite{2011Ap&SS.335..257O,2012ApJ...744...60G} have pioneered a new method for studying galactic atmospheres: UV absorption-line spectroscopy
\cite{2013ApJ...777...59T,2013ApJ...763..148S,2013ApJS..204...17W,2013ApJS..204...17W,2014ApJ...796..136B,2016ApJ...833...54W,2016A&A...590A..68R,2016ApJ...817..111D,2017ApJ...849..106S,2017A&A...607A..48R,2018MNRAS.476.4965M,2018MNRAS.478.3890B,2018ApJ...866...33L,Burchett_2019ApJ...877L..20B,2020MNRAS.497..498C}.
Lines of sight toward bright quasars can intercept the atmospheres of galaxies at projected distances less than $\sim 100$~kpc from those lines of sight \cite{Sargent+1980ApJS...42...41S}.  Absorption lines produced by neutral hydrogen clouds with redshifts essentially identical to those of the projected galaxies are commonplace, along with coinciding lines from heavier elements in a variety of ionization stages (\cite{1992ApJS...80....1S,2017ARA&A..55..389T}).  Those observations show that not all of the atmospheric gas in halos of mass $\sim 10^{12} \, M_\odot$ is at the gravitational temperature but instead is distributed over phases that can differ by two orders of magnitude in temperature and density.  

The total amount of this \textit{multiphase gas} is comparable to the combined mass of all the galaxy's stars and perhaps also the mass of ambient atmospheric gas closer to the halo's gravitational temperature \citep{Tumlinson_2011Sci...334..948T}.  Its existence was already evident from ground-based optical spectroscopy of quasars, particularly of H~I, Mg II, and C IV absorption lines left by cool clouds around galaxies of greater redshift \cite{Sargent+1980ApJS...42...41S,1992ApJS...80....1S,2001ApJ...556..158C,2010ApJ...714.1521C,Rudie+2019ApJ...885...61R,Tripp_2008ApJS..177...39T,2013ApJ...770..130Z,Zhu_2014MNRAS.439.3139Z}.  However, COS studies of nearer galactic atmospheres have given us a more detailed picture of the different components, their flow patterns, and their relationships to the stellar mass of the central galaxy
\cite{Werk_2014ApJ...792....8W, Werk2016_ApJ...833...54W,Prochaska_2019ApJS..243...24P,2016ApJ...817..111D,Stocke_2013ApJ...763..148S,Keeney_2017ApJS..230....6K,2020MNRAS.497..498C}.
Section \ref{sec:CGM_Pressure} will consider what those observations tell us about atmospheric pressures around galaxies like the Milky Way.

\subsubsection{Dispersion Measure Constraints \label{Section:DispMeasure}}

An even newer window on galactic atmospheres is opening with increasingly sophisticated observations of fast radio bursts (FRB, discovered in 2007 by \cite{2007Sci...318..777L}) from galaxies at cosmological distances \cite[e.g., ][]{2019Natur.572..352R,2019ApJ...885L..24C,2020ApJ...899..161L}.  As a radio burst passes through intervening galactic atmospheres on its way to Earth, the column density of free electrons disperses the burst, delaying the arrival of longer-wavelength radio waves.  Simultaneous observations of FRBs at different wavelengths therefore tell us the dispersion measure
\begin{equation}
    \int \frac {n_e(r)} {1+z} \, dr
\end{equation}
where the integral is over the comoving coordinate $r$ along the line of sight to the FRB's origin and the factor of $1+z$ accounts for time dilation and redshift effects \citep{2003ApJ...598L..79I,2018ApJ...852L..11S,2019ApJ...872...88R}.
  The majority of that signal comes from the cosmological mean electron density, which is consistent with the cosmic baryon fraction inferred from CMB observations \cite{Macquart+2020}.  However, lines of sight passing through especially dense galactic atmospheres can add to that signal, with detectable consequences \cite{McQuinn2014ApJ...780L..33M}.  One early search for an enhanced signal has already provided interesting constraints on the electron column density at a projected radius $\sim 25$~kpc from a galaxy with a stellar mass similar to the Milky Way's \cite{2019Sci...366..231P}. The current interpretation of those observations is that not all of the baryons associated with a given halo are within the virial radius. Within a few years those constraints should rapidly improve.

\subsection{The Milky Way Revisited}

Decades elapsed before the $10^6$~K galactic ``corona" proposed by Lyman Spitzer was directly detected.  In the meantime, emission from $10^4$~K neutral hydrogen clouds was discovered above the galactic disk, a discovery enabled by the motions of the clouds, which have higher velocities relative to the Sun than H~I clouds in the disk \citep{Muller_HVCs_1963CRAS..257.1661M}.  Those clouds became known as the \textit{high velocity clouds} or HVCs \citep{1997ARA&A..35..217W}.  They have column densities as great as $\sim 10^{19} \, {\rm cm^{-2}}$, but their gas masses are hard to estimate because their heights above the disk are hard to pin down.  Bracketing them with absorption-line observations of halo stars at differing distances \citep[e.g.,][]{1993ApJ...416L..29D} has shown that many HVCs are within $\sim 10$~kpc, but some (including the Magellanic Stream) are tens of kiloparsecs away \citep{1993ApJ...416L..29D,Putman_2012ARA&A..50..491P}.  Closer to the disk, UV absorption-line studies of hot stars in the galactic halo show that the clouds with singly-ionized elements are distributed with scale heights above the disk of $\sim 0.3$ kpc \citep{1988ApJ...330..942V}, while more highly ionized elements can be found at heights $\sim 10$ times greater \citep{1987ApJ...314..380S}. Diffuse C~IV and O~III line emission from this halo gas was first observed using a UV telescope aboard the Space Shuttle \citep{1990ApJ...350..242M}. 

\subsubsection{X-ray Emission and Absorption by Halo Gas}

Unambiguous direct detection of the Milky Way's $\sim 10^6$~K CGM gas is not an easy task, in part because the Sun sits in a local hot-gas bubble difficult to differentiate from hot gas at greater distances.  The two most abundant elements, H and He, are both nearly fully ionized at $\sim 10^6$~K, so most of our information about the Milky Way's hot gas comes from  highly ionized oxygen, the next most abundant element, through the O~VII and O~VIII lines. Detections of X-ray absorption shadows coincident with foreground molecular gas clouds have gradually established that some of the O~VIII emission is coming from hot gas in the Milky Way's halo, well beyond the local bubble \citep{1991Sci...252.1529S,2007ApJ...658.1081G,2007PASJ...59S.141S}.  

More recent X-ray studies of both emission and absorption along multiple lines of sight have helped to constrain the radial distribution of that gas in the Milky Way's hot atmosphere \citep{Gupta_2012ApJ...756L...8G,MillerBregman_2013ApJ...770..118M,MillerBregman_2015ApJ...800...14M,Fang_2015ApJS..217...21F}.  Joint constraints placed by O~VII absorption lines along lines of sight to quasars, as observed from our location 8~kpc from the galaxy's center, indicate that the atmosphere's ambient density declines with radius approximately as $n_e \propto r^{-1.7}$ out to at least 100~kpc from the center \citep{MillerBregman_2013ApJ...770..118M}.  Emission-line constraints, which are sensitive primarily to the inner $\sim 40$~kpc of the atmosphere, indicate a slightly shallower decline ($\propto r^{-1.4}$) at smaller radii \citep{MillerBregman_2015ApJ...800...14M}.  The ratio of O~VII to O~VIII emission is consistent with a temperature $T \approx 2 \times 10^6$~K \citep{HenleyShelton_2013ApJ...773...92H} that may decline slowly with radius \citep{MillerBregman_2015ApJ...800...14M}.  Integrating the best-fitting density models over volume gives an atmospheric mass less than half the total baryonic mass originally associated with the halo's dark matter.  And even when the Galaxy's stellar mass is added, up to half of the baryonic mass associated with the Milky Way's halo is still considered ``missing" \citep[e.g.,][]{MillerBregman_2013ApJ...770..118M,MillerBregman_2015ApJ...800...14M,Voit_pNFW_2019ApJ...880..139V,Faerman_2020ApJ...893...82F}

\subsubsection{Ram-Pressure Stripping and Dispersion Measure}

\begin{figure}[!t]
    \centering
    \includegraphics[width=5.4in,trim=0.1in 0.0in -0.1in 0.0in,clip]{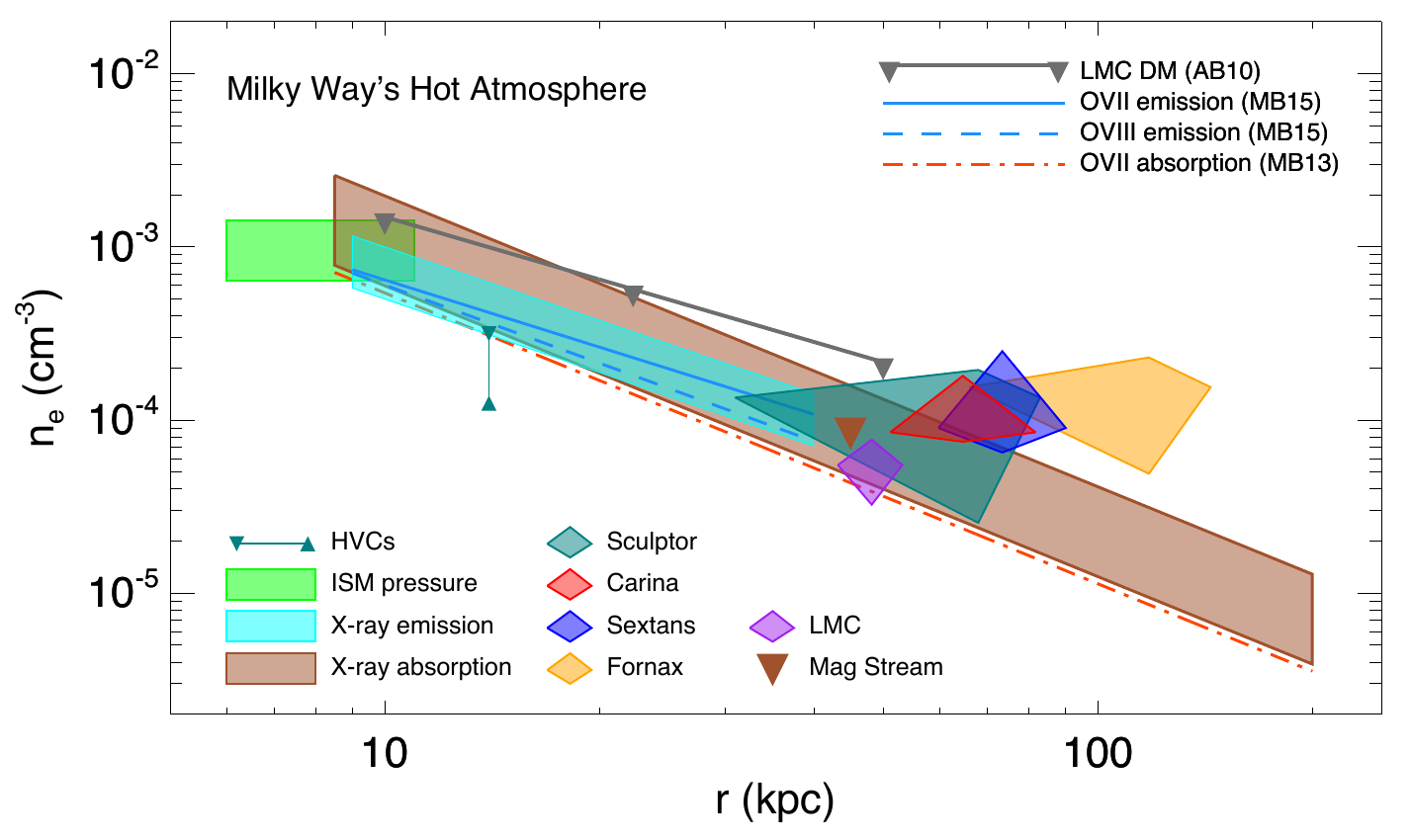}
    \caption{Constraints on the radial electron density profile of the Milky Way's hot atmosphere (adapted from \citep{Voit_pNFW_2019ApJ...880..139V}).  Long brown and blue strips show X-ray constraints, while diamond-like polygons show constraints from ram-pressure stripping.  Inverted gray triangles show upper limits from dispersion-measure observations of pulsars in the Large Magellantic Cloud \citep{AndersonBregman_2010ApJ...714..320A}. Green triangles show constraints derived from observations of high-velocity clouds \citep{Putman_2012ARA&A..50..491P}. (See \citep{Voit_pNFW_2019ApJ...880..139V} and \S \ref{sec:MilkyWayCGM} for citations and deeper discussions of the data sources.) 
    \label{Figure:MW_density}}
\end{figure}

Complementary estimates of the Milky Way's atmospheric density can be derived by gauging how effectively the ram pressure experienced by dwarf galaxies moving through the Milky Way's atmosphere strips cooler gas out of the dwarfs.  Those density estimates agree with the X-ray estimates to within a factor $\sim 2$--3, depending on the elemental abundances assumed in the X-ray modeling \cite{Voit_pNFW_2019ApJ...880..139V}.  Most of the ram-pressure stripping constraints apply to radii $\sim 50$--100~kpc, where they imply $n_e \sim 10^{-4} \, {\rm cm^{-3}}$.  At smaller radii, dispersion-measure observations toward radio pulsars in the Large Magellanic Cloud place an upper limit of $\lesssim 7 \times 10^{19} \, {\rm cm^{-2}}$ on the electron column density out to 50~kpc \citep{AndersonBregman_2010ApJ...714..320A}.  Figure \ref{Figure:MW_density} shows both kinds of constraints, along with electron-density profiles derived for the Milky Way from X-ray emission and absorption observations.

\subsection{Basics of Atmospheric Structure} 

The rest of this article will treat all of these galactic atmospheres similarly, from the atmospheres of galaxy clusters down to those around galaxies like the Milky Way, assuming that the volume-filling component is close enough to being hydrostatic for gravitational compression to make its temperature similar to the halo's gravitational temperature $T_\phi$. That assumption is reasonable for galaxy clusters and groups but less well justified for lower-mass galactic halos in which gas motions appear to approach speeds comparable to the halo's circular velocity. In a nearly hydrostatic state, the atmosphere's structure is determined by the distribution of specific entropy among its gas parcels \citep[e.g.,][]{Voit+02,Voit+03,Voit_2005RvMP...77..207V}.  Consequently, the structure of a galactic atmosphere reflects its thermodynamic history, which is shaped by all of the heating events that have added entropy to its gas and all of the cooling events that have reduced its specific entropy. 

\subsubsection{Convective Stability}

An atmosphere in which specific entropy is not constant on equipotential surfaces cannot remain static.  To see why, consider a gas parcel of entropy $K$ within an atmospheric layer of entropy $\bar{K}$.  In pressure balance, a parcel with $K > \bar{K}$ will be less dense than its surroundings, because $n \propto (P/K)^{3/5}$. Buoyancy will therefore push a high-entropy parcel to greater altitude.  Similarly, a parcel with $K < \bar{K}$ will be denser than its surroundings, so gravity will pull a low-entropy parcel to lower altitude.  Convection therefore sorts atmospheric gas according to specific entropy until the entropy gradient is aligned with the gravitational potential gradient.\footnote{In stellar astrophysics, this condition is known as the \textit{Schwarzschild criterion} \citep{SchwarzschildCriterion_1906WisGo.195...41S}.}

\subsubsection{Entropy Profile and Hydrostatic Temperature}

Once convection has sorted the atmosphere's gas parcels so that $K$ increases monotonically with altitude, a gravitationally confined atmosphere's structure depends primarily on the shape of the confining potential well and the distribution of specific entropy in its atmospheric gas.  Section \ref{sec:HSE} showed that the atmosphere's temperature in this configuration is $\sim T_\phi$.  This section makes that connection more precise, by showing how the relationship between $T$ and $T_\phi$ arises from the atmosphere's entropy gradient.

The temperature profile of a hydrostatic atmosphere in a spherical potential is related to its pressure profile through the equation
\begin{equation}
    \frac {d \ln P} {d \ln r} \: = \: -2 \frac {T_\phi (r)} {T(r)} 
    \; \; .
    \label{eq:HSE_ln2}
\end{equation}
In other words, the power-law slope of the pressure gradient determines the constant of proportionality relating an atmosphere's hydrostatic temperature at radius $r$ to the gravitational temperature $T_\phi(r) = \mu m_p v_{\rm c}^2(r) /2k$ at the same radius (see \S \ref{sec:HSE}).  Noting that $P \propto T^{5/2} K^{-3/2}$, we can rewrite the relationship given in equation (\ref{eq:HSE_ln2}) as 
\begin{equation}
  \frac {d} {d \ln r} \left( \frac {T} {K^{3/5}} \right) \: = \: 
       - \frac {4} {5} \frac {T_\phi(r)} {K^{3/5}(r)}
       \; \; .
\end{equation}
Given a known entropy profile $K(r)$, we can then directly integrate to obtain the hydrostatic temperature profile 
\begin{equation}
 T(r) \: = \: \frac{4}{5} \int^{r_{\rm b}}_r 
                   \left[ \frac {K(r)} {K(r^\prime)} \right]^{3/5} T_{\phi}(r^\prime) 
                   \, \frac {d r^\prime} {r^\prime} 
                  \: + \:  
        \left[ \frac {K(r)} {K(r_{\rm b})} \right]^{3/5} T_{\rm b}  
    \; \; .
    \label{eq:Tprofile}
\end{equation}
At the bounding radius $r_{\rm b}$, the temperature boundary condition $T_{\rm b}$ is jointly determined by $K(r_{\rm b})$ and the boundary pressure $P_{\rm b} = P(r_{\rm b})$.  We will assume that $T_{\rm b} \lesssim T_\phi (r_{\rm b})$, because a hydrostatic atmosphere with $T_{\rm b} \gg T_\phi$ is confined primarily by external pressure, not so much by gravity. 

The form of equation (\ref{eq:Tprofile}) reveals how an atmosphere's entropy profile determines its hydrostatic temperature profile.  Gas temperature at a given radius is a weighted combination of the boundary temperature $T_b$ and the value of $T_\phi$ in the vicinity of $r$, with the weights depending on $K(r)$.  In an atmosphere with a substantial entropy gradient, temperatures at $r \ll r_{\rm b}$ reflect the local value of $T_\phi$ because the boundary term in equation (\ref{eq:Tprofile}) becomes negligible. The constant of proportionality relating $T$ and $T_\phi$ is then closely linked to the entropy gradient. That relationship is 
\begin{equation}
    T \: = \: \left( \frac {4} {3 \alpha_K} \right) T_\phi 
      \: = \: \left( \frac {2 \mu m_p} {3 k \alpha_K}  \right) v_{\rm c}^2
      \; \; .
      \label{eq:HSE_temp}
\end{equation}
for the idealized case of an atmosphere with a power-law entropy profile $K \propto r^{\alpha_K}$ in a potential well with constant $v_{\rm c}$.

Notice that heating and cooling processes that change the normalization of $K(r)$ without changing its slope \textit{do not change the atmosphere's hydrostatic temperature} as long as $T_\phi$ does not change.  In a typical galactic atmosphere, the contribution of atmospheric gas to the gravitational potential is minor, meaning that atmospheric reconfigurations cause only small changes in $T_\phi$.  Significant changes in an atmosphere's hydrostatic temperature therefore result from heating and cooling processes that \textit{change the entropy gradient.}  

By now it should be clear that the shape of a hydrostatic atmosphere's entropy profile $K(r)$ determines its temperature and the entropy profile's normalization determines the atmosphere's density.  Heating causes the atmosphere to expand.  Cooling causes it to contract.  If heating and cooling are gradual compared to the time required for a sound wave to propagate, the atmosphere's temperature remains close to the local value of $(4/3 \alpha_K) T_\phi$ during both heating-driven expansion and cooling-driven contraction.   

Central heating can raise the atmosphere's temperature by flattening the overall entropy gradient (reducing $\alpha_K$).  Central cooling can lower the atmosphere's temperature by making the entropy gradient steeper (increasing $\alpha_K$).  But heating can also lower an atmosphere's hydrostatic temperature if its distribution with radius increases the overall entropy gradient.\footnote{For example, heating restricted to the atmosphere's outer layers causes them to expand, reducing the weight pressing down on the lower layers, which then expand adiabatically and drop to lower temperature.}

\subsubsection{Thermodynamic Phase Diagrams}

Astronomers customarily think of a galaxy's atmospheric gas in terms of temperature and density, rather than pressure and entropy, because $T$ and $n$ are more directly measured with observations than $P$ or $K$.  However, pressure and entropy more directly reflect the thermodynamic history of an atmosphere and how its structure responds to heating and cooling.  Radiative cooling lowers the entropy of a gas parcel, but not necessarily its temperature.  Gradual heating raises the entropy of a gas parcel, but not necessarily its temperature.  Passing through a shock front causes a sudden jump in both entropy and temperature, but the temperature of the shocked gas might then decrease adiabatically as buoyancy pushes it upward toward layers of lower pressure.

\begin{figure}[!t]
\centering
\includegraphics[width=5.4in,trim=0.1in 0.0in -0.1in 0.0in,clip]{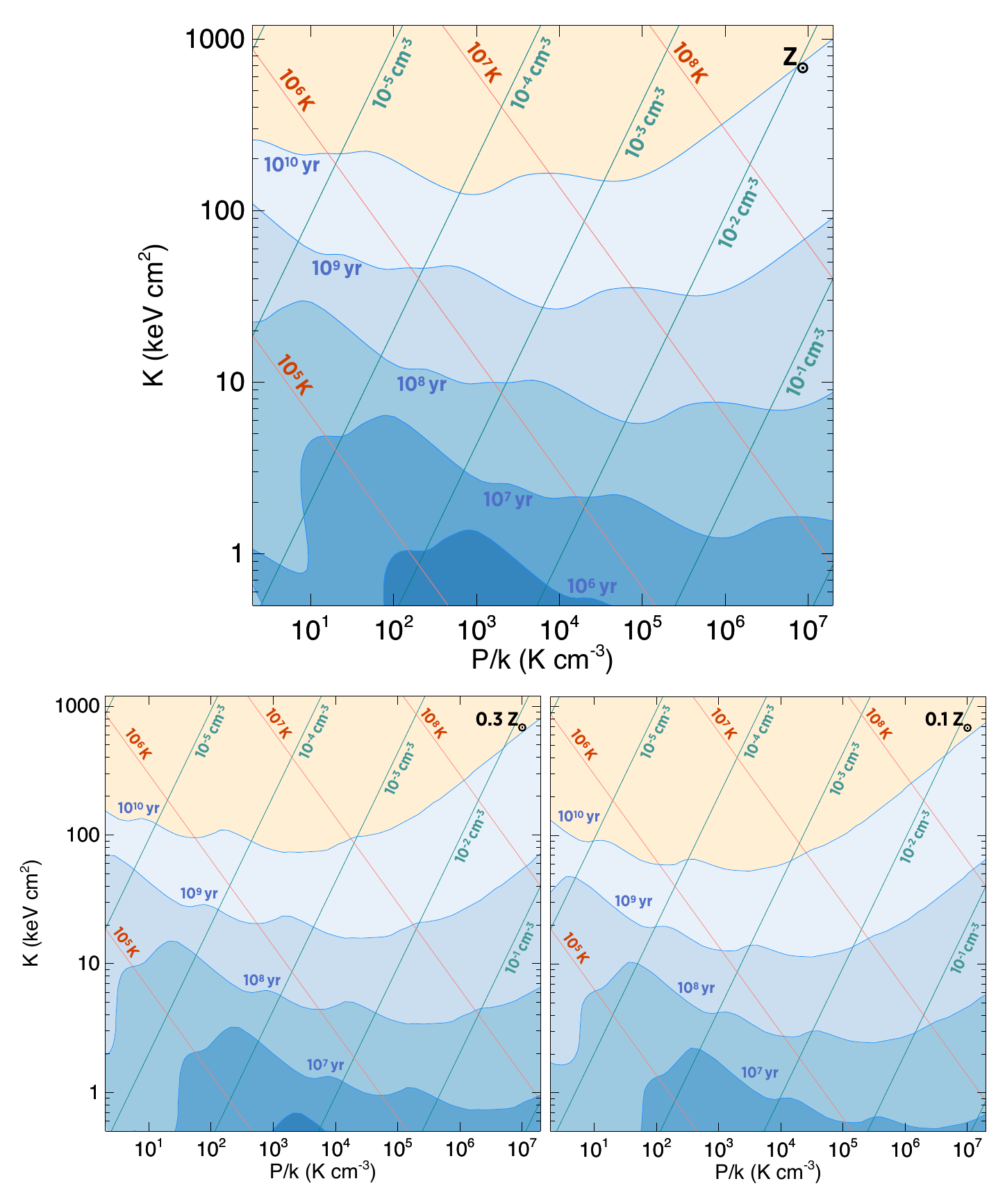}
\caption{Contours of constant cooling time in the pressure-entropy plane for plasma in coronal ionization equilibrium with solar abundances ($Z_\odot$, top panel), 0.3 times solar abundances (0.3 $Z_\odot$, bottom left panel), and 0.1 times solar abundances (0.1 $Z_\odot$, bottom right panel).  Labeled diagonal lines show temperature and electron density. \label{Figure:P-Kplane}}
\end{figure}

One can become more accustomed to thinking of atmospheric gas in terms of pressure and entropy through the use of phase diagrams in which $P$ and $K$ are the principal axes.  Figure \ref{Figure:P-Kplane} shows some examples, so let us unpack them. The horizontal axes showing $P/k$ are in units of ${\rm K \, cm^{-3}}$. The vertical axes showing $K$ are in units of ${\rm keV \, cm^2}$, because X-ray astronomers usually report temperatures in units of keV.\footnote{ 1 keV corresponds to $1.16 \times 10^7$K.}  The diagonal lines rising from left to right are lines of constant density; the ones declining from left to right are lines of constant temperature. 

In the $P$--$K$ plane, the effects of hydrodynamics and thermodynamics are orthogonal.  A gas parcel experiencing only adiabatic hydrodynamical motions shifts horizontally but not vertically in such a diagram, because adiabatic compression and expansion do not change $K$.  On the other hand, isobaric gains and losses of heat energy shift the parcel vertically, changing $K$ and not $P$.  Radiative cooling shifts a gas parcel down. If it sinks toward lower layers of greater pressure, it will also shift to the right.  Shock heating shifts a parcel up but also to the right, because both $P$ and $K$ sharply increase. 

These diagrams also include contours of constant cooling time, with shading showing how cooling time rises from the bottom of each plot (darker blue) toward the top (lighter blue and then orange).  Gas in the darker regions toward the bottom cools on a timescale that is short compared to a typical dynamical time in the CGM (which is $\gtrsim 100 \, {\rm Myr}$ at $r \gtrsim 30 \, {\rm kpc}$). Gas that has reached the orange region is unlikely to cool on an astrophysically interesting time scale, because it has $t_{\rm cool} > 10 \, {\rm Gyr}$. 

Note that for a wide range of CGM temperatures ($10^{5.5-7.5}$~K), the contours of constant cooling time are nearly horizontal. Within that temperature range, the primary radiative cooling mechanisms are changing from collisionally-excited line emission at the lower end (approximately following $\Lambda \propto T^{-0.8}$ in the range $10^{5.5-6.5}$~K) to bremsstrahlung ($\Lambda \propto T^{1/2}$) at greater $T$.  Expressing this dependence on temperature as $\Lambda \propto T^\lambda$ allows the scaling of cooling time to be written as
\begin{equation}
    t_{\rm cool}  \: \propto \:  T^{1-\lambda} n^{-1}  
                  \: \propto \:  P^{-(2 \lambda + 1)/5} K^{3(2-\lambda)/5}
    \label{eq:tcool_rels}
\end{equation}
showing why $t_{\rm cool}$ depends much more strongly on $K$ than on $P$.  In the heart of the CGM temperature range, the dependence of radiative cooling on temperature corresponds to $\lambda \approx -0.5$. Equation \ref{eq:tcool_rels} shows that the pressure dependence of $t_{\rm cool}$ is negligible for this value of $\lambda$, whereas the dependence of $t_{\rm cool}$ on $K$ is strong.  In other words, adiabatic compression does not significantly lower the cooling time of a gas parcel, because its temperature rises almost as much as its density, while $\Lambda (T)$ declines.  As a result, its cooling time depends almost exclusively on $K$. 

\subsubsection{Consistency in Cooling-Time Structure}

Before moving on to more detailed analyses of how galactic atmospheres come to be as they are, we will conclude Section~\ref{sec:AmbientCGM} with one more figure illustrating two important patterns that motivate the approach we will be taking.  Figure~\ref{figure:Consistency} shows the atmospheric structure of four representative galaxies overlaid on a $P$--$K$ phase diagram:

\begin{figure}[!t]
\centering
\includegraphics[width=4.5in]{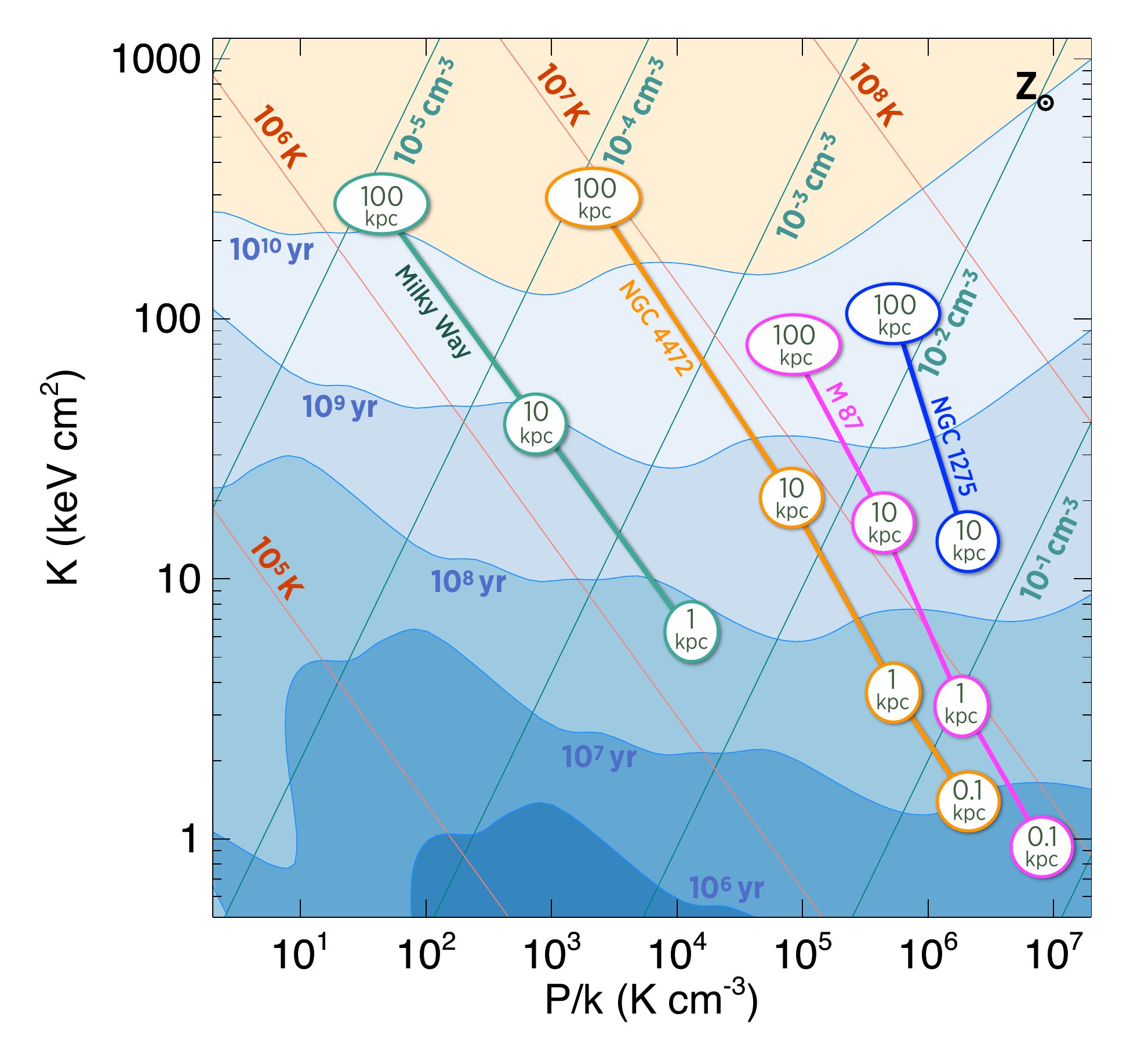}
\caption{Consistency of cooling time structure in the CGM of massive galaxies. The four galaxies schematically represented in this thermodynamic phase diagram include an actively star-forming central galaxy in a massive galaxy cluster (NGC~1275 in the Perseus Cluster), the central galaxy of a less massive galaxy cluster without much star formation (M87 in the Virgo Cluster), a non-central massive elliptical galaxy with no star formation (NGC 4472 in the Virgo Cluster), and an average spiral galaxy that continually forms stars (our Milky Way).  All of them have $t_{\rm cool} \sim 10^9$~yr at $\sim 10$~kpc and $t_{\rm cool} \sim 10^{10}$~yr at $\sim 100$~kpc, even though the gas pressures there differ by more than three orders of magnitude.
\label{figure:Consistency}}
\end{figure}

\begin{itemize}

    \item NGC 1275 (in blue) is the central galaxy of the Perseus Cluster, a relatively massive and nearby cluster of galaxies.  Outflows from its AGN are currently producing large cavities with sizes indicating a kinetic energy output $\gtrsim 10^{44} \, {\rm erg \, s^{-1}}$.  Its $\sim 3 \times 10^7$~K atmosphere is multiphase with a spectacular emission-line nebula (at $10^4$~K) extending tens of kpc from the center. It also contains abundant molecular gas (at $< 10^2$~K). Cooling-time estimates shown at 10~kpc and 100~kpc are derived from observations by \citep{Fabian_2006MNRAS.366..417F}.   Unlike most large elliptical galaxies, this BCG is forming stars at a rate $\sim 50 \, M_\odot \, {\rm yr^{-1}}$ \citep{Mittal_2015MNRAS.450.2564M}.  
    
    \item M87 (in magenta) is the central galaxy of the Virgo Cluster, which is closer but less massive than the Perseus Cluster.  Its central AGN is also producing cavities but with less kinetic power ($\sim 10^{43} \, {\rm erg \, s^{-1}}$).  Its $\sim 1.5 \times 10^7$~K atmosphere is multiphase, but with a smaller emission-line nebula than in NGC 1275 and little molecular gas. There is no current star formation.  Cooling-time estimates shown at 0.1~kpc, 1~kpc, 10~kpc, and 100~kpc are derived from observations by \citep{Werner+2012MNRAS.425.2731W}.
        
    \item NGC 4472 (in orange) is another member of the Virgo Cluster.  Its stellar mass is similar to M87's, but it is not centrally located in the cluster's halo.  Its central AGN is putting out a kinetic power $\sim 10^{41} \, {\rm erg \, s^{-1}}$.  Its $\sim 1 \times 10^7$~K atmosphere does not have any obvious multiphase gas.  There is no detectable star formation.  Cooling-time estimates shown at 0.1~kpc, 1~kpc, 10~kpc, and 100~kpc are derived from observations by \citep{Russell_M87_2015MNRAS.451..588R}.
    
    \item The Milky Way (in green) has a much smaller stellar mass than the other three galaxies and is the only one of the four with an actively star-forming galactic disk.  Cooling-time estimates shown at 10~kpc and 100~kpc are based on the observations shown in Figure~\ref{Figure:MW_density}.  The estimate for 1~kpc comes from an extrapolation that extends the best-fitting density profile from \citep{MillerBregman_2013ApJ...770..118M}.
    
\end{itemize} 
\noindent Here are the two patterns to notice:
\begin{itemize}

    \item  The dependence of cooling time and specific entropy on radius in each of these galactic atmospheres is quite similar, even though pressure differs by three orders of magnitude, density differs by two orders of magnitude, and temperature differs by an order of magnitude. We therefore need to understand how these four very different galaxies come to have such similar radial profiles of cooling time and entropy.
    
    \item  The ambient cooling time inside of 10 kpc in each of these galactic atmospheres is $\lesssim 1$~Gyr and can be as short as $\sim 10$~Myr at the smallest observable radii.  As Lyman Spitzer noted, such a short cooling time suggests a long-term regulation mechanism that somehow provides enough heat to offset radiative cooling, over timescales no longer than $\sim t_{\rm cool}$.
    
\end{itemize}
Our approach will therefore be to start with the heating mechanism that originally generates the entropy profiles of these atmospheres (\S \ref{section:CosAtmo}), consider how radiative cooling alters those entropy profiles (\S \ref{sec:Cooling}), and identify the heating mechanisms capable of offsetting radiative cooling (\S \ref{sec:Heating}).  Then we will be ready to analyze the observational clues indicating how those processes interact (\S \ref{sec:Weather}) and to assess theoretical ideas about how those interactions lead to self-regulation (\S \ref{sec:Balance}).

\section{Cosmological Accretion \label{section:CosAtmo}}

A galaxy acquires its gaseous atmosphere mainly through cosmological accretion, which occurs preferentially along the large scale dark-matter filaments connecting cosmological halos.  Accreting matter generally enters a halo at a speed similar to the halo's circular velocity, which can be supersonic compared with the sound speed of the infalling gas.  Deceleration of the infalling gas therefore produces shock fronts that dissipate the kinetic energy of infall into heat that raises the CGM entropy level.  Without radiative cooling, this process of entropy generation results in an atmosphere that is approximately hydrostatic, with a radial density profile similar to the total mass density profile of the dark-matter halo \citep{Bertschinger_1985,Voit_2005RvMP...77..207V}.

\subsection{Cosmological Entropy Scale}

A simple calculation provides a useful estimate of the CGM entropy level produced by cosmological accretion and how it scales with halo mass and cosmic time (based on the definition of halo mass in \S \ref{sec:boundaries}). Gas at the halo's gravitational temperature and distributed in space like its dark matter has a specific entropy similar to  
\begin{equation}
    K_{200 {\rm c}} \: \equiv \: 
        \frac {kT_{200 {\rm c}}} {\bar{n}_{e,200c}^{2/3}} 
    \: \propto \: M_{200 {\rm c}}^{2/3} H^{-2/3}  
    \label{eq:K200}
\end{equation}
where $T_{200 {\rm c}} \equiv T_\phi (r_{200 {\rm c}}) = (\mu m_p/2k) (GM_{200 {\rm c}} / r_{200 {\rm c}})$ is the halo's gravitational temperature and 
\begin{equation}
    \bar{n}_{e,200 {\rm c}} \: \equiv \: \frac {200 f_{\rm b} \rho_{\rm cr}} {\mu_e m_p}
        \: \approx \: 1.4 \times 10^{-4} \,  {\rm cm^{-3}} \, \cdot \,  E^2(z)
\end{equation}
is the mean electron density corresponding to a total mass density $200 \rho_{\rm cr}$ \citep[e.g.,][]{vkb05}.  In the latter expression, $\mu_e m_p$ is the mean mass per electron, $f_{\rm b}$ is the baryonic mass fraction of the universe, and $E(z) \equiv H(z) / H_0$, with $H_0$ representing the current value of $H(z)$.

Equation \ref{eq:K200} shows that the cosmologically-generated CGM entropy scale $K_{200 {\rm c}}$ depends almost entirely on the product of halo mass and cosmological time $t$, because $H(z)$ in a $\Lambda$CDM universe slowly evolves from $\approx (2/3)t^{-1}$ early in time to $\approx t^{-1}$ at the current time.  For a massive elliptical galaxy, one finds
\begin{equation}
    K_{200 {\rm c}} \: \approx \: 110 \, {\rm keV \, cm^2} 
        \left( \frac {M_{200 {\rm c}}} {10^{13} \, M_\odot} \right)^{2/3}
        [E(z)]^{-2/3}
        \; \; .
\end{equation} 
Comparing this entropy scale with the cooling-time contours in Figure \ref{Figure:P-Kplane} shows that the current cosmological entropy scale of the atmosphere in a halo of mass $M_{200 {\rm c}} \sim 10^{13} \, M_\odot$ corresponds to a cooling time of $\sim 10$~Gyr.  However, both $M_{200 {\rm c}}$ and $t$ were smaller earlier in time.  Therefore, CGM gas that entered the halo much earlier in time (and passed through an accretion shock with greater density) has a lower entropy level than this typical value and also a shorter cooling time.

\subsection{Spherical Accretion}

A slightly less simple calculation gives the entropy level expected in the outer parts of an idealized halo of mass $M_{\rm halo}(t)$ that grows by continuous accretion of spherically symmetric shells of matter \citep{2001ApJ...546...63T, Voit+03}.  At first, each concentric shell that will end up in the halo expands almost as fast as the rest of the universe. But the excess matter density interior to the shell slows it down until the shell's expansion halts at a turnaround radius $r_{\rm ta}$.  After halting, the shell falls inward at increasing speed until passing through an accretion shock at $r_{\rm acc} \approx r_{\rm ta} / 2$.  The shell's dark matter is unimpeded by that shock front, which affects only the gas.  The dark matter continues to plunge inward and proceeds to oscillate back and forth through the center, if strict spherical symmetry is imposed.  The shell's gas, on the other hand, rapidly decelerates at the shock front, where the shock turns its kinetic energy into heat and raises the entropy of the accreted gas.

\begin{figure}[!t]
\centering
\includegraphics[scale=0.6]{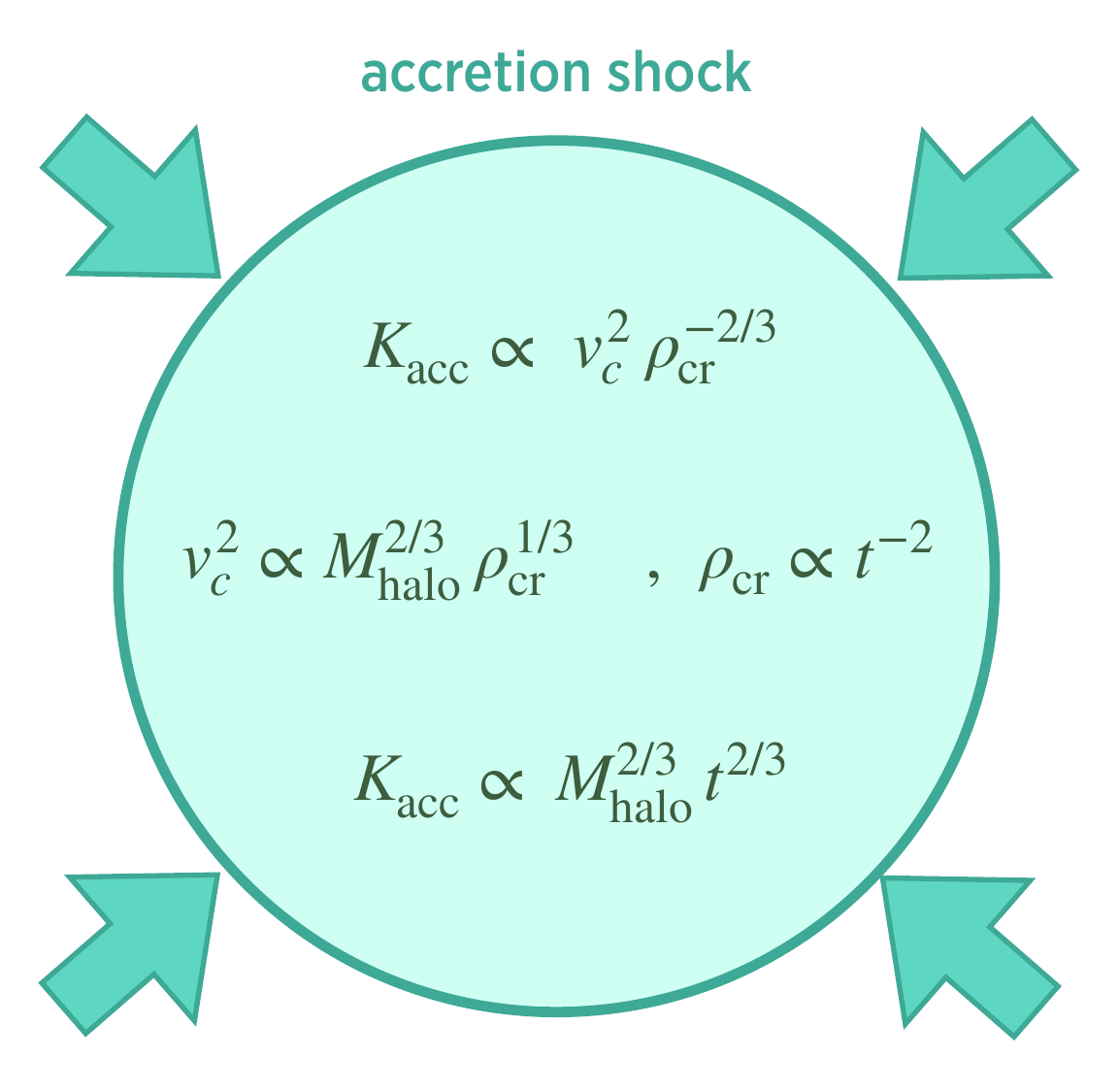}
\caption{The entropy gained by gas falling through an accretion shock depends on the mass of the halo $M_{\rm halo}$ and the age of the universe $t$ at the time of infall. 
\label{fig:SphericalAccretion}
}
\end{figure}

The amount of entropy generated in this spherical accretion shock can be calculated directly from the usual shock jump conditions.  Here we will assume that the thermal energy of the incoming gas is negligible compared to its kinetic energy.\footnote{Corresponding to a shock with a Mach number much greater than unity} In that limit, the jump condition $\rho_2 = 4 \rho_1$ relates the density $\rho_2$ of the post-shock gas to the density $\rho_1$ of the pre-shock gas in a medium in which $P \propto K \rho^{5/3}$.  If 100\% of the incoming kinetic energy is thermalized by the shock front, then $3kT_2/2 = \mu m_p v_{\rm acc}^2 / 2$, where $v_{\rm acc}$ is the infall speed and $T_2$ is the post-shock gas temperature. Combining these conditions gives 
\begin{equation}
    K_{\rm acc} = \frac {\mu m_p v_{\rm acc}^2} {3} 
                    \left( \frac {\mu_e m_p} {4 \rho_1} \right)^{2/3} 
\end{equation}
for the entropy generated at a strong accretion shock, as long as $v_{\rm acc}$ is the infall speed in the frame of the post-shock gas (see Figure \ref{fig:SphericalAccretion}).

Relating $K_{\rm acc}$ to $M_{\rm halo}$ and $t$ requires expressions for $\rho_1$ and $v_{\rm acc}$.  The total mass accretion rate $\dot{M}_{\rm halo} = 4 \pi r_{\rm acc}^2 \rho_1 v_{\rm acc} / f_{\rm b}$ determines the preshock density, and the infall speed comes from the accreting shell's orbital dynamics.  Dark energy is not yet dominant enough to have significantly affected the shell's motion, meaning that $v_{\rm acc}^2 \approx (GM_{\rm halo}/r_{\rm acc}) \xi$, where $\xi=2(1- r_{\rm acc} / r_{\rm ta})$.  Putting these pieces together gives
\begin{equation}
 \begin{split}
    K_{\rm acc} \: &\approx \: \frac {\mu m_p} {3} 
                \left ( \frac{\pi G^2 \xi^2 \mu_e m_p} {f_{\rm b}} 
                                    \right )^{2/3} 
                \left( \frac{d \ln M_{\rm halo}} { d \ln t} \right )^{-2/3}  
                                (M_{\rm halo} t)^{2/3}
                \\ 
        &\approx \: 284 \, {\rm keV \, cm^2} 
            \left( \frac{d \ln M_{\rm halo}}{ d \ln t} \right )^{-2/3} 
            \left( \frac {M_{\rm halo}} {10^{13} \, M_\odot} \right)^{2/3}
            (H_0 t)^{2/3}
 \end{split}
 \label{eq:Kacc}
\end{equation}
where the lower line of the equation has assumed $r_{\rm acc} \approx r_{\rm ta} / 2$.

The outcome of this exercise shows why the cosmological entropy of a galactic atmosphere is $\propto (M_{\rm halo} t)^{2/3}$.  Figure \ref{fig:SphericalAccretion} provides a more succinct summary, but without the constant of proportionality.  The quantity $K_{\rm acc}$ is somewhat greater than $K_{200c}$ in part because it reflects the entropy currently being generated at the accretion shock but also because the accreting gas has been assumed to enter the halo as a smooth, uniform flow.  In fact, cosmological accretion in a $\Lambda$CDM universe is sporadic, hierarchical, and inhomogeneous, meaning that the mass-weighted mean density of accreting gas is somewhat greater than $\rho_1$, making the actual amount of entropy generated by accretion shocks somewhat less than $K_{\rm acc}$.

\subsection{The Cosmological Baseline Entropy Profile}
\label{sec:Kbaseline}

Numerical simulations of cosmological structure formation account for all the messiness of entropy generation associated with hierarchical mergers of halos with inhomogeneous atmospheres.  Without radiative cooling, the outcome of this complex process turns out to be simple to state.  The cosmological entropy profile of a galactic atmosphere is nearly a power law in radius for $r \gtrsim 0.2 r_{200 {\rm c}}$, with
\begin{equation}
 \begin{split}
    K_{\rm base}(r) &\approx 1.32 \, K_{200 {\rm c}} 
          \left( \frac {r} {r_{200 {\rm c}}} \right)^{1.1}
          \\
         &\approx 144 \, {\rm keV \, cm^{2}} 
         \left( \frac {M_{200c}} {10^{13} \, M_\odot} \right)^{2/3}
                [E(z)]^{-2/3}
          \left( \frac {r} {r_{200 {\rm c}}} \right)^{1.1}
    \label{eq:Kbase}
 \end{split}
\end{equation}      
The same basic result emerges from different simulations using entirely different hydrodynamical methods, with only $\sim 10$\% differences in the normalization constant and power-law index \citep{vkb05}.  

\begin{figure}[!t]
  \centering
  \includegraphics[width=3.0in]{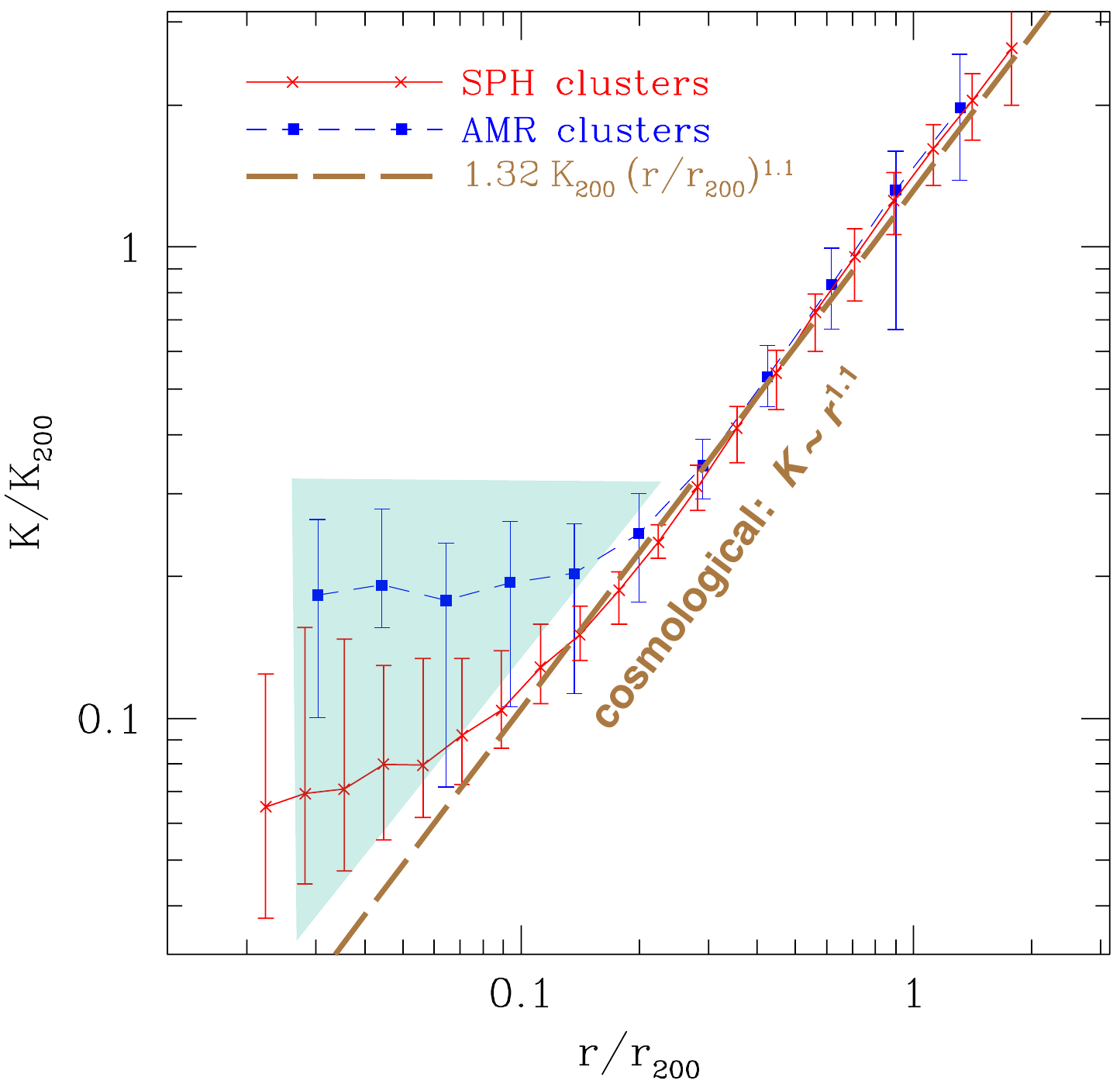}
  \caption{The cosmological baseline entropy profile from \citet{vkb05}.  Numerical simulations of cosmological structure formation produce atmospheric entropy profiles following the power law $K \propto r^{1.1}$ at $r \gtrsim 0.2 r_{200 {\rm c}}$, independent of the simulation method.  Scaling them by $K_{200 {\rm c}}$ and $r_{200 {\rm c}}$ removes their dependences on halo mass, demonstrating that cosmological structure formation without radiative cooling and galaxy formation would make atmospheres with nearly self-similar structure.  At small radii ($r \lesssim 0.2 r_{200 {\rm c}}$), different hydrodynamical methods, such as smoothed particle hydrodynamics (SPH, red crosses) and adaptive mesh refinement (AMR, blue squares), can produce differing results (shaded region) that arise from differences in their treatments of mixing \citep{Mitchell+09}.
  \label{figure:entropyprofiles}}
\end{figure}

At smaller radii ($r \lesssim 0.2 r_{200 {\rm c}}$), mixing driven by hierarchical merging causes the cosmological entropy profile to become flatter (see Figure \ref{figure:entropyprofiles}). The results of mixing in simulations depend somewhat on the hydrodynamical method being applied \citep{Mitchell+09}.  Recent improvements to the mixing behavior of smoothed particle hydrodynamics algorithms \citep[e.g., ][]{Price2008} have reduced those discrepancies.  However, the entropy profiles at $r \lesssim 0.2 r_{200 {\rm c}}$ in both real and simulated galactic atmospheres depend much more on radiative cooling and feedback than on cosmological structure formation \citep[e.g.,][]{Voit+02}.

The simplicity of the result in Figure \ref{figure:entropyprofiles} is useful because it provides a well-defined baseline for assessing the effects of radiative cooling and feedback from galaxies on the gas surrounding them.  Additional heating of the atmosphere raises its entropy above the cosmological baseline.  Central cooling allows the entropy profile to fall below the cosmological baseline.  However, the effects of cooling can also be counterintuitive. Cooling and condensation of the lowest entropy gas can preferentially remove low-entropy gas from the atmosphere, thereby raising the mean entropy of the remaining atmospheric gas \citep{KnightPonman_1997MNRAS.289..955K,Bryan_2000ApJ...544L...1B,vb01}.

\subsection{Observed Entropy Profiles}
\label{sec:Kobserved}

\begin{figure}[!t]
  \centering
  \includegraphics[width=4.5in]{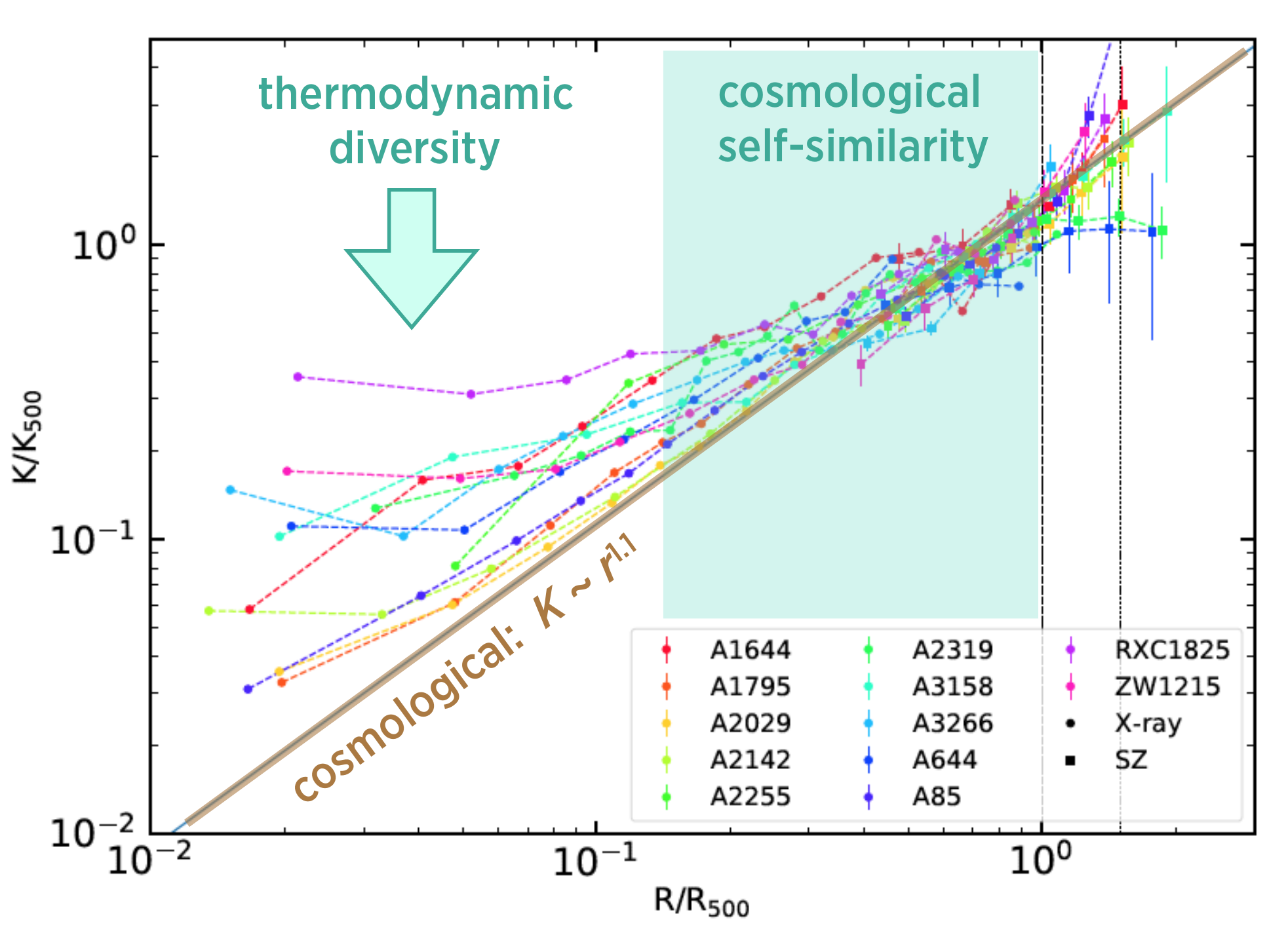}
  \caption{Comparisons of observed entropy profiles of galaxy clusters with the cosmological baseline profile (adapted from \citet{Ghirardini_2019A&A...621A..41G} and reproduced with permission from the European Southern Observatory). Each entropy profile is scaled according to a radius $r_{500 {\rm c}}$ and an entropy $K_{500 {\rm c}}$ derived from the cluster's mass.  (The quantity $K_{500 {\rm c}}$ is analogous to $K_{200 {\rm c}}$ but is defined with respect to $r_{500 {\rm c}}$ instead of $r_{200 {\rm c}}$, which is marked with a vertical dotted line.)  In the range $0.15 r_{500 {\rm c}} < r < r_{500 {\rm c}}$, galaxy cluster atmospheres are nearly self-similar and are close to the cosmological baseline profile produced in numerical simulations, but their entropy profiles are much more diverse at $r < 0.15 r_{500 {\rm c}}$, reflecting each cluster's unique thermodynamic history.
  \label{figure:ClusterEntropyProfiles}}
\end{figure}

X-ray observations of massive galaxy clusters show that the entropy profiles of their  atmospheres do indeed approach the cosmological baseline profile beyond $r \gtrsim 0.1 r_{200 {\rm c}}$ \citep{pcn99,Cavagnolo+09,Pratt2010_REXCESSentropy,Ghirardini2017A&A...604A.100G,Ghirardini_2019A&A...621A..41G}.  At $r_{500 {\rm c}}$ ($\approx 0.6 r_{200 {\rm c}}$), they are quite similar to the baseline profile (see Figure \ref{figure:ClusterEntropyProfiles}).  Beyond that radius, diminishing X-ray surface brightness makes the observations more difficult.  Some studies \citep{2012MNRAS.427L..45W,2018A&A...614A...7G,2019SSRv..215....7W} suggest that the entropy profiles of clusters fall below the baseline profile at $r > r_{500 {\rm c}}$, but at least part of that discrepancy may be due to clumping of the X-ray emitting gas, which makes the apparent mean gas density seem greater than the actual mean gas density \citep{2011ApJ...731L..10N, 2017MNRAS.469.1476S}.  Correcting for clumping tends to bring the observed entropy at $r_{500 {\rm c}} < r < r_{200 {\rm c}}$ closer to the predicted baseline profile \citep{2017MNRAS.469.1476S}.  Thermodynamic properties throughout most of a galaxy cluster's atmosphere are therefore consistent with entropy generation via cosmological accretion shocks. 

Entropy profiles of galaxy clusters are more diverse inside of $0.15 r_{500 {\rm c}}$ ($\approx 0.1 r_{200 {\rm c}}$), with deviations from the baseline reflecting the thermodynamic history of the cluster core.  They cannot be adequately represented with a single power law, because many galaxy clusters have nearly isentropic cores \citep{Cavagnolo+09}. Early attempts to account for this deviation from a single power-law profile applied the simplest parametric model that could work \citep{Donahue+05,Donahue+06}, which simply added a constant to the power law relation that applies at larger radii: 
\begin{equation}
    K(r) = K_0 + K_{100} \left( \frac {r} {100~{\rm kpc}} \right)^{\alpha_K}
    \; \; .
\end{equation} 
The $K_0$ parameter usually characterizes the excess entropy at small radii, relative to a pure power-law entropy profile.  Negative values of the $K_0$ parameter are uncommon but can arise if $K(r)$ has a steeper power-law slope at small radii than at large radii or a discontinuity like the ones at the outer boundaries of X-ray coronae (\S \ref{sec:Coronae}).

This three-parameter model can be adequately fit within $r_{2500}$ to a large majority of the galaxy cluster observations in the \textit{Chandra} data archive, giving $\alpha_K = 1.2 \pm 0.4$ and $K_{100} = 126 \pm 45 \, {\rm keV \, cm^2}$ \citep{Cavagnolo+09}.  For comparison, the cosmological baseline profile on the mass scale of a galaxy cluster is
\begin{equation}
    K_{\rm base}(r) \approx 110 \, {\rm keV \, cm^2} 
         \left( \frac {M_{200c}} {10^{15} \, M_\odot} \right)^{0.3}
                [E(z)]^{-2/3}
          \left( \frac {r} {100 \, {\rm kpc}} \right)^{1.1}
\end{equation}      
which has similar values of $\alpha_K$ and $K_{100}$.  That regularity contrasts with  the best-fitting values of excess entropy ($K_0$), which are quite diverse (see Figure \ref{figure:ClusterEntropyProfiles}), ranging from highs of $\gtrsim 300 \, {\rm keV \, cm^2}$ down to low values indistinguishable from zero.  Self-similarity among galaxy-cluster entropy profiles therefore breaks down near the entropy scale ($\sim 100 \, {\rm keV \, cm^2}$) at which $t_{\rm cool}$ is similar to the age of the universe \citep{vb01}. 

\begin{figure}[!t]
  \centering
  \includegraphics[width=5.2in]{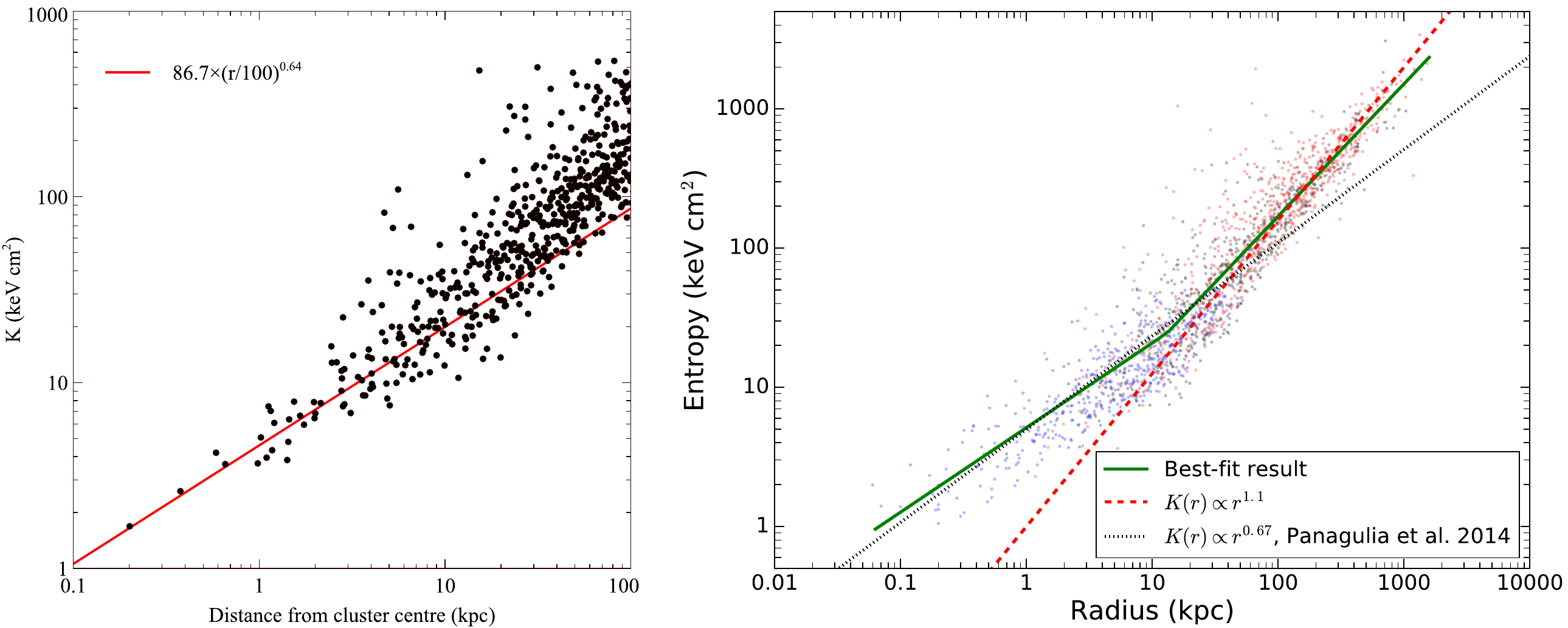}
  \caption{Entropy profiles at small radii in galaxy clusters and groups.  The left panel shows a compilation of entropy-profile measurements from \citet{Panagoulia_2014MNRAS.438.2341P}.  The right panel shows a larger compilation from \citet{Babyk2018ApJ...862...39B}.  At small radii, the entropy measurements appear to converge toward the ``universal" entropy profile expressed in equation (\ref{eq:Kinner}).  However, many clusters have central entropy levels that are considerably greater (see Figure \ref{sec:Kobserved}).  This inner entropy profile should therefore be interpreted as a universal \textit{lower limit} on the ambient entropy of a galactic atmosphere.
  \label{figure:UniversalEntropy}}
\end{figure}

More recent analyses, focusing on high quality data that enable full deprojection of the central temperature profile, indicate that the deviations of cluster atmospheres from self-similarity inside of $\sim 100$~kpc are sometimes better fit with a broken power law than by adding a constant to a single power law  \citep{Panagoulia_2014MNRAS.438.2341P, Hogan_2017_tctff,Babyk2018ApJ...862...39B} (see Figure \ref{figure:UniversalEntropy}). Instead of a universal entropy floor at a well-defined minimum value of $K_0$ \citep{Kaiser_1991ApJ...383..104K,EvrardHenry_1991ApJ...383...95E,pcn99}, there appears to be a ``universal" inner power-law profile \citep{Panagoulia_2014MNRAS.438.2341P,Babyk2018ApJ...862...39B}:
\begin{equation}
  K_{\rm inner}(r) \approx 19 \, {\rm keV \, cm^2}           
            \left( \frac {r} {10 \, {\rm kpc}} \right)^{2/3}
            \label{eq:Kinner}
\end{equation}
about which this article will have more to say in \S \ref{sec:tctff_profiles}.


\section{Radiative Cooling \label{sec:Cooling}}

Galaxy formation starts with gravitational assembly of a cosmological halo.  It can continue as long as radiative cooling allows the halo's baryonic gas to shed the entropy it has gained during the process of assembly.  That needs to happen on a timescale shorter than the age of the universe, and observations show that the atmospheres of many massive galaxies do indeed have $t_{\rm cool} \sim 1$~Gyr at $r \sim 10$~kpc, as Lyman Spitzer postulated for the Milky Way in 1956 (see Figure \ref{figure:Consistency}).

This section considers what would happen if radiative cooling operated without producing a feedback response.  If cooling had been unopposed, then there would be a continuous flow of cooling gas, called a \textit{cooling flow}, into the massive galaxy at the center of every large cosmological halo.  Observations show that reality is more complex, because such a cooling flow triggers a disruptive feedback response that limits further cooling.  However, considering the case of pure cooling helps to illustrate why halos of mass $\lesssim 10^{13} \, M_\odot$ tend to be dominated by individual central galaxies, while those of mass $\gtrsim 10^{13} \, M_\odot$ tend to contain groups and clusters with multiple large galaxies.




\subsection{Cooling Flows \label{sec:CoolingFlows}}


Shortly after the atmospheres of galaxy clusters were first characterized with spatially resolved X-ray spectroscopy in the mid-1970s \citep{1976MNRAS.175P..29M,1977ApJ...211L..63S,Mushotzky_1984PhST....7..157M}, several groups identified a puzzling issue that later became known as the ``cooling flow problem" \citep{1973ApJ...184L.105L,Gull1975MNRAS.173..585G,cb77,fn77,mb78}. Analyses of the X-ray observations showed that $t_{\rm cool} \ll 10 \, {\rm Gyr}$ in the atmospheres of many BCGs out to distances tens of kiloparsecs from the center. During the next couple of decades, the leading interpretation of this finding was that the central gas in those galaxy clusters must be continually flowing inward at rates $\gtrsim 10^{2} \, M_\odot \, {\rm yr}^{-1}$
\cite{1976ApJ...203..569L,fn77,Thomas_1987MNRAS.228..973T,1984Natur.310..733F,Nulsen_1986MNRAS.221..377N,Bertschinger_1989ApJ...340..666B}. 
In such a ``cooling flow", gravitational compression continually replaces the thermal energy lost to radiation, keeping the ambient temperature near the gravitational temperature until the flow speed approaches the local circular velocity. Then the flow becomes unstable to inhomogeneous condensation.
However, this interpretation became increasingly problematic as searches for condensed gas at the center of the flow continued to come up empty handed.  Optical and radio observations seeking the ultimate mass sink failed to find either $\gtrsim 10^{2} \, M_\odot \, {\rm yr}^{-1}$ of star formation or the amounts of cold molecular gas ($\gtrsim 10^{12} \, M_\odot$) that would have accumulated over $\sim 10^{10} \, {\rm yr}$ \citep[e.g., ][]{om89,mo89,1990ApJ...355..401G,1994A&A...283..407B,1994A&A...281..673M}.  

\subsubsection{Steady Cooling Flow Models}

A cooling flow's characteristics can be expressed quite simply in terms of specific entropy.  Multiplying equation (\ref{eq:tcool_K}) for an atmosphere with no heat input by $r$ and then dividing it by $v_r = dr/dt$ leads to  
\begin{equation}
    \alpha_K
         = \frac {d \ln K} {d \ln r} 
         = \frac {r} {v_r} \frac {d \ln K} {dt}
         = \frac {t_{\rm flow}} {t_{\rm cool}}
\end{equation}
where $t_{\rm flow} \equiv | r / v_r |$, with $v_r < 0$. The mass inflow rate of a pure cooling flow is therefore
\begin{equation}
    \dot{M}_{\rm cool}  
         \: = \: \frac {4 \pi r^3 \rho} {\alpha_K t_{\rm cool}}
         \: = \:\frac {8 \pi \mu m_p} {3 \alpha_K} 
                    \left( \frac {n_i} {n_e} \right)
                    \left[ (kT)^2 \Lambda \right] 
                    \left( \frac {r} {K} \right)^3
          \; \; .
          \label{eq:Mdot_cool}
\end{equation}
Equation (\ref{eq:Mdot_cool}) shows that the inflow rate depends on only two things, the atmosphere's gas temperature (which is $\sim T_\phi$) and its entropy profile $K(r)$. 

A steady subsonic cooling flow (with constant $\dot{M}_{\rm cool}$) in an isothermal potential well (with constant $T_\phi$) has a nearly constant temperature. Gas temperature remains constant in such a cooling flow because the $P dV$ work going into compression of inflowing gas fully compensates for the thermal energy lost to radiation.  The flow is ``cooling" despite the lack of temperature change because it is continually losing entropy.  Its temperature is given by equation (\ref{eq:HSE_temp}), because a subsonic flow is close to hydrostatic equilibrium.  The resulting steady flow in this idealized case must therefore have an entropy profile $K \propto r$, because all factors in equation (\ref{eq:Mdot_cool}) other than the $(r/K)^3$ factor are constant.  As a result, the cooling flow's gas temperature is $T \approx (4/3) T_\phi$, and its gas density profile is $\rho \propto r^{-3/2}$ \citep{1984Natur.310..733F,Bertschinger_1989ApJ...340..666B,Voit_2011ApJ...740...28V}. 

The cores of galaxy clusters with short central cooling times are generally not isothermal, because $T_\phi (r)$ rises with radius from $\sim 10$~kpc to $\gtrsim 100$~kpc.  Equation (\ref{eq:HSE_temp}) therefore calls for a corresponding radial rise in gas temperature.\footnote{The positive temperature gradients observed in galaxy clusters with short central cooling times are often cited as evidence for cooling of the central gas, but that is a misconception.  A static, thermally balanced galaxy-cluster atmosphere with the same entropy gradient as a subsonic cooling flow would have the same temperature profile.  The primary reason for the positive temperature gradients observed in galaxy cluster cores is the increase in $T_\phi$ from $\lesssim 10^7$~K near the BCG to several times $10^7$~K at $r > 100$~kpc.}  Observations show that the resulting rise in atmospheric gas temperature over that range in radius is modest, rarely exceeding $T \propto r^{0.3}$ \citep{VoigtFabian_2004MNRAS.347.1130V,Donahue+06}. The factor in square brackets in equation (\ref{eq:Mdot_cool}) is therefore a rising function of radius.  For $\dot{M}_{\rm cool}$ to be constant, the entropy profile of a steady subsonic cooling flow must be $K \propto r (T^2 \Lambda)^{1/3}$, giving a density profile $\rho \propto r^{-3/2} (T/\Lambda)^{1/2}$.\footnote{If the cooling is primarily through bremsstrahlung radiation ($\Lambda \propto T^{1/2}$), then these relations reduce to $K \propto r \, T^{5/6}(r)$ and $\rho \propto r^{-3/2} \, T^{1/4}(r)$.}  The characteristic entropy-profile slope of a pure cooling flow ($1 \lesssim \alpha_K \lesssim 1.3$) is therefore nearly indistinguishable from the $\alpha_K \approx 1.1$ slope of the cosmological baseline profile, even though its physical origin is completely different \citep{Voit_2011ApJ...740...28V}.  

To obtain the natural cooling-flow rate of a cosmological atmosphere, one can substitute a cluster's cosmological baseline profile from equation (\ref{eq:Kbase}) into equation (\ref{eq:Mdot_cool}) while setting $\alpha_K = 1.1$ and $T = 1.2 \, T_\phi$.  Doing so yields
\begin{equation}
    \dot{M}_{\rm cool} \approx 400 \, M_\odot \, {\rm yr}^{-1} 
                \left( \frac {kT_\phi} {\rm keV} \right)^{1/2} H(z) \, \Lambda_{-23} 
                \left( \frac {r} {r_{200 {\rm c}}} \right)^{0.1}
        \; \; ,
        \label{eq:Mdot_cool_base}
\end{equation}
where $\Lambda_{-23} \equiv \Lambda(T) / (10^{-23} \, {\rm erg \, cm^3 \, s^{-1}})$. In other words, the natural cooling-flow rate in a galaxy cluster would be several hundred solar masses per year, if there were not enough energetic feedback to compensate for radiative cooling.  However, the entropy profiles observed near the centers of almost all galaxy clusters are inconsistent with steady-state cooling.\footnote{The Phoenix Cluster may be an interesting exception \cite{McDonald_Phoenix_2019ApJ...885...63M}.}  Substituting the observed ``universal" inner entropy profile from equation (\ref{eq:Kinner}) into equation (\ref{eq:Mdot_cool}) gives the much more modest cooling-flow rate
\begin{equation}
    \dot{M}_{\rm cool} 
            \approx 13 \, M_\odot \, {\rm yr}^{-1} 
                \left( \frac {kT_\phi} {\rm keV} \right)^2 \Lambda_{-23}
                \left( \frac {r} {10 \, {\rm kpc}} \right)
            \; \; .
        \label{eq:Mdot_cool_inner}
\end{equation}
Furthermore, many galaxy clusters have inner entropy levels exceeding the one used to obtain equation (\ref{eq:Mdot_cool_inner}), implying even smaller cooling-flow rates at small radii.  Notice also that the dependence of $\dot{M}$ on $r$ in equation (\ref{eq:Mdot_cool_inner}), which arises because $K(r) \propto r^{2/3}$ at small radii, is inconsistent with a steady-state cooling flow.

\subsubsection{Problems with Steady Cooling Flows}

The apparent linear dependence of $\dot{M}_{\rm cool}$ on radius represented in equation (\ref{eq:Mdot_cool_inner}) was one of the first problems with the steady cooling-flow hypothesis to be recognized.  It was identified in analyses of \textit{Einstein} and ROSAT observations \citep[e.g., ][]{Thomas_1987MNRAS.228..973T,1988ApJ...335..688W,1994MNRAS.269..409A,1994MNRAS.269..589W,1995ApJ...438L..71P}, falsifying the simplest steady cooling flow models and spurring development of revised cooling-flow models in which the gas was inhomogeneous \cite{Nulsen_1986MNRAS.221..377N,Thomas_1987MNRAS.228..973T,Fabian1994}.  In an inhomogeneous cooling flow, it was thought, inflowing gas blobs of greater density would condense out of the flow at larger radii, resulting in a declining mass-flow rate as the flow moved inward \cite{WhiteSarazin_analytical_1987ApJ...318..612W,1988ApJ...331..102S,1989ApJ...345...22S}. 
However, this idea did not survive closer theoretical scrutiny, for reasons that \S \ref{sec:BuoyancyDamping} will discuss in detail.

A more serious problem arose as optical, infrared, and radio observations of BCGs stubbornly refused to yield any evidence for massive cooling flows in the form of commensurate star formation rates \cite{jfn87,1989AJ.....98.2018M} or huge stockpiles of cold gas \citep{1990ApJ...360...20M,1995AJ....109...26O,1990ApJ...355..401G,1994A&A...281..673M,1994ApJ...422..467O,1994A&A...281..673M}. As observational technology and sensitivity improved toward the end of the 1990s and into the 21st century, molecular gas and star formation were found to be closely correlated with the magnitude of those supposed cooling flows. But the observed amounts of cold gas and rates of star formation turned out to be vastly smaller than those expected from pure cooling flow models \citep{1994ASSL..190..169E,1997MNRAS.284L...1J,1998ApJ...494L.155F,2000MNRAS.318.1232W_VibH2,2000A&A...354..439K,2001MNRAS.324..443J,2003ApJ...594L..13E,2002MNRAS.337...49E,2003ApJ...594L..13E,2003AA...412..657S_CO,2006A&A...454..437S,2006ApJ...647..922E}.

Decisive falsification of the steady cooling-flow hypothesis came from spectroscopic X-ray observations with the XMM-\textit{Newton} Reflection Grating Spectrometer \citep{pet_etal_01,pet_etal_03}. Those spectra showed that several hundred solar masses of gas per year could not be condensing out of the hot gas in cluster cores.  Gas that goes from $\gtrsim 10^7$~K to $\ll 10^7$~K radiates a predictable series of emission lines that would allow a direct measurement of the steady cooling rate, if they were present.  The lack of strong O~VII and Fe~XVII emission lines, representative of gas dropping below $\sim 10^7$~K, convincingly ruled out simple steady cooling-flow models for galaxy clusters.

\subsection{Coolability of Halo Gas \label{sec:Coolability}}

While steady cooling flows do not happen in present-day galaxy clusters, extending some fundamental cooling-flow ideas to galactic halos of all sizes has provided useful insights into galaxy formation.   For example, \citet{ReesOstriker1977MNRAS.179..541R, Binney1977ApJ...215..483B,Silk1977ApJ...211..638S}, and \citet{wr78},  identified a natural maximum mass scale for formation of individual galaxies by considering the conditions that allow gas within a cosmological halo to have a cooling time significantly shorter than the age of the universe.  Those considerations show that most of the gas associated with a halo of mass $\gtrsim 10^{13} \, M_\odot$ cannot cool fast enough to form galaxies.  Consequently, the upper limit on an individual galaxy's stellar mass is $\lesssim 10^{12} \, M_\odot$, once the universe's baryonic mass fraction ($f_{\rm b} \approx 0.16$) is taken into account.  Here we reframe the classic arguments leading to that conclusion in terms of specific entropy to illuminate the deep links between CGM entropy and galaxy evolution.  

\subsubsection{Quantifying Coolability}

The simplest measure of a galactic atmosphere's ability to cool on a cosmological timescale is the dimensionless quantity
\begin{equation}
    C \equiv \left[ H(z) \, t_{\rm cool} \right]^{-1}
\end{equation}
which we will call \textit{coolability}.  Gas with $C \gg 1$ is easily able to cool in time to fuel star formation in the halo's central galaxy, while gas with $C \ll 1$ remains in the CGM because it cools so slowly.  The coolability of CGM gas is set by the specific entropy ($\sim K_{200 {\rm c}}$) it acquires as it passes through an accretion shock and by the temperature ($\sim T_{200 {\rm c}}$) the atmosphere approaches as it relaxes toward hydrostatic equilibrium.  Figure~\ref{Figure:P-Kplane} shows that CGM coolability depends much more strongly on entropy than on temperature, because $t_{\rm cool}$ at constant $K$ changes by less than a factor of $\sim 2$ in the range $10^{5.5} \, {\rm K} \lesssim T \lesssim 10^{7.5} \, {\rm K}$.

\begin{figure}[!t]
  \centering
  \includegraphics[
  width=5.3in]{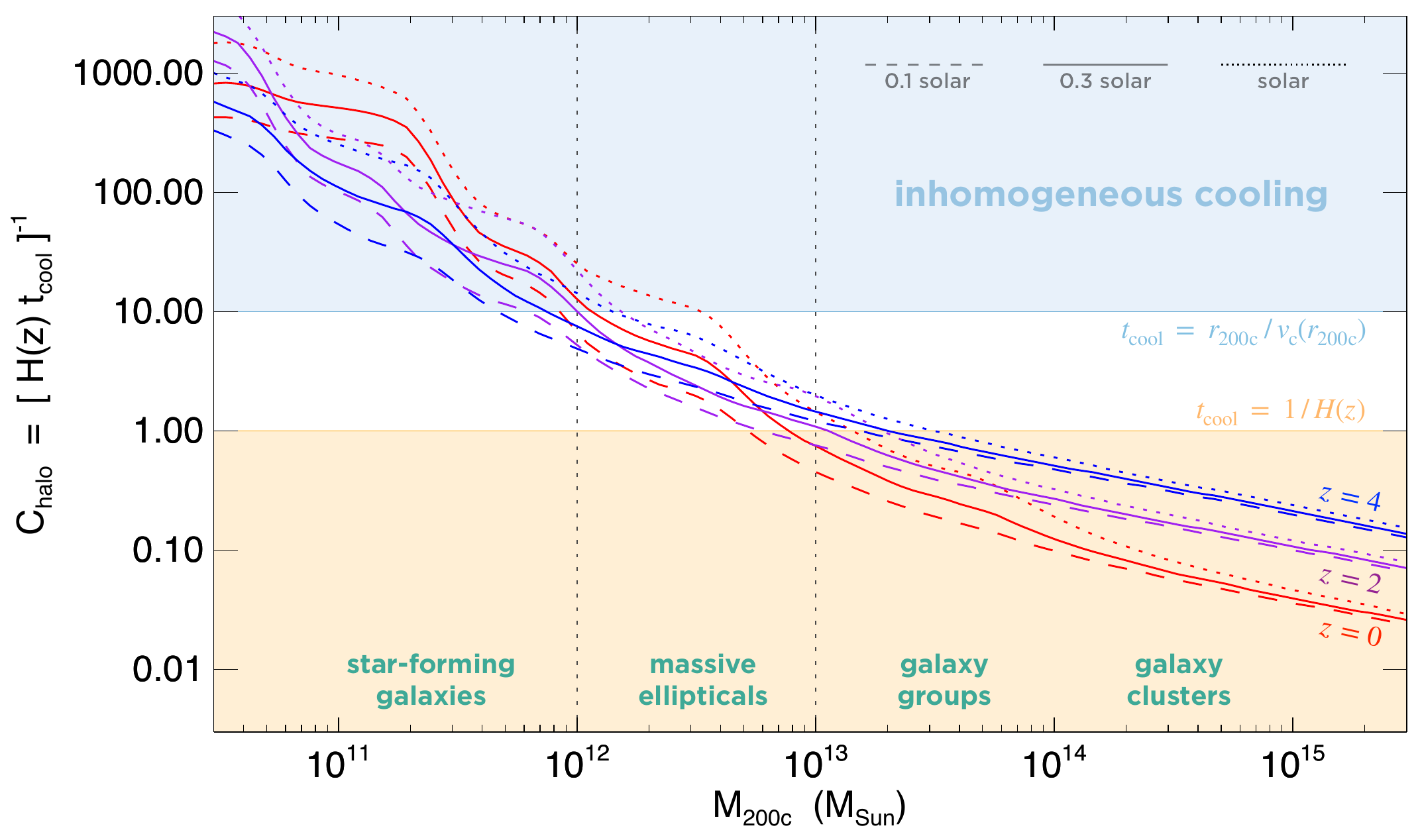}
  \caption{Coolability of cosmological galactic atmospheres as a function of halo mass and redshift.  Lines trending from upper left to lower right show the quantity $C_{\rm halo} \equiv [ H(z) \, t_{\rm cool} (M_{200 {\rm c}},z) ]^{-1}$ calculated according to equation (\ref{eq:coolability}) for CGM gas with heavy-element abundances relative to hydrogen of 0.1 (dashed lines), 0.3 (solid lines), and 1.0 (dotted lines) times their solar proportions. Line colors indicate redshifts of $z =0$ (red), $z = 2$ (purple), and $z = 4$ (blue).  The typical cooling time of cosmologically shocked CGM gas in the orange shaded region is long compared to the age of the universe at redshift $z$.  Cooling of newly accreted CGM gas therefore does not supply many star-forming clouds to central galaxies in halos of mass $\gtrsim 10^{13} \, M_\odot$.  At the other end of the halo-mass scale, in the blue shaded region, cosmologically shocked CGM gas has a typical cooling time less than the time to fall a distance $\sim r_{200 {\rm c}}$ at a speed $\sim v_{\rm c}$.  CGM cooling in halos of mass $\lesssim 10^{12} \, M_\odot$ is therefore expected to be inhomogeneous and able to continually supply the central galaxy with star-forming gas clouds.
  \label{fig:Coolability}}
\end{figure}

Diffuse gas therefore enters a cosmological halo with a characteristic coolability that we will represent as
\begin{equation}
    C_{\rm halo}(M_{200 {\rm c}},z) \: = \:
        \left( \frac {2 n_i} {n} \right)
        \frac { (k T_{200 {\rm c}})^{1/2} \Lambda (T_{200 {\rm c}}) }
              { 3 H(z) K_{200 {\rm c}}^{3/2} }
              \; \; ,
    \label{eq:coolability}
\end{equation}
based on setting its cooling time equal to $t_{\rm cool} (K_{200 {\rm c}},T_{200 {\rm c}})$. Figure~\ref{fig:Coolability} shows the dependence of $C_{\rm halo}$ on halo mass and CGM abundance for $z=0$, 2, and 4.  The dependence of $C_{\rm halo}$ on $z$ at fixed halo mass is weak in the $10^{12-13} \, M_\odot$ mass range because $K_{200 {\rm c}} \propto [ M_{200 {\rm c}} / H(z) ]^{2/3}$ and also because $T^{1/2} \Lambda(T)$ is approximately constant at the relevant gas temperatures. Consequently, the halo mass above which accretion shocks produce enough entropy to make $t_{\rm cool}$ greater than the current age of the universe remains near $10^{13} \, M_\odot$ during most of cosmic time.  In halos of greater mass, newly accreting CGM gas (in the orange shaded region) does not contribute to star formation in the central galaxy (or any other galaxy in the halo). As a result, the universe's biggest galaxies are unlikely to make more than $\sim 10^{12} \, M_\odot$ of stars, given a cosmic baryon fraction of $f_{\rm b} \approx 0.16$.

At the other end of the halo mass scale, where $C_{\rm halo} > 10$, cooling of the CGM is rapid and generally inhomogeneous.  The ambient gas in those halos can radiate away the entropy imparted by cosmological accretion on a timescale $t_{\rm cool} < 0.1 \, H^{-1}(z)$, which is shorter than the freefall time from the virial radius ($\sim r_{200 {\rm c}}$).  Under those conditions, thermal instability will amplify any density contrasts present in the CGM, thereby causing the medium to develop complex multiphase structure (see \S \ref{sec:ThermalInstability} and \S \ref{sec:BuoyancyDamping}).  Clouds that are cooler and denser than the rest of the CGM then sink into the central galaxy, supplying the fuel needed to sustain star formation.

Figure \ref{fig:Coolability} captures the essence of many more elaborate assessments of halo gas cooling.  For example, \citet{BirnboimDekel_2003MNRAS.345..349B} have shown that radiative cooling without energetic feedback prevents accretion shocks from persisting near the borders of low-mass halos.  Without feedback to compensate for radiative cooling, the halo's ambient atmosphere collapses inward as it sheds entropy, allowing infalling gas to plummet far inside the virial radius before experiencing an accretion shock.  Even when feedback is added, radiative cooling is still able to shed the entropy generated through accretion on timescales shorter than a freefall time in halos of mass $\lesssim 10^{12} \, M_\odot$ at essentially all redshifts relevant to galaxy evolution \citep{DekelBirnboim2006MNRAS.368....2D}.  Cold, narrow streams of gas therefore carry much of the accreting gas through the virial radius and toward the central galaxy in cosmological simulations of those halos \citep{Keres_2005MNRAS.363....2K,Dekel_2009Natur.457..451D,FaucherGiguere_2011MNRAS.417.2982F}.

Outside of the blue shaded region in Figure~\ref{fig:Coolability}, in halos of mass exceeding $\sim 10^{12} \, M_\odot$, one might expect the central galaxies to be transitioning away from rapid star formation and into quiescence as inhomogeneous cooling subsides.  Analyses of large galaxy surveys show that the properties of central galaxies in halos with a total mass $\sim 10^{12-13}\, M_\odot$ do indeed align with that expectation \citep[e.g., ][]{Behroozi+2013ApJ...770...57B}.  However, the calculation shown in Figure~\ref{fig:Coolability} is too simplistic to support that conclusion because it does not account for lower entropy gas that accreted into those halos earlier in time and has since migrated toward the center.  That central gas now has a cooling time substantially less than the current age of the universe and seems like it should be capable of cooling and flowing into the central galaxy. Yet it does not.  Instead, an energetic feedback response from the central galaxy limits net cooling and formation of cold clouds. Sections \ref{sec:Heating}, \ref{sec:Weather}, and \ref{sec:Balance} will discuss how that feedback process might operate.

\subsubsection{Individual Galaxies vs.~Galaxy Clusters}

A more robust implication of the calculation summarized in Figure~\ref{fig:Coolability} is that $\sim 10^{13} \, M_\odot$ should be the mass scale at which halos go from being dominated by individual galaxies to containing groups and clusters of large galaxies.  Part of the reason for that transition is cosmological entropy generation, which prevents a large proportion of the ambient gas in a cluster-scale halo of mass $\sim 10^{14-15} \, M_\odot$ from cooling and fueling star formation in the central galaxy.  Another part of the reason is the large orbital speeds of galaxies in massive halos, which inhibit galaxy-galaxy mergers. 

In order for two galaxies to merge and form a single larger galaxy, the interactions between them need to dissipate some of their original orbital energy.  Typically, the dissipation happens through tidal forces that reduce the orbital energy of the galaxies, relative to one another, while increasing the orbital energy of the stars within each galaxy.  That process can be effective as long as the relative orbital speed of the two-galaxy system is comparable to the stellar orbital speeds within those galaxies.  However, once the relative speeds of galaxies become greater than their internal orbital speeds, tidal dissipation becomes much less efficient.  The central galaxies of some galaxy clusters appear to have ``cannibalized" some of their satellites \citep[e.g.,][]{Massey_A3827_2015MNRAS.449.3393M}, but most of the galaxies in a cluster remain in orbit around it and do not end up merging into a single huge galaxy at the halo's center.


\subsection{The Overcooling Problem \label{sec:Overcooling}}

We have seen that Figure~\ref{fig:Coolability} implies that a large proportion of the baryons associated with halos of mass $\lesssim 10^{13} \, M_\odot$ should be able to cool and form stars.  Sophisticated numerical simulations of cosmological galaxy formation arrive at the same basic conclusion \citep[e.g.,][]{Katz_1992ApJ...391..502K,Benson+2003ApJ...599...38B,Benson2010PhR...495...33B,SomervilleDave_2015ARA&A..53...51S,NaabOstriker2017ARA&A..55...59N}.  Without feedback capable of limiting radiative cooling, the circumgalactic gas in those halos would ``overcool" and produce many more stars than are observed in today's galaxies.  This discrepancy has been called the \textit{overcooling problem}.

\subsubsection{Stellar Baryon Fraction}

Analyses of large galaxy surveys \citep{Moster_2010ApJ...710..903M,Behroozi+2010ApJ...717..379B} have shown that the fraction of a halo's baryons in the form of stars ($f_*$) is limited to less than 40\% of the theoretical maximum. The observed stellar mass fraction is a strong function of halo mass
\cite{2006MNRAS.368..715M,2009ApJ...696..620C,2012ApJ...744..159L} but is nearly independent of redshift for $z<4$ \cite{Behroozi+2013ApJ...762L..31B}.  In contrast to what Figure~\ref{fig:Coolability} would seem to imply, $f_*$ rises with $M_{\rm halo}$ up to $M_{\rm halo} \sim 10^{12} \, M_\odot$. indicating that star formation enabled by cooling of a cosmological atmosphere must be less efficient in halos of mass $\ll 10^{12} \, M_\odot$ than in halos of greater mass. Therefore, observed galaxy populations cannot be reconciled with the $\Lambda$CDM cosmological model without an explanation for the strong dependence of $f_*$ on $M_{\rm halo}$.\footnote{Alternatively, the $\Lambda$CDM model might not be correct \cite{McGaugh+2010ApJ...708L..14M,McGaughPhysRevLett.106.121303}.}  

One method for arriving at that result is known as \textit{abundance matching} (recently reviewed by Wechsler and Tinker \citep{WechslerTinker2018ARA&A..56..435W}).  Large galaxy surveys provide the mean number density of galaxies as a function of both galactic stellar mass ($M_*$) and cosmological redshift $z$.  Numerical simulations of cosmological structure formation provide the average number density of halos of mass $M_{200 {\rm c}}$ predicted at each redshift by the $\Lambda$CDM model.  One can then map each stellar mass $M_*$ onto a halo mass $M_{200 {\rm c}}$ corresponding to the same mean number density in a large cosmological volume.  Dividing each value of $M_*$ by the corresponding value of $f_{\rm b} M_{200 {\rm c}}$ gives estimates for both $f_*(M_{200 {\rm c}})$ and $f_*(M_*)$.\footnote{Note that the numerical value of $f_*$ depends upon the definition of $M_{\rm halo}$.  For example, $f_*$ defined with respect to $M_{500c}$ can be $\sim 50$\% greater than $f_*$ defined with respect to $M_{200c}$.}

\begin{figure}[!t]
  \centering
  \includegraphics[width=5.3in,
  trim=0.15in 0.0in 0.0in 0.0in]{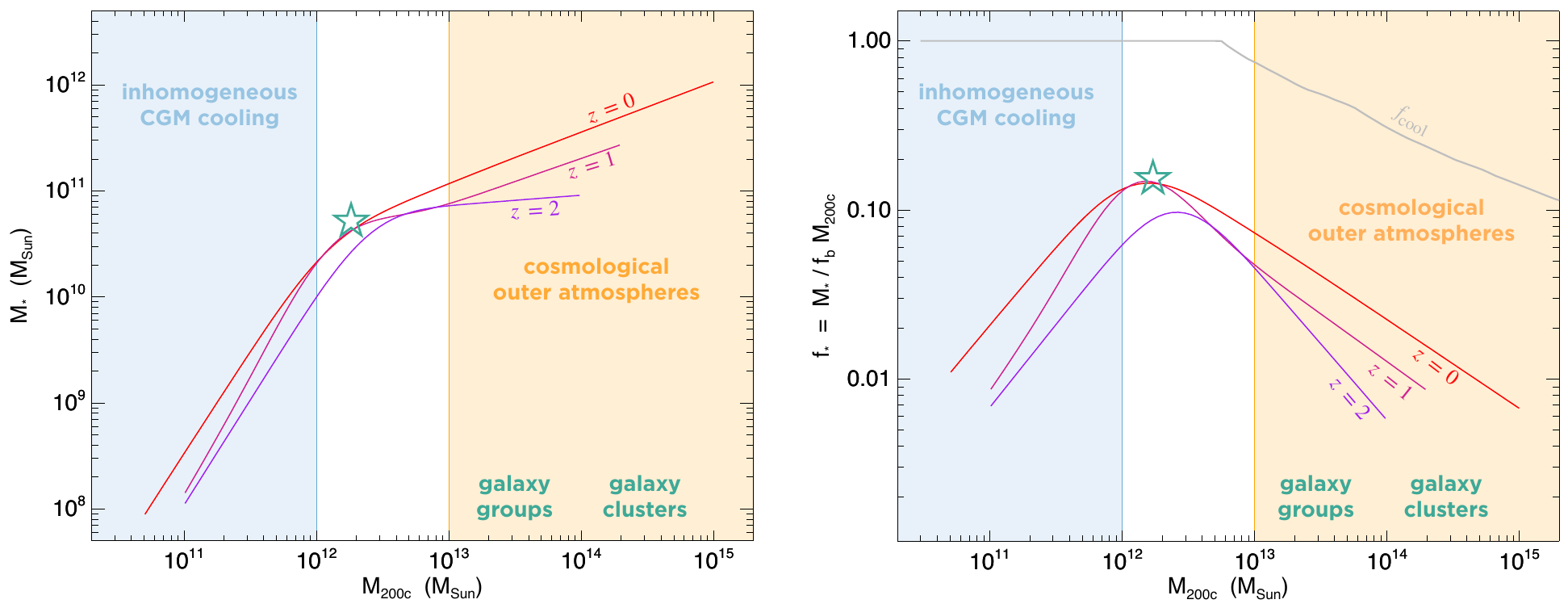}
  \caption{Stellar mass ($M_*$) of a halo's central galaxy as a function of halo mass ($M_{200 {\rm c}}$).  The left panel (based on \cite{Behroozi2019MNRAS.488.3143B}) shows the result of mapping $M_*$ onto $M_{200 {\rm c}}$ through abundance matching.  The right panel shows the stellar baryon fraction $f_*$ obtained from dividing $M_*$ by the baryonic mass associated with the halo.  Lines show the relationships at $z = 0$ (red), $z = 1$ (violet red), and $z = 2$ (purple).  A green star indicates the Milky Way's current properties.  Throughout most of cosmic time, the stellar baryon fraction peaks near a halo mass $\sim 10^{12} \, M_\odot$.  Star formation in lower mass halos is much less efficient, even though the cooling time of their cosmologically shocked CGM gas is short enough for inhomogeneous cooling.  Above a halo mass $\sim 10^{13} \, M_\odot$, accretion shocks boost the cooling time of recently accreted gas above $H^{-1}(z)$, and so the properties of the outer atmospheres in those halos depend more directly on cosmology than on radiative cooling. A gray line labeled $f_{\rm cool}$ in the right panel shows the fraction of cosmologically shocked gas with a cooling time less than the age of the universe, estimated from equation (\ref{eq:CoolingRadius}). 
  \label{fig:SF_Efficiency} }
\end{figure}

Figure~\ref{fig:SF_Efficiency} shows the relationships between $M_*$, $f_*$, and $M_{200 {\rm c}}$ that emerge from abundance matching.  Galactic star formation appears to be most efficient in halos of mass $\sim 10^{12} \, M_\odot$, which contain galaxies with a stellar mass similar to the Milky Way's.  However, the stellar baryon fraction in those halos appears to be no greater than $f_* \approx 0.2$ and declines significantly toward both smaller halo masses and greater halo masses.  

The observed characteristics of individual dwarf galaxies do seem to indicate that they reside in halos that have managed to turn only a small fraction of the available gas into stars \citep[]{DekelSilk1986ApJ...303...39D}.   Based on abundance-matching analyses, star formation in low-mass dwarf galaxies is far less efficient than in Milky-Way sized halos, in the sense that $f_* \ll 1$ \cite{Behroozi+2013ApJ...762L..31B}. The most commonly cited explanation for this inefficiency is that supernova explosions during early episodes of star formation release enough energy to drive a galactic wind capable of ejecting most of the halo's gas \citep[e.g.][]{Larson_1974MNRAS.169..229L,ChevalierClegg1985Natur.317...44C,DekelSilk1986ApJ...303...39D}.  Section~\ref{sec:Heating} will look more closely at the plausibility of gas ejection from a galaxy's halo by supernova explosions.

\subsubsection{Cooling Radius}

In halos of greater mass, star formation becomes less efficient than it is at $M_{200 {\rm c}} \sim 10^{12} \, M_\odot$.  That happens largely because of the greater mean entropy and longer cooling time of the CGM in those halos, which limits the total amount of gas that can cool.  However, that feature alone does not account for the declining inefficiency of star formation as the stellar mass of a galaxy approaches $\sim 10^{11} \, M_\odot$. The amount of cosmologically-accreted gas capable of cooling in a halo of mass $\sim 10^{13} \, M_\odot$ is closer to $\sim 10^{12} \, M_\odot$, and yet most of this available gas does not end up in the central galaxy.

An estimate for the radius within which significant radiative cooling of CGM gas is possible can be derived from the cosmological baseline entropy profile.  The estimate applies the approximate relation $t_{\rm cool} \propto K^{3/2}$ for gas at typical CGM temperatures, which gives
\begin{equation}
    t_{\rm cool} (r) 
        \: \approx \: \frac {1.5} {H(z) C_{\rm halo}} 
                \left( \frac {r} {r_{200 {\rm c}}} \right)^{1.65}
\end{equation}
when combined with equation (\ref{eq:Kbase}).  Setting $t_{\rm cool} = H^{-1} (z)$ then defines a \textit{cooling radius} 
\begin{equation}
    r_{\rm cool} \approx 0.74 \, C_{\rm halo}^{0.6} \, r_{200 {\rm c}}
    \label{eq:CoolingRadius}
\end{equation}
within which the cooling time of a cosmological atmosphere is less than the current age of the universe.  For a galaxy cluster at $z \approx 0$, which has a halo coolability $C_{\rm halo} \approx 0.03$, the cosmological cooling radius is $r_{\rm cool} \approx 0.1 \, r_{200 {\rm c}}$.  And for a galaxy group at $z \approx 0$, with a halo coolability $C_{\rm halo} \approx 0.2$, one finds $r_{\rm cool} \approx 0.3 \, r_{200 {\rm c}}$.

This particular definition of $r_{\rm cool}$ is not universal.  A halo's cooling radius is often defined in the literature with respect to an estimate of the halo's age that is smaller than $H^{-1}(z)$.  However, the amount of gas within $r_{\rm cool}$ in the cosmological atmosphere of a galaxy cluster or group is a few times $\sim 10^{12} \, M_\odot$, implying a cooling rate of several hundred solar masses per year (see also equation \ref{eq:Mdot_cool_base}).

And yet, the galaxies at the centers of these massive halos typically form stars at rates one to two orders of magnitude smaller \citep[e.g., ][]{McDonald_Phoenix_2019ApJ...885...63M}, and many are forming no new stars at all.  This lack of ``overcooling" at the high end of the galaxy mass range has proven harder to explain than at the low-mass end, because supernova energy alone cannot drive CGM gas out of those much deeper potential wells (see the review by \citet{Benson2010PhR...495...33B}).  

Feedback from AGNs appears to be required, for reasons we will discuss in Section~5. However, feedback energy input into the CGM does not necessarily need to be continuous. Early energy input that raises the entropy and lowers the density of CGM gas lengthens its cooling time and moves its cooling radius inward. One consequence of that heating and expansion of the CGM is a long-term reduction of the ambient medium's natural cooling-flow rate. In some present-day galaxies, steady cooling flows might therefore supply star-forming gas to the galactic disk at a rate determined by how effectively \textit{prior} feedback episodes have heated and lifted CGM gas \citep{Qu_2018ApJ...856....5Q,Stern_2020MNRAS.492.6042S,Stern_2021ApJ...911...88S}.

\section{Heating of Galactic Atmospheres  \label{sec:Heating}}



Recognition that galactic atmospheres must be heated started with Lyman Spitzer's speculations about the Milky Way's ``corona" in 1956.  Identifying the heat source later became a central concern of galaxy evolution studies as astronomers began to grapple with the cooling-flow problem in galaxy clusters and the CGM overcooling problem in smaller halos. The most obvious energy sources are the supernova explosions that follow star formation and the AGN outbursts that happen when a galaxy's gas accretes onto its central black hole.  This section assesses how the energy released by those events compares with radiative cooling of the CGM and shows that supernovae may be able to provide enough energy to prevent overcooling in low-mass halos but cannot succeed in high-mass halos, where AGN heating appears to be both necessary and sufficient.

However, this section's assessment of energy sources leaves an important question unanswered.  Observations show that these feedback mechanisms must eventually tune themselves to the state of the CGM, at least in high-mass halos (see Figure \ref{figure:Consistency}), so that time-averaged heating approximately balances radiative cooling. Section~\ref{sec:Weather} outlines the observations showing how interplay between heating and cooling produces phenomena akin to weather in galactic atmospheres.  Section~\ref{sec:Balance} then returns to the question of balance and interprets the ``weather patterns" that bring tuning about.


\subsection{Stellar Heating}
\label{sec:SN_heating}


Two types of supernova explosions can contribute to CGM heating.  One is prompt. The other is delayed.  Within $\sim 100$~Myr of a star-formation episode, stars with birth masses exceeding $\sim 8 \, M_\odot$ exhaust their nuclear fuel and undergo core collapse, causing explosions that expel $\sim 10^{51} \, {\rm erg}$ of kinetic energy (e.g. \cite{2018MNRAS.480..800H,2018MNRAS.477.1578H}).  Core-collapse supernovae therefore provide a prompt feedback response to star formation.  Over a longer time period, some of the stellar population's white dwarfs produce supernova explosions ejecting similar amounts of kinetic energy per event \cite{2018MNRAS.477.1578H,2018ApJ...854...52S}.  The white-dwarf supernova rate declines more gradually than the core-collapse rate, over a period of many Gyr.  Both types of supernova play important roles in CGM heating. 


 \subsubsection{Prompt Supernova Feedback}
 
 The amount of energy that core-collapse supernovae release can be expressed in terms of the specific supernova energy $\epsilon_{\rm SN}$ per unit mass of the parent stellar population.  Uncertainties in both the distribution of initial stellar masses and the mean energy release per supernova event preclude a precise determination of $\epsilon_{\rm SN}$.  Standard assumptions lead to an approximate value 
 \begin{equation}
     \epsilon_{\rm SN} 
        \: \approx \: \frac {10^{51} \, {\rm erg}} {100 \, M_\odot}
        \: \approx \: (700 \, {\rm km \, s^{-1}})^2
        \: \approx \: 3 \, {\rm keV}/\mu m_p
 \end{equation}
 corresponding to a single core-collapse supernova per $100 \, M_\odot$ of star formation \citep[e.g.,][]{Maoz_2011MNRAS.412.1508M}. 
 
Supernova heat input can temporarily fend off overcooling if it matches or exceeds the total amount of atmospheric radiative cooling.  Importantly, if the feedback response produces more energy then the galaxy's atmosphere sheds through radiative cooling, then it can expand the galaxy's atmosphere, causing its mean density and pressure to decline. Atmospheric expansion requires an energy input that scales with the gravitational binding energy of the CGM, and significant expansion of the ambient CGM reduces the future amount of heat input needed to prevent CGM overcooling. Feedback regulation of star formation therefore does not require precise heating-cooling balance at all times.


\subsubsection{Mass Loaded Galactic Winds}

The most prevalent analytical framework for characterizing prompt supernova feedback in a low-mass galactic halo assumes that supernovae drive a wind that expels gas from the central galaxy at a rate
\begin{equation}
    \dot{M}_{\rm wind} = \eta_M \dot{M}_*
    \; \; .
\end{equation}
Here, $\dot{M}_*$ is the galaxy's star formation rate and $\eta_M$ is known as the \textit{mass loading factor} of the wind (see \cite{SomervilleDave_2015ARA&A..53...51S} and references therein).
Models that rely on gas ejection to prevent excessive star formation often assume that a large fraction of the halo's gas passes through the central galaxy, where some of it fuels star formation and produces supernovae.  If the resulting galactic wind can eject the bulk of the accreted gas from the halo, then not much additional energy is needed to compensate for radiative cooling of the CGM gas that remains.

However, explosive expulsion of gas that has fallen all the way into the central galaxy requires far more specific energy than gentler modes of CGM expansion. To illustrate the point, consider the NFW potential well of a cosmological dark matter halo,
\begin{equation}
    \phi_{\rm NFW} (r) \: = \: - 4.625 \, v_{\rm max}^2  
            \left( \frac {r_{\rm s}} {r} \right) \, 
            \ln \left( 1 + \frac {r} {r_{\rm s}} \right)
        \; \; ,
        \label{eq:phi_NFW}
\end{equation}
obtained from the mass density profile $\rho_M (r) \propto r^{-1} ( 1 + r / r_{\rm s} )^{-2}$  \citep{nfw97}.  This density profile is $\propto r^{-2}$ at the scale radius $r_{\rm s}$, and the halo's circular velocity reaches a maximum value equal to $v_{\rm max}$ at $r_{\rm max} \approx 2.16 \, r_{\rm s}$.  Unbinding of gas that has reached the bottom of the potential well requires imparting a specific energy $\sim 4.6 \, v_{\rm max}^2$.\footnote{Adding a comparable specific energy to gas within the potential well of a singular isothermal sphere, $\phi_{\rm SIS} (r) = v_{\rm c}^2 \ln (r / r_{200 {\rm c}})$, can achieve a similar outcome by lifting galactic gas from $\sim 0.01 r_{200 {\rm c}}$ to $\sim r_{200 {\rm c}}$, because $\ln (100) \approx 4.6$.}  In contrast, gentle expansion of the entire atmosphere by a factor of $\sim 2$ in radius requires much less specific energy\footnote{Because $\phi_{\rm SIS}(2r) - \phi_{\rm SIS}(r) \approx 0.7 v_{\rm c}^2$} but still manages to reduce the ambient CGM gas density by an order of magnitude. We will discuss the consequences of CGM lifting in \S \ref{sec:Lifting}, after considering the energetics of CGM ejection in more detail.

\begin{figure}[!t]
  \centering
  \includegraphics[width=5.3in,
  trim=0.15in 0.0in 0.0in 0.0in]{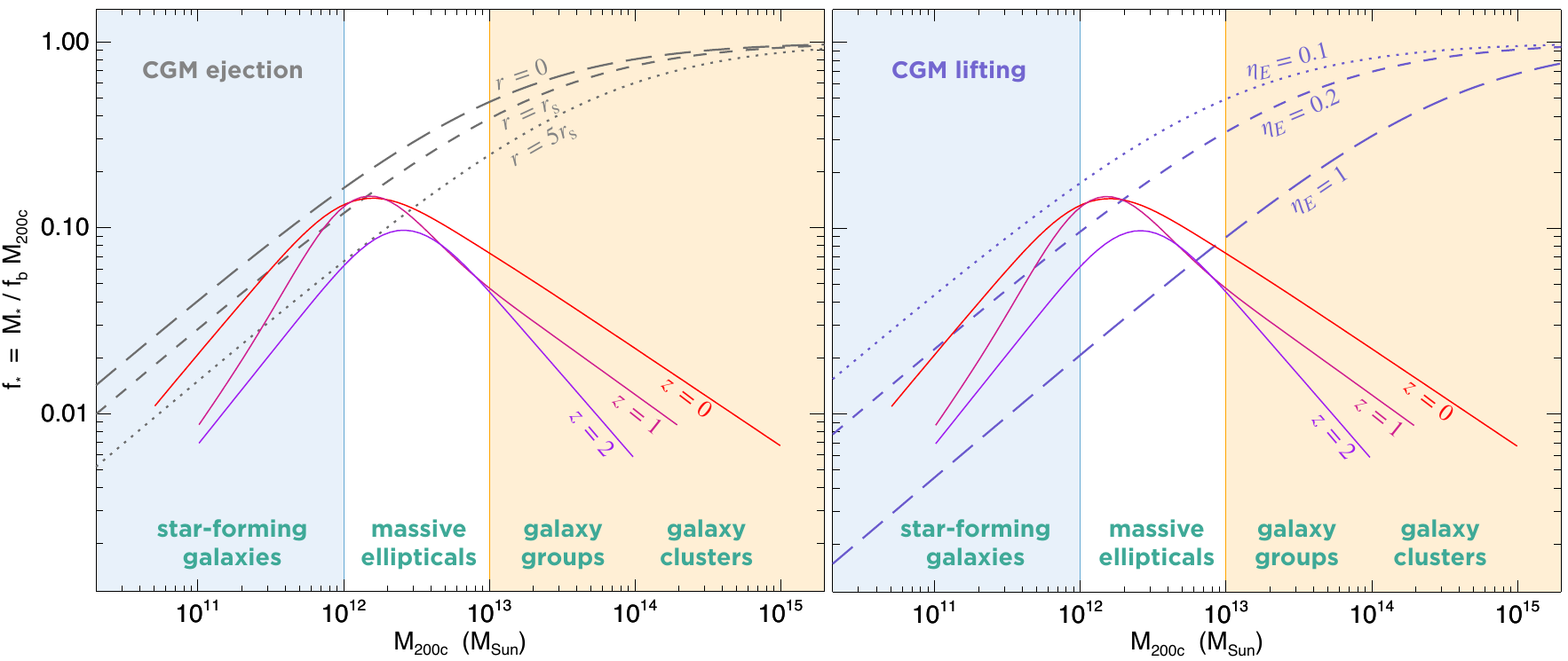}
  \caption{Lower limits on the amount of star formation needed to mitigate overcooling through prompt supernova feedback.  The left panel shows limits derived by requiring a galaxy's core-collapse supernova energy to unbind the halo's remaining baryons.  A long-dashed line illustrates the minimum amount of star formation necessary to unbind gas from the vicinity of the central galaxy ($r = 0$), the short-dashed line represents unbinding of gas from the scale radius $r_{\rm s}$, and the dotted line is for unbinding of gas from $5r_{\rm s}$.  All of the lines apply to halos at $z=0$ and assume maximal supernova feedback efficiency ($\eta_E = 1$).  In the right panel are analogous limits derived from the amount of star formation needed to lift the halo's gas by a factor $\sim 2$ in radius, assuming different efficiencies for supernova energy conversion: $\eta_E = 1$ (long-dashed line), $\eta_E = 0.2$ (short-dashed line), and $\eta_E = 0.1$ (dotted line). The solid lines in both panels are identical to the observed $f_*(M_{200 {\rm c}})$ relationships shown in Figure \ref{fig:SF_Efficiency} and demonstrate that supernova energy from the central galaxy cannot significantly alter the CGM in halos more massive than a few times $10^{12} \, M_\odot$.}
  \label{fig:EjectionLifting}
\end{figure}

An approximate lower limit on the amount of star formation needed to eject the halo's accreted gas by cycling it through mass loaded galactic winds can be obtained by equating the supernova energy input $\eta_E \epsilon_{\rm SN} f_*$ and the specific energy $4.6 \, v_{\rm max}^2 (1 - f_*)$ required to unbind leftover galactic gas, giving  
\begin{equation}
    f_* \: \gtrsim \: \frac {4.6 \, v_{\rm max}^2} 
        {\eta_E \epsilon_{\rm SN} + 4.6 \, v_{\rm max}^2}
        \; \; .
    \label{eq:fstar_limit_ejection}
\end{equation}
The parameter $\eta_E$, sometimes called the \textit{energy loading factor}, is the fraction of supernova energy that goes into driving the galactic wind.  A long-dashed gray line in the left panel of Figure~\ref{fig:EjectionLifting} illustrates the lower limit on $f_*$ expressed in equation (\ref{eq:fstar_limit_ejection}), assuming $\eta_E \approx 1$ and $z \approx 0$.\footnote{It is more stringent earlier in time because $v_{\rm max}^2 \propto M_{\rm halo}^{2/3} H^{2/3}(z)$.}  Comparing that limit with the red line indicating the observed value of $f_*$ at $z = 0$ shows that ejection of the leftover gas fraction $1-f_*$ from the vicinity of the central galaxy ($r < r_{\rm s}$) is only marginally plausible for low-mass halos with $M_{200 {\rm c}} \lesssim 10^{12} \, M_\odot$ because it requires $\eta_E \approx 1$.  The similarity in slope between the red line and the gray line among lower-mass galaxies does indeed suggest a strong link between supernova energy output and the observed relation between stellar mass and halo mass, but detailed modeling has shown that the required conversion efficiency of supernova energy to wind energy is difficult to achieve (see \S \ref{sec:RadiativeLosses}).

\subsubsection{Preventative Wind Feedback}

Ejection of a halo's accreted gas requires less supernova energy if much of that gas never manages to make it into the central galaxy.  Accreting gas may encounter substantial resistance from the ambient CGM during infall. If that gas settles into the CGM at a greater radius, it is less tightly bound to the halo and more easily expelled from it.  Phenomena that prevent accreting gas from entering a halo's central galaxy are sometimes called \textit{preventative feedback} and can be parameterized in terms of the fraction $\zeta$ of cosmologically accreting gas that reaches the central galaxy.  The gas supply into the central galaxy in that case is $\zeta f_{\rm b} \dot{M}_{200 {\rm c}}$ \citep{DaveFinlatorOppenheimer_2012MNRAS.421...98D}. Strong preventative feedback therefore corresponds to $\zeta \ll 1$.   

Dotted and short-dashed lines in the left panel of Figure~\ref{fig:EjectionLifting} show lower limits analogous to the one expressed in equation (\ref{eq:fstar_limit_ejection}) but for gas at greater altitudes in an NFW potential well.\footnote{In a typical galactic halo, the radius $5 r_{\rm s}$ corresponds approximately to $\sim 0.2$--$0.5 \, r_{200 {\rm c}}$.}  Unbinding of CGM gas from those greater altitudes is less taxing on the supernova energy supply but requires reinterpretation of the mass-loading parameter $\eta_M$.  If $\dot{M}_{\rm wind}(r)$ is an increasing function of radius outside of the central galaxy, then $\eta_M$ is no longer a single-valued wind parameter fully determined by what is happening inside of the galaxy.  Instead, it depends on how a galactic wind couples with the CGM, a topic we will return to in \S \ref{sec:Implications}.

The primary implication of the left panel in Figure~\ref{fig:EjectionLifting} is that prompt supernova feedback in high-mass halos ($M_{200 {\rm c}} \gg 10^{12} \, M_\odot$) is energetically incapable of unbinding a significant proportion of the halo's gas, independent of these nuances in modeling.  A more powerful energy source is needed to prevent that CGM gas from fueling too much star formation, regardless of how the CGM is distributed in radius. 

\subsubsection{Uncertain Radiative Losses}
\label{sec:RadiativeLosses}

Radiative losses further complicate matters.  The efficiency parameter $\eta_E$ for supernova energy transfer to the CGM is difficult to determine through theoretical modeling.  Numerical simulations of galactic winds in low-mass halos have become an enormous enterprise during the last three decades, not only because they are so important to understanding galaxy evolution, but also because the results depend heavily on numerical resolution. Many astrophysical processes come into play (see \cite{SomervilleDave_2015ARA&A..53...51S} and \cite{NaabOstriker2017ARA&A..55...59N} for reviews).  The modeling is complicated because core-collapse supernovae explode in complex multiphase environments, with intricately interwoven gaseous components spanning large ranges in density, temperature, and size scale.  That complexity provides many channels for supernova energy to dissipate within gas that has a short cooling time, resulting in radiative losses that can substantially suppress $\eta_E$.    

There is still no consensus on what value of $\eta_E$ is appropriate for modeling of galactic winds.  However, the answer clearly depends on the spatial distribution of supernova explosions.  When supernovae are clustered, they create large bubbles of hot gas within a galactic disk that then erupt into the galaxy's halo.  Those eruptions transport supernova energy into the CGM more efficiently than supernovae happening at random locations in a galactic disk, as originally shown by \cite{MacLowMcCray1988ApJ...324..776M} and more recently reaffirmed by \cite{Fielding_2018MNRAS.481.3325F, 2017ApJ...841..101L,2020ApJ...890L..30L}.
However, current modeling indicates that $\eta_E \sim 0.1$--0.2 in galaxies like the present-day Milky Way even if supernovae are clustered \cite{2020ApJ...890L..30L}, implying that supernova-driven winds are unlikely to unbind most of the gas that a $\sim 10^{12} \, M_\odot$ halo accretes.

\subsubsection{CGM Ejection or CGM Lifting?}
\label{sec:Lifting}

Given the likelihood that only a relatively modest fraction of galaxy's core-collapse supernova energy output goes into the CGM, it is worth considering what that amount of SN energy is able to do.  The right panel of Figure~\ref{fig:EjectionLifting} shows a breakdown.  Each of the blue lines in the panel illustrates a limit
\begin{equation}
    f_* \: \gtrsim \: \frac {0.5 \, v_{\rm max}^2} 
        {\eta_E \epsilon_{\rm SN} + 0.5 \, v_{\rm max}^2}
    \label{eq:fstar_limit_lifting}
\end{equation}
that comes from equating $\eta_E \epsilon_{\rm SN} f_*$ and the specific energy needed to lift a gas mass $(1 - f_*) f_{\rm b} M_{200 {\rm c}}$ by a factor of $\sim 2$ in radius.\footnote{Assuming an NFW-like  potential well}  Lifting the whole CGM by this factor reduces its density and raises its cooling time by an order of magnitude.  

Feedback that gradually lifts the CGM therefore seems more likely to succeed in halos of mass $\gtrsim 10^{12} \, M_\odot$ than feedback that explosively ejects the CGM. The short-dashed line in the right panel of Figure~\ref{fig:EjectionLifting} shows that transferring only $\sim 20$\% of the available supernova energy to the CGM can lift it by a factor of $\sim 2$ in present-day halos of mass $\lesssim 10^{12.5} \, M_\odot$. Comparing that finding with the estimates in the left panel of Figure~\ref{fig:EjectionLifting} demonstrates that slowing star formation by gently lifting the CGM through gradual heating is much less demanding on the SN energy budget than is explosive unbinding of gas that has reached the halo's central galaxy.

\subsubsection{Delayed Supernova Feedback}

After a galaxy's period of prompt supernova feedback has ceased, its aging stellar population can still provide some heat as white-dwarf supernovae (SNIa) continue to explode.  Among elliptical galaxies belonging to galaxy clusters, the observed rate for a stellar population of age $\sim 10$~Gyr is $\sim 3 \times 10^{-14} \, {\rm \, yr^{-1}} \, M_\odot^{-1} (t / 10 \, {\rm Gyr})^{-1.3}$ \cite{FriedmannMaoz_2018MNRAS.479.3563F}. Among isolated elliptical galaxies, which tend to be less massive, the observed specific SNIa rate for a stellar population of similar age is $\sim 2 \times 10^{-14} \, {\rm \, yr^{-1}} \, M_\odot^{-1} (t / 10 \, {\rm Gyr}^{-1})$ \cite{MaozGraur_2017ApJ...848...25M}. 

The total SNIa energy output, integrated over time, is only $\sim 10$\% of the energy output from core-collapse supernovae, but it can be transferred to the CGM more efficiently, because SNIa tend to explode in gaseous environments that are more diffuse and homogeneous, thereby minimizing radiative losses. This delayed version of supernova feedback does not contribute much to lifting or ejection of a massive galaxy's CGM but can help to sweep gas out of a mature galaxy and into the CGM, if other forms of feedback have already lowered the CGM's pressure \cite{Voit_2015Natur.519..203V,Voit_2020ApJ...899...70V}.  To assess the efficacy of supernova sweeping, one can compare the specific energy $\epsilon_*$ of gas shed by an old stellar population with the depth of the galaxy's gravitational potential. Multiplying the specific supernova rate by $10^{51} \, {\rm erg}$ per SN Ia and dividing by the specific stellar mass loss rate ($t_*^{-1}$) gives 
\begin{equation}
    \epsilon_* \approx \frac {2 \, {\rm keV}} {\mu m_p} 
                       \frac {t_*} {200 \, {\rm Gyr}}
\end{equation}
for an old stellar population with an age $\sim 10$~Gyr. The 200~Gyr timescale chosen for scaling $t_*$ corresponds to 0.5\% of the stellar mass per Gyr but depends in detail on the stellar initial mass function \citep[e.g.,][]{LeitnerKravtsov_2011ApJ...734...48L}.

This amount of SNIa energy is enough to sweep ejected stellar gas out of a massive galaxy, because $\epsilon_*^{1/2} \approx 560 \, {\rm km \, s^{-1}}$ exceeds the circular velocity of even the most massive galaxies.  However, it is unable to expel that gas from a massive galaxy's halo if $v_{\rm max} \gtrsim (\epsilon_* / 4.6)^{1/2} \approx 260 \, {\rm km \, s^{-1}}$.  Without another energy source for feedback, gas swept out of the central galaxy accumulates in the galaxy's CGM. Sweeping of gas out of the galaxy then becomes progressively more difficult, because the accumulating CGM must be lifted along with the gas continually shed by the aging stars.  The resulting buildup of expelled stellar gas ultimately boosts the central gas density enough for radiative cooling to exceed SNIa heating, initiating a cooling flow toward the galaxy's center, where its supermassive black hole resides \citep{MathewsLoewenstein_1986ApJ...306L...7M,LoewensteinMathews_1987ApJ...319..614L,Ciotti_1991ApJ...376..380C}.

\subsection{Black Hole Heating}
\label{sec:BH_Heating}

Accretion of gas onto a galaxy's central supermassive black hole appears to be the backstop that fuels feedback when supernovae fail to keep pace with radiative cooling of atmospheric gas.  Consider that a galaxy with $M_* \sim 10^{11} \, M_\odot$ typically has a central black hole of mass $M_{\rm BH} \sim 10^8 \, M_\odot$.  The black hole's rest-mass energy ($\sim 10^{62} \, {\rm erg}$) therefore greatly exceeds the galaxy's collective SN energy output ($\sim 10^{60} \, {\rm erg}$).  According to the theory of accretion disks, a fraction $f_{\rm acc} \sim 0.1$ of the accreting rest-mass energy escapes in the form of radiation and kinetic outflows before the majority becomes trapped in the black hole. The accretion energy available to offset radiative cooling of the galaxy's atmosphere is therefore an order of magnitude greater than the galaxy's total supernova energy output.  

\subsubsection{Energy Requirements}
\label{sec:EnergyRequirements}

Unfortunately, the fraction of the black hole's rest-mass energy that becomes thermalized in the galaxy's atmosphere remains highly uncertain.  We will therefore parameterize our ignorance about the thermalization fraction in terms of an efficiency factor $f_{\rm BH}$, defining it so that 
\begin{equation}
    \dot{E}_{\rm BH} = f_{\rm BH} \dot{M}_{\rm BH} c^2
\end{equation}
is the atmospheric heating rate resulting from a mass accretion rate $\dot{M}_{\rm BH}$ onto the central black hole. If this heating rate exceeds radiative cooling of the galaxy's atmosphere, then the excess heating expands the galaxy's atmosphere, lifts its CGM, and reduces its radiative losses.

Equating the total amount of heating and the energy required for CGM lifting ($\sim 0.5 v_{\rm max}^2 f_{\rm b} M_{\rm 200c}$) yields the relation
\begin{equation}
    \frac {M_{\rm BH}} {M_{\rm 200c}} 
      \: \sim \: \frac {0.5 v_{\rm max}^2} {c^2} \frac {f_{\rm b}} {f_{\rm BH}}
      \: \sim \:  \frac {k T_\phi} {\mu m_p c^2} \frac {f_{\rm b}} {f_{\rm BH}}
       \; \; .
       \label{eq:MBH_Mhalo}
\end{equation}
Typically, the stellar velocity dispersion of the central galaxy in a halo of mass $\lesssim 10^{13} \, M_\odot$ is related to $v_{\rm max}$ through $2 \sigma_v^2 \sim v_{\rm max}^2$, and so it is illuminating to rewrite equation (\ref{eq:MBH_Mhalo}) as
\begin{equation}
    M_{\rm BH} \sim 2 \times 10^8 \, M_\odot 
        \left( \frac {\sigma_v} {200 \, {\rm km \, s^{-1}}} \right)^5
        \left( \frac {v_{\rm max}^2} {2 \sigma_v^2} \right)^{2.5}
        \left( \frac {f_{\rm BH}} {10^{-2.5}} \right)^{-1} E(z)
        \; \; .
\end{equation}
This relationship is quite similar to the observed low-redshift $M_{\rm BH}$--$\sigma_v$ relation \citep{KormendyHo2013ARAA..51..511K}, given $f_{\rm BH} \sim 10^{-2.5}$ and $v_{\rm max}^2 \sim 2 \sigma_v^2$.  Gentle lifting of a massive galaxy's CGM through AGN feedback therefore requires $f_{\rm BH} \sim 10^{-2.5}$. A much greater value of $f_{\rm BH}$ would completely unbind the atmosphere, and a much smaller value would be insufficient to lift it.

\begin{figure}[!t]
\centering
\includegraphics[width=5.5in]{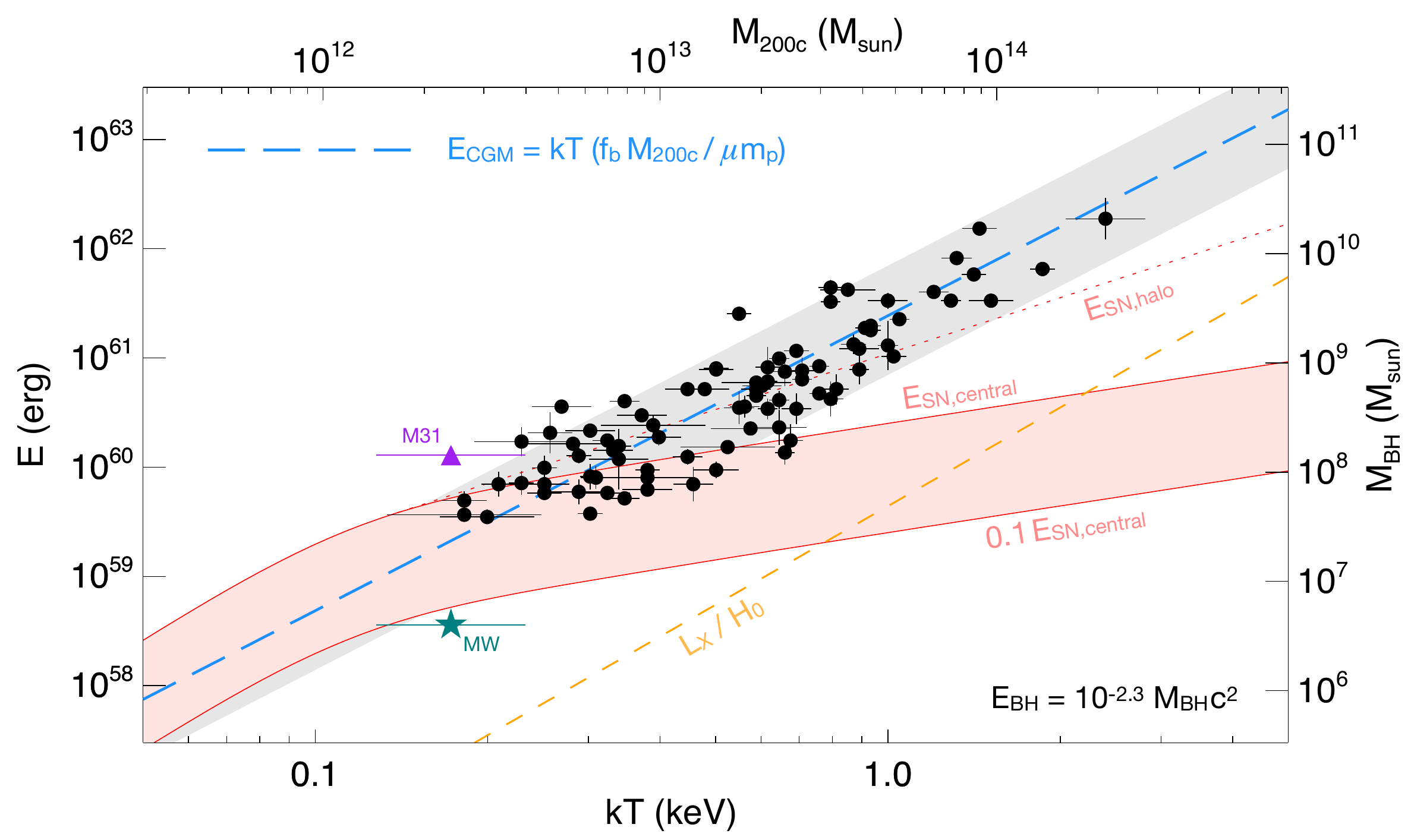}
\caption{Energy requirements for CGM lifting. A dashed blue line shows the CGM energy scale $E_{\rm CGM} \equiv kT (f_{\rm b} M_{200c} / \mu m_p)$.  Significant lifting of the CGM requires feedback energy of this magnitude.  Black points show the data set from \citet{Gaspari2019}, in which the temperature $kT$ of hot CGM gas near the galaxy correlates more closely with $M_{\rm BH}$ than any other property (including $\sigma_v$). They are scaled to the left-hand axis by setting $E_{\rm BH} = 10^{-2.3} M_{\rm BH} c^2$ and to the top axis by setting $M_{200c} = 5 \times 10^{13} \, M_\odot \, (kT/ 1 \, {\rm keV})^{1.7}$. The gray region shows the $10^{\pm 0.5}$ range around the $E_{\rm CGM}$ line.  Most of the points fall within that region, indicating that $f_{\rm BH} \sim 10^{-2.3}$ is necessary for AGN feedback to lift the CGM.  The pink region indicates the amount of energy available for CGM lifting from supernovae in a halo's central galaxy, assuming $0.1 \leq \eta_E \leq 1$.  A dotted line shows the total SN energy produced in the entire halo.  The purple triangle and green star show the M31 and the Milky Way black holes, respectively.  A short-dashed orange line labeled $L_X / H_0$ represents an estimate of the cumulative radiative losses from hot halo gas, which are generally much less than the energy required for CGM lifting.
\label{figure:ECGM_EBH}}
\end{figure}

A recent comparison of central black hole masses with galactic atmospheric temperatures by \citet{Gaspari2019} strongly indicates a close link between black hole mass and the energy required for CGM lifting (see Figure \ref{figure:ECGM_EBH}).  Among a large array of X-ray and optical observables, Gaspari et al.~found that $M_{\rm BH}$ correlates most closely with atmospheric temperature $T_{\rm CGM}$ in the vicinity of the galaxy, which reflects the depth of the galaxy's potential well.  That correlation, with an intrinsic scatter of 21\% in mass, is even tighter than the $M_{\rm BH}$--$\sigma_v$ relation in this particular sample of massive galaxies.  The observed $M_{\rm BH}$--$T_{\rm CGM}$ correlation in the figure tracks the energy scale 
\begin{equation}
    E_{\rm CGM} \equiv kT (f_{\rm b} M_{200c} / \mu m_p)
\end{equation}
required for lifting {\rm all} of the baryons associated with a galactic halo by a factor $\sim 2$ in radius, when an efficiency factor $f_{\rm BH} \approx 10^{-2.3}$ is applied to it.

The importance of atmospheric radiative losses relative to the CGM lifting energy $E_{\rm CGM}$ can be assessed by comparing a galactic atmosphere's current luminosity with the time-averaged power $E_{\rm CGM} H_0$ required to lift the halo's baryons.  At halo masses $\sim 10^{14} \, M_\odot$, representing the transition from galaxy groups to galaxy clusters, that time-averaged power is
\begin{equation}
    \langle \dot{E}_{\rm BH}\rangle  
        \: \sim \: 2 \times 10^{44} \, {\rm erg \, s^{-1}} 
            \left( \frac {M_{\rm 200c}} {10^{14} \, M_\odot} \right)^{5/3}
        \; \; .
\end{equation}
However, X-ray observations of galaxy clusters and groups \citep{Sun+09,Lovisari_2015A&A...573A.118L} indicate
\begin{equation}
    L_X \approx 10^{43} \, {\rm erg \, s^{-1}} \left( \frac {M_{\rm 200c}} {10^{14} \, M_\odot} \right)^2
    \; \; .
\end{equation}
This mismatch shows that the average AGN feedback power required for significant lifting of a galaxy group's atmosphere outstrips by an order of magnitude the power currently required to compensate for the atmosphere's radiative losses.  Therefore, the observed deficit of baryons in groups out to $r_{\rm 500c}$ (see \S \ref{sec:group_baryons}) implies that time-averaged AGN feedback in groups must exceed the current $L_X$ by a large factor in order to account for the observed structure of group atmospheres. 

Galaxy clusters, on the other hand, do not exhibit significant baryon deficits, implying that time-averaged feedback power roughly compensates for radiative losses.  Only losses from the central region with $t_{\rm cool} \lesssim H_0^{-1}$ need to be replenished, because radiative losses at larger radii are too slow to change atmospheric structure.  Those central regions typically have $L_X \lesssim 10^{45} \, {\rm erg \, s^{-1}}$.  Sustaining such a power output for $\sim 10$~Gyr requires accretion of up to $\sim 6 \times 10^{10} \, M_\odot$, given $f_{\rm BH} \sim 10^{-2.5}$, a total mass that is comparable to the maximum masses of the most massive black holes to have so far been discovered.

These observational findings corroborate an important conclusion that \citet{BoothSchaye_2010MNRAS.405L...1B} drew from simulations of AGN feedback.  In their simulations, the mass of a galaxy's central supermassive black hole ends up correlating closely with the mass of its galaxy's halo, because accretion onto the black hole must release an amount of energy comparable to $E_{\rm CGM}$ in order to regulate its own growth.  That result is independent of the efficiency factor assumed for rest-mass energy conversion over four orders of magnitude in efficiency. However, there is a secondary dependence on the concentration of halo mass toward the center of the potential, because the potential wells of highly concentrated halos are deeper for a given mass.  As a result, the relationship emerging from the simulations ($M_{\rm BH} \propto M_{\rm halo}^{1.55 \pm 0.05}$) is slightly shallower than expected for a population of self-similar isothermal halos ($M_{\rm BH} \propto M_{\rm halo}^{5/3}$, \citep[see][]{SilkRees1998AA...331L...1S}).  The blue dashed line in Figure~\ref{figure:ECGM_EBH} shows a relationship with $M_{\rm BH} \propto M_{200 {\rm c}}^{1.6}$, nearly identical to the one found by \citet{BoothSchaye_2010MNRAS.405L...1B}.

\subsubsection{Bipolar Outflows}
\label{sec:bipolar}

\begin{figure}[!t]
\centering
\includegraphics[width=5.3in]{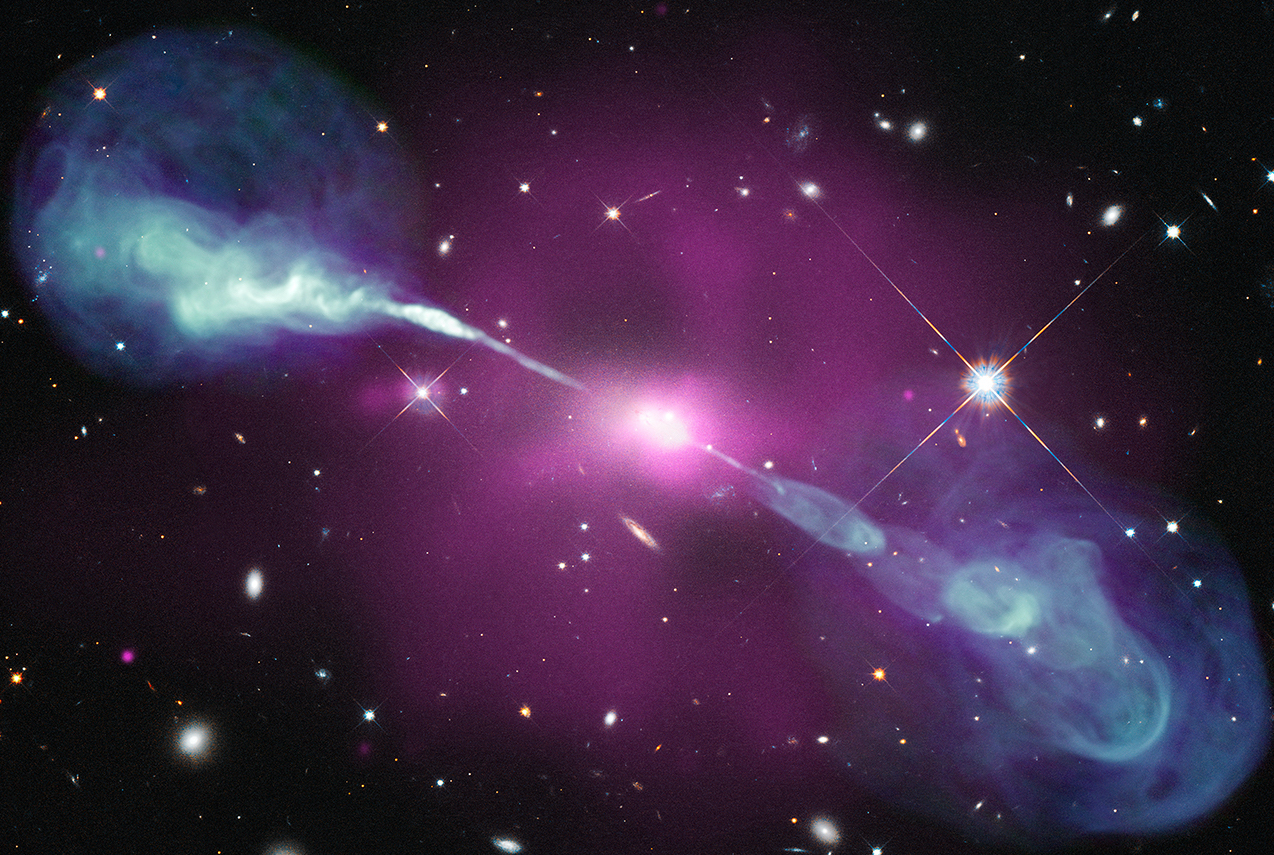}
\caption{Energetic bipolar outflows from the galaxy Hercules A.  Radio observations (blue) show synchrotron emission from relativistic electrons.  X-ray observations (purple) reveal the surrounding atmosphere.  The outflows emerge from the galaxy as narrow jets which then broaden into lobes filled with relativistic plasma as they push against the galaxy's atmosphere.
\label{figure:HerculesA}}
\end{figure}

Observations have long suggested that the central black hole of a massive galaxy does indeed pump energy into the surrounding environment through energetic bipolar outflows extending to hundreds of kiloparsecs (for reviews, see \citep{McNamaraNulsen2012NJPh...14e5023M,McNamara_2014ApJ...785...44M,HardcastleCroston_2020NewAR..8801539H}).  Figure \ref{figure:HerculesA} shows the radio galaxy Hercules A, a spectacular example of the phenomenon.  Accretion of gas onto the central black hole is thought to be the power source, resulting in relativistic jets emerging along the spin axis of the accretion disk \cite{Begelman_RMP_1984}.  Those jets terminate in radio lobes where pressure resistance from the galaxy's atmosphere becomes comparable to the diverging momentum flux of the jets.  However, the power carried by the jets is difficult to estimate from radio observations alone.

\subsubsection{Bubble Calorimetry}
\label{sec:calorimetry}

\begin{figure}[!t]
\centering
\includegraphics[width=5.2in]{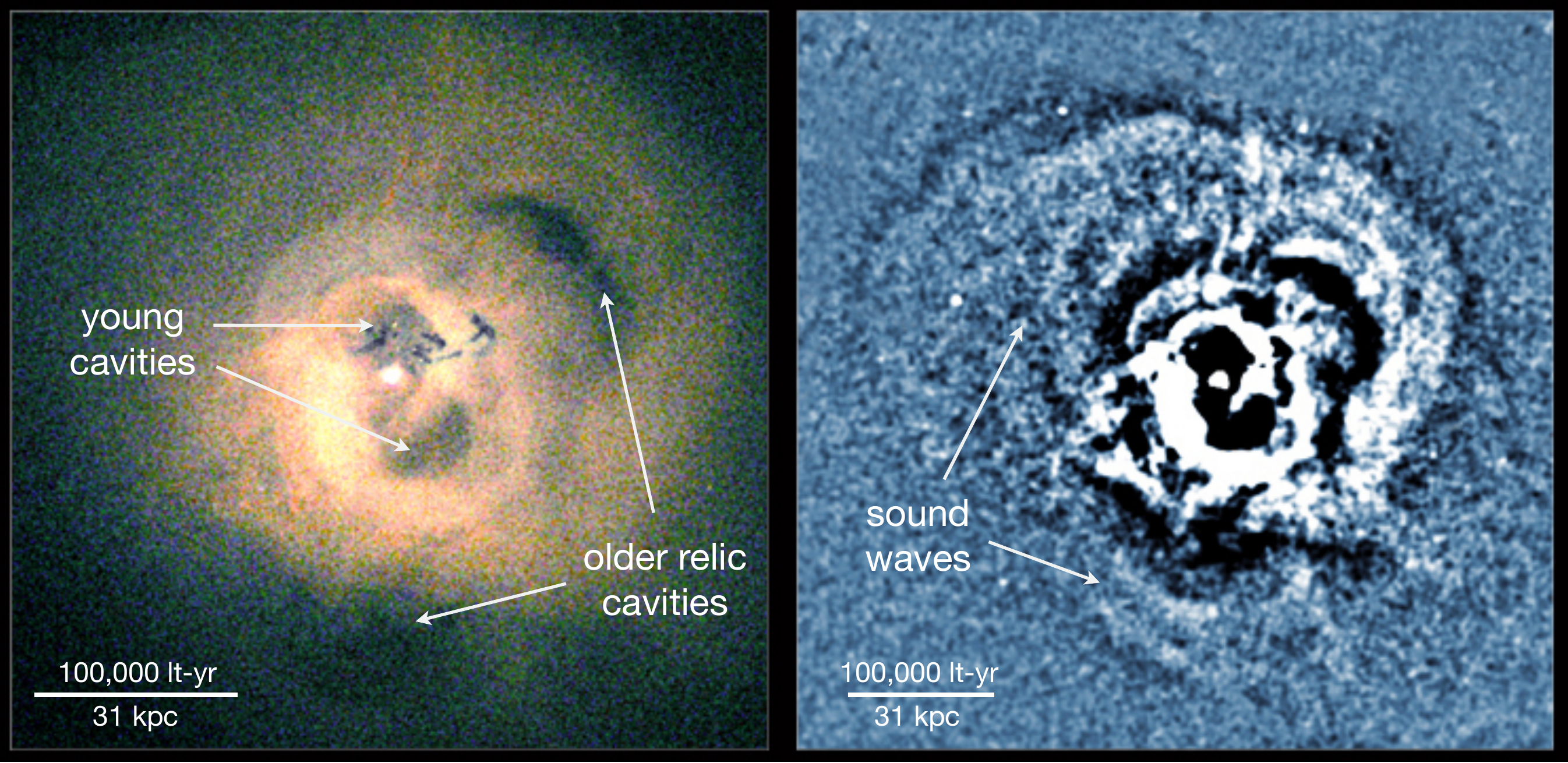}
\caption{Early \textit{Chandra} observations of X-ray cavities and sound waves in the Perseus Cluster (adapted from \citet{Fabian03}).  The left panel shows multiple generations of X-ray cavities.  On the right, an edge-sharpened version of the same image reveals sound waves propagating outward from the cavities.
\label{figure:PerseusCavities}}
\end{figure}

The first accurate assessments of AGN jet power came from \textit{Chandra} observations of the cavities that the jets excavate in galactic atmospheres as they dissipate energy into the lobes.  Earlier X-ray observatories had provided tantalizing evidence for cavities in a few nearby galaxy clusters \citep{Boehringer+1993MNRAS.264L..25B,Carilli_1994MNRAS.270..173C,HUangSarazin_1998ApJ...496..728H,OwenEilek_1998ApJ...493...73O,mn07}. \textit{Chandra's} superior sensitivity and resolving power showed that such cavities are common (see Figure \ref{figure:PerseusCavities}).

\begin{figure}[!t]
  \centering
    \includegraphics[width=0.45\linewidth]{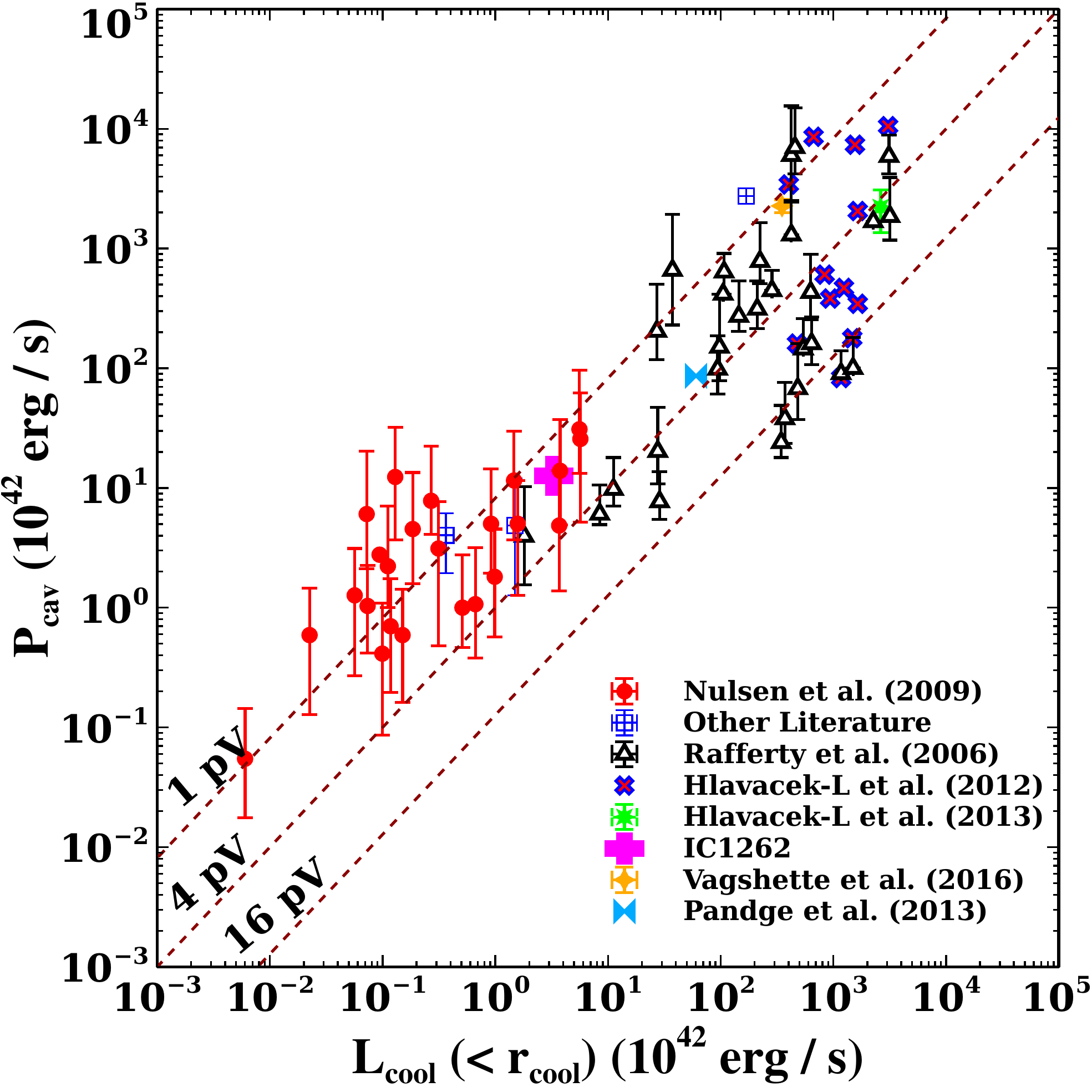}
\hspace{0.1in}
    \includegraphics[width=0.47\linewidth]{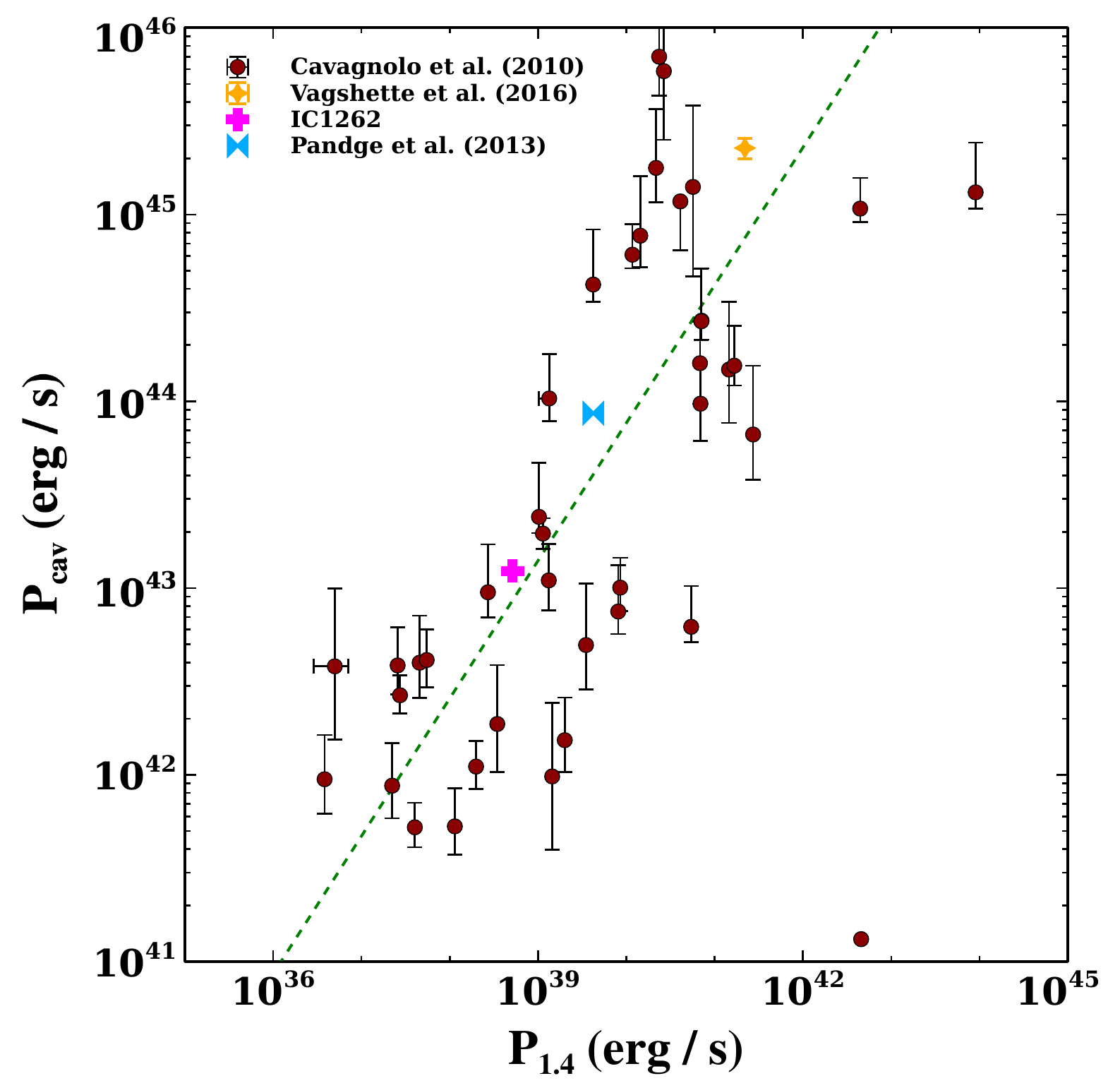}
  \caption{Relationships between radiative losses ($L_{\rm cool}$), radio power at 1.4~GHz ($P_{1.4}$), and the AGN feedback power ($P_{\rm cav}$) inferred from X-ray cavities (compiled by \citep{Pandge2019}).  The left panel shows how $P_{\rm cav}$ inferred from a cavity inflation energy $4PV$ compares with X-ray luminosity from within the cooling radius $r_{\rm cool}$ at which $t_{\rm cool} \approx H_0^{-1}$. The middle red dotted line shows the locus along which the inferred $P_{\rm cav}$ would equal $L_{\rm cool}$. Along the other two red dotted lines cavity inflation energies of $1PV$ and $16PV$ would be needed for inferred $P_{\rm cav}$ to equal $L_{\rm cool}$.  The right panel shows how inferred cavity power correlates with radio power.  A green dotted line shows the approximate relation $P_{\rm cav} / P_{1.4} \approx 5800 \, (P_{1.4} / 10^{40} \, {\rm erg \, s^{-1}})^{-0.3}$ found by \citet{Cavagnolo+10}.  
  \label{figure:corr}}
\end{figure}

Shortly after \textit{Chandra} began returning images of cavities in galaxy clusters, \citet{Churazov+02,Churazov_2000A&A...356..788C,Churazov+01} pioneered their use as calorimeters for assessing the mechanical energy output of their AGNs. That technique was then broadly applied \citep[e.g.,][]{Birzan+04, Birzan+08,mn07,McNamaraNulsen2012NJPh...14e5023M}.  Modern X-ray observations enable estimates of cavity volume $V$, the pressure $P$ of adjacent gas, and the timescale $t_{\rm buoy} \sim r/v_{\rm c}$ for buoyancy to lift the cavities away from their observed locations.  Assuming the cavities to be filled with relativistic plasma of energy density $u = 3P$ leads to a total energy content $3PV$, with an additional $PV$ going into the work required to subsonically inflate them.  Combining those quantities gives an AGN cavity power estimate $P_{\rm cav} \approx 4 PV / t_{\rm buoy}$, equivalent to 
\begin{equation}
    P_{\rm cav} \sim 10^{44} \, {\rm erg \, s^{-1}} 
      \left( \frac {P} {10^{-10} \, {\rm erg \, cm^{-3}}} \right)
      \left( \frac {v_{\rm c}} {10^3 \, {\rm km \, s^{-1}}} \right)
      \left( \frac {r} {10 \, {\rm kpc}} \right)^2
      f_V
      \; \; ,
\end{equation}
where $f_V$ is the volume filling factor of cavities at radius $r$.  In other words, the AGN power output indicated by the cavities is similar in magnitude to radiative losses from a galaxy cluster's core ($L_{\rm x,core} \sim 10^{43-44} \, {\rm erg \, s^{-1}}$).  Furthermore, cavity power estimates correlate with radiative losses over several orders of magnitude in inferred power, as shown in the left panel of Figure~\ref{figure:corr}.  However, note that the cavity power estimated from $4PV$ in galaxy groups (at $L_{\rm cool} \sim 10^{42} \, {\rm erg \, s^{-1}}$) typically \textit{exceeds} radiative losses from within the radius $r_{\rm cool}$ at which $t_{\rm cool}$ approximately equals the universe's age.  

Apparently, AGN feedback in groups is overheating the group's atmosphere, as would seem to be necessary for gradual lifting of that atmosphere through thermally-driven expansion.  The same inference follows from large surveys of radio galaxies (see \citep{HeckmanBest_2014} for a review).  In principle, one can estimate the time-averaged kinetic jet power in galaxies of a given stellar mass by determining their time-averaged radio power at 1.4~GHz and applying the observed relationship between radio power and cavity power, shown in the right panel of Figure~\ref{figure:corr}.  In practice, the dependence of time-averaged radio power on stellar mass is difficult to estimate, because the fraction of galaxies currently exhibiting high radio power rises strongly with stellar mass. Galaxies with $M_* \lesssim 10^{11.3} \, M_\odot$ must therefore experience large temporal fluctuations in AGN feedback power \citep{Best_2005}.  Averaging over large samples can mitigate those uncertainties. \citet{Best_2007} have shown that applying the observed relationships of radio power to cavity power results in estimates of total feedback heating that exceed radiative cooling by at least an order of magnitude in groups with $kT \lesssim 1 \, {\rm keV}$.

    \subsubsection{Cooling Time and Radio Power}
\label{sec:tcool_Prad}

Prior to the \textit{Chandra} mission, the circumstantial evidence indicating a close connection between central cooling and AGN fueling was not taken seriously enough. For example, \citet{1990AJ.....99...14B} showed that BCGs in clusters with short cooling times were highly likely to be radio sources, with a weak correlation between central cooling time and radio power. However, the observed radio power ($L_R \sim 10^{40-41}$ erg s$^{-1}$) was considered insignificant compared to the observed radiative losses. Also, the bipolar morphology of many radio sources suggested that their energy was not being distributed in a manner that could offset radiative losses far from the jet axis.

After \textit{Chandra} observations established that AGN outflows were providing enough power to offset radiative losses, astronomers began to reexamine the correlations between short central cooling time, radio power, and the presence of X-ray cavities. Among nearby galaxy clusters, \citet{DunnFabian2006} found detectable X-ray cavities only in the ones with central cooling times less than 3~Gyr.  In a larger sample extending to $z = 0.2$, \citet{Cavagnolo+08} found that \textit{all} of the galaxy clusters with strong central radio sources ($P_{1.4} > 10^{40} \, {\rm erg \, s^{-1}}$) also had central entropy levels $< 30 \, {\rm keV cm^2}$, corresponding to $t_{\rm cool} \lesssim 1 \, {\rm Gyr}$.  \citet{Cavagnolo+10} later showed that radio power strongly correlates with the kinetic power inferred from X-ray cavities (see the right panel of Figure~\ref{figure:corr}).

\begin{figure}[!t] 
  \centering 
    \includegraphics[width=0.95\textwidth]{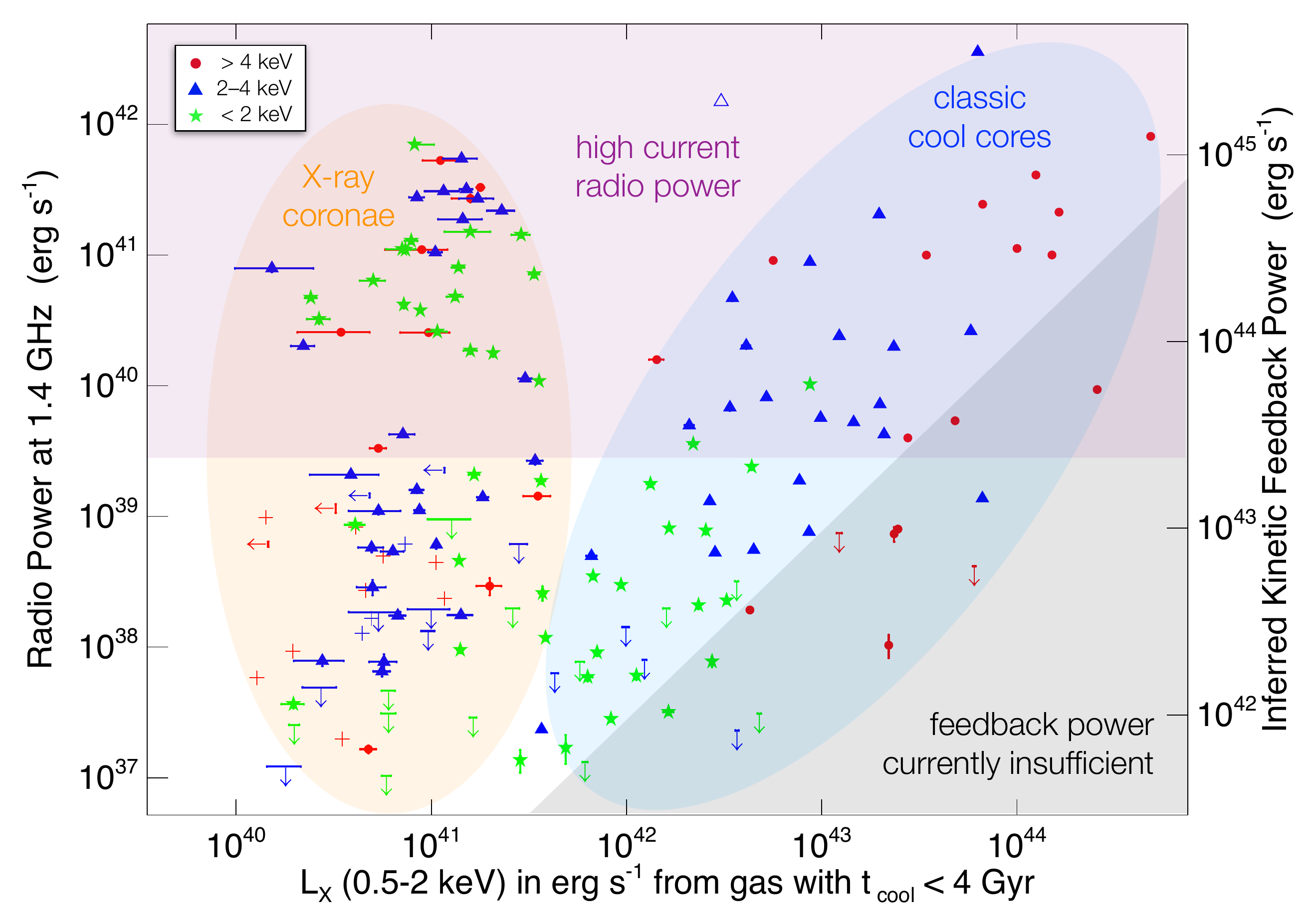}
  \caption{Relationships between radio power and radiative losses in galaxy groups and clusters, adapted from \citet{Sun09}.  The horizontal axis shows X-ray luminosity in the 0.2--2~keV band from inside the radius where $t_{\rm cool} \approx 4 \, {\rm Gyr}$.  The vertical axis shows the radio power $P_{1.4} = P_\nu(1.4 \, {\rm GHz}) \times 1.4 \, {\rm GHz}$.  The right-hand axis shows the cavity power one infers by applying the approximate relation $P_{\rm cav} = 5800 \, P_{1.4} \, (P_{1.4} / 10^{40} \, {\rm erg \, s^{-1}})^{-0.3}$ from \citep{Cavagnolo+10}.  In the gray triangle at the lower right, radiative losses currently exceed the inferred kinetic power.  Red circles show massive galaxy clusters with $kT > 4 \, {\rm keV}$, green stars show galaxy groups with $kT < 2 \, {\rm keV}$, and blue triangles show low-mass galaxy clusters with intermediate temperatures.  All of the strong radio sources with $P_{1.4} > 10^{39.5} \, {\rm erg \, s^{-1}}$ are surrounded by gas with $t_{\rm cool} < 4 \, {\rm Gyr}$. In classic cool cores, radio power correlates with radiative losses, and the inferred kinetic power typically exceeds $L_X$. However, radio power among sources within X-ray coronae spans over 4 orders of magnitude and does not correlate with radiative losses, suggesting that AGN feedback in that population fluctuates between low-power states and high-power states that vastly exceed the power required to offset radiative losses.
  \label{figure:Sun2009}}
\end{figure}

The tight link between high current radio power and short central cooling time strongly implies that the AGN power is responding to a need for feedback.  However, \citet{Sun09} found that the nature of the link depends on the AGN's environment.  Figure~\ref{figure:Sun2009}, adapted from Sun's paper, illustrates several relationships between radio power and central radiative losses within a large sample of nearby galaxy groups and clusters.  Position along the horizontal axis indicates X-ray luminosity from atmospheric gas with a cooling time $< 4$~Gyr, which we will call $L_{\rm cool}$.  In the ``classic cool core" population with $L_{\rm cool} \gtrsim 10^{42} \, {\rm erg \, s}$, radio power correlates with $L_{\rm cool}$ and indicates a total kinetic power that generally exceeds $L_{\rm cool}$.  The few exceptions in which kinetic power is insufficient suggest that the AGN engine is rarely dormant in that population.  

In contrast, radio power does not appear to correlate with $L_{\rm cool}$ in the ``X-ray coronae" population with $L_{\rm cool} \lesssim 10^{41.5} \, {\rm erg \, s}$. The most powerful examples demonstrate that kinetic power sometimes greatly exceeds $L_{\rm cool}$.  Also, there is no apparent correlation between radio power and halo mass, as indicated by the X-ray temperature of gas outside the corona, whereas both $L_{\rm cool}$ and radio power tend to correlate with halo mass in the classic cool core population.

Intriguingly, the transitional range in $L_{\rm cool}$ between X-ray coronae and classic cool cores coincides with the SNIa power from a massive elliptical galaxy's old stellar population.  Multiplying a total stellar mass $\sim 10^{11.5} \, M_\odot$ by the specific SNIa rate ($\sim 3 \times 10^{14} \, M_\odot^{-1} \, {\rm yr}^{-1}$) and SNIa energy ($\sim 10^{51} \, {\rm erg}$) gives a total power $\sim 10^{41.5} \, {\rm erg \, s^{-1}}$.  If $L_{\rm cool}$ exceeds this power output, then AGN feedback is continually needed to offset radiative cooling and apparently becomes tuned to $L_{\rm cool}$.  But if SNIa power exceeds $L_{\rm cool}$, then the tuning mechanism must be different.  

\begin{figure*}[t]
    \centering
    \includegraphics[width=5.3in]
        {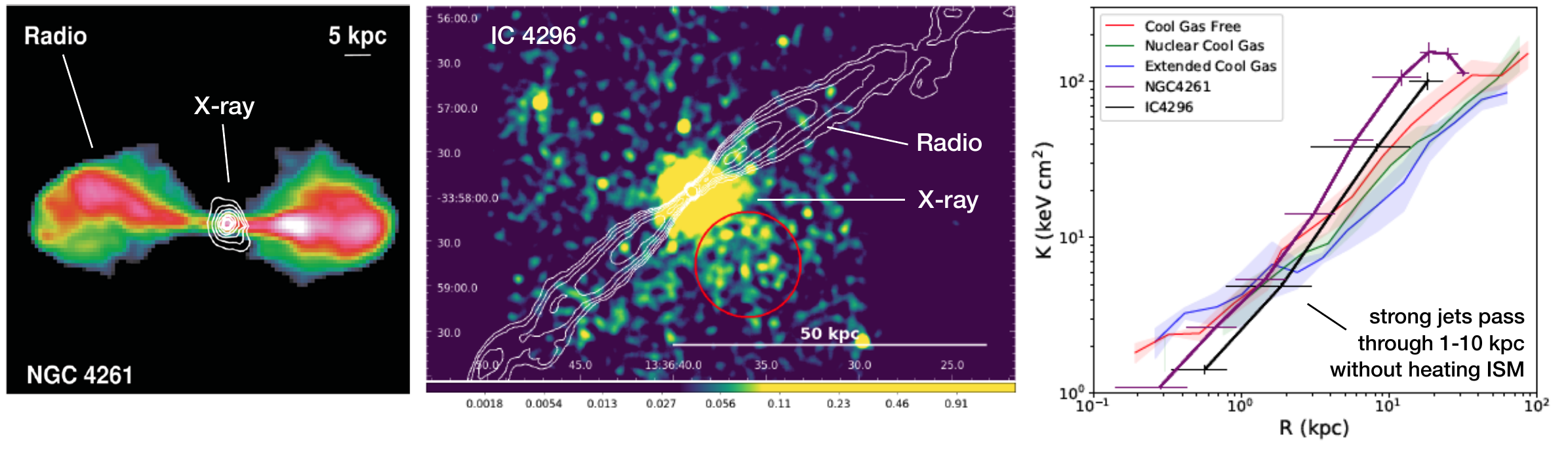}
    \caption{Strong narrow jets in NGC 4261 and IC 4296. The jets in NGC 4261 (left panel) have a kinetic power $\sim 10^{44} \, {\rm erg \, s^{-1}}$ and thermalize $\sim 25$--40 kpc from the center \cite{Werner+2012MNRAS.425.2731W}. The jets in IC 4296 (center panel) have a kinetic power $\sim 10^{44} \, {\rm erg \, s^{-1}}$ and thermalize $\sim 50$--250 kpc from the center \cite{Grossova_2019MNRAS.488.1917G}.  But the entropy profiles of those galaxies (right panel) from 1-10 kpc remain similar to those of other massive ellipticals with jets two orders of magnitude less powerful \cite{Grossova_2019MNRAS.488.1917G}. In fact, their unusually low entropy at $< 1$~kpc suggests that cooling of the central gas is currently fueling these strong outbursts.  The morphologies of those jets indicate that they penetrate to large distances because they are narrow and limit cooling at 1-10 kpc because their heat input expands the CGM and lowers the confining pressure it applies to the ISM \cite{Voit_2020ApJ...899...70V}.}
    \label{fig:narrow_jets}
\end{figure*}

One possibility is the black-hole feedback valve mechanism proposed in \citep{Voit_2020ApJ...899...70V}, which \S \ref{sec:Valve} will discuss in more detail.  That paper shows how SNIa heating in galaxies with a large central velocity dispersion ($\sigma_v \gtrsim 240 \, {\rm km \, s^{-1}}$) can lead to a quasi-steady state with a modest cooling flow at small radii ($\lesssim 1$~kpc) and a slow SNIa-heated outflow at larger radii ($\sim 1$--10~kpc).  Such a state requires a low confining CGM pressure at $\sim 10$--20~kpc, kept low by occasional outbursts of AGN feedback that limit accumulation of CGM gas.  Observations show that at least some of the X-ray coronae with large radio power meet the necessary requirements and a few exhibit powerful bipolar jets (see Figure  \ref{fig:narrow_jets}) that do not significantly disrupt the gas with $t_{\rm cool} < 4$~Gyr \citep{Werner+2012MNRAS.425.2731W,Grossova_2019MNRAS.488.1917G,Frisbie_2020ApJ...899..159F}.  Instead of tuning jet power to match radiative losses, AGN feedback in these systems appears to be regulating the CGM pressure that confines the SNIa-heated atmospheric gas and determines the magnitude of the cooling flow at small radii.

\subsubsection{Weak Shocks}
\label{sec:shocks}

The X-ray and radio observations have not been forthcoming about exactly how the AGN accomplishes self-regulation. It is one thing to show that AGN power output can compensate for radiative cooling in CGM gas and is correlated with radiative losses. Showing exactly how and where the AGN feedback energy is dissipating is more difficult.  

At least some of that dissipation occurs in weak shock fronts.  High-resolution X-ray observations of numerous nearby galaxy groups and clusters have revealed the presence of weak shocks in the vicinity of the jets and cavities produced by AGN feedback \citep[e.g.,][]{Fabian03,2005ApJ...635..894F,Fabian_2006MNRAS.366..417F,Forman+07,Blanton_2009ApJ...697L..95B,Randall+11,Liu_2019MNRAS.484.3376L}.  The Mach numbers of those shocks are typically $\lesssim 1.5$ \citep{Eckert_2021Univ....7..142E}.

A simple scaling argument shows why the shocks in classic cool cores tend to be weak.  The speed $v_{\rm sh}$ of a spherical shock at radius $r$ driven by continuous power input at the rate $\dot{E}_{\rm AGN}$ is determined by
\begin{equation}
    v_{\rm sh}^3 \sim \frac {\dot{E}_{\rm AGN} \, r} {M_{\rm gas} (r)} 
    \label{eq:vsh3}
    \; \; ,
\end{equation}
where $M_{\rm gas}(r)$ is the gas mass within radius $r$.  The shock's Mach number (${\cal M} \sim v_{\rm sh} \sqrt{\mu m_p /kT}$) is therefore
\begin{equation}
    {\cal M} \: \sim \: \left( \frac {1} {v_{\rm c}} \frac {\mu m_p} {kT} \right)^{1/3} v_{\rm sh}
             \: \sim \: \left[ \frac {\dot{E}_{\rm AGN}} {L_{\rm rad}(r)} \frac {t_{\rm ff}} {t_{\rm cool}}                  \right]^{1/3}
             \; \; ,
\end{equation}
where $L_{\rm rad} \sim (M_{gas} / \mu m_p) kT / t_{\rm cool}$ is the radiative luminosity from gas within $r$ and we have assumed $kT \sim \mu m_p v_{\rm c}^2$.  In the atmospheres of massive galaxies, the ratio $t_{\rm cool}/t_{\rm ff}$ is rarely observed to be less than $\sim 10$ (see \S \ref{sec:Evidence}), meaning that $\dot{E}_{\rm AGN}$ needs to exceed $L_{\rm rad}(r)$ by a large factor in order to produce a shock with ${\cal M}$ significantly greater than unity.  In fact, it needs to be an order of magnitude greater than $L_{\rm rad}(r)$ to produce even a weak shock.  In other words, frequent weak shocks can gently maintain a nearly steady state in the core of a galaxy cluster, but sporadic heating by strong shocks would be excessively disruptive (see also \citep{Gaspari+2014ApJ...783L..10G}).

Another way to characterize those shocks is in terms of the entropy jump across the shock front,
\begin{equation}
    \Delta K \: \approx \: \frac {\mu m_p v_{\rm sh}^2} {3(4n_1)^{2/3}} \: - \: 0.16 K_1
    \; \; ,
\end{equation}
where $\Delta K = K_2 - K_1$ is the difference between the preshock entropy ($K_1$) and the postshock entropy ($K_2$), $n_1$ is the preshock electron density, and $v_{\rm sh}$ is the difference in flow speed across the shock \citep{Voit+03}.  Ignoring the correction term containing $K_1$ and substituting the approximate shock speed from equation (\ref{eq:vsh3}) gives
\begin{equation}
    \Delta K \sim \frac {\mu m_p} {3(kT)^2} \left( \frac {\dot{E}_{\rm AGN}} {8 \pi \mu_e m_p} \right)^{2/3}
        \left[ \frac {K_1(r)} {r^{2/3}} \right]^2
        \label{eq:DeltaK1}
\end{equation}
for a gas density profile approximately following $n_e \propto r^{-1}$.  Entropy profiles of classic cool cores are observed to follow $K(r) \approx 19 \, {\rm keV \, cm^2} \, (r/10 \, {\rm kpc})^{2/3}$ \citep{Panagoulia_2014MNRAS.438.2341P,Babyk2018ApJ...862...39B}, often with a central entropy excess ($K_0$) at small radii \citep{Donahue+06,Cavagnolo+09} (see \S \ref{sec:Kobserved}).  Plugging that relation into equation (\ref{eq:DeltaK1}) yields
\begin{equation}
    \Delta K \sim 2.7 \, {\rm keV \, cm^2} \, 
        \left( \frac {\dot{E}_{\rm AGN}} {10^{43} \, {\rm erg \, s^{-1}}} \right)^{2/3} 
                            \left[ \frac {kT(r)} {1 \, {\rm keV}} \right]^{-2}
        \label{eq:DeltaK2}
    \; \; ,
\end{equation}
linking the entropy jump produced by a single shock-heating event to the AGN kinetic power and the ambient gas temperature. 

The relationship in equation (\ref{eq:DeltaK2}) allows constraints on AGN power to be inferred from the observed entropy profiles of massive galaxies \citep{vd05}. High-mass systems with large AGN power ($\sim 10^{45} \, {\rm erg \, s^{-1}}$), corresponding to the red dots in the upper right corner of Figure \ref{figure:Sun2009}, have $kT \sim 2$~keV at $< 10$~kpc.  Equation (\ref{eq:DeltaK2}) therefore implies that the outbursts needed to match radiative cooling at larger radii should produce central entropy jumps of $\sim 15 \, {\rm keV~ cm^2}$.  The  observed central entropy excesses\footnote{Relative to extrapolations of the power-law profiles at larger radii} in massive cool-core clusters are similar to that estimate and are therefore consistent with intermittent shock heating by outbursts capable of offsetting radiative cooling throughout the core \citep{vd05,Cavagnolo+09}.  Among smaller systems with lower AGN power ($\sim 10^{42} \, {\rm erg \, s^{-1}}$), corresponding to the green stars at $L_{\rm cool} \sim 10^{42} \, {\rm erg \, s^{-1}}$ in Figure \ref{figure:Sun2009}, equation (\ref{eq:DeltaK2}) implies an entropy jump $\sim 2 \, {\rm keV \, cm^2}$ for $kT \sim 0.5$~keV.  Observations of those systems indeed indicate $K_0 \sim 2 \, {\rm keV \, cm^2}$, again consistent with the outbursts required to balance radiative cooling at larger radii \citep{Werner+2012MNRAS.425.2731W,Werner+2014MNRAS.439.2291W,Voit+2015ApJ...803L..21V,Voit_2020ApJ...899...70V}. 

However, most of the shocks produced by those outburst events become subsonic at radii not much greater than $\sim 10$--20~kpc.  For example, setting $v_{\rm sh}^2 \sim kT / \mu m_p$ in equation (\ref{eq:vsh3}) leads to the estimate
\begin{eqnarray}
    r_{\rm ss} & \sim & \frac {\dot{E}_{\rm AGN}^{1/2}} {2 \pi \mu_e m_p} 
                        \frac {(\mu m_p)^{3/2}} {(kT)^3} K^{3/2} \\
     & \sim & 10 \, {\rm kpc} \, \left( \frac {\dot{E}_{\rm AGN}} {10^{43} \, {\rm erg \, s^{-1}}} \right)^{1/2}
                             \left[ \frac {kT(r)} {1 \, {\rm keV}} \right]^{-3}
\end{eqnarray}
for the radius $r_{\rm ss}$ at which shocks consistent with the observed $K_0$ values go subsonic, given the $K(r)$ profiles observed in classic cool cores.  As the shock decays into a sound wave beyond $r_{\rm ss}$, the high entropy gas behind the shock front forms a bubble that subsequently rises buoyantly and subsonically.\footnote{The shocked atmospheric gas might not rise very far if $\Delta K \lesssim K_1$, but much of the gas in the bubble may have considerably greater entropy, if it consists of relativistic plasma from the AGN or ejecta that have passed at high speed through a reverse shock. The observed contrasts between X-ray cavities and their surroundings indicate that they do indeed have much greater specific entropy.}

There are exceptions, though.  A few notable sources in classic cool-core clusters, including Hydra A and MS0735+74 (see Figure \ref{Figure:MS0735}), have been able to drive shock fronts to much larger radii \citep{McNamara_2005Natur.433...45M,Nulsen_Hydra_2005ApJ...628..629N,2007ApJ...660.1118G} and indicate that AGN feedback events can produce total energies as large as $\sim 10^{61} \, {\rm erg}$, but those events appear to be uncommon \citep[e.g., ][]{2002ApJ...568..163N}.  Among the lower-mass X-ray coronae there are examples of powerful ($\sim 10^{44} \, {\rm erg \,s^{-1}}$), highly collimated outflows, including those from NGC 4261 and IC 4296, that reach radii of tens of kiloparsecs \citep{Grossova_2019MNRAS.488.1917G,Frisbie_2020ApJ...899..159F}.  Equation (\ref{eq:DeltaK2}) predicts $\Delta K > 10 \, {\rm keV \, cm^2}$ for the shocks driven by such outflows, but the X-ray observations indicate $K_0 < 1 \, {\rm keV \, cm^2}$, with no evidence for deviation from a power-law entropy profile at small radii (see Figure \ref{fig:narrow_jets}).  Apparently, narrow AGN outflows with sufficient momentum flux can drill cleanly through X-ray coronae, channeling substantial feedback power into the CGM without producing strong shock fronts on kiloparsec scales (see also Figure \ref{Figure:NCG1265_corona}).

\subsubsection{Turbulence, Mixing, Conduction, and Cosmic Rays}
\label{sec:turbulence}

\begin{figure}[!t]
\centering
\includegraphics[width=5.3in]{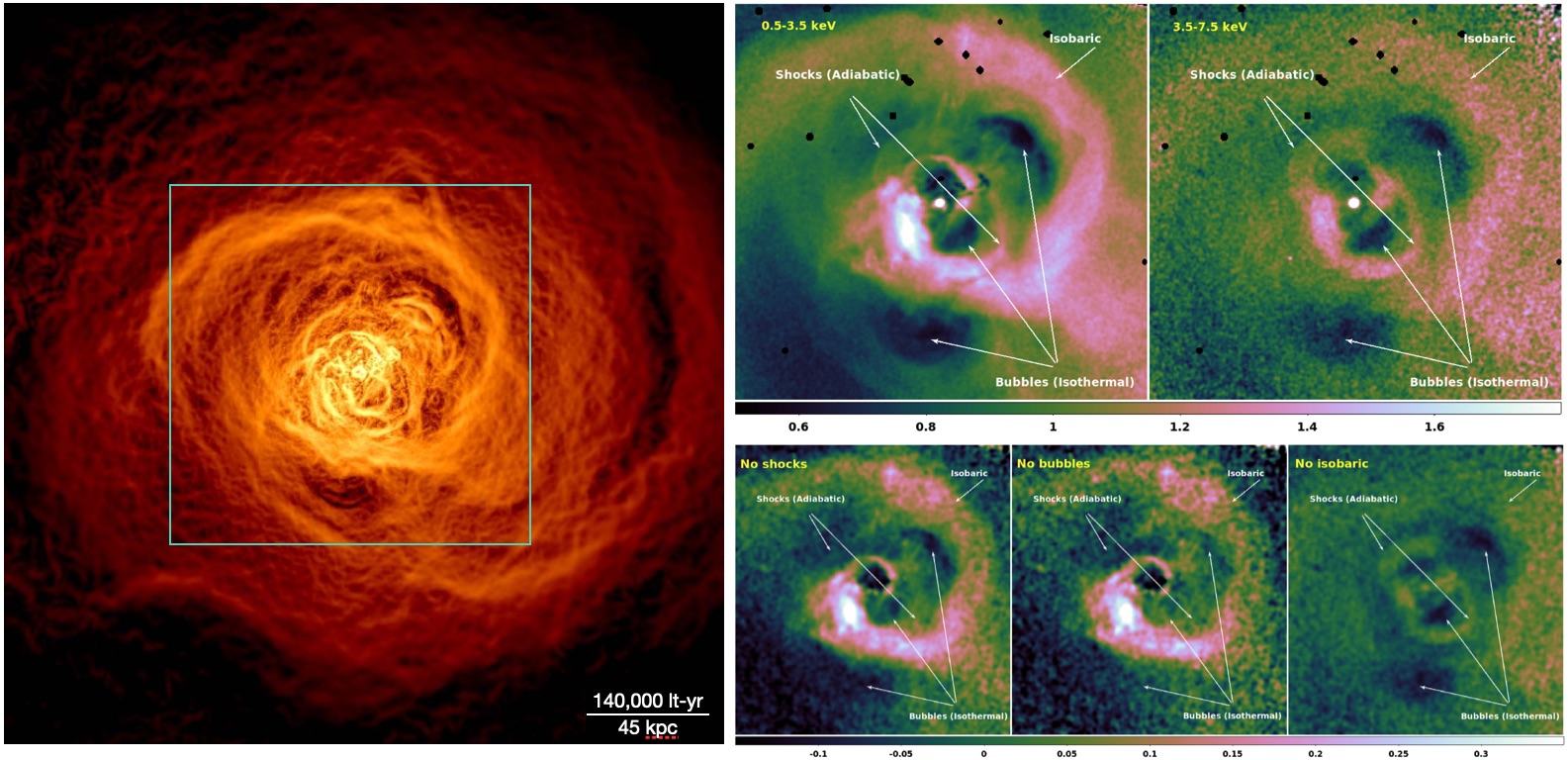}
\caption{Waves and turbulence in the Perseus Cluster's atmosphere. \textit{Left panel:} X-ray surface-brightness substructure revealed by gaussian gradient filtering, performed by \citet{Sanders+2016} on a deep \textit{Chandra} observation.  The bubbles inflated by AGN feedback in the cluster core drive internal gravity waves, sound waves, and turbulence, all of which contribute to the observed surface-brightness substructure. A square outlines the region shown in the panels to the right. \textit{Upper right panels:} Central region of the Perseus Cluster in a soft X-ray band (0.5-3.5~keV, left) and a hard X-ray band (3.5-7.5~keV, right).  \textit{Lower right panels:} Three subtractions of the hard band image from the soft band image, weighted differently by \citet{Churazov_2016MNRAS.463.1057C} so as to distinguish bubbles and internal gravity waves from weak shocks and sound waves. The subtraction on the left is weighted so that the shock fronts cancel, emphasizing the gravity waves and the bubbles.  The subtraction in the middle is weighted so that the older bubbles largely cancel.  The subtraction on the right is weighted so that the internal gravity waves cancel, emphasizing the shock fronts and sound waves.
\label{figure:PerseusOutburst}}
\end{figure}

Thermalization of kinetic AGN power input becomes harder to sort out after a subsonic bubble forms.  A high-entropy bubble rises buoyantly through the ambient gas, lifting some of it and later allowing it to fall. As denser gas around the bubble falls, the falling gas drives internal gravity waves, sound waves, and turbulence (see Figure~\ref{figure:PerseusOutburst}).  All of those phenomena can transport energy away from the rising bubble and can eventually thermalize it \citep{Churazov+01,Begelman+01}. Also, cosmic rays coming from within the bubble can propagate out of the bubble via either streaming or diffusion before thermalizing  \citep{GuoOh_2008MNRAS.384..251G,Ruszkowski_2017ApJ...844...13R,JacobPfrommer_2017MNRAS.467.1478J,Butsky_2020ApJ...903...77B}.

The thermalization and transport rates for all of those heating channels depend on highly uncertain features of the ambient medium, such as its viscosity and thermal conduction coefficients and the MHD wave interactions that govern cosmic-ray propagation and thermalization.  Theorists are actively working to reduce those uncertainties, making nonthermal energy transport and dissipation a lively area of current CGM research \citep[e.g.,][]{Butsky_2018ApJ...868..108B,Su_2020MNRAS.491.1190S,Ji_2020MNRAS.496.4221J,Bustard_2021ApJ...913..106B}, but not one we can review in detail.  Instead, we will briefly comment on turbulence because the upcoming XRISM mission\footnote{https://www.nasa.gov/content/goddard/xrism-x-ray-imaging-and-spectroscopy-mission} will soon vastly expand the observational constraints on that mode of heating and mixing in cluster cores. 

Turbulent dissipation is a plausible mechanism for heating of atmospheric gas beyond where weak shocks are observed. \citet{Zhuravleva_2014Natur.515...85Z} have argued that thermalization of the turbulent energy inferred from surface-brightness substructure may be sufficient to compensate for radiative losses, but the argument is not conclusive because of large observational uncertainties. Turbulence may also help to transport heat in a manner similar to thermal conduction. In the presence of a radial entropy gradient, turbulence can mix gas parcels that differ in specific entropy, allowing low-entropy gas to become higher-entropy gas without passing through shocks. \citet{HillelSoker_2016MNRAS.455.2139H} argue that the heating via turbulent transport and mixing is significantly greater than what comes from turbulent dissipation.

\begin{figure}[!t]
\centering
\includegraphics[width=5.3in]{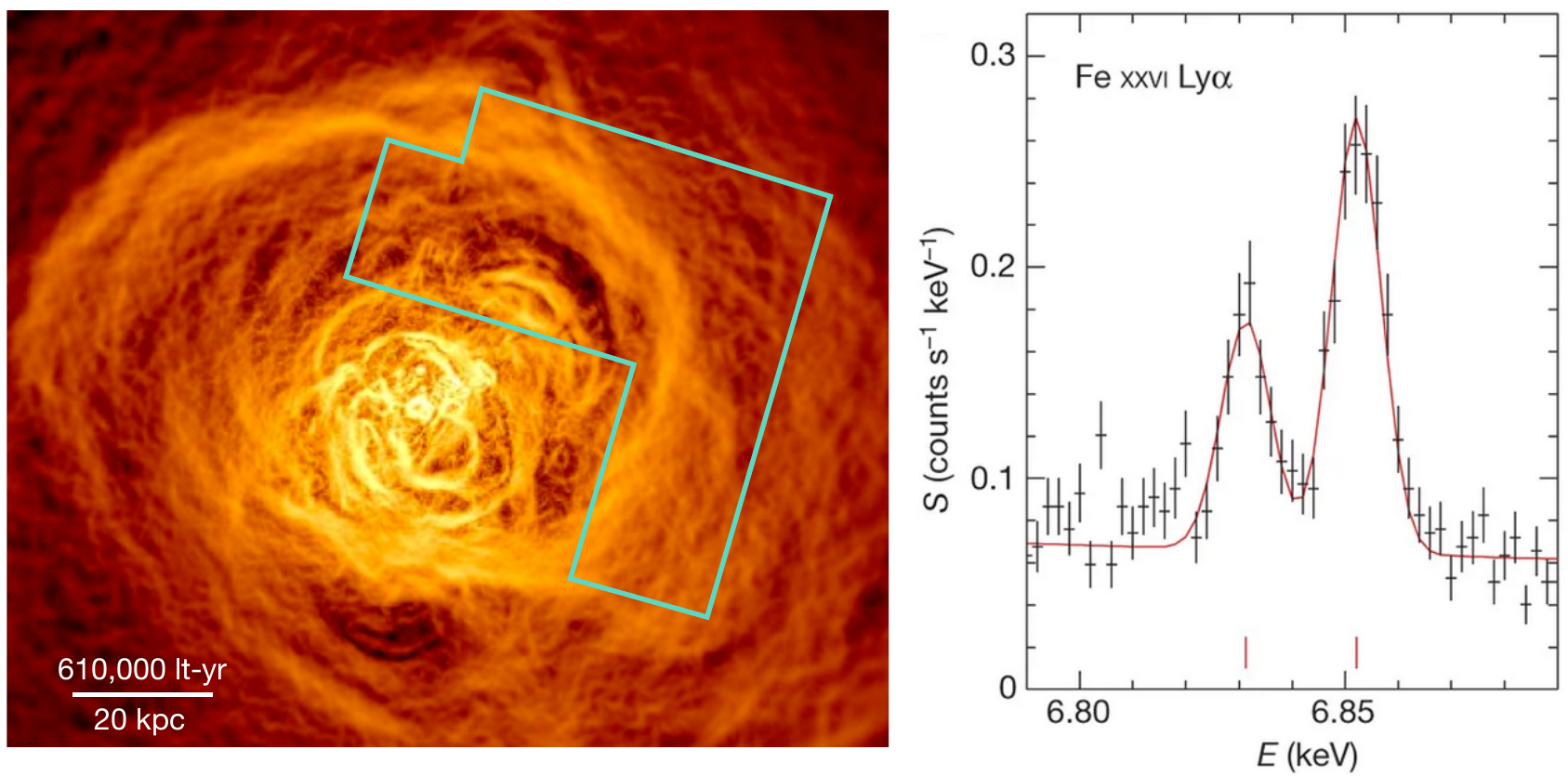}
\caption{Velocity dispersion of the Perseus Cluster's core atmosphere, measured with the Hitomi satellite's Soft X-ray Spectrometer (SXS) \citep{HitomiPerseus2016}. \textit{Left Panel:} Region of the Perseus Cluster observed with the SXS to obtain the spectrum on the right.  This aperture excludes the central AGN and the highest contrast bubbles around it. \textit{Right Panel:} SXS X-ray spectrum of the Ly$\alpha$ doublet from hydrogen-like iron ions (Fe XXVI) in that region. A red line shows the best fitting pair of Gaussian line profiles, indicating a line-of-sight velocity dispersion of $\sigma_v = 164 \pm 10 \, {\rm km \, s^{-1}}$. Other X-ray emission lines in the spectrum indicate the same dispersion.  A similar spectrum of the central region (including the AGN)} indicates a slightly greater velocity dispersion there ($\sigma_v = 187 \pm 13 \, {\rm km \, s^{-1}}$).   \label{figure:HitomiSpectrum}
\end{figure}

The most compelling evidence that the X-ray emitting plasma in a galaxy cluster is indeed turbulent comes from the short-lived\footnote{RIP Hitomi (17 Feb 2016 - 26 Mar 2016)} \textit{Hitomi} satellite (Figure \ref{figure:HitomiSpectrum}).  It carried an X-ray spectrograph with high spectral resolution and arcminute-scale angular resolution, well suited to obtaining spectra of extended thermal X-ray sources. The \textit{Hitomi} observation of the Perseus Cluster provided a brief but rich spectroscopic view of the hot-gas velocity field in a single galaxy cluster \citep{HitomiPerseus2016}, indicating line-of-sight turbulent velocities of $\sim 160 \, {\rm km \, s^{-1}}$ that correspond to an equivalent pressure support no more than $\sim 10\%$ of the thermal pressure of the gas \citep{HitomiPerseus2016,2018PASJ...70....9H}.

This amount of turbulence appears to be too small to provide enough heating outside of central $\sim60$ kpc or so \citep{2018MNRAS.478L..44B,Sanders+2016}. Sound waves remain a viable channel for AGN energy transport in Perseus, but the sound-wave dissipation rate depends on an unknown bulk viscosity \citep{2017MNRAS.464L...1F}.  Also, comparisons by \citet{HillelSoker_2020ApJ...896..104H} of the velocity structure of the optical H$\alpha$-emitting filaments with simulations of various heating mechanisms suggest that turbulent dissipation is not the primary heating source in galaxy clusters, leaving open important questions about how AGN energy propagates and thermalizes beyond the central $\sim 10$--20~kpc, where weak shocks are typically observed. 

\section{Circumgalactic Weather \label{sec:Weather}}

Supermassive black holes clearly release enormous amounts of energy into the atmospheres of massive galaxies. They seem to be highly disruptive, yet the atmospheres surrounding them have persisted for many Gyr and are frequently observed to have central cooling times considerably less than 1~Gyr at $\lesssim 10$~kpc (see Figure \ref{figure:Consistency}).  Together, those findings indicate that the energy output from black hole accretion is self-regulating, diminishing when the central cooling time rises above a few hundred Myr and rising when the central cooling time falls below a few hundred Myr.

Observations of the phenomena produced by this feedback loop suggest certain resemblances to terrestrial weather. Like Earth's clouds, gas clouds in the atmospheres of massive galaxies exhibit complex morphologies and turbulent velocities indicating chaotic dynamics with unpredictable and sometimes surprising details.  But the overall ``weather" patterns observed among massive galaxies follow broader systematic trends which are presumably the emergent properties of self-regulation.

\subsection{Multiphase Gas in Galaxy Cluster Cores}

Our most vivid and complete views of the weather in galactic atmospheres come from multiwavelength observations of the huge central galaxies (the BCGs) in massive galaxy clusters.  That is where the volume-filling hot gas around massive galaxies is most easily observed with X-ray telescopes, revealing cavities, shock fronts, sound waves, and turbulence. Frequently those atmospheres also contain cooler multiphase gas spanning wide ranges in both temperature ($30 \, {\rm K} \lesssim T \lesssim 10^8 \, {\rm K}$) and density ($10^{-2} \, {\rm cm^{-3}} \lesssim n 
\lesssim 10^4 \, {\rm cm^{-3}}$) and extending up to $\sim 100$~kpc from the center.  This section describes the observed properties of the multiphase gas in the cores of galaxy clusters and \S \ref{sec:multiphase_tcool} discusses how those gas properties are linked to the cooling time of the ambient hot gas.

\subsubsection{Brightest Cluster Galaxy (BCG) Nebulae}

Astronomers have known for decades that the central galaxies of some galaxy clusters display spectacular emission-line nebulae.  One of the most famous examples is the nebula associated with NGC 1275, at the center of the Perseus Cluster. Studies of its unusual emission-line spectrum go back as least as far as Humason's in 1932 \citep{Humason_1932_NGC1275}.  Those emission lines led Paul Seyfert to include NGC 1275 among the original ``Seyfert galaxies" in 1943 \citep{Seyfert_1943}.  Minkowski's 1957 image showed that the nebulosity extends over tens of kiloparsecs \citep{Minkowski_1957_NGC1275}.  The nebula's morphology, along with the velocity field inferred from the Doppler shifts of its emission lines and the neighboring radio emission, led the Burbidges to speculate in 1965 that NGC 1275 was undergoing a violent outburst \citep{Burbidges_1965_NGC1275}. Their energy estimate for that outburst, up to $\sim 10^{59} \, {\rm erg}$, was remarkably similar to what we infer today from the Perseus Cluster's X-ray cavities (see \S \ref{sec:calorimetry}). 

\begin{figure}[!t]
\centering
\includegraphics[width=5.3in]{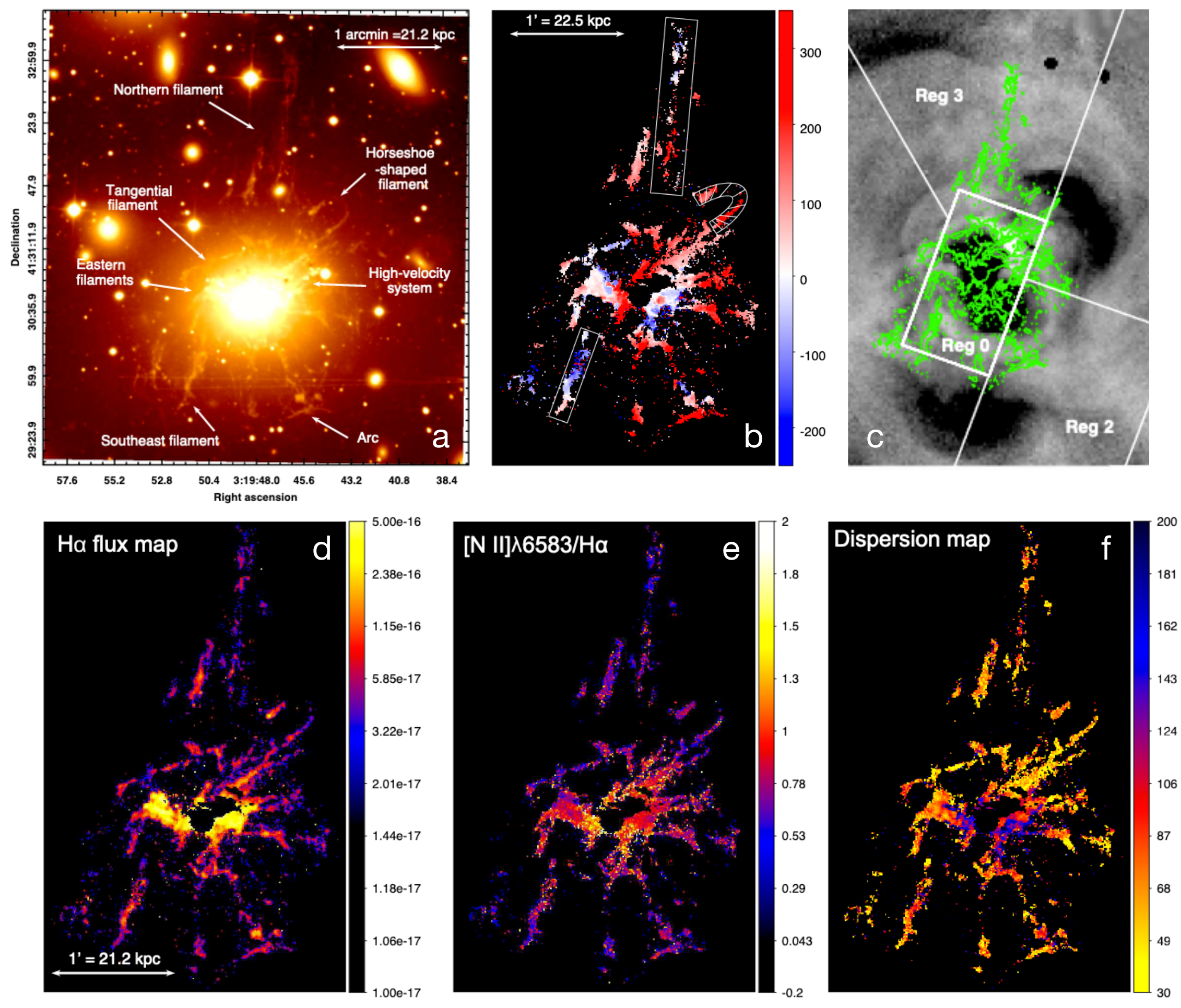}
\caption{Six views of the emission-line nebula in NGC 1275 (the Perseus Cluster's BCG) from \citet{Gendron-Marsolais_2018MNRAS.479L..28G}: (a) Narrow-band image including the H$\alpha$ emission line, with major morphological features annotated. (b) Line-of-sight velocity field of the nebula, relative to the galaxy's center of mass, measured by the SITELLE Fourier transform spectrometer on the Canada-France-Hawaii Telescope. (c) Nebular surface-brightness contours superimposed on a \textit{Chandra} X-ray image showing some of the Perseus Cluster's cavities. (d) H$\alpha$ surface-brightness map from SITELLE. (e) Map showing the intensity of the low-ionization [N II] 658.3 nm forbidden line, relative to H$\alpha$. (f) Line-of-sight nebular velocity dispersion measured with SITELLE. 
\label{figure:NGC1275_nebula_SITELLE}}
\end{figure}

Classic BCG nebulae, like the one in NGC1275 (Figure \ref{figure:NGC1275_nebula_SITELLE}), feature low-ionization forbidden lines ([OII], [NII], [OI], [SII]) that are unusually strong compared to the ones observed in the more typical ``H II region" ionization nebulae energized by UV light from young stars. In the classic Baldwin-Phillips-Terlevich (BPT) diagrams used for classifying extragalactic nebulae  \citep{BPT_diagrams}, the observed ratios of forbidden-line to hydrogen recombination-line intensity place BCG nebulae near the boundary between normal star-forming galaxies and the LINER (Low Ionization Nuclear Emission Region) class usually associated with low-luminosity AGNs \citep{Heckman_LINERs_1980A&A....87..152H}. As the Burbidges recognized in 1965 \citep{Burbidges_1965_NGC1275}, those emission-line ratios imply more heat input per ionizing photon than UV light from a young stellar population can supply, and therefore indicate the presence of at least one additional heat source (see also \citep{jfn87,vd97}). Also, some of the line-emitting filaments have no obvious nearby star formation \citep{Hatch_2006MNRAS.367..433H,Sparks_2009ApJ...704L..20S,Johnstone_2012MNRAS.425.1421J}.

\begin{figure}[!t]
\centering
\includegraphics[width=5.3in]{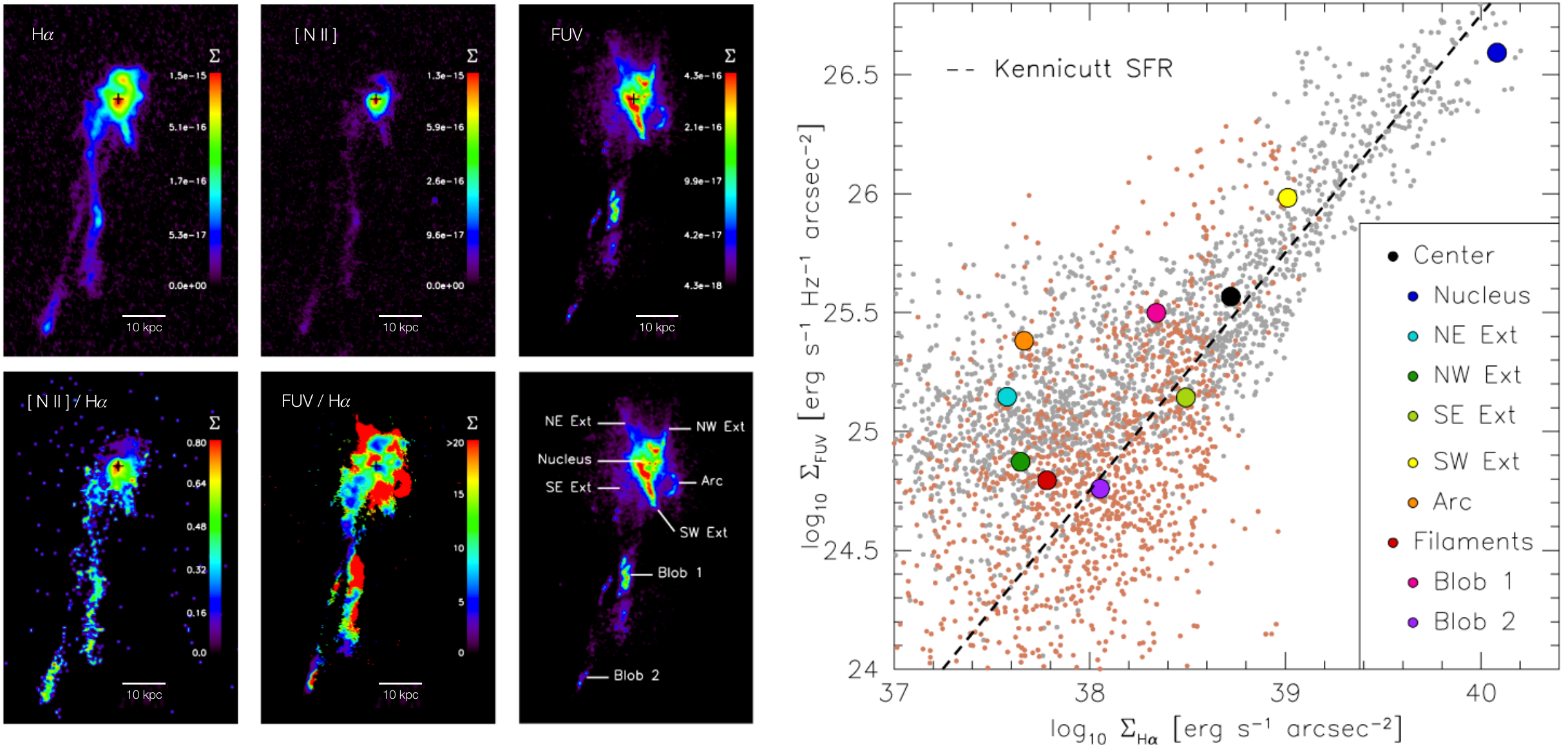}
\caption{Spatial relationships between nebular emission-line intensity and far-UV emission from young stars observed in the BCG of galaxy cluster Abell 1795, from \citet{McDonald_A1795_2009ApJ...703L.172M}.  The six panels on the left, proceeding clockwise from the top left, show H$\alpha$ surface brightness ($\Sigma_{{\rm H}\alpha}$), [N II] surface brightness, far-UV surface brightness from young stars ($\Sigma_{\rm FUV}$), a labeled FUV map showing particular morphological features, a map of the FUV/H$\alpha$ surface-brightness ratio, and a map of the [N II]/H$\alpha$ line ratio.  The right hand panel shows how $\Sigma_{{\rm H}\alpha}$ correlates with $\Sigma_{\rm FUV}$.  A dashed line shows the general trend in $\Sigma_{\rm FUV}/\Sigma_{{\rm H}\alpha}$ found by \citet{Kennicutt_1998ARA&A..36..189K} among normal star-forming galaxies.  The nebula in Abell~1795 generally follows the trend, but local deviations from it become large at low surface-brightness levels. 
\label{figure:Abell1795_MMTF}}
\end{figure}

Even though young stars cannot be the sole energy source, the overall luminosity of H$\alpha$ emission from a BCG nebula still strongly correlates with other indicators of star formation, both within a particular BCG nebula (see Figure~\ref{figure:Abell1795_MMTF}) and across the entire BCG population. Those indicators include excess blue light \citep{jfn87,Cardiel+95,Cardiel+98,Bildfell+08}, excess UV light \citep{Donahue_2010ApJ...715..881D,Hicks_2010ApJ...719.1844H,Odea_2010ApJ...719.1619O,Hoffer+2012ApJS..199...23H}, and far-infrared luminosity \citep{ODea+08,Edge_2010A&A...518L..47E,Rawle+2012ApJ...747...29R}.  In BCGs with modest star formation rates ($\dot{M}_* \sim 1$--$10 \, M_\odot \, {\rm yr}^{-1}$), the overall scaling relations among star-formation properties are typical of those found in star-forming disk galaxies.  And the scaling relations in BCGs with larger star-formation rates are more like those in luminous starburst galaxies \citep{Donahue+2011ApJ...732...40D}.

Exactly what powers the line emission from BCG nebulae remains an open question.\footnote{Frustratingly so to the co-authors of this review, whose first joint publication proposed an unsuccessful (or at best incomplete) hypothesis more than three decades ago \cite{vd90}}  Photoelectric heating by recently formed, hot, massive stars obviously contributes but sometimes is not the dominant energy source. Other physical processes that have been considered include X-ray photons emitted by the hot gas \citep{jfn87,vd90,VoitDonahueSlavin_1994ApJS...95...87V}, extreme UV radiation from intermediate temperature mixing layers surrounding the filaments \citep{BegelmanFabian_1990MNRAS.244P..26B}, suprathermal particle heating such as electron or proton mediated thermal conduction \citep{Sparks_1992ApJ...399...66S,Ferland_2009MNRAS.392.1475F}, and cosmic rays \citep[e.g.,][]{Ruszkowski_2018ApJ...858...64R}.  Then there are variants that posit combinations of slow shocks with some of these other energy sources \citep[e.g.,][]{McDonald_2012ApJ...746..153M}. 

Nearly all of the proposed non-stellar power sources boil down to the same generic solution: energy likely flows from the ambient plasma into the colder, more condensed nebular gas, with a flux similar to the product of energy density and sound speed in the ambient plasma \citep{vd97}.  Hot thermal plasma and relativistic plasma are both adjacent to the nebulae.  As long as those energy sources infuse the nebular gas with more thermal energy per ionization event than photoelectric heating from young stars, then the optical spectroscopic signatures predicted by speculative models of those mechanisms are nearly indistinguishable from each other. Complementary constraints are therefore needed to distinguish among these possibilities.


\subsubsection{Dust and Warm H$_2$ in BCG Nebulae}
\label{sec:multiphase_dust}

At first glance, the nebular gas in BCGs would seem to be a natural byproduct of atmospheric cooling.  There is plenty of hot gas with $t_{\rm cool} < 1 \, {\rm Gyr}$ in the cores of those clusters.  And gas cooling through the $\sim 10^4$~K temperature range produces UV and optical emission lines.\footnote{However, the observed emission-line luminosities are much greater than one would expect from a pure cooling flow, even at the inflated cooling rates inferred from simple steady cooling-flow models of the X-ray observations \citep[e.g., ][]{Heckman_1989}.}  

One clue to the origin of the nebular gas in BCGs comes from observations showing that it contains dust particles \citep{Sparks_1989ApJ...345..153S,DonahueVoit_1993ApJ...414L..17D} along with H$_2$ and CO \citep{1994ASSL..190..169E,2000ApJ...545..670D}. Those molecules would not form under the prevailing conditions without dust particles to catalyze molecule formation. But the dust particles themselves are not expected to survive much longer than $\sim 1$~Myr in the neighboring X-ray emitting gas because of sputtering \citep{DraineSalpeter_1979ApJ...231...77D}.  Instead of coming from the hot gas, the dust is probably supplied by outflows from aging stars.  Its presence therefore demonstrates that BCG nebulae are part of a complex cycle, one that probably includes mixing of dusty gas ejected from stars with dust-poor gas condensing out of the hot atmosphere \citep{VoitDonahue2011ApJ...738L..24V}. 

\begin{figure}[!t]
\centering
\includegraphics[width=4.5in]{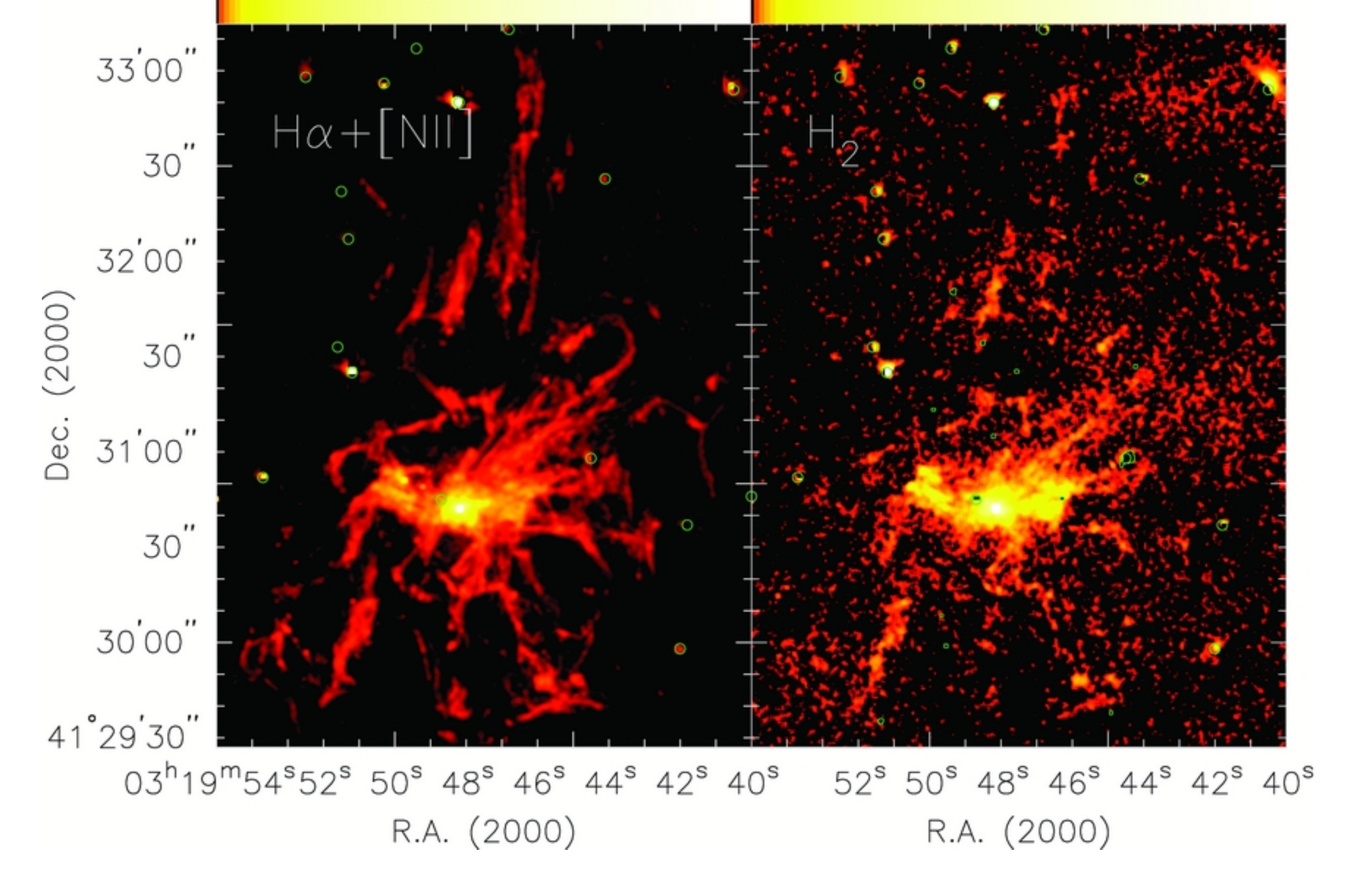}
\caption{Spatial relationships between optical H$\alpha$+[N II] line emission (left, \citep{2001AJ....122.2281C}) and vibrationally-excited H$_2$ molecules (right, \citep{2012ApJ...744..112L}) in the NGC 1275 nebula at the center of the Perseus Cluster. 
\label{figure:H2EmissionLines}}
\end{figure}

In fact, prominent optical nebulosity in and around a BCG reliably indicates that a BCG also contains a much larger amount of cold, dusty molecular gas revealed by radio observations of carbon monoxide and other molecules \citep{Edge_2001MNRAS.328..762E, 2017ApJ...844L..17W, 2019ApJ...879..103F, 2019MNRAS.489..349R, 2019A&A...631A..22O,2020A&A...640A..65C, 2021ApJ...909L..29D}.  The optical nebulae are likely to be thin surface layers on more massive molecular gas clouds.  For example, there is nearly a one-to-one match between BCGs that have H$\alpha$ emission and those that have collisionally-excited infrared line emission from H$_2$  \citep{1994ASSL..190..169E,1993ApJ...410L..19K,1997MNRAS.284L...1J,1998ApJ...494L.155F,2000ApJ...545..670D}. Also, infrared images of the vibrationally-excited molecular hydrogen closely correspond with the morphologies of the H$\alpha$ filaments (see Figure \ref{figure:H2EmissionLines}). 


\begin{figure}[!t]
  \centering
    \includegraphics[width=0.47\linewidth]{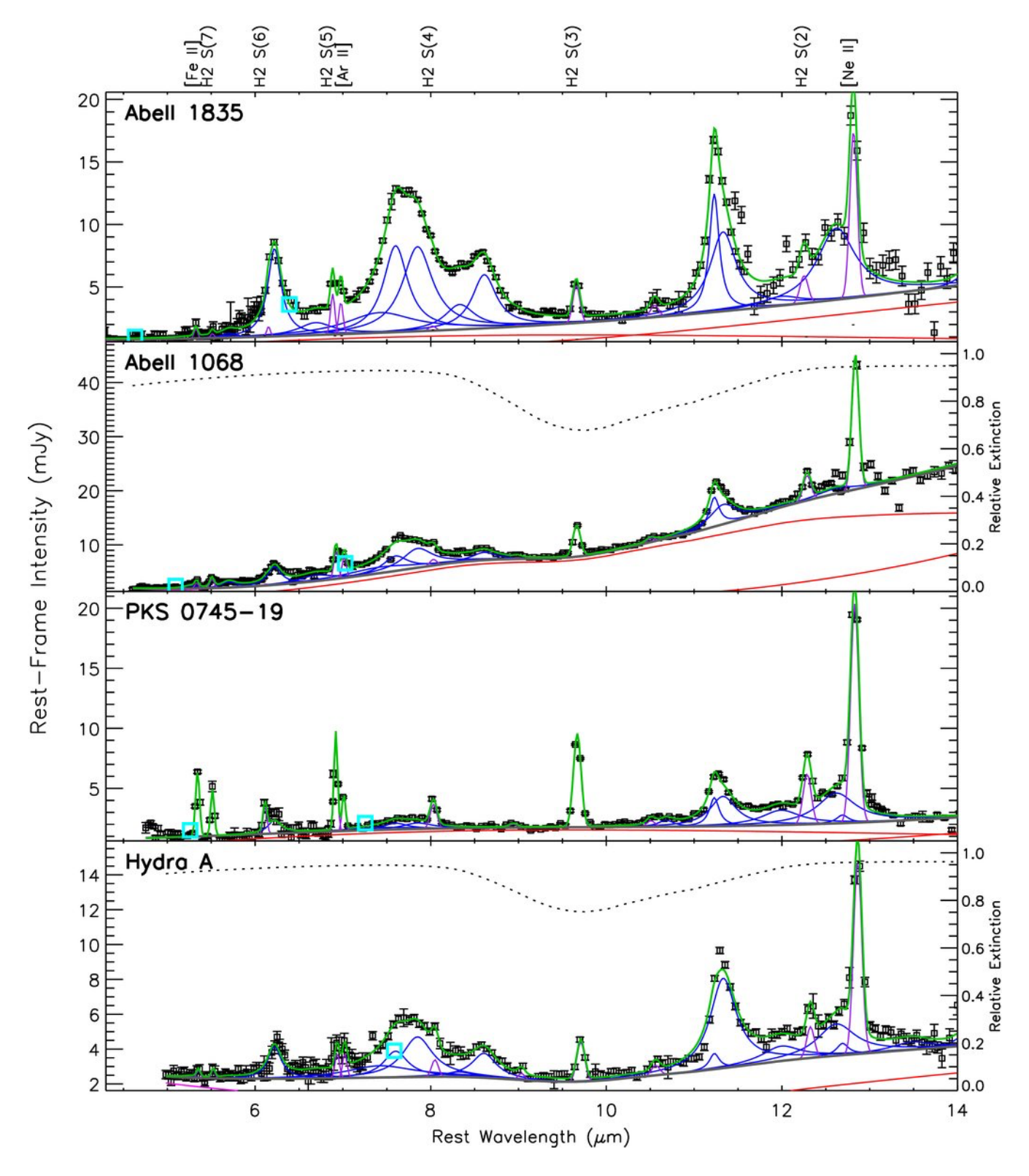}
  \hspace{0.1in}
    \includegraphics[width=0.47\linewidth]{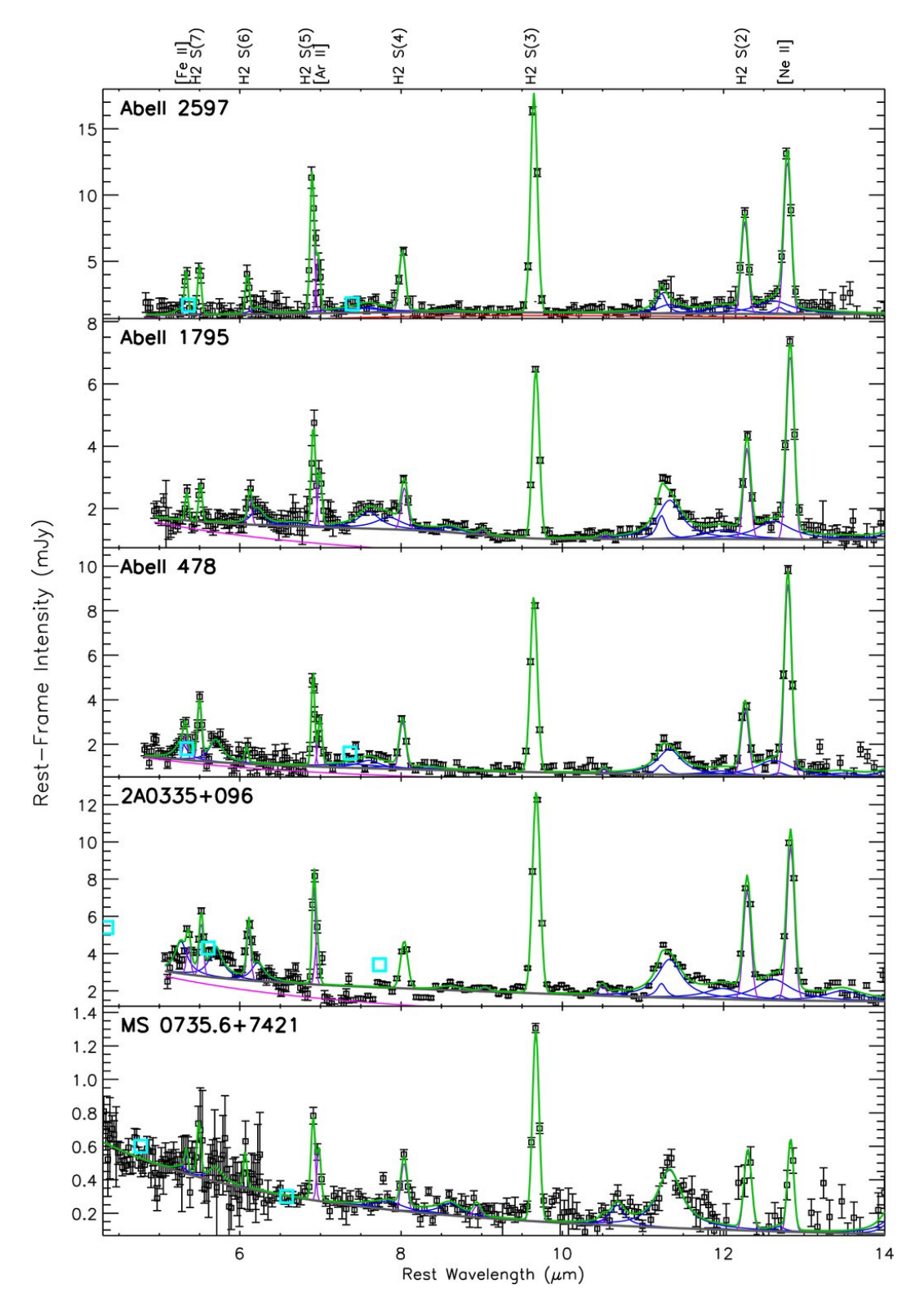}
\caption{Spitzer IRS Spectra of BCGs at 4.3--14 $\mu$m in nine cool-core galaxy clusters (from \citep{Donahue+2011ApJ...732...40D}).  Colored lines represent various components of the best-fitting spectrum: thermal dust emission (red), stellar continuum emission (magenta), the sum of dust and stellar continuum emission (gray), broad emission features from PAHs (blue), and unresolved fine-structure and molecular hydrogen emission lines (violet) labeled at the top.  A green line represents the combined spectral model, plotted over the observed rest-frame flux intensities. In the two cases where a silicate extinction feature may be present, a dotted line near the top of the frame, scaled to the right-hand axis, represents the extinction curve, which affects all components similarly.  Cyan squares represent complementary broad-band Spitzer/IRAC photometry. 
  \label{figure:SpitzerSpectra}}
\end{figure}

Mid-infrared spectra of BCG nebulae, observed at $\sim 5$--$30 \, \mu$m with the \textit{Spitzer} Space Telescope, are rich in spectral features demonstrating that the molecular gas in BCGs is quite similar to star-forming gas in lower-mass galaxies \citep{Donahue+2011ApJ...732...40D}.  Figure~\ref{figure:SpitzerSpectra} shows nine examples. Old stars dominate the optical continuum emission from BCGs but not the infrared spectra of BCGs with molecular gas.  Instead, most of the mid-IR continuum emission is from dust grains at a range of temperatures. 

Perhaps more surprising than the presence of dust is the presence of polycyclic aromatic hydrocarbons (PAHs), which are more fragile than dust grains and more easily destroyed by fast electrons and energetic photons, plenty of which are nearby. Nevertheless, the PAH emission features observed with \textit{Spitzer} are indistinguishable from those observed in less massive star-forming galaxies.  They also exhibit the same correlations with the overall star-formation rate, as do the thermal dust emission and fine-structure line emission \citep{Donahue+2011ApJ...732...40D}.

One might therefore be tempted to conclude that the cool clouds in BCGs are identical to those in other star-forming galaxies, except for being embedded within a gigantic population of old stars.  But at least one feature of the infrared spectra from those clouds is very different from those of star-forming galaxies and is still not understood: the prominent lines of rotationally-excited molecular hydrogen. For example, H$_2$ S(3) emission features near 10~$\mu$m are unusually prominent in all the spectra in Figure~\ref{figure:SpitzerSpectra}, regardless of whether the star formation rate is high or low. Furthermore, the mid-infrared emission line ratios of H$_2$ are inconsistent with gas at a single temperature, just like the vibrationally-excited H$_2$ lines around 2~$\mu$m. These two characteristics indicate that the H$_2$ gas in BCG nebulae spans a wide temperature range and is energized by processes other than star formation.  The James Webb Space Telescope (JWST) will be well suited to mapping that H$_2$ emission and helping us understand what is powering it.

\subsubsection{Molecular Gas in BCGs}
\label{sec:multiphase_H2}

\begin{figure}[!t]
\centering
\includegraphics[width=5.3in]{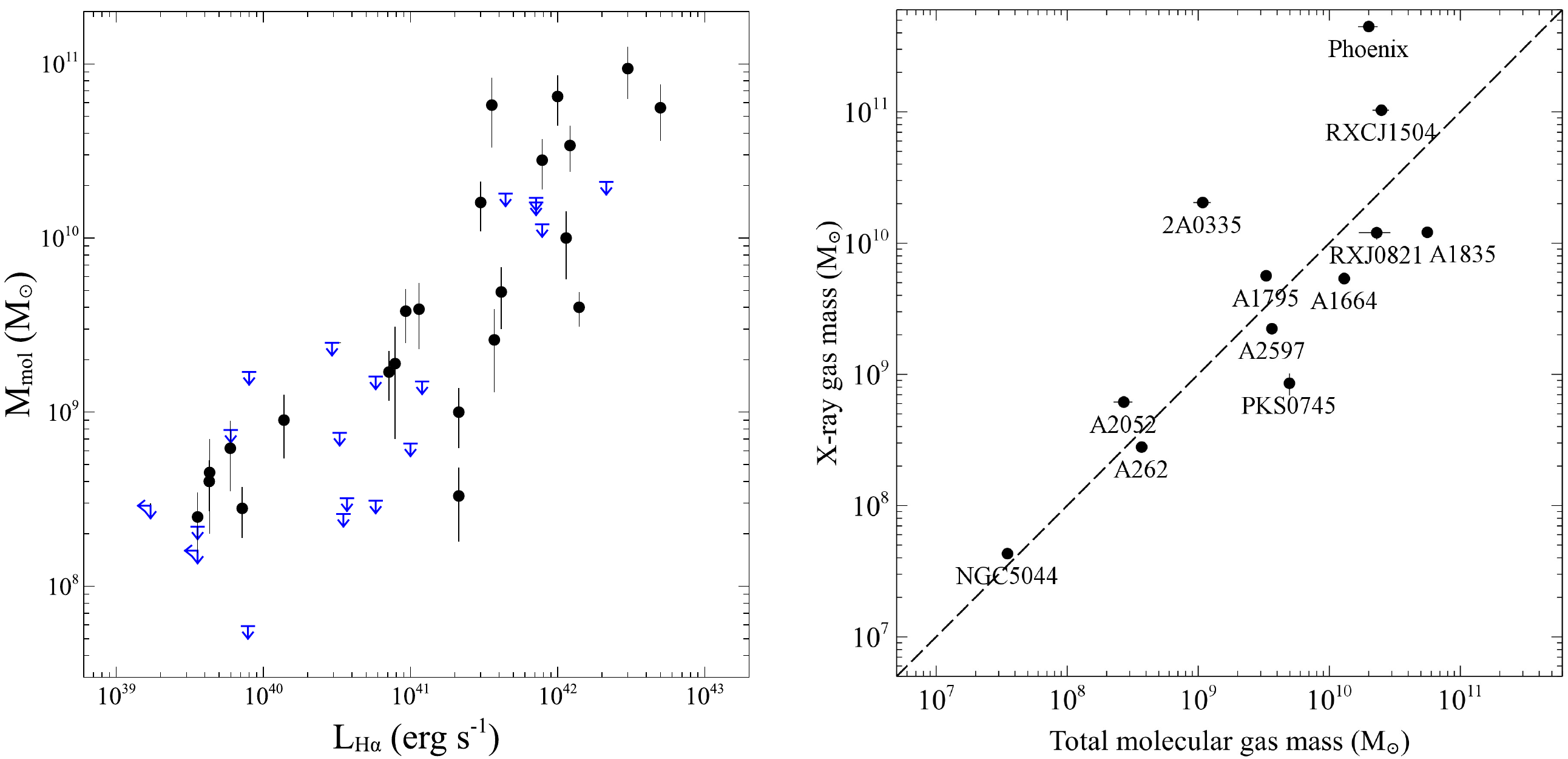}
\caption{Correlations of molecular gas mass observed in BCGs with H$\alpha$ luminosity (left) and cospatial X-ray emitting gas mass (right).  The left panel (from \citet{Pulido_2018ApJ...853..177P}) shows that the molecular gas mass inferred from CO emission ranges up to $10^{11} \, M_\odot$ and is highly correlated with the H$\alpha$ luminosity of the BCG nebula.  The right panel (from \citet{Russell_2019MNRAS.490.3025R}) shows that those molecular gas masses are comparable to the mass of the hot atmosphere within a similar radius.
\label{figure:BCG_H2_mass}}
\end{figure}

The mass of molecular gas associated with a BCG can be as large as $10^{11} \, M_\odot$, exceeding the entire stellar mass of the Milky Way galaxy (Figure~\ref{figure:BCG_H2_mass}).  It correlates strongly with the total H$\alpha$ luminosity of the BCG nebula and therefore also with the BCG's star formation rate.  In the most prodigious examples, star formation rates approach $10^3 \, M_\odot \, {\rm yr}^{-1}$, rivaling the most actively star-forming galaxies in the observable universe \citep{McDonald_Phoenix_2012Natur.488..349M}.  Most giant elliptical galaxies currently do not form any stars at all and stopped forming them billions of years ago.  This intriguing subpopulation of BCGs therefore provides some of the most important clues about how the baryon cycles of giant galaxies operate. The next two sections will consider those clues, first by looking at the connection between atmospheric cooling time and multiphase gas (\S \ref{sec:multiphase_tcool}) and then by examining the dynamics of the multiphase gas (\S \ref{sec:MultiphaseDynamics}).

\subsection{Multiphase Gas and Cooling Time}
\label{sec:multiphase_tcool}

Connections between atmospheric cooling time and multiphase gas in BCGs have been noted since the 1980s \citep{1981ApJ...250L..59H,1983ApJ...272...29C,Heckman_1989}.  For example, \citet{1985ApJS...59..447H} recognized that BCG nebulae appeared only in galaxy clusters with a cooling time less than the age of the universe ($\sim H_0^{-1}$), seemingly consistent with the notion that gas with a shorter cooling time would cool, condense, and collect at the center of the atmosphere.  

But as the spatial resolution of X-ray observations improved, the critical value of $t_{\rm cool}$ associated with multiphase gas and star formation in BCGs shifted to shorter timescales.  Using the color gradients of BCG starlight to trace star formation, \citet{Rafferty+08} showed that young stars were present only in BCGs with an atmospheric cooling time $\lesssim 1$~Gyr at 12~kpc from the center.  A contemporaneous study by \citet{Cavagnolo+08} of a larger set of BCGs (the ACCEPT sample \citep{Cavagnolo+09}) showed that detectable H$\alpha$ luminosity was present only in atmospheres with a central entropy level $K_0 \lesssim 30 \, {\rm keV \, cm^2}$, indicating $t_{\rm cool} \lesssim 1$~Gyr (see Figure~\ref{Figure:P-Kplane}).  In both cases the transition to a multiphase atmosphere is sharp, linking the presence of multiphase gas more closely with the central galaxy's dynamical time than with a cosmological cooling time scale.

\begin{figure}[!t]
  \centering
  \includegraphics[width=5.3in]{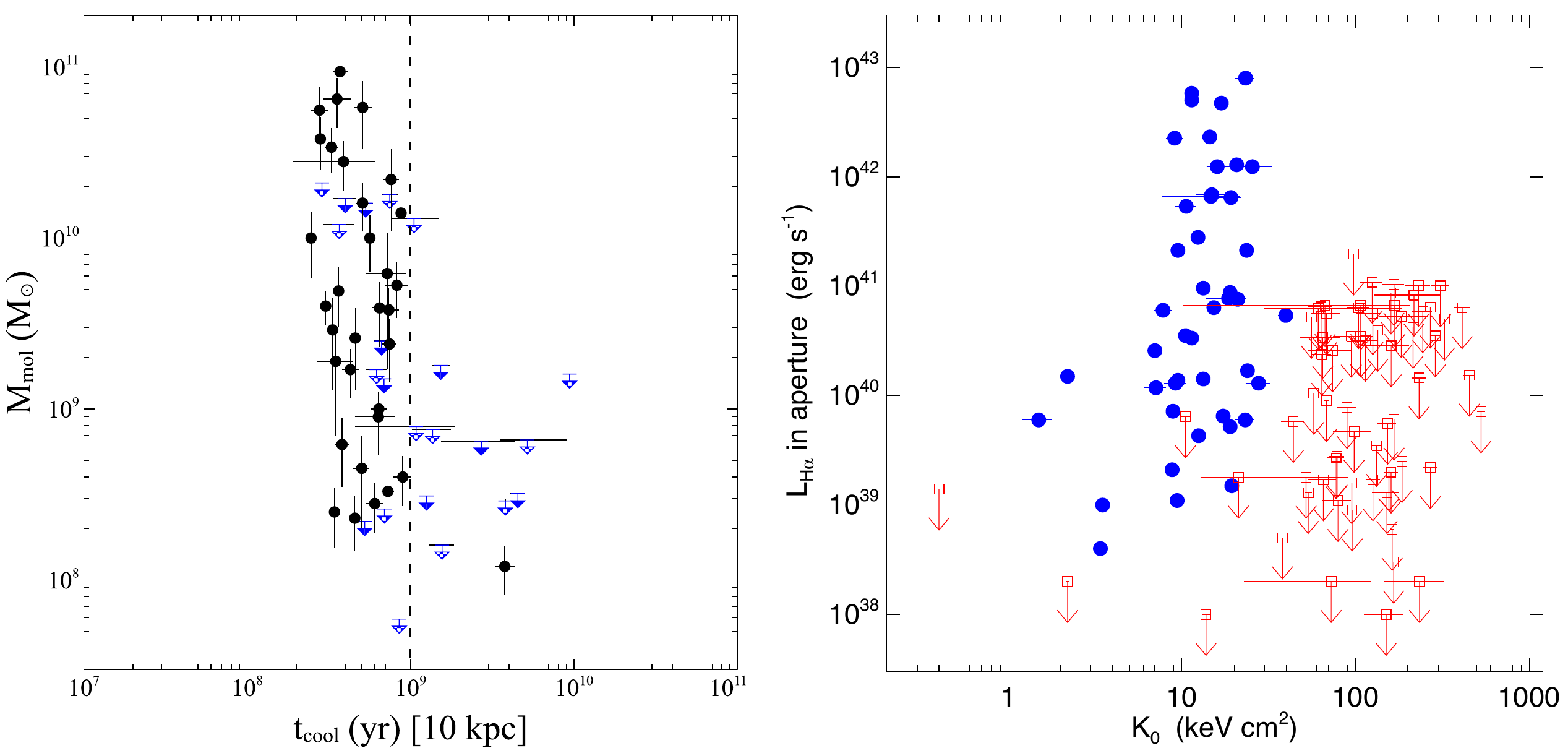}
\caption{Thresholds for the incidence of multiphase gas in BCGs. \textit{Left Panel:} Dependence of molecular gas mass on cooling time at 10~kpc (from \citet{Pulido_2018ApJ...853..177P}).  \textit{Right Panel:} Dependence of H$\alpha$ luminosity on central entropy level $K_0$ from the ACCEPT sample (update of \citep{Cavagnolo+08} presented in \citep{VoitDonahue2015ApJ...799L...1V}). The threshold values of $t_{\rm cool} (10 \, {\rm kpc}) \lesssim 1 \, {\rm Gyr}$ and $K_0 \lesssim 30 \, {\rm keV \, cm^2}$ are equivalent.
  \label{figure:tc_K0_threshold}}
\end{figure}

Figure~\ref{figure:tc_K0_threshold} shows two updated versions of those findings.  The left panel illustrates the relationship between molecular gas mass and the cooling time of gas 10~kpc from a BCG's center \citep{Pulido_2018ApJ...853..177P}.  The right panel illustrates how a BCG's H$\alpha$ luminosity is related to $K_0$.  In each case, the threshold is strikingly abrupt, more like a binary switch than a gradual transition.  Exceptions with low $t_{\rm cool}$ and low $K_0$ and no sign of multiphase gas are rare but potentially very interesting \citep{Martz_2020ApJ...897...57M}.  More importantly, the sharp thresholds found in studies like these imply a very strong link between star formation in a massive galaxy and the thermodynamics of its atmosphere.

\subsection{Dynamics of Multiphase Gas}
\label{sec:MultiphaseDynamics}

Now it is time to turn to the weather. Section~\ref{sec:BH_Heating} has already shown that energetic outflows fueled by black-hole accretion pump energy into the inner regions of a galaxy cluster's atmosphere.  They drive shocks that heat the atmosphere from below and inflate high-entropy bubbles that buoyantly rise.  Both phenomena can result in convective flows capable of entraining lower-entropy atmospheric gas and lifting it to greater altitudes.  Interactions between the central AGN and the atmosphere above it therefore produce observable outflows of multiphase gas, along with turbulence and circulation traced by the multiphase gas. In some respects, the atmospheric dynamics driven by AGN feedback in a massive galaxy resemble terrestrial weather patterns driven by solar heating of Earth's surface. The analogy should not be taken too literally but can provide useful conceptual guidance about how a BCG's baryon cycle operates.

\subsubsection{Multiphase Outflows}
\label{sec:MultiphaseOutflows}

\begin{figure}[!t]
  \centering
  \includegraphics[width=5.3in]{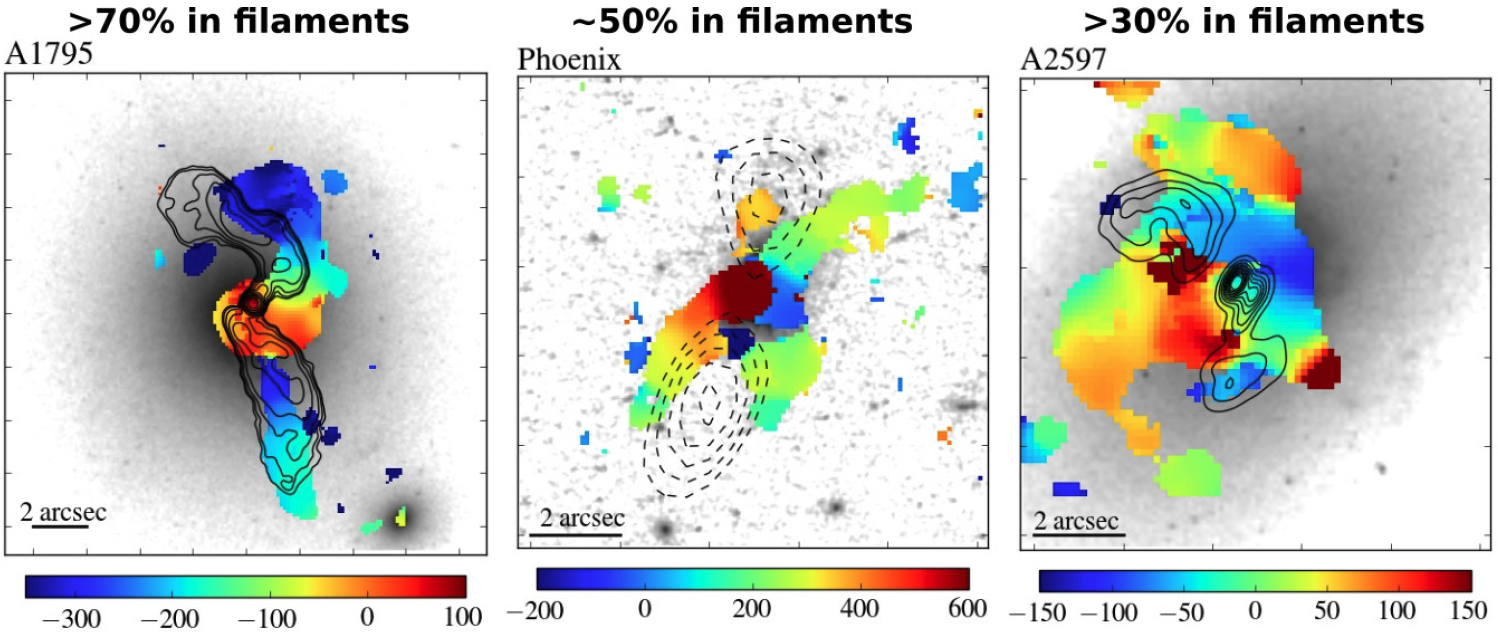}
\caption{ALMA velocity maps of molecular gas in BCGs (from \citet{Russell_2019MNRAS.490.3025R}). Line-of-sight velocity centroids are shown in color.  Grayscale \textit{Hubble} images beneath the ALMA maps show optical starlight.  Dotted lines show locations of X-ray cavities in \textit{Chandra} observations.  Solid contours show radio emission observed with the VLA.  Generally, molecular gas that is more filamentary exhibits clearer morphological and dynamical correspondences with X-ray cavities and radio outflows.
  \label{fig:Russell_montage_velocity}}
\end{figure}

Some of the most puzzling observations of multiphase gas dynamics in and around BCGs have come from the ALMA observatory, which can map CO emission from molecular gas with unprecedented spatial and spectral resolution \citep{McNamara_2014ApJ...785...44M,Russell_A1664_2014ApJ...784...78R,Russell_2019MNRAS.490.3025R,Olivares_2019A&A...631A..22O}.  One surprise has been the unexpectedly small velocity dispersion of the overall population of molecular gas clouds.  Much of a BCG's molecular gas mass has a velocity dispersion roughly half that of the BCG's stars, indicating that those dense gas clouds do not orbit ballistically in the BCG's gravitational potential.  Instead, the molecular clouds are either short lived, unable to significantly accelerate before disintegrating, or they are dynamically coupled to the hot atmosphere by drag forces, perhaps mediated by magnetic fields. Another surprise has been that molecular gas clouds with unusually large speeds spatially coincide with outflows of much hotter gas that appear to be lifting the molecular gas out of the BCG \citep{McNamara_2014ApJ...785...44M,Russell_A1664_2014ApJ...784...78R,Russell_2019MNRAS.490.3025R,Russell_2016MNRAS.458.3134R,Vantyghem_2A0335_2016ApJ...832..148V,Vantyghem_RXJ1504_2018ApJ...863..193V} (see Figure \ref{fig:Russell_montage_velocity}).

These features are surprising because the molecular clouds are $\gtrsim 10^5$ times more dense than the hot gas that appears to be pushing them around.  This density contrast is more than two orders of magnitude greater than that between rocks and air near Earth's surface.  And yet the BCG's ``air" seems to be levitating its ``rocks."

One plausible explanation for the link between uplift and molecular gas has emerged from theoretical considerations.  In numerical simulations of AGN feedback in galaxy cluster cores, rising high-entropy bubbles are able to induce condensation of lower-entropy ambient gas pulled up in their wakes \citep[e.g.,][]{Revaz_2008A&A...477L..33R,LiBryan2014ApJ...789..153L,Prasad_2015ApJ...811..108P}.  Condensation can happen if uplift is able to raise the lower-entropy gas to an altitude at which its cooling time is shorter than the freefall time back to its original altitude \citep[e.g.,][]{McNamara_2016ApJ...830...79M,Voit_2017_BigPaper}.  Then the entropy contrast between the uplifted gas and its surroundings has a chance to become large before descent of the uplifted gas returns it to a layer of equivalent entropy.  The terrestrial analogy here is to a thunderstorm, in which raindrops condense as updrafts of humid gas adiabatically cool below their dewpoint.  

In a BCG nebula, molecule formation may happen as a consequence of uplift, meaning that the low-density outflows are not necessarily accelerating the molecular clouds themselves. However, condensation of previously hot gas is unlikely to be all that is happening, given the presence of dust grains in the putative condensates.  Dust can grow in condensed gas that is sufficiently cold and dense, but rapid incorporation of refractory elements into solid-state grains requires some pre-existing dust grains to act both as sites for grain growth through nucleation and as catalysts for molecule formation.  The presence of dust therefore implies that the nebular gas has been seeded with dust grains that originated in outflows from the BCG's stars and have survived long enough in the galaxy's atmosphere to act as catalysts for further dust formation \citep{VoitDonahue2015ApJ...799L...1V}.  In that context, it is worth noting that any dust grains managing to survive within the hot gas greatly increase its radiative cooling rate \citep{SilkBurke_1974ApJ...190...11S} and can speed the process of condensation until sputtering destroys them.

\subsubsection{Multiphase Turbulence}
\label{sec:MultiphaseTurbulence}
Asymmetric outflows and uplift inevitably drive a certain amount of atmospheric turbulence \citep[e.g.,][]{WittorGaspari_2020MNRAS.498.4983W}.  Observations of galaxy-cluster atmospheres indicate that the turbulence is subsonic, with a turbulent Mach number in the $\sim 0.1$--0.3 range.  Direct X-ray observations of turbulent speeds are rare and will remain so until the XRISM mission is successfully launched (\S \ref{sec:turbulence}). Indirect constraints on turbulent speeds for a larger number of cluster atmospheres come from analyses of surface-brightness fluctuations and assume that those irregularities stem from electron-density fluctuations produced by turbulence \citep{Zhuravleva_2014Natur.515...85Z,Zhuravleva_2015MNRAS.450.4184Z,Zhuravleva_2016MNRAS.458.2902Z,Zhuravleva_2018ApJ...865...53Z}.  Those indirect turbulence measurements are consistent with the direct observations, but the observational uncertainties are large, because disturbances other than turbulence can also produce surface-brightness fluctuations.

\begin{figure}[!t]
  \centering
  \includegraphics[width=5.3in]{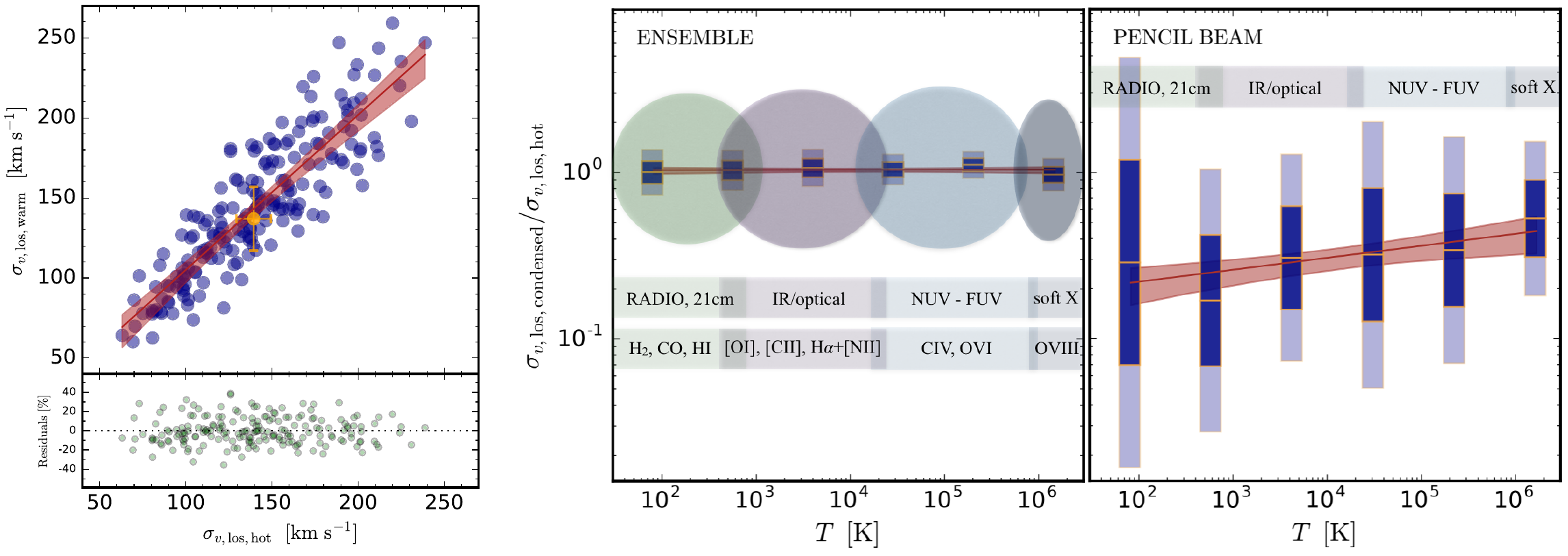}
\caption{Line-of-sight velocity dispersion ($\sigma_{v,{\rm los}}$) relationships among different gas phases in a simulated galaxy-cluster core (adapted from \citet{Gaspari_2018ApJ...854..167G}). \textit{Left panel:} Correlation of $\sigma_{v,{\rm los}}$ in the warm phase ($\sim 10^4$~K gas) and the hot phase ($\sim 10^7$~K) for the entire ensemble of warm clouds.  Blue points show instantaneous relationships at different moments during a single simulation.  A red line shows the best linear fit with a shaded uncertainty region, with residuals along the bottom. Velocity dispersions in both phases rise and fall together as AGN feedback fluctuates.  An orange circle with error bars shows the relationship observed in the Perseus cluster core, using Hitomi (Figure~\ref{figure:HitomiSpectrum}) and SITELLE (Figure~\ref{figure:NGC1275_nebula_SITELLE}). \textit{Right Panels:} Ratios of cool phase to hot phase velocity dispersion as a function of cool-phase temperature for the entire ensemble of cool clouds (left) and along pencil beams through the simulation volume (right).  Restricting the aperture to a pencil beam generally reduces that ratio, because the beam's intersections with cool clouds often result in a sampled length scale much smaller than the sampled path length through the hot gas.
  \label{fig:GaspariSnowGlobes}}
\end{figure}

It is also tempting to infer the level of turbulence in the hot atmosphere from observations of denser multiphase gas components embedded within it.  Observations show that line-of-sight speeds measured from Doppler shifts of the H$\alpha$ line tracing $10^4$~K gas correlate well with speeds inferred from Doppler shifts of molecular lines tracing much denser gas \citep{Tremblay_A2597_2018ApJ...865...13T,2022arXiv220107838O}.  Likewise, the velocity dispersions of differing gas phases observed in idealized numerical simulations of AGN feedback exhibit similarly strong correlations \citep[][see Figure~\ref{fig:GaspariSnowGlobes}]{Gaspari_2018ApJ...854..167G}.  In simulations, at least, the velocity dispersion of the ensemble of molecular clouds associated with a BCG nebula reflects the velocity dispersion of the hot gas, which is about half the stellar velocity dispersion.  Those findings indicate a close dynamical connection between the molecular clouds and the hot gas, despite the great difference in gas density.

Simulations also demonstrate that the vertical motions associated with turbulence promote multiphase condensation in essentially the same way that bulk uplift promotes it \citep{Gaspari+2013MNRAS.432.3401G,Gaspari_2015A&A...579A..62G,Gaspari_2017MNRAS.466..677G}.  Turbulent motions strong enough to lift a parcel of low-entropy gas to an altitude at which it can cool before descending back to its original altitude can stimulate development of a multiphase medium.  For that to happen, the turnover time of a large turbulent eddy needs to be comparable to the cooling time.  Turbulence that is too strong shreds and mixes incipient condensates before they develop much contrast \citep{BanerjeeSharma_2014MNRAS.443..687B}, and turbulence that is too weak fails to overcome the buoyancy effects that suppress condensation, which will be discussed in \S \ref{sec:BuoyancyDamping}.  In models of stratified galactic atmospheres, the level of turbulence most favorable for promoting multiphase condensation has a velocity dispersion approximately half that of the galaxy's stars \citep{Gaspari+2013MNRAS.432.3401G,Voit_2018ApJ...868..102V}, similar to what is observed.

When measuring turbulence and making line-width comparisons among different gas phases, it is important to recognize that the observed line widths may depend strongly on aperture and gas temperature.  A pencil beam through a cluster core intercepts volume-filling hot gas through the entire core, but its intersections with cold gas clouds may sample much smaller length scales.  Consequently, the turbulent velocity dispersion of the cold gas along a pencil beam can be substantially smaller than that of the hot gas along the same line of sight, as shown in the right panel of Figure~\ref{fig:GaspariSnowGlobes}. 

\begin{figure}[!t]
  \centering
  \includegraphics[width=5.3in]{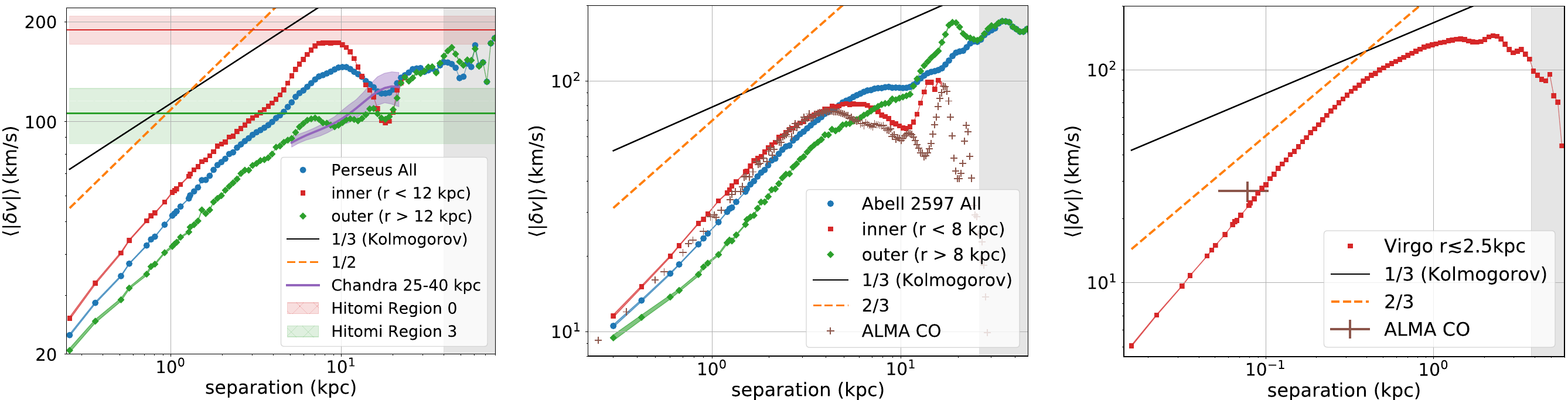}
\caption{Velocity structure functions of BCG nebulae in the Perseus Cluster (left), the cluster Abell 2597 (middle), and the Virgo Cluster (right), measured from H$\alpha$ lines by \citep{Li_2020ApJ...889L...1L}. Red, green, and blue points show the mean absolute velocity difference $\langle | \delta v | \rangle$ as a function of transverse separation $l$ in the plane of the sky for the inner part, the outer part, and the full BCG nebula, respectively.  In each panel, a solid gray line shows the power-law slope $\langle | \delta v | \rangle \propto l^{1/3}$ characteristic of Kolmogorov turbulence, and a dotted orange line shows $\langle | \delta v | \rangle \propto l^{2/3}$.  All three BCG nebulae have velocity structure functions considerably steeper than the Kolmogorov slope, indicating that the turbulence does not follow Kolmogorov scaling.
  \label{fig:Li_Velocity_Structure}}
\end{figure}

One method for probing how turbulent velocities depend on length scale in a galaxy cluster core is to measure the velocity structure function of H$\alpha$ emission lines from a BCG nebula. Figure~\ref{fig:Li_Velocity_Structure} shows some examples. The function $\langle | \delta v (l) | \rangle$ is the mean of the absolute velocity differences among lines of sight separated by a projected distance $l$ on the plane of the sky.  Its peak indicates the scale on which turbulence is being driven and is consistent with a driving scale comparable to the radii at which X-ray cavities are observed.  Its slope toward smaller separations reflects how the kinetic energy imparted by the cavities cascades toward smaller scales and dissipates.  In each of the cases depicted, the velocity structure function's slope becomes steeper than the slope expected from Kolmogorov scaling, indicating that turbulence is not steadily cascading through a Kolmogorov hierarchy of eddies.  Instead, the steeper slope may be indicating a gravity-wave cascade resulting from nonlinear entropy perturbations (see \citep{WangRuszkowski_2021MNRAS.504..898W} and \S \ref{sec:NonlinearCoupling}).

\subsubsection{Multiphase Circulation}
\label{sec:Circulation}

\begin{figure}[!t]
  \centering
  \includegraphics[width=5.3in]{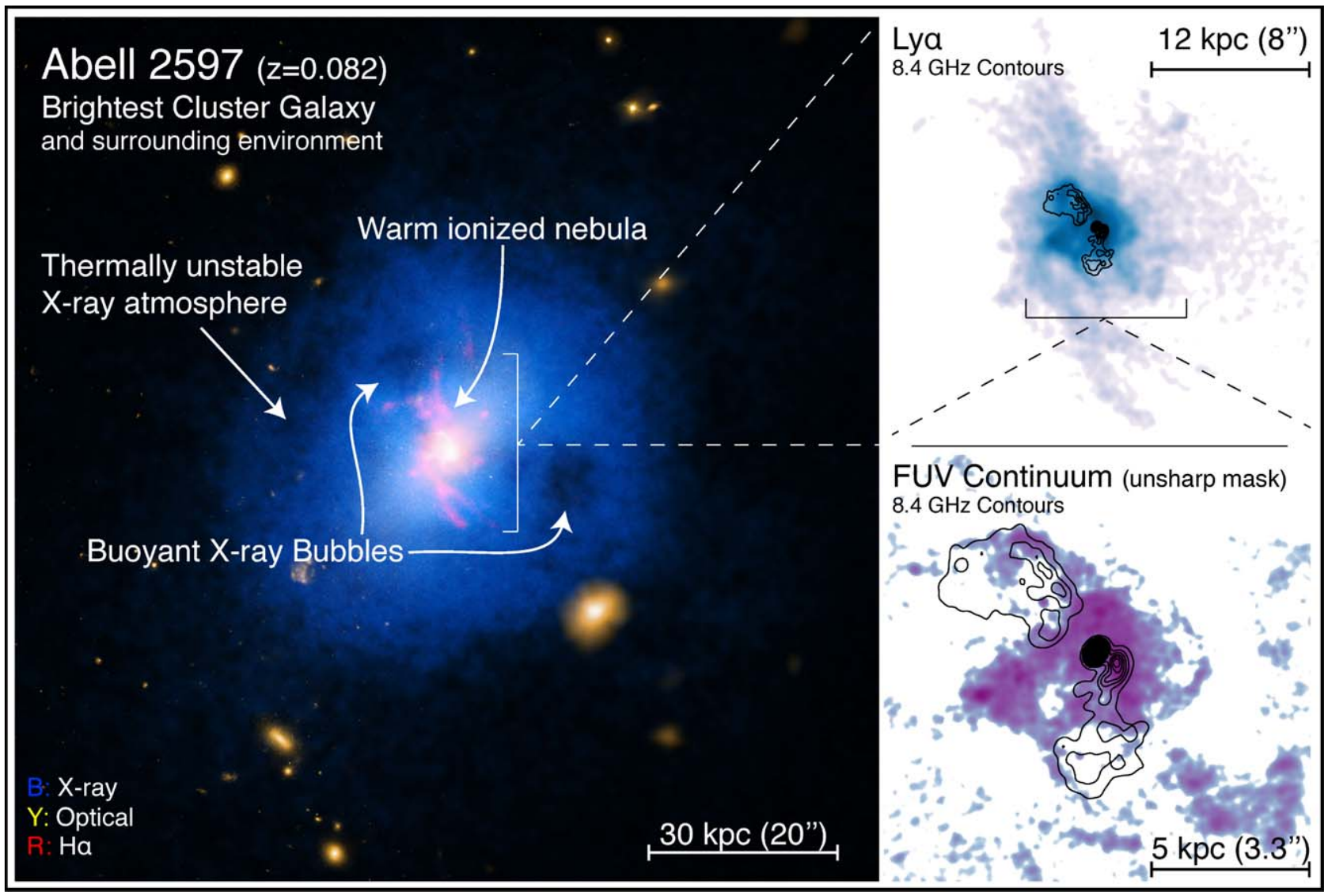}
\caption{Multiwavelength views of the circulation cycle in and around the BCG of galaxy cluster Abell 2597 (from \citet{Tremblay_A2597_2018ApJ...865...13T}).  The left panel shows X-ray emission from hot gas (blue) and H$\alpha$ emission from the BCG nebula (red) superimposed on galactic starlight (yellow).  The upper right panel zooms closer into the BCG, showing Ly$\alpha$ emission from the BCG nebula (grayscale) and radio emission from relativistic electrons in the AGN outflow (contours).  The lower right panel zooms in even closer, showing how far-UV emission from young stars (purple) traces the edges of the cavities inflated by the radio-emitting plasma (contours). 
  \label{fig:Tremblay18_Fig1}}
\end{figure}

Both uplift and turbulence are features of a broader ``weather" pattern in and around BCGs.  Bipolar outflows from the central AGN drive the pattern by inflating buoyant cavities (see Figure~\ref{fig:Tremblay18_Fig1}).  As those cavities rise, they lift lower-entropy gas outward along the outflow axis.  Equatorial gas must then circulate inward to replace the uplifted gas.  Meanwhile, cold clouds associated with the uplifted gas can rain back down toward the center once the hydrodynamic phenomena that lifted them subside. All of these bulk motions can generate sound waves, internal gravity waves, and turbulence.

Many aspects of the overall circulation cycle are evident in multiwavelength observations of the BCG in the galaxy cluster Abell 2597 \citep{Tremblay2016,Tremblay_A2597_2018ApJ...865...13T}.  In the image on the left side of Figure~\ref{fig:Tremblay18_Fig1}, one can see two obvious cavities in the X-ray emitting atmosphere, approximately 15~kpc from the cluster's center.  A plume of H$\alpha$ emission from the BCG nebula trails the cavity to the upper left. The association between the right-hand cavity and the nebular gas is less obvious, but the image on the upper right shows that Ly$\alpha$ emission from the nebular gas is aligned with radio emission from relativistic plasma ejected by the central AGN, presumably in an outburst more recent than the one that excavated the outer cavities.  On the lower right is an image showing a more obvious relationship between the nebular gas and the AGN outflow. Far-UV starlight tracing recent star formation outlines the edges of radio-emitting plasma blobs within the central 5~kpc, and star formation implies the presence of molecular gas clouds.  The morphological correspondence between star formation and the AGN outflow is therefore consistent with the possibility that uplift stimulates both multiphase gas condensation and star formation.  

\begin{figure}[!t]
  \centering
  \includegraphics[width=5.3in]{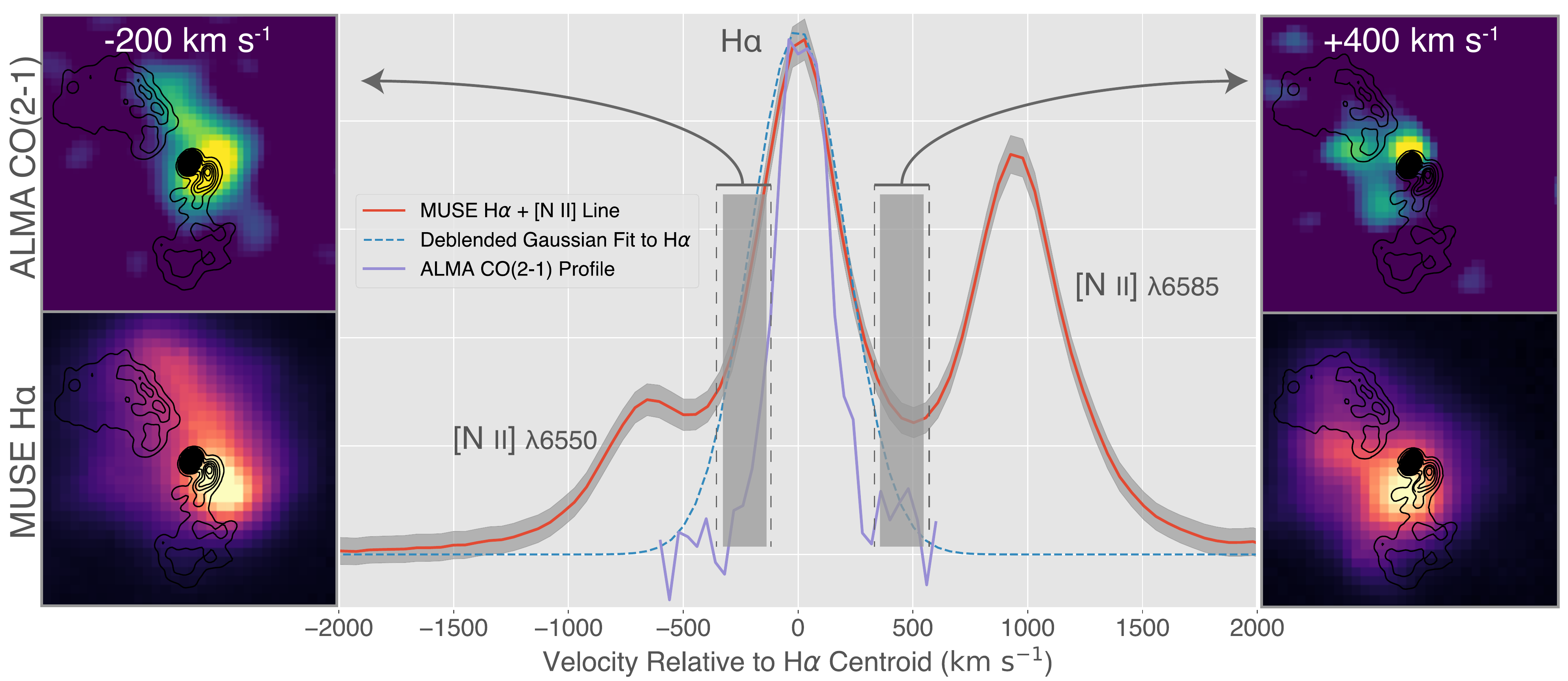}
\caption{Mapping of the morphological and kinematic relationships between multiphase gas components and the AGN outflow at the center of Abell 2597 (from \citep{Tremblay_A2597_2018ApJ...865...13T}).  The middle panel shows both optical MUSE and radio ALMA spectra of the central region. A solid red line represents the H$\alpha$+[N II] emission-line blend, and a dotted blue line represents a Gaussian fit to just the H$\alpha$ component.  A purple line shows the CO (2--1) emission line observed with ALMA. Images along the sides of the middle panel show how the morphologies of fast-moving gas components are related to the radio-emitting plasma (contours).  To the left are gas components moving toward us at approximately 200 km s$^{-1}$, relative to the line centroid.  To the right are gas components moving away from us at approximately 400 km s$^{-1}$ in the same inertial frame.
  \label{fig:Tremblay18_musealmaspec}}
\end{figure}

ALMA observations of Abell 2597's central region reveal the morphologies and motions of the molecular gas more directly.  Figure~\ref{fig:Tremblay18_musealmaspec} focuses on the gas components moving at unusually high speeds. Its central panel shows the overall profiles of the H$\alpha$ and CO (2--1) lines from the central region.  Their cores are narrower than the stellar velocity dispersion ($\sim$ 250--450 km s$^{-1}$), with a Gaussian velocity dispersion of $240 \pm 11 \, {\rm km \, s^{-1}}$ for the H$\alpha$ gas and $107 \pm 6 \, {\rm km \, s^{-1}}$ for the molecular gas.  However, both those lines have non-Gaussian wings produced by higher velocity gas.  Mapping just the high-velocity gas, as in the side panels, shows that both the H$\alpha$ and molecular components trace the edges of the AGN outflow, signaling a dynamical connection between AGN activity and acceleration of multiphase gas, presumably including uplift.  A precise uplift speed is difficult to infer from these data, given the uncertain projection corrections needed to account for motion in the plane of the sky.

\begin{figure}[!t]
  \centering
  \includegraphics[width=5.3in]{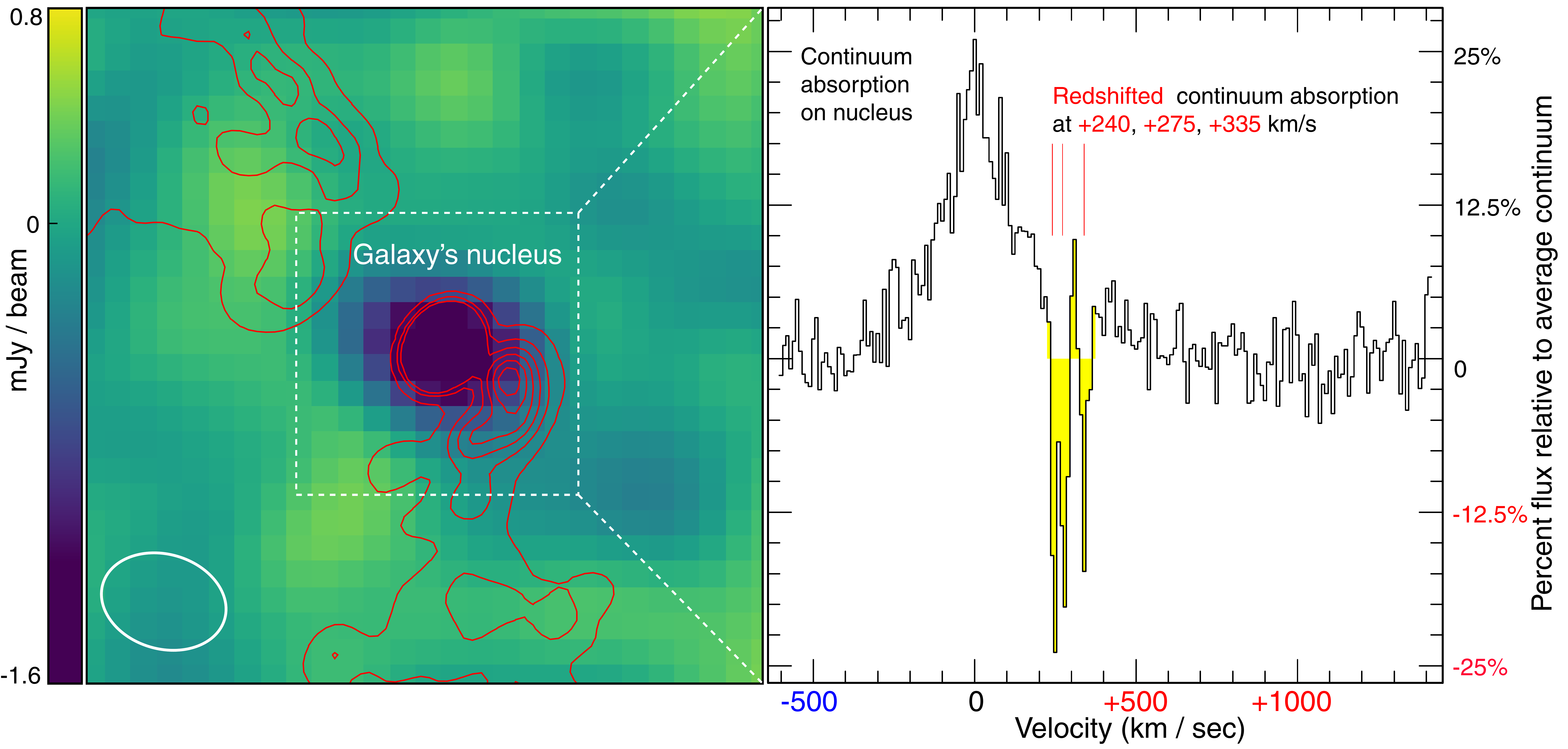}
\caption{ALMA absorption-line observations revealing three molecular clouds falling toward the central black hole in Abell 2597 (from \citet{Tremblay2016,Tremblay_A2597_2018ApJ...865...13T}).  On the left is a narrow-band CO emission intensity map with superimposed radio continuum emission contours in red.  A white ellipse represents the ALMA beam. On the right is a CO spectrum of the portion of the image containing the AGN, showing three distinct absorption lines corresponding to molecular clouds falling inward at 240 km s$^{-1}$, 275 km s$^{-1}$, and 335 km s$^{-1}$.
  \label{fig:Tremblay_A2597_inflow}}
\end{figure}

 The most remarkable gas motions revealed by the ALMA observations of Abell 2957 require no projection corrections, because they are measured from absorption along the line of sight to the central black hole.  Using the radio continuum emission from the black hole's vicinity as a backlight, \citet{Tremblay2016} have measured the motions of three molecular clouds between us and the black hole.  Figure~\ref{fig:Tremblay_A2597_inflow} shows the result.  All three CO (2--1) absorption-line detections come from clouds moving toward the black hole at speeds ranging from 
240 km s$^{-1}$ to 335 km s$^{-1}$.  Assessing the probability of seeing three such clouds along the same line of sight implies that they are likely to be less than a few hundred parsecs from the AGN, meaning that they are a plausible fuel source for future AGN outbursts.

\begin{figure}[!t]
  \centering
  \includegraphics[width=5.3in]{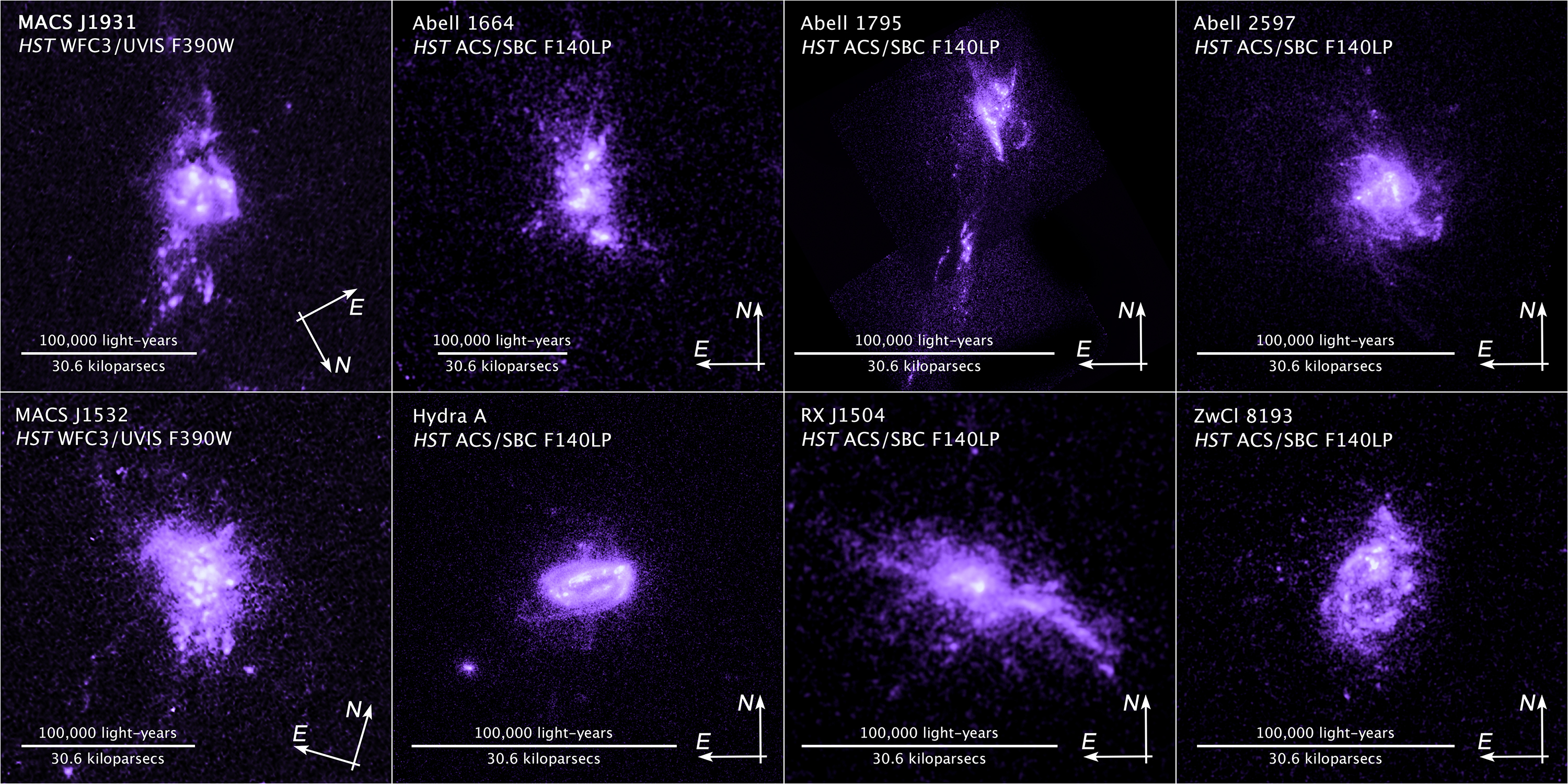}
\caption{Ultraviolet images from \textit{Hubble} showing eight examples of extended star formation in the cores of galaxy clusters, including Abell 2597 (from \citep{Donahue_UV_2015ApJ...805..177D,Tremblay_UV_2015MNRAS.451.3768T}).  Except for Hydra A, star formation is not restricted to an organized rotating disk. Instead, it is chaotically spread over regions approaching 100,000 light-years across, often aligned with an AGN outflow, suggesting that star formation in BCGs is part of a circulation cycle driven by AGN feedback.
  \label{fig:UV_BCG_gallery}}
\end{figure}

Together, the observations in Figures \ref{fig:Tremblay18_Fig1} through \ref{fig:Tremblay_A2597_inflow} are consistent with a circulation cycle that is driven by AGN feedback and also provides fuel for AGN feedback.  The BCG in Abell~2597 is one of the most comprehensively observed examples of its type.  Many other BCGs exhibit similar ``weather" phenomena, displaying extended multiphase gas and star formation aligned with AGN outflows (Figure~\ref{fig:UV_BCG_gallery}), but the overall picture remains incomplete.  Detailed dynamical observations will be necessary to unravel these weather patterns, with ALMA, MUSE, and soon JWST, poised to make major contributions during the coming decade.

\section{Achieving Balance \label{sec:Balance}}

Sections \ref{sec:Heating} and \ref{sec:Weather} presented ample circumstantial evidence favoring a self-regulating feedback loop that keeps the atmospheres of massive galaxies in a quasi-steady state but did not address a key question:  How does that feedback loop close?  Figure~\ref{figure:Consistency} shows that galactic atmospheres from the most massive galaxy clusters down to the Milky Way tend to have similar cooling-time structure, with $t_{\rm cool} \sim 1$~Gyr at 10 kpc and cooling times as short as $\sim 30$~Myr at $\sim 1$~kpc.  In the most massive halos, the feedback keeping their inner atmospheres in a quasi-steady state must correct itself on timescales shorter than those cooling timescales. In Milky-Way scale halos, a similar feedback loop may be regulating the atmosphere's structure, but it is also possible that those halos currently harbor cooling flows with mass-inflow rates determined by earlier episodes of strong feedback (e.g., \citep{Stern_2019MNRAS.488.2549S}).

Fueling of the AGN engine that supposedly maintains approximate thermal balance in massive galactic halos depends on conditions within just a few parsecs of the central black hole.  Somehow, the black hole's fuel supply on that small scale must be sensitive to the cooling time of gas at distances at least as large as 10~kpc, because the incidence of strong AGN feedback is so closely tied to the cooling time there (\S \ref{sec:tcool_Prad}).  But the difference in those length scales is enormous.

Then there is the incidence of multiphase gas, which is likewise closely linked to the ambient cooling time at $\sim 10$~kpc (\S \ref{sec:multiphase_tcool}).  That cooler gas may be the cause of strong feedback, an outcome of strong feedback, or both.  However, accumulation of multiphase gas needs to be self-limiting, because an AGN feedback loop that continually increases its own fuel supply would lead to a runaway increase in AGN power.

This section focuses on a particular class of potential solutions to these puzzles.  As Binney and Tabor \citep{TaborBinney1993MNRAS.263..323T,BinneyTabor_1995MNRAS.276..663B} pointed out in the 1990s, catastrophic cooling of hot gas at the center of a galactic atmosphere is likely to fuel accretion onto a supermassive black hole, producing outbursts of energy that make the atmosphere convectively unstable and promote multiphase condensation.  During the past decade, many numerical simulations have demonstrated that those outburst events can be self-regulating, for reasons we will now explore in detail.

\subsection{The Precipitation Hypothesis \label{section:hypothesis}}

\begin{figure}[!tbp]
  \centering
    \includegraphics[width=\textwidth]{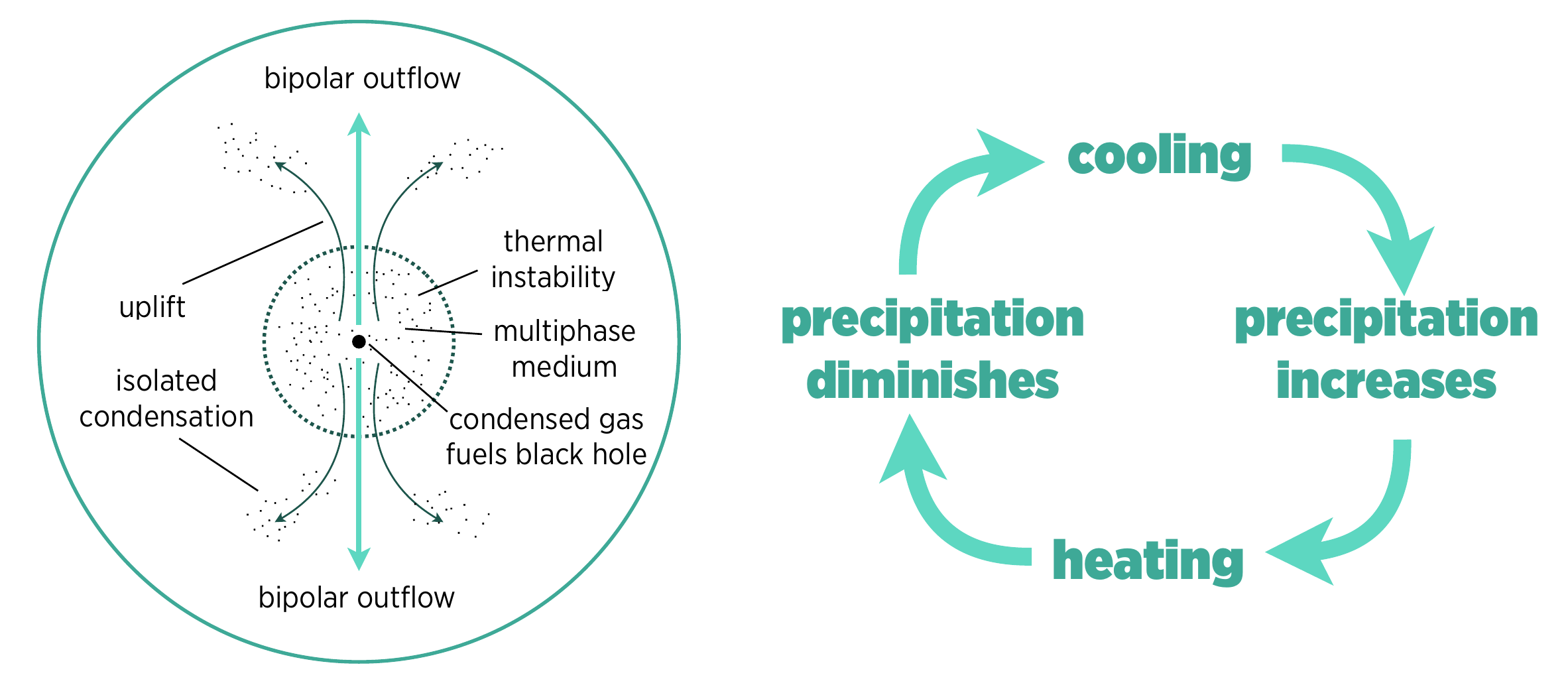}
  \caption{A schematic illustration of the precipitation hypothesis for self-regulating AGN feedback. 
  \label{figure:hypothesis}}
\end{figure}

The authors of this article have a favorite name for the generic idea common to this class of solutions, which includes \textit{cold feedback} \citep{ps05} and \textit{chaotic cold accretion} \citep{Gaspari+2013MNRAS.432.3401G}.  We call it the \textit{precipitation hypothesis}. According to that hypothesis, development of a multiphase medium in the vicinity of a massive galaxy's central black hole is critical to closing the AGN feedback loop.  

Figure~\ref{figure:hypothesis} schematically illustrates the fundamental idea.  The envisioned feedback loop proceeds as follows: Suppose the galaxy's hot atmosphere is initially homogeneous and that accretion of ambient gas onto the central black hole is too slow for AGN feedback to balance cooling.  Then the specific entropy of the ambient gas will decrease and a cooling flow will develop. As long as that flow remains undisturbed, it can remain nearly homogeneous, with a temperature near the local gravitational temperature (see \S \ref{sec:CoolingFlows}), until either the inflow speed ($\sim r / t_{\rm cool}$) approaches the local circular velocity or angular momentum causes the inflow to settle into a rotationally supported disk. Then the cooling flow becomes more susceptible to multiphase condensation and production of cold, dense clouds. However, atmospheric disturbances produced by feedback outbursts, cosmological infall, or orbiting satellite galaxies can push the cooling flow into multiphase condensation long before its flow speed approaches $v_{\rm c}$.  

As multiphase condensation starts to happen in an atmosphere without much rotational support, the coldest, densest clouds begin to rain toward the center.  The rain of cold clouds can supply accretion fuel to the central black hole much more rapidly than homogeneous accretion of the ambient gas can supply it.  As this ``precipitation" causes AGN feedback power to increase, the resulting outflows disturb the ambient medium and can temporarily promote more precipitation by lifting lower entropy gas to greater altitudes, where it is more unstable to multiphase condensation.  Ultimately, the feedback loop prevents a runaway, because heating, uplift, and mixing all raise the specific entropy and cooling time of the ambient medium until the precipitation finally diminishes and AGN feedback power declines. 

\subsubsection{Boosting of the Bondi Limit}

Development of a multiphase medium can sharply boost AGN fueling because the limiting black-hole accretion rate depends strongly on the specific entropy of the accreting gas.  Hermann Bondi's classic calculation of steady pressure-limited accretion onto a massive object \citep{Bondi_1952MNRAS.112..195B} results in
\begin{equation}
    \dot{M}_{\rm Bondi} \: = \: 4 \pi \lambda_{\rm Bondi} \, (\mu_e m_p) \,  (\mu m_p)^{3/2} (G M_{\rm BH})^2 K^{-3/2}
    \label{eq:Mdot_Bondi}
\end{equation}
when expressed in terms of specific entropy $K$ and black hole mass $M_{\rm BH}$.  The calculation assumes that accretion is spherically symmetric, with no change in specific entropy. For gas with $P \propto \rho^{5/3}$ the appropriate numerical correction factor is $\lambda_{\rm Bondi} \approx 1/4$. Low-entropy precipitation can therefore accrete much more rapidly onto a central black hole than can the ambient medium, at least in principle, because it is more easily compressed.  

To assess the significance of that change in the pressure-limited accretion rate, we can scale equation (\ref{eq:Mdot_Bondi}) to values of $M_{\rm BH}$ and $K$ typical of galaxy-cluster cores, obtaining 
\begin{equation}
    f_{\rm BH} \dot{M}_{\rm Bondi} c^2 \: = \: 
      3 \times 10^{41} \, {\rm erg \, s^{-1}} 
        \left( \frac {f_{\rm BH}} {0.005} \right)
        \left( \frac {M_{\rm BH}} {10^9 \, M_\odot} \right)^2
        \left( \frac {K} {10 \, {\rm keV \, cm^2}} \right)^{-3/2}
    \label{eq:Edot_Bondi}
\end{equation}
where $f_{\rm BH}$ is the efficiency factor discussed in \S \ref{sec:BH_Heating} for black-hole energy transfer to a galactic atmosphere. If all the factors in parentheses are of order unity, then the resulting power output is marginally sufficient to offset radiative cooling in small groups, but cannot do so in the cores of the most strongly cooling galaxy clusters \citep{Rafferty+2006ApJ...652..216R,Hardcastle_2007MNRAS.376.1849H}.  Those cluster cores radiate $L_{\rm cool} \sim 10^{45} \, {\rm erg \, s^{-1}}$ and exhibit cavities implying similar amounts of feedback power (see Figure \ref{figure:Sun2009}).  

Among the factors on the right-hand side of equation (\ref{eq:Edot_Bondi}), only a change in the entropy factor can plausibly account for such a large AGN power output.  Accretion of a gas component with specific entropy below 0.1~keV~cm$^2$ can supply it, and the entropy levels of BCG nebulae are considerably smaller than that.  Accumulation of precipitating gas in a BCG nebula and its associated molecular clouds can therefore strongly boost AGN feedback output, as envisioned by the precipitation hypothesis.\footnote{Some cosmological numerical simulations of AGN feedback have provided this boost artificially through a boost factor that raises the Bondi limit by multiple orders of magnitude \citep[e.g.,][]{Springel_2005MNRAS.361..776S,BoothSchaye_2009MNRAS.398...53B}, but those artificial boosts are becoming increasingly unnecessary as numerical resolution improves.} 

\subsubsection{Overcoming Angular Momentum}
\label{sec:OvercomingAngMom}

While development of precipitation in a galactic atmosphere can boost the Bondi limit on steady accretion, it does not guarantee a boosted black hole accretion rate.  At least some of the cool clouds raining toward the galactic center must also have low enough specific angular momentum to descend into the black hole's accretion disk.  Both hydrodynamical drag against a headwind of ambient gas and collisions with other cool clouds can reduce a cloud's specific angular momentum \citep{ps10,Hobbs_2011MNRAS.413.2633H,Gaspari+2013MNRAS.432.3401G,Gaspari_2015A&A...579A..62G,Prasad_2017MNRAS.471.1531P}.  But demonstrating that cool clouds forming at kiloparsec scales can then fall to within a few parsecs of the central black hole is a formidable computational challenge. Though progress is being made, exactly how angular momentum transfer allows the cool clouds to accrete onto the central black hole remains largely an unsolved problem (e.g..,\citep{Angles-Alcazar_2017MNRAS.470.4698A}).

\citet{Gaspari+2013MNRAS.432.3401G,Gaspari_2015A&A...579A..62G,Gaspari_2017MNRAS.466..677G} have provided some of the most convincing numerical demonstrations of how precipitation can couple with black hole fueling, through a mechanism they call \textit{chaotic cold accretion}. Their idealized simulations begin with a black hole embedded in the atmosphere of a typical galaxy group, but without any multiphase gas.  Adiabatic accretion in that homogeneous atmosphere unsurprisingly proceeds at the Bondi rate. When radiative cooling is allowed, the accretion rate can rise by two orders of magnitude, but only if the ambient gas has no initial angular momentum.  Organized rotation results in an accretion rate similar to the Bondi rate of the ambient gas, because the trajectories of the condensing clouds circularize at radii determined by their specific angular momenta (see also \citep{LiBryan2012ApJ...747...26L}). However, including turbulence raises the accretion rate closer to the case of pure cooling with no net angular momentum.  And remarkably, adding heating that balances the average radiative losses from each gas layer does not significantly reduce that accretion rate.  Accretion remains boosted by two orders of magnitude above the ambient medium's Bondi rate even when there are no net losses of thermal energy.

\indent The key to enabling strongly boosted accretion is a turbulent velocity dispersion that exceeds the atmosphere's typical rotation speed \citep{Gaspari_2015A&A...579A..62G,Prasad_2017MNRAS.471.1531P}.  Stochastic cloud-cloud collisions then continually replenish the portion of phase space that has low specific angular momentum and would otherwise be depleted by infall.  The result is a rapid and continual cascade of cold clouds toward the central black hole.\footnote{Perhaps similar to the ones observed by \citet{Tremblay2016} and discussed in \S \ref{sec:Circulation}.}  In idealized numerical experiments, chaotic cold accretion can establish a quasi-steady inhomogeneous accretion flow within $\sim 30$~Myr.  But if rotational motions dominate turbulent ones, most of the condensing gas instead collects into an orbiting disk or torus, where star formation can deplete it \citep{Li_2015ApJ...811...73L,Prasad_2018ApJ...863...62P}.\footnote{Recent numerical simulations by \citet{WangRuszkowski_2020MNRAS.493.4065W} have demonstrated that there may be another way for precipitating clouds in a magnetized medium to shed angular momentum: Magnetic braking can occur if trailing strands of gas magnetically tied to the cloud substantially increase the drag forces it experiences.}

\subsection{Thermal Instability}
\label{sec:ThermalInstability}

If the precipitation hypothesis is correct, then feedback acting on a galactic atmosphere tends to drive the ambient medium toward a state marginally susceptible to precipitation.  Development of a multiphase medium is inevitable if accretion of homogeneous ambient gas fails to yield enough feedback to offset radiative losses, allowing its entropy and cooling time to decline.  And after a multiphase medium develops, feedback capable of reducing precipitation through net heating of the ambient medium pushes it in the opposite direction, toward the marginal state.  The question then becomes:  What are the characteristics of that marginal state?

Answers to that question are often framed in terms of thermal instability.  The relevant astrophysical literature is vast and potentially confusing, because many different phenomena go by the name ``thermal instability."  We will therefore try to be careful and precise when defining what we mean when using that term.  Importantly, a galactic atmosphere can formally be thermally unstable without producing multiphase condensation, because of nonlinear buoyancy effects to be discussed in \S \ref{sec:BuoyancyDamping}.  This section will focus just on thermal stability of \textit{linear} perturbations, following \citep{Voit_2017_BigPaper}.

\subsubsection{Thermal Instability without Gravity}

Without gravity, both the pressure and specific entropy of the ambient medium can be uniform. Its thermal stability then depends only on whether a perturbation in its specific entropy grows or decays with time.  If ${\cal L}$ is the net cooling rate per unit mass, then equation (\ref{eq:tcool_K}) gives
\begin{equation}
    \frac {d} {dt} \ln K = - \frac {2} {3} \frac {\mu m_p} {k} \frac {\cal L} {T}
     \; \; .
\end{equation}
In general, both the entropy $K$ of the perturbation and the background entropy $\bar{K}$ may be changing with time.  The fractional entropy contrast of the perturbation is then $\delta \ln K = \ln (K/\bar{K})$, and it evolves with time according to 
\begin{equation}
  \frac {d} {dt} ( \delta \ln K ) 
                        \: =  \: - \left( \frac  {2} {3} \frac {\mu m_p} {k} \right)
                                       \left. \frac {\partial ({\cal L}/T)} {\partial \ln K} \right|_A \delta \ln K
                        \; \; ,
\end{equation}
where $A$ is an arbitrary thermodynamic quantity that remains constant.  The perturbation's amplitude therefore grows monotonically with time if
\begin{equation}
      \left. \frac {\partial ({\cal L}/T) } {\partial \ln K} \right|_A  \: < \: 0 \; \; ,
\end{equation}
equivalent to the thermal instability condition originally derived by \citet{Balbus86}.  If the background medium is also thermally balanced (${\cal L} = 0$), then this condition reduces to 
\begin{equation}
      \left. \frac { \partial {\cal L} } {\partial \ln K} \right|_A  \: < \: 0 \; \; ,
\end{equation}
equivalent to the thermal instability condition originally derived by \citet{Field65}.  Setting $A = P$ and defining 
\begin{equation}
  \omega_{\rm ti} \equiv - \left( \frac  {2} {3} \frac {\mu m_p} {k} \right)
                                       \left. \frac {\partial ({\cal L}/T)} {\partial \ln K} \right|_P
            \; \; ,
\end{equation}
then yields the thermal instability timescale $\omega_{\rm ti}^{-1}$ for isobaric multiphase condensation. 

Separating ${\cal L}$ into a cooling rate per unit mass $C$ and a heating rate per unit mass $H$, so that ${\cal L} = C - H$, reveals how the timescale for perturbation growth compares with the timescale on which the ambient medium heats or cools.  If all of the cooling is radiative, then
\begin{equation}
    \omega_{\rm ti} = \frac {1} {t_{\rm cool}}
   					 \, \left[  \, \left( \frac {6 - 3 \lambda} {5} \right)
   					    \, + \, \frac {H} {C} \left. \frac {\partial \ln (H/T)} {\partial \ln K} \right|_P \, \right]
    \label{eq:omega_ti}
\end{equation}
for $\Lambda(T) \propto T^\lambda$ (see equation \ref{eq:tcool_rels}).  Without heat input, the second term inside the square brackets vanishes, resulting in $\omega_{\rm ti}^{-1} = 5 \, t_{\rm cool} / (6 - 3 \lambda)$.  Isobaric perturbation growth therefore happens for $\lambda < 2$ and progresses exponentially. However, those perturbations do not get much of an opportunity to develop into multiphase condensation, because the timescale for perturbation growth is similar to the timescale for cooling of the background medium \citep[e.g.,][]{Malagoli_1987ApJ...319..632M,bs89}.  
  
Adding some heat gives the perturbations more time to develop contrast by reducing cooling of the background medium while low-entropy perturbations condense within it.  If heating per unit volume is constant, then the second term in the square brackets of equation (\ref{eq:omega_ti}) vanishes, yielding the same timescale for perturbation growth as in the no-heating case. Temporal variations of the heating rate in such an environment therefore do not affect the growth in contrast of an isobaric perturbation.  More generally, variations in heating that occur on a timescale less than or comparable to $t_{\rm cool}$ allow thermal instability to progress into multiphase condensation on a timescale $\sim t_{\rm cool}$, as long as time-averaged heating sufficiently delays global cooling of the ambient medium.

\subsubsection{Thermal Instability with Gravity}

Gravity fundamentally alters how thermal instability proceeds. In the presence of gravity, the pressure of a static background medium cannot be uniform, and condensation couples interestingly with buoyancy \citep{Defouw_1970ApJ...160..659D,Cowie_1980MNRAS.191..399C,Nulsen_1986MNRAS.221..377N,Malagoli_1987ApJ...319..632M,Loewenstein_1989MNRAS.238...15L,bs89}.  For example, an isobaric perturbation with lower entropy than its surroundings is also denser and accelerates in the direction of gravity.  If the background medium is isentropic, with constant $\bar{K}$, then the perturbation's entropy contrast can steadily increase, but its motion through the background medium may then initiate hydrodynamic instabilities capable of shredding it \citep{Nulsen_1986MNRAS.221..377N}.  And if the background medium has an entropy gradient, then a thermally unstable perturbation can excite internal gravity waves.

Consider what happens to an entropy perturbation in a spherically symmetric and otherwise hydrostatic background medium with $d \ln \bar{K} / d \ln r = \alpha_K$.  Let $\boldsymbol\xi$ be a vector field describing the perturbation in terms of displacements away from the atmosphere's equilibrium state.  The perturbation's equation of motion is then
\begin{equation}
    \ddot{\boldsymbol\xi} = - \nabla \phi - \frac {\nabla ( \bar{P} + \delta P) }
                                                  {\bar{\rho} + \delta \rho}
            \; \; ,
\end{equation}
where $\bar{P}$ and $\bar{\rho}$ represent the pressure and density of the unperturbed state, while $\delta P$ and $\delta \rho$ represent the corresponding perturbation amplitudes.  Simplifying the equation by retaining only terms of linear order and applying the equilibrium condition $\nabla \bar{P} = - \bar{\rho} \, \nabla \phi$ gives
\begin{equation}
    \ddot{\boldsymbol\xi} \: = \:
       \left[ \frac {3} {5} ( \delta \ln K) + \frac {2} {5} (\delta \ln P) \right] \nabla \phi 
         \: - \: \frac {\bar{P}} {\bar{\rho}} \nabla (\delta \ln P) 
            \; \; ,
        \label{eq:ddot_xi}
\end{equation}
where $\delta \ln P \equiv \ln (P/\bar{P})$ and we have used the relation $\rho \propto (P/K)^{3/5}$.  

In an adiabatic medium, entropy perturbations are directly related to displacements through 
$\delta \ln K =  - \, \boldsymbol\xi \cdot \nabla \ln \bar{K}$.  The radial part of the perturbation's equation of motion therefore yields
\begin{equation}
    \ddot{\xi}_r + \omega_{\rm buoy}^2 \xi_r  
      \: = \:
    \frac {3 } {5} \left( \frac {2g} {3} - c_s^2 \frac {\partial} {\partial r} \right) (\delta \ln P) 
            \; \; ,
        \label{eq:gmodes_stable}
\end{equation}
where $\omega_{\rm buoy} \equiv [(3/5) \nabla \phi \cdot \nabla \ln \bar{K}]^{1/2} = (6 \alpha_K / 5)^{1/2} t_{\rm ff}^{-1}$ is the Brunt-V{\"a}is{\"a}l{\"a} frequency for buoyant oscillations, $c_s = (5P/3\rho)^{1/2}$ is the adiabatic sound speed, $g = G M_r / r^2$ is the local gravitational acceleration, and $\xi_r$ is the radial displacement.  The solutions of this equation with frequencies similar to $\omega_{\rm buoy}$ are internal gravity waves.  

In a non-adiabatic medium, internal gravity waves couple with thermal instability because 
\begin{equation}
    \frac {d} {dt} (\delta \ln K) \: = \: \frac {\partial} {\partial t} (\delta \ln K) 
                                          + \dot{\boldsymbol\xi} \cdot \nabla \ln \bar{K}
                                \: = \: \omega_{\rm ti} (\delta \ln K)
                    \label{eq:Kcoupling}
\end{equation}
as long as isentropic pressure perturbations are stable.  Wavelike solutions proportional to $e^{-i \omega t}$ then have the property
\begin{equation}
    \left( 1 - i \frac {\omega_{\rm ti}} {\omega} \right) \delta \ln K =  - \frac {\alpha_K \xi_r} {r}
        \; \; .
\end{equation}
In this relationship, the $i \omega_{\rm ti} / \omega$ term represents a phase difference that enables heating and cooling to pump internal gravity waves with $\omega \gg \omega_{\rm ti}$.  

\begin{figure}[t]
  \centering
    \includegraphics[width=4.5in]{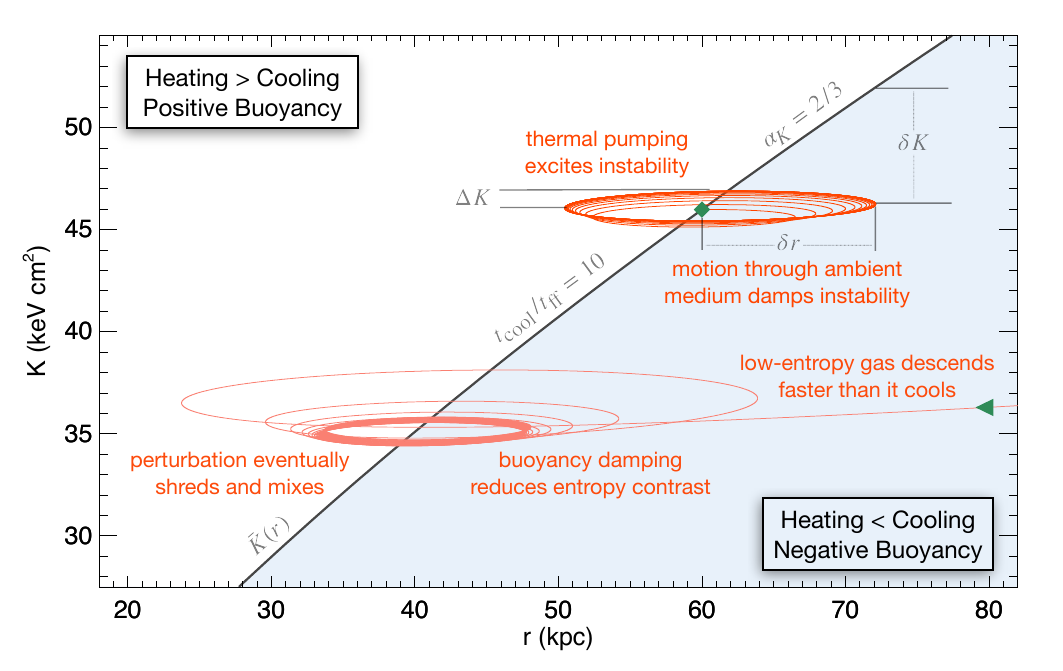}
  \caption{Growth and damping of thermal instability in a stratified, thermally balanced medium (from \citep{Voit_2021ApJ...908L..16V}).  A charcoal line in the figure shows a median entropy profile $\bar{K}(r) \propto r^{2/3}$ along which $t_{\rm cool} / t_{\rm ff}$ = 10. Cooling exceeds heating in the blue region below that line and heating exceeds cooling above it.  Thermal pumping excites internal gravity waves in this medium, because $\alpha_K^{1/2} (t_{\rm cool}/t_{\rm ff}) \approx \omega_{\rm buoy} t_{\rm cool} \gg 1$.  Wave growth saturates when energy gained through thermal pumping equals energy dissipated into the ambient medium.  The red line starting with the green diamond at 60~kpc shows a perturbation trajectory with an amplitude that grows toward saturation, and the red line entering the figure from the right (marked by the green triangle) shows a perturbation trajectory that begins with a large amplitude and decays to the saturation amplitude.  
  \label{figure:DampingDuo}}
\end{figure}

Figure~\ref{figure:DampingDuo} provides a more intuitive illustration of what happens. Its upper trajectory represents a perturbation introduced by pushing a fluid element at 60~kpc (green diamond) toward greater altitudes, into the blue shaded region.  Cooling exceeds heating in that region, and so the perturbation's trajectory bends downward on a timescale $\sim t_{\rm cool}$.  While that is happening, gravity acts on this low-entropy perturbation to pull it inward on a timescale $\sim t_{\rm ff}$.  The perturbation's trajectory then passes through the median entropy profile because inward acceleration is more rapid than entropy loss. After crossing the median profile, the trajectory is in the unshaded region where heating exceeds cooling and represents a high-entropy perturbation that buoyancy forces push outward.  Gravity-wave oscillations ensue, and thermal pumping continually increases their amplitude until nonlinear saturation limits their growth, for reasons we will discuss next. 

\subsection{Buoyancy Damping}
\label{sec:BuoyancyDamping}

\begin{figure}[t]
  \centering
    \includegraphics[width=\textwidth]{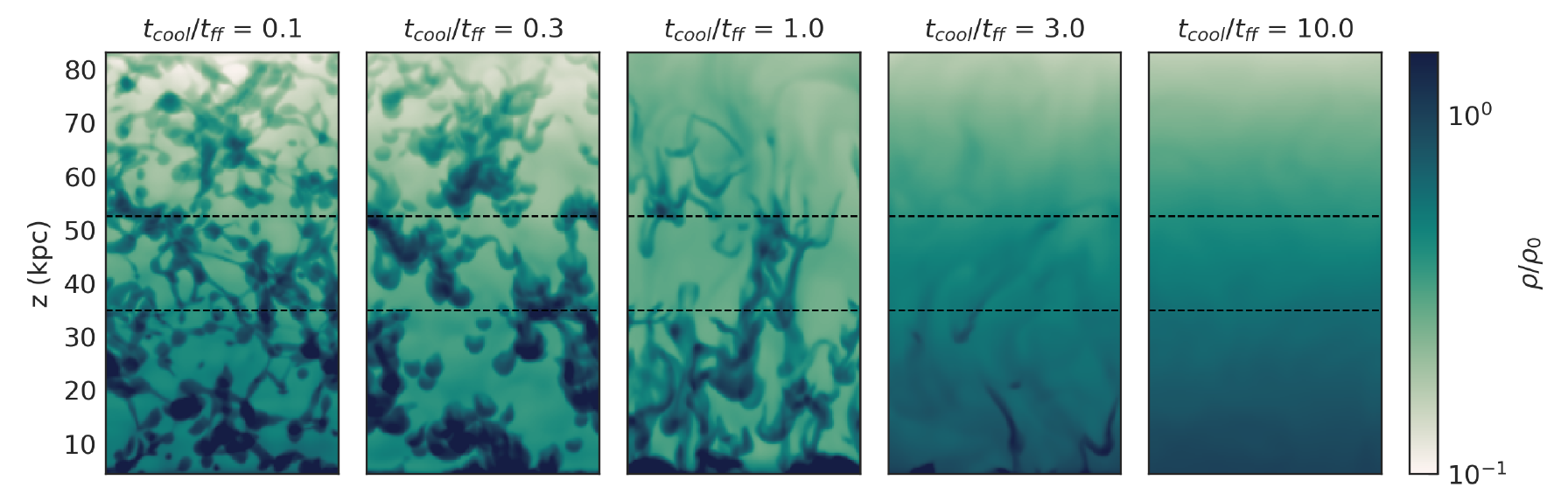}
  \caption{Damped thermal instability as a function of $t_{\rm cool}/t_{\rm ff}$ in simulations by \citet{Butsky_2020ApJ...903...77B}.  In the simulation environment, a uniform heating rate per unit mass balances average radiative losses within each horizontal layer of the atmosphere.  Toward the left are environments with $t_{\rm cool} < t_{\rm ff}$, in which multiphase condensation forms before buoyancy effects can damp thermal instability.  Toward the right are environments with $t_{\rm cool} > t_{\rm ff}$, in which buoyancy effects damp multiphase condensation.  Perturbations are initialized with $\delta \ln \rho \sim 0.02$ in each case, and the panels show the state of each simulation at approximately 4 times the median cooling time of gas between the horizontal dashed lines. 
  \label{figure:Butsky_tctff}}
\end{figure}

Numerical simulations show that oscillatory thermal instability saturates before producing multiphase condensation in thermally balanced atmospheres with $\alpha_K \sim 1$ and $t_{\rm cool} / t_{\rm ff} \gg 1$.  \citet{McCourt+2012MNRAS.419.3319M} found that growth of thermally unstable perturbations saturates at a fractional density amplitude $\delta \ln \rho \sim \alpha_K^{1/2} (t_{\rm ff} / t_{\rm cool})$.  Figure~\ref{figure:Butsky_tctff} shows a more recent demonstration from \citep{Butsky_2020ApJ...903...77B}. McCourt et al.~attributed the saturation amplitude to nonlinear mode coupling but were not specific about how that would occur.  However, the saturation mechanism is important to understand because it determines the characteristics of a galactic atmosphere that is marginally unstable to precipitation.  

Our preferred term for that mechanism is \textit{buoyancy damping} \citep{Voit_2017_BigPaper}.  Its fundamental features are present in heuristic analyses of nonlinear perturbation damping (\S \ref{sec:HeuristicDamping}).  But explicit consideration of nonlinear mode coupling is necessary to identify how the kinetic energy introduced by thermal instability propagates and ultimately rethermalizes in a medium with $t_{\rm cool} / t_{\rm ff} \gg 1$ (\S \ref{sec:NonlinearCoupling}).

\subsubsection{Heuristic Nonlinear Damping}
\label{sec:HeuristicDamping}

The early analyses of \citet{Cowie_1980MNRAS.191..399C} and \citet{Nulsen_1986MNRAS.221..377N} treated nonlinear damping of thermal instability heuristically by adding a damping term to a thermally unstable perturbation's equation of motion.  Scaling the damping term to mimic the effects of hydrodynamic drag on an oscillating thermally unstable gas blob results in saturation of thermal instability when kinetic energy losses owing to drag become comparable to kinetic energy gains through thermal pumping.  And drag on a bobbing blob of radial thickness $\sim k_r^{-1}$ causes it to lose kinetic energy on a timescale $\omega_{\rm D}^{-1} \sim | k_r \dot{\xi}_r |^{-1}$. 

According to this line of reasoning, buoyant oscillations of a thermally unstable gas blob should saturate with a fractional amplitude
\begin{equation}
    | \delta \ln \rho | \: \sim \: | \delta \ln K | 
                        \: \sim \: \alpha_K \frac { | \xi_r | } {r} 
                        \: \sim \: \frac {\alpha_K} {k_r r} \left( \frac {\omega_{\rm ti}} {\omega_{\rm buoy}} \right)
                        \: \sim \: \frac {\alpha_K^{1/2}} {k_r r} \left( \frac {t_{\rm ff}} {t_{\rm cool}} \right)
            \; \; .
            \label{eq:SaturationAmplitude}            
\end{equation}
Blobs of radial thickness comparable to $r$ therefore saturate with the greatest amplitudes, and the magnitude of their density contrast aligns with the findings of numerical simulations such as those in Figure \ref{figure:Butsky_tctff}.  In an environment with $\alpha_K^{1/2} (t_{\rm cool}/t_{\rm ff}) \gg 1$, the entropy contrast $\delta K$ between a saturated blob and its surroundings stems mainly from its radial displacement, while the actual entropy changes the blob experiences (see Figure \ref{figure:DampingDuo}) have an amplitude $| \Delta K | \sim (t_{\rm ff} / t_{\rm cool}) | \delta K |$.

However, the overall picture remains incomplete, because the oscillations of a bobbing gas blob decay by exciting other internal gravity waves, which are also thermally unstable.  Saturation of an entire ensemble of gravity waves therefore requires the kinetic energy they receive from thermal pumping to dissipate into some other form of energy \citep{Voit_2017_BigPaper}, and that is what we will look at next.

\subsubsection{Nonlinear Gravity-Wave Coupling}
\label{sec:NonlinearCoupling}

In Earth's atmosphere and oceans, internal gravity waves decay through a mechanism known as \textit{parametric subharmonic instability} (see \citep{StaquetSommeria_2002AnRFM..34..559S} for a review).  Their generic dispersion relation is
\begin{equation}
    \omega^2 = \left( \frac {k^2 - k_r^2} {k^2} \right) \omega_{\rm buoy}^2
\end{equation}
where $k$ is the total wavenumber and $k_r$ is the radial wavenumber.\footnote{This relation can be derived from equation (\ref{eq:gmodes_stable}) with the help of the equations for conservation of mass and horizontal momentum.}  Internal gravity waves with a propagation component in the radial direction therefore oscillate more slowly than those propagating perpendicular to gravity. 

This difference in frequency enables a primary gravity wave to resonate with a pair of secondary waves with frequencies whose sum equals that of the primary wave.  In other words, the secondary waves are subharmonic and form a wave triad that saps energy from the primary wave at a rate $\sim \omega_{\rm buoy} | k_r \xi_r |$.
The primary wave therefore decays, with its energy cascading through nonlinear coupling into lower frequency waves of increasingly large wavenumber that propagate ever more vertically.  And its decay rate reproduces the saturation amplitude represented in equation (\ref{eq:SaturationAmplitude}).

Ultimately, the gravity-wave cascade ends up producing small-scale Kolmogorov turbulence as the gravity waves start to break.  Increasingly vertical propagation eventually leads to wavefronts with density inversions that overturn like ocean waves do.  The resulting turbulence then thermalizes the wave energy, removing it from the gravity-wave ensemble.

Interestingly, the gravity-wave cascade has a power spectrum distinctly different from that of Kolmogorov turbulence.  Expressed in terms of a velocity structure function (see \S \ref{sec:MultiphaseTurbulence}), the power spectrum of parametric subharmonic instability has $\langle | \delta v | \rangle \propto k^{-1}$, whereas Kolmogorov turbulence has $\langle | \delta v | \rangle \propto k^{-1/3}$.  The velocity power spectra summarized in Figure \ref{fig:Li_Velocity_Structure} therefore suggest that the velocity fields observed in galaxy cluster cores may be more consistent with a gravity-wave cascade than with Kolmogorov turbulence.  Recognition of this possibility is a relatively recent development\footnote{See the Erratum to \citep{Voit_2017_BigPaper}} that has not yet been investigated with numerical simulations (but see \citep{WangRuszkowski_2021MNRAS.504..898W}) and deserves more attention.

\subsection{Drivers of Multiphase Condensation}
\label{sec:CondensationDrivers}

\begin{figure}[t]
  \centering
    \includegraphics[width=\textwidth]{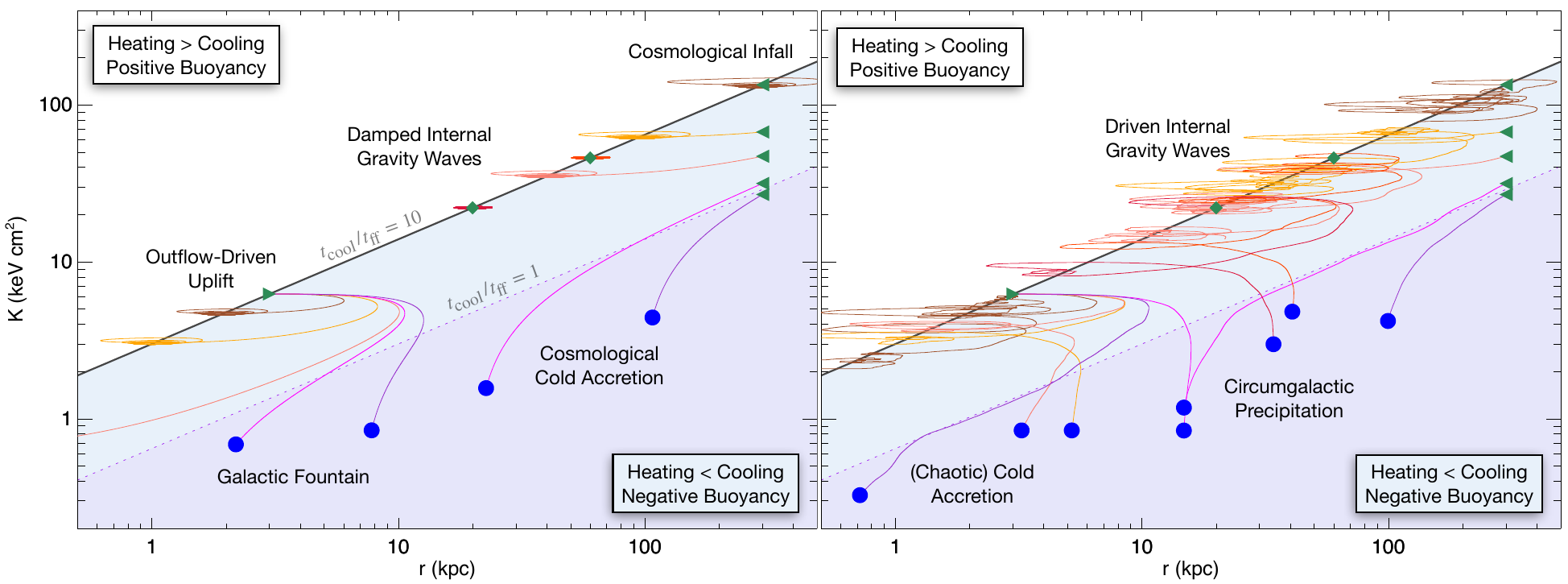}
  \caption{Pathways to multiphase condensation illustrated as schematic perturbation trajectories in the $K$--$r$ plane (from \citep{Voit_2021ApJ...908L..16V}).  In each panel, purple shading beneath the dotted line labeled $t_{\rm cool} / t_{\rm ff} = 1$ shows the region in which buoyancy damping cannot suppress multiphase condensation, and the charcoal line labeled $t_{\rm cool} / t_{\rm ff} = 10$ shows the atmosphere's median entropy profile, which has a power-law slope $K \propto r^{2/3}$.  The panel on the left depicts perturbation trajectories in a static atmosphere, most of which converge to the saturation amplitude determined by buoyancy damping. The panel on the right shows how atmospheric disturbances can alter those trajectories, through momentum fluctuations capable of counteracting buoyancy damping.  Disturbing a galactic atmosphere therefore promotes multiphase condensation (symbolized by blue dots), allowing it to happen in environments with median $t_{\rm cool} / t_{\rm ff}$ ratios significantly greater than unity, depending on the strength of those disturbances. 
  \label{figure:BlobOrbitSummary}}
\end{figure}

Based on the results summarized in the previous section, it would be reasonable to conclude that multiphase condensation happens in stratified galactic atmospheres only if $t_{\rm cool} / t_{\rm ff} \lesssim 1$.  The significance of this tipping point was recognized and elucidated decades ago by \citet{Hoyle_1953ApJ...118..513H} and has since been incorporated into many semi-analytic models of galaxy formation and galactic atmospheres  \citep[e.g.,][]{WhiteFrenk1991ApJ...379...52W,SomervilleDave_2015ARA&A..53...51S}.  However, both observations and simulations of galactic atmospheres show that multiphase condensation can happen in stratified atmospheres with ambient $t_{\rm cool} / t_{\rm ff}$ ratios more than an order of magnitude greater \citep[e.g.,][see also \S \ref{sec:Evidence}]{Voit_2015Natur.519..203V,Li_2015ApJ...811...73L}.\footnote{A caveat: At least some of the multiphase gas observed in BCG nebulae is dusty, meaning that it cannot all result from condensation of the hot ambient medium, which is essentially dust-free. One potential dust source is the red-giant winds from the BCG's aging stellar population, which can catalyze additional dust formation if they are able to seed the condensing gas with dust before sputtering destroys all of the dust grains (e.g., \citep{VoitDonahue2011ApJ...738L..24V,VoitDonahue2015ApJ...799L...1V}). }
Those findings imply the presence of disturbances capable of suppressing buoyancy damping.  Figure \ref{figure:BlobOrbitSummary} schematically illustrates some of the possibilities and the rest of this section discusses them.

\subsubsection{Uplift}

We have already discussed the first type of disturbance, bulk uplift, in \S \ref{sec:MultiphaseOutflows} and \S \ref{sec:MultiphaseDynamics}.  Adiabatic lifting of low-entropy ambient gas that has collected near the atmosphere's center lowers its local $t_{\rm cool} / t_{\rm ff}$ ratio because $t_{\rm ff}$ increases while $t_{\rm cool}$ remains relatively constant, at least at first.  But as the uplifted gas approaches altitudes at which $t_{\rm cool} / t_{\rm ff} \sim 1$, its entropy and cooling time start to decline on a timescale comparable to the dynamical time.  The purple, magenta, and salmon colored trajectories starting at 3~kpc in the left panel of Figure \ref{figure:BlobOrbitSummary} show some examples.  Radial uplift by a factor $\sim 3$ in this idealized galactic atmosphere, which has a median timescale ratio $t_{\rm cool} / t_{\rm ff} = 10$, induces condensation of the perturbations represented by the purple and magenta trajectories and barely misses causing the salmon-colored trajectory to end in condensation.  

This pathway to multiphase condensation was envisioned by early ``galactic fountain" models of the Milky Way's atmosphere \citep{ShapiroField_1976ApJ...205..762S,Bregman_1980ApJ...236..577B} and has since been realized in many numerical simulations of AGN feedback \citep[e.g.,][]{Revaz_2008A&A...477L..33R,LiBryan2014ApJ...789..153L,Prasad_2015ApJ...811..108P}.  But inducing condensation of uplifted gas becomes more difficult as an atmosphere's median $t_{\rm cool} / t_{\rm ff}$ ratio increases, because uplift to a larger radius is required \citep{Voit_2017_BigPaper}.  Drag forces that limit the descent of uplifted gas clouds can help to promote multiphase condensation \citep{McNamara_2016ApJ...830...79M}, but not if hydrodynamic instabilities shred the clouds and mix them with the ambient medium faster than they can cool \citep{Nulsen_1986MNRAS.221..377N}.

\begin{figure}[t]
  \centering
    \includegraphics[width=5.0in]{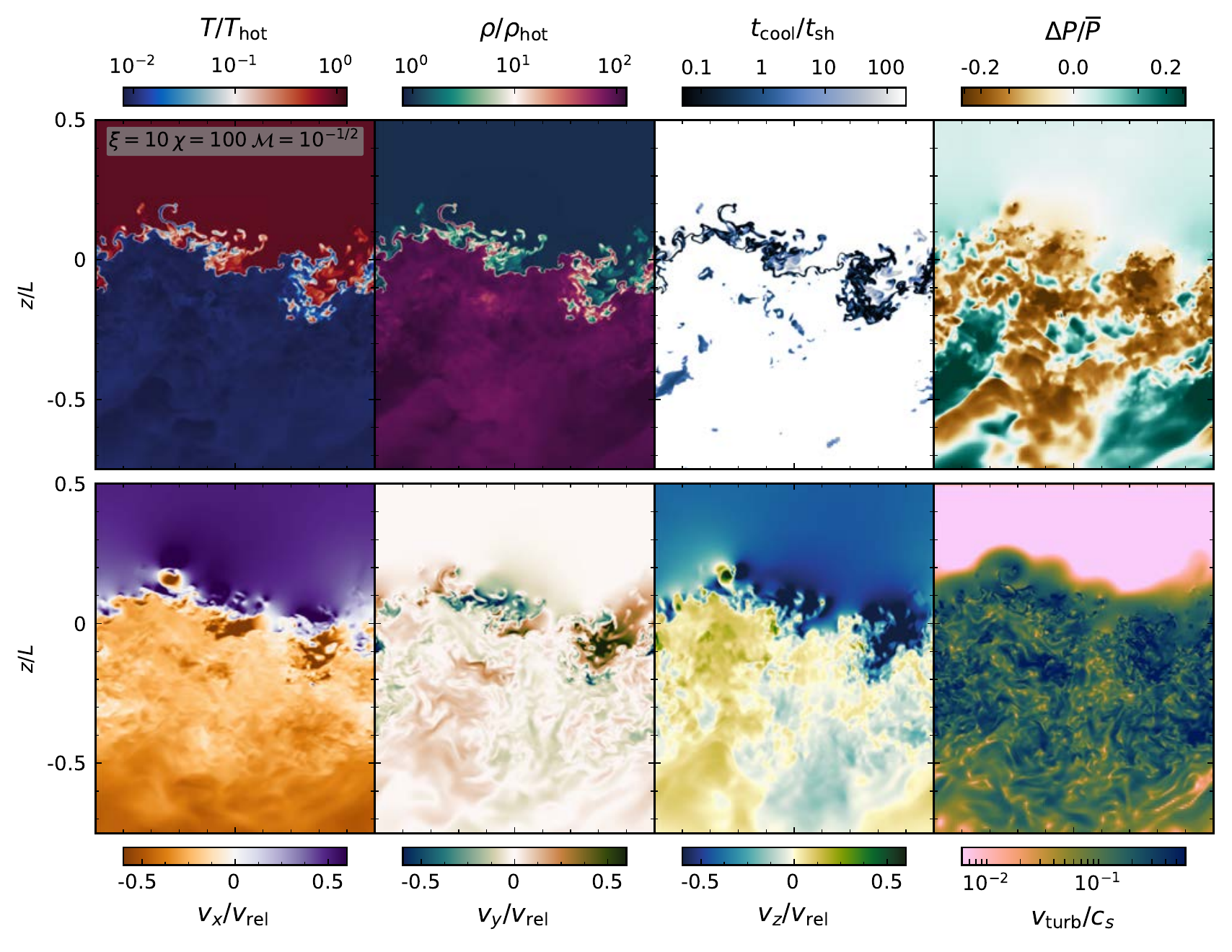}
  \caption{Fractal structure of a turbulent radiative mixing layer produced by shear flow at the interface between a cold dense cloud and a more diffuse medium that is 100 times hotter (from \citet{Fielding_2020ApJ...894L..24F}). Panels across the top depict contrasts in temperature ($T$), density ($\rho$), the ratio of cooling time to shearing time ($t_{\rm cool}/t_{\rm sh}$), and pressure ($\Delta P / \bar{P}$).  Panels across the bottom depict the three-dimensional velocity field and its turbulent Mach number ($v_{\rm turb}/c_{\rm s}$).
  \label{figure:MixingLayer}}
\end{figure}

The complex role of mixing has recently received considerable attention \citep[e.g.,][]{GronkeOh_2018MNRAS.480L.111G,GronkeOh_2020MNRAS.492.1970G,GronkeOh2020MNRAS.494L..27G,Ji_2018MNRAS.476..852J,TanOhGronke_2021MNRAS.502.3179T,Sparre_2019MNRAS.482.5401S,Fielding_2020ApJ...894L..24F}.  Whether or not hydrodynamic instabilities can destroy cool clouds experiencing a headwind depends on a competition between hydrodynamic mixing and radiative cooling.  Turbulent mixing folds thermal energy into the cool gas and produces an intermediate-temperature boundary layer, perhaps with fractal structure (see Figure \ref{figure:MixingLayer}), that maximizes radiative cooling.  If enhanced radiative cooling exceeds turbulent thermal energy transport, then it can preserve the cool cloud.  Under some circumstances, particularly if the cool clouds are large, mixing can even enable mass transfer from the hot phase to the cool phase, enhancing multiphase condensation and allowing momentum transfer to accelerate the cool clouds without destroying them. 

\subsubsection{Turbulence \label{section:turbulence}}

We have also already discussed a potential role for turbulence in initiating multiphase condensation (in \S \ref{sec:MultiphaseTurbulence} and \S \ref{sec:OvercomingAngMom}).  Turbulent momentum impulses can counteract buoyant damping by lifting lower-entropy gas to greater altitudes and can succeed in stimulating multiphase condensation if the local $t_{\rm cool} / t_{\rm ff}$ ratio in some of the turbulently levitated gas parcels approaches unity.  Then those parcels can condense before they descend.\footnote{Proponents of turbulent stimulation of condensation consider bulk uplift a subcategory of turbulent levitation, while proponents of stimulated condensation via uplift consider turbulent levitation a subcategory of uplift.}

The trajectories in the right panel of Figure \ref{figure:BlobOrbitSummary} derive from a heuristic model of turbulent levitation \citep{Voit_2018ApJ...868..102V}.  Their initial conditions are identical to those of the trajectories on the left, but they are exposed to random momentum impulses intended to mimic turbulence.  Two trajectories in the left panel are identical to the ones in Figure \ref{figure:DampingDuo}.  One of them starts at the green diamond 60~kpc from the center.  The other starts at the green triangle with an entropy level of $50 \, {\rm keV \, cm^2}$ at 300~kpc.  Both of them converge to the saturation amplitude, but their counterparts in the right panel of Figure \ref{figure:BlobOrbitSummary} end in multiphase condensation $\sim 40$~kpc from the center, because dynamical noise in that model is sufficient to disrupt buoyancy damping.  In essence, they represent internal gravity waves that have been driven past the amplitude at which gravity is a rapid enough restoring force.

The uplift trajectories starting at 3~kpc are also significantly altered.  Now the orange and salmon colored trajectories starting there end in condensation, while the purple and magenta trajectories still condense but do so at different distances from the center.  In that respect the heuristic model qualitatively mimics simulations of chaotic cold accretion, in which turbulence is essential for feeding the central black hole with cold clouds (\S \ref{sec:OvercomingAngMom}).

Consequently, the median value of $t_{\rm cool} / t_{\rm ff}$ in an atmosphere able to harbor multiphase condensation depends on the amplitudes of gravity waves driven within it.  The heuristic model depicted in Figure \ref{figure:BlobOrbitSummary} suggests that an ambient medium with $t_{\rm cool} / t_{\rm ff} = 10$ can be driven into multiphase condensation with disturbances resulting in a one-dimensional turbulent velocity dispersion roughly $1/3$ the circular velocity of the potential well, equivalent to about $1/2$ the velocity dispersion of ballistic particles.  The same amount of driven turbulence stimulates chaotic cold accretion in numerically simulated atmospheres with a similar median $t_{\rm cool} / t_{\rm ff}$ ratio \citep{Gaspari+2013MNRAS.432.3401G,Gaspari_2015A&A...579A..62G}.  And those results align with observations showing that the turbulent velocities of BCG nebulae and their molecular clouds are typically about half the stellar velocity dispersion of the BCG (see \S \ref{sec:MultiphaseTurbulence}).

\citet{Gaspari_2018ApJ...854..167G} have argued that the ratio $t_{\rm cool} / t_{\rm eddy}$, where $t_{\rm eddy}$ is the turnover time for large turbulent eddies, is more closely connected with multiphase condensation, advocating for a condensation criterion $t_{\rm cool} / t_{\rm eddy} \approx 1$.  Faster turbulence will certainly inhibit condensation by mixing and heating the perturbations faster than they can condense \citep{BanerjeeSharma_2014MNRAS.443..687B}, while slower turbulence will fail to lift perturbations quickly enough to counteract buoyancy damping.  However, the criterion $t_{\rm cool} / t_{\rm eddy} \approx 1$ is not general enough to account for all forms of multiphase condensation.  For example, internal gravity waves are a form of radial motion capable of stimulating condensation without the help of turbulent eddies.  The primary barrier to multiphase condensation in the entropy-stratified atmospheres of massive galaxies is buoyancy damping, and the phenomena capable of surmounting it depend most critically on the atmosphere's median value of $\omega_{\rm buoy} t_{\rm cool} \sim \alpha_K^{1/2} (t_{\rm cool} / t_{\rm ff})$.

\subsubsection{Cosmological Infall}

Another potential driver of multiphase condensation in galactic atmospheres is cosmological infall of low-entropy gas.  Whether or not the infalling gas can condense depends on its initial entropy level, as shown by the trajectories starting with green triangles at 300~kpc in Figure \ref{figure:BlobOrbitSummary}.  If a gas blob with $t_{\rm cool} / t_{\rm ff} \lesssim 1$ enters an atmosphere with $t_{\rm cool} / t_{\rm ff} \gg 1$, it is likely to condense.  But if $t_{\rm cool} / t_{\rm ff} \gg 1$ within the entering gas blob, it does not condense unless some sort of disturbance hinders its descent and eventual settling into an ambient gas layer of equivalent entropy. Similarly, gas stripped out of satellite galaxies can feed low-entropy perturbations into a central galaxy's atmosphere, and the same condensation criteria apply to those perturbations.

\begin{figure}[t]
  \centering
    \includegraphics[width=4.5in]{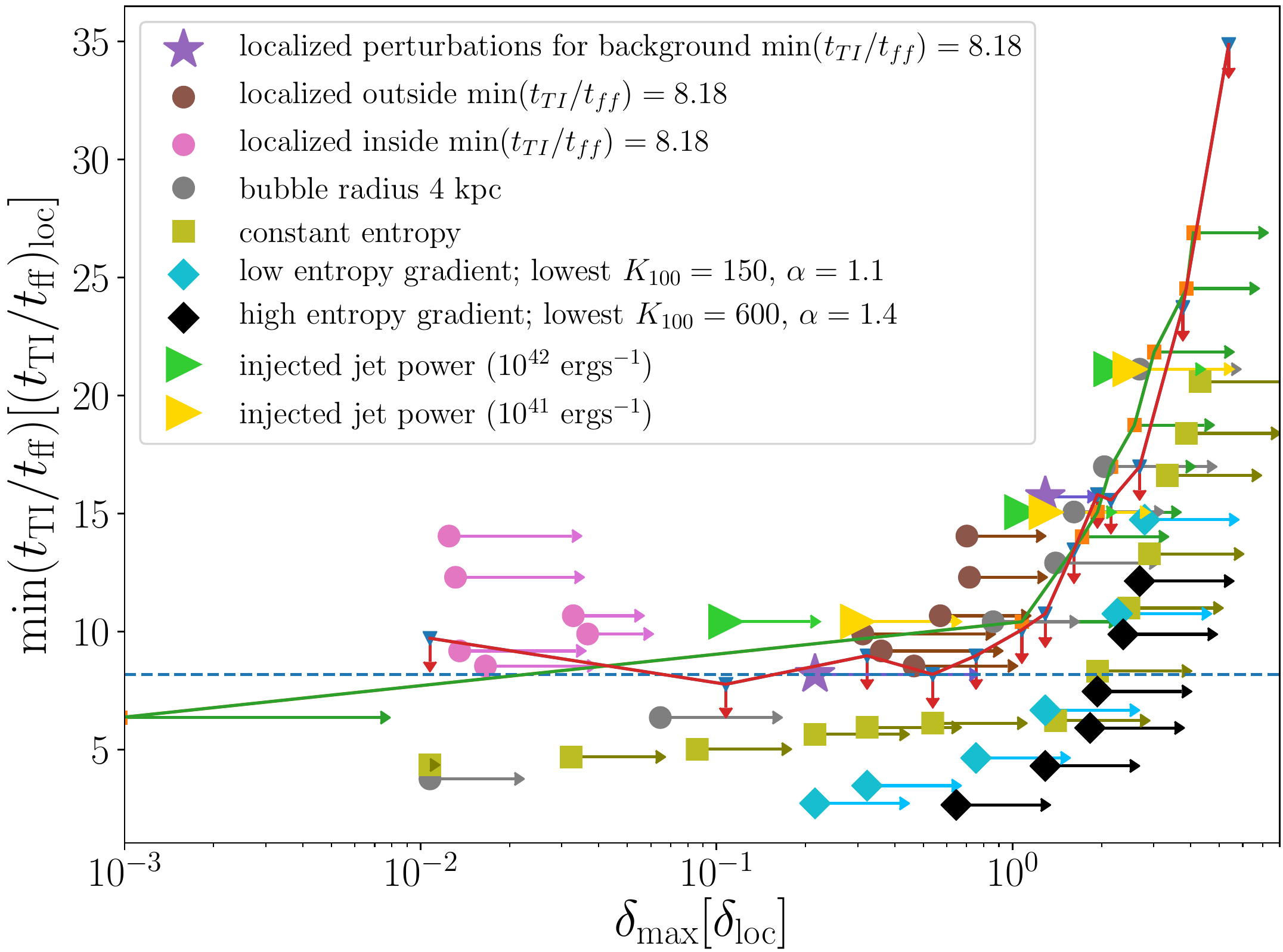}
  \caption{The ``condensation curve" from \citet{Choudhury_2019MNRAS.488.3195C} illustrating criteria for multiphase condensation that depend jointly on the fractional amplitude of initial density perturbations ($\delta_{\rm max}$) and the minimum ratio of thermal instability time to freefall time ($t_{\rm TI}/t_{\rm ff}$) in the ambient galactic atmosphere.   Generally, $t_{\rm TI} \approx t_{\rm cool}$ and perturbations with large amplitudes condense regardless of the ambient atmosphere's typical $t_{\rm cool}/t_{\rm ff}$ value. But condensation of perturbations with initially linear amplitudes depends on the ambient $t_{\rm cool}/t_{\rm ff}$ value and other atmospheric characteristics and usually requires $\min (t_{\rm cool}/t_{\rm ff}) \lesssim 10$ if there are no other atmospheric disturbances.  (See their paper for details.)
  \label{figure:CondensationCurve}}
\end{figure}

\citet{Choudhury_2019MNRAS.488.3195C} have recently analyzed the joint dependence of multiphase condensation on initial perturbation amplitude and a galactic atmosphere's median value of $t_{\rm cool} / t_{\rm ff}$.  Figure \ref{figure:CondensationCurve} shows the ``condensation curve" resulting from their analysis.  It represents the outcomes of idealized simulations of spherical galactic atmospheres in which heating balances the average radiative losses in each atmospheric layer.  The parameter $\delta_{\rm max}$ is the maximum fractional amplitude of the initial density perturbations, which are isobaric, and the parameter $\min(t_{\rm TI}/t_{\rm ff})$ is the ambient atmosphere's minimum initial ratio of thermal instability time ($t_{\rm TI} = \omega_{\rm ti}^{-1} \approx t_{\rm cool}$) to freefall time.  Except for the motions induced by the initial perturbations, these atmospheres are undisturbed.  Perturbations with great enough initial amplitudes always condense.  However, the values of $\delta_{\rm max}$ required for condensation are greater in atmospheres with greater $\min(t_{\rm TI}/t_{\rm ff})$ and are typically several times greater than unity in atmospheres with $\min(t_{\rm TI}/t_{\rm ff}) \gtrsim 10$.  

Condensation of perturbations with initial amplitudes less than unity can happen in atmospheres with $\min(t_{\rm TI}/t_{\rm ff}) \lesssim 10$, but the critical amplitude for condensation then depends on the ambient atmosphere's structure.  In particular, the critical amplitude for condensation depends on the atmosphere's entropy gradient, with larger entropy gradients (black diamonds) requiring larger initial amplitudes than shallower entropy gradients (blue diamonds and green squares).  Those results are qualitatively consistent with suppression by buoyancy damping, because they imply that multiphase condensation depends more critically on $\omega_{\rm buoy} t_{\rm cool} \sim \alpha_K^{1/2} (t_{\rm cool} / t_{\rm ff})$ than on $t_{\rm cool} / t_{\rm ff}$ alone.

\subsubsection{Convection}

Yet another route to multiphase condensation is to flatten the ambient atmosphere's entropy gradient, thereby reducing $\omega_{\rm buoy} t_{\rm cool}$ without necessarily changing $\min (t_{\rm cool} / t_{\rm ff})$.  Binney and Tabor \citep{TaborBinney1993MNRAS.263..323T,BinneyTabor_1995MNRAS.276..663B} recognized that centralized heating by AGN feedback would promote multiphase condensation by flattening the atmosphere's entropy gradient. Production of multiphase gas through this mechanism plays a major role in the \textit{cold feedback} mechanism envisioned by \citet{ps05}.  However, excessive central heating inverts the atmosphere's central entropy gradient, causing convection that can stimulate runaway condensation instead of well regulated precipitation.

Once convection begins, buoyancy damping is no longer a barrier to multiphase condensation, regardless of the atmosphere's median $t_{\rm cool} / t_{\rm ff}$ ratio, because the atmosphere no longer supports internal gravity waves.  Incipient condensates start to descend as their density contrast increases and they never pass through an atmospheric layer of equivalent entropy.  Hydrodynamical instabilities may be able to shred those condensates before they are able to reach a large density contrast, but numerical simulations generally show that centralized heat input tends to promote multiphase condensation.

\begin{figure}[t]
  \centering
    \includegraphics[width=\textwidth]{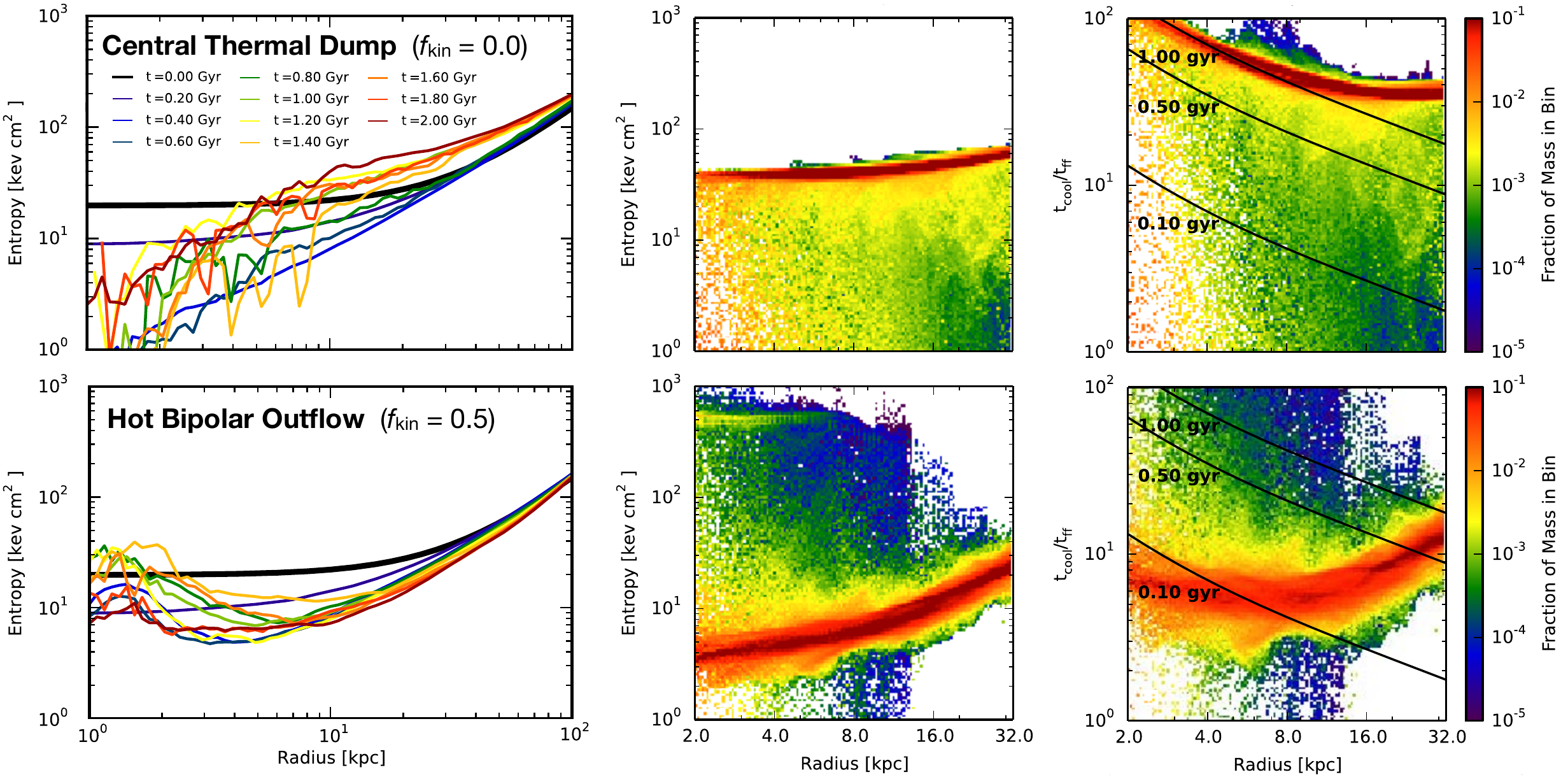}
  \caption{Contrasting outcomes of centralized thermal energy input (top row) and bipolar kinetic energy input (bottom row) in idealized numerical simulations of AGN feedback (adapted from \citep{Meece_2017ApJ...841..133M}).  The upper left panel shows a time series of median entropy profiles from a simulation with purely thermal feedback ($f_{\rm kin} = 0.0$) introduced at the center.  It does not achieve a steady state because it stimulates convection that flattens the global entropy gradient (upper middle panel) and eliminates buoyancy damping, allowing multiphase condensation to proceed despite a large median $t_{\rm cool}/t_{\rm ff}$ ratio (upper right panel).  Changing only the feedback mode leads to a very different outcome.  The lower left panel shows a time series from a simulation that injects feedback as a bipolar outflow in which half the energy is kinetic ($f_{\rm kin} = 0.5$). That simulation settles into a self-regulated steady state with a significant median entropy gradient (lower middle panel) more closely resembling observations of cool core clusters.  Buoyancy damping therefore limits condensation and allows AGN feedback to self-regulate with $t_{\rm ff} / t_{\rm cool} \sim 10$ (lower right panel).
  \label{figure:Convection}}
\end{figure}

Figure \ref{figure:Convection} illustrates how dramatically centralized heating and the convection it drives can boost multiphase condensation and interfere with self-regulation of AGN feedback.  It shows two simulations of an idealized galaxy cluster core from \citet{Meece_2017ApJ...841..133M} that differ only in how they inject AGN feedback energy.  The top row of panels shows a simulation with no kinetic feedback ($f_{\rm kin} = 0.0$).  All of the feedback energy is dumped into the center in thermal form, driving convection that transports thermal energy outward.  In the bottom row of panels, half of the feedback energy is kinetic ($f_{\rm kin} = 0.5$), injected as a bipolar outflow of hot gas.  Instead of driving convection, the bipolar outflow drills through the atmosphere's central region and thermalizes the feedback energy at greater altitudes, adding heat without flattening the global entropy gradient.

The simulation with centralized heating fails to reach a self-regulated state because its AGN feedback mode is unable to shut off multiphase condensation and AGN fueling.  Its median entropy profile at $\sim 10$~kpc therefore continually rises during the 2~Gyr period depicted. In the upper middle panel showing the entropy distribution at each radius 1.5~Gyr into the simulation, one can see a flat entropy core at $\sim 40 \, {\rm keV \, cm^2}$ extending to $\gtrsim 10$~kpc.  It also contains abundant low-entropy gas because flattening of the entropy profile has suppressed buoyancy damping.  At the upper right is the distribution of $t_{\rm cool} / t_{\rm ff}$ at each radius, with black lines tracing contours of constant $t_{\rm cool}$, as labeled.  Both the median cooling time and median entropy at small radii in this simulation substantially exceed the observed values in cool-core clusters.

In contrast, the simulation with kinetic feedback settles into a quasi-steady state within 1~Gyr.  Median entropy levels fluctuate in the central few kiloparsecs because of fluctuations in jet power, but the overall entropy profile remains remarkably stable and closely resembles X-ray observations of the entropy profiles in cool-core clusters.  AGN feedback is tightly self-regulated in this simulation because the significant and persistently positive entropy gradient limits multiphase condensation.  The entropy distribution in the lower middle panel shows that there is some low-entropy gas significantly below the median at $\lesssim 10$~kpc, but not nearly as much as in the centrally heated simulation, and there is no low-entropy gas at larger radii.  Also, the lower right panel shows that $t_{\rm cool} / t_{\rm ff} \sim 10$ for most of the gas within 30~kpc of the center.

These features are consistent with other comparisons of kinetic feedback with central thermal energy dumps in massive galaxies.  For example, \citet{Weinberger_2017MNRAS.465.3291W} directly compared the central-heating feedback algorithm used in the original Illustris simulations of cosmological galaxy formation with the updated kinetic feedback algorithm implemented in the IllustrisTNG simulations.  The spatially distributed heating that happens in IllustrisTNG results in galactic atmospheres more similar to observed ones, because it is less explosive. It therefore allows the entropy of ambient gas at $\sim 10$~kpc to remain near $10 \, {\rm keV \, cm^2}$, as in the lower panels of Figure \ref{figure:Convection}, rather than rising toward $100 \, {\rm keV \, cm^2}$, as in the upper panels.  Other simulation comparisons concur \citep{Gaspari+2014ApJ...783L..10G,Choi_2015MNRAS.449.4105C}, finding that kinetic feedback more successfully self-regulates, resulting in atmospheric X-ray luminosities more similar to observations.\footnote{One exception is the RomulusC simulation \citep{Tremmel_RomulusC_2019MNRAS.483.3336T}.  In that cosmological numerical simulation of AGN feedback in a $10^{14} \, M_\odot$ halo, centralized heat input ends up producing a bipolar outflow, because of an orbiting disk of dense gas around the AGN that redirects the hot outflow along its rotation axis.  When that happens, AGN feedback produces outcomes resembling those seen in simulations of kinetic AGN feedback \citep{2021MNRAS.504.3922C}.}

\subsubsection{Angular Momentum}

Rotation is a fifth way to promote multiphase condensation, but it has not been studied as systematically in numerical simulations of AGN feedback.  Angular momentum suppresses buoyancy damping by suppressing buoyancy itself, as rotational motions approach the local circular velocity.  However, multiphase condensation enabled by rotation does not boost AGN feedback unless some other mechanism can remove enough angular momentum from the condensates for them to sink inward and accrete onto the central black hole.

\begin{figure}[t]
  \centering
    \includegraphics[width=\textwidth]{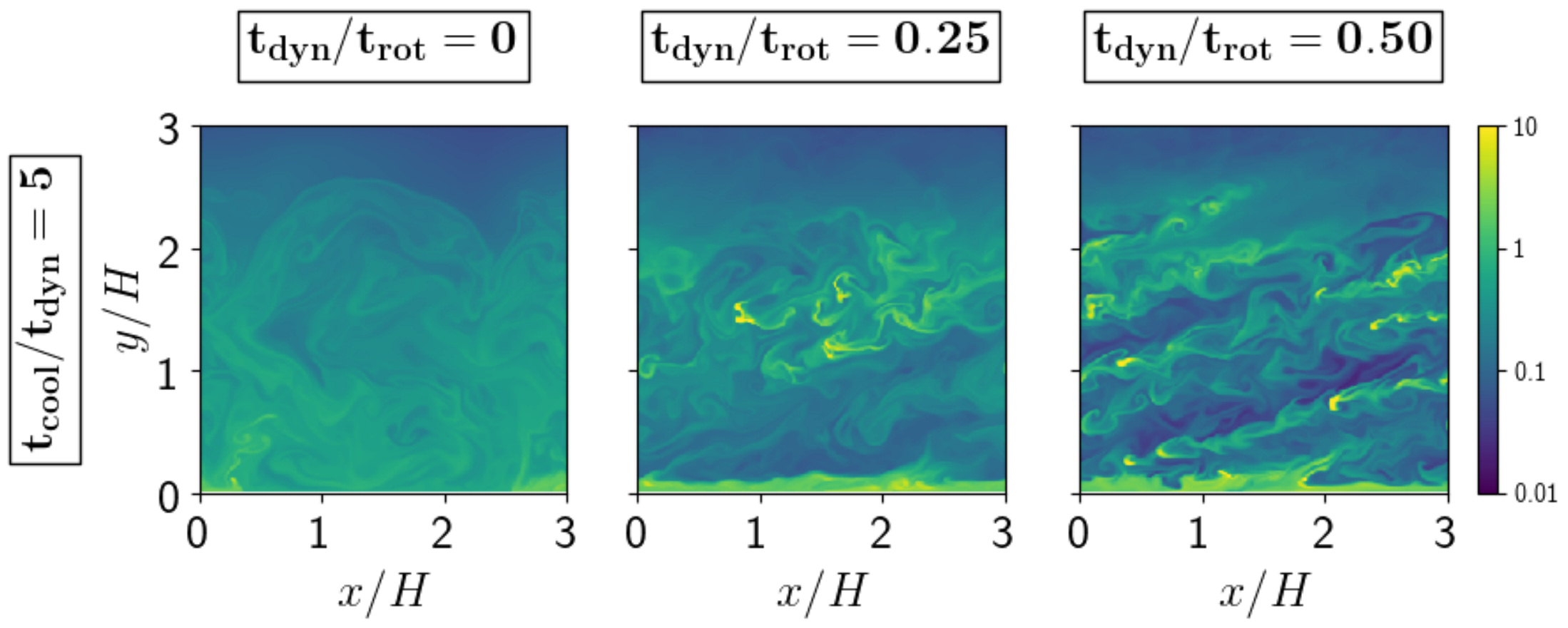}
  \caption{Fields of density contrasts  ($\rho/\rho_0$) from simulations of multiphase condensation in a rotating galactic atmosphere (from \citep{Sobacchi_2019MNRAS.486..205S}).  Rotation can promote multiphase condensation because it reduces the effective gravitational acceleration $g_{\rm eff}$ that determines the local dynamical time $t_{\rm dyn} \propto g_{\rm eff}^{-1/2}$, thereby lowering the frequency of buoyant oscillations and suppressing buoyancy damping.  However, the effects of Coriolis forces on an incipient condensate make rotation even more effective at promoting condensation.    Multiphase condensation in an atmosphere with a given value of $t_{\rm cool} / t_{\rm dyn}$ becomes increasingly likely as the atmosphere's ratio of dynamical time to rotational time increases, as shown in the figure's three panels, which depict density contrasts at a time $t = 7 t_{\rm cool}$ from the start of each simulation.
  \label{figure:Rotation}}
\end{figure}

Simulations by \citet{Sobacchi_2019MNRAS.486..205S} have shown that rotation can enhance multiphase condensation beyond what one would infer from reduction of the effective gravitational acceleration ($g_{\rm eff}$) in a rotating frame (see Figure \ref{figure:Rotation}).  The effect stems from Coriolis forces.  A condensing gas blob above the midplane of a galactic atmosphere rotating in a spherical potential will sink inward in the direction of the effective gravitational force.  But as the blob moves inward, the effective gravitational force acting on it changes direction, deflecting its descent and reducing the buoyancy effects that would otherwise damp condensation.  The result is a condensation criterion that depends on: (1) the effective dynamical time ($t_{\rm dyn} \propto g_{\rm eff}^{-1/2}$) through the $t_{\rm cool} / t_{\rm ff}$ ratio, (2) the ratio of dynamical time to rotation time ($t_{\rm dyn} / t_{\rm rot}$), and (3) the angle between the rotation axis and the direction of effective gravitational acceleration. 

\subsection{The Precipitation Limit}
\label{sec:PrecipitationLimit}

Taken as a whole, this set of astrophysical models implies that realistic galactic atmospheres become marginally susceptible to multiphase condensation and precipitation with median $t_{\rm cool} / t_{\rm ff}$ ratios significantly greater than unity.  The classic criterion $t_{\rm cool} / t_{\rm ff} \approx 1$ applies only to dynamically quiet, non-rotating, and homogeneous atmospheres with a significant entropy gradient, in which buoyancy damping can cause thermal instability to saturate.\footnote{The CGM model of \citet{MallerBullock_2004MNRAS.355..694M} is an important precursor to current precipitation-limited CGM models but differs from them in several key respects.  First, feedback is not assumed to maintain the CGM in a state of time-averaged thermal balance.  Second, the cooling time of the hot phase is assumed to be comparable to the halo's age and so is not determined by a local $t_{\rm cool}/t_{\rm ff}$ ratio.  Third, the atmosphere is assumed to be isentropic, which determines its structure and disables buoyancy damping.}  But realistic galactic atmospheres can be disturbed in many different ways, and moderate disturbances of atmospheres with $1 \lesssim t_{\rm cool} / t_{\rm ff} \lesssim 10$ drive them into precipitation \citep{Voit_2017_BigPaper,Voit_2018ApJ...868..102V}.

\begin{figure}[!t]
  \centering
    \includegraphics[width=4.0in]{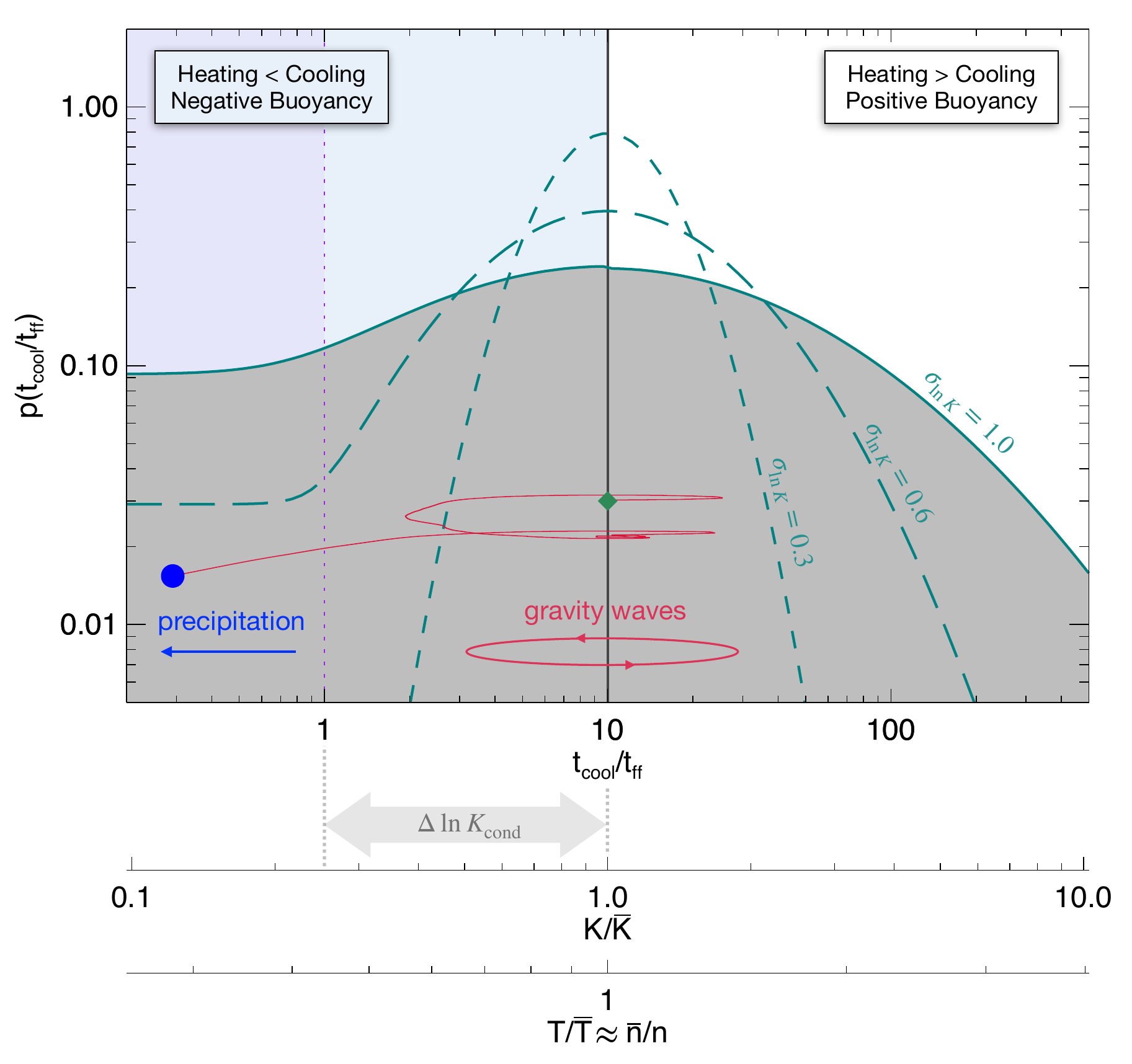}
  \caption{Schematic illustration of the distribution of $t_{\rm cool} / t_{\rm ff}$ in a galactic atmosphere that is marginally susceptible to precipitation (from \citep{Voit_2021ApJ...908L..16V}).  The teal lines show probability distribution functions that are lognormal at $t_{\rm cool} / t_{\rm ff} > 1$, with a median at $t_{\rm cool} / t_{\rm ff} = 10$.  Their widths are determined by the standard deviation of entropy fluctuations around the median.  A solid line represents $\sigma_{\ln K} = 1.0$, a long-dashed line represents $\sigma_{\ln K} = 0.6$, and a short dashed line represents $\sigma_{\ln K} = 0.3$.  The difference between the median entropy and the entropy level at which $t_{\rm cool} / t_{\rm ff} = 1$ is $\Delta \ln K_{\rm cond}$.  If $\sigma_{\ln K} \ll \Delta \ln K_{\rm cond}$ then those perturbations simply oscillate as internal gravity waves, and there is no precipitation.  But as $\sigma_{\ln K}$ approaches $\Delta \ln K_{\rm cond}$, the tail of the distribution extends below $t_{\rm cool} / t_{\rm ff} = 1$, and perturbations in the tail precipitate.  
  \label{figure:tctff}}
\end{figure}

Figure \ref{figure:tctff} schematically illustrates how a marginally unstable galactic atmosphere's median $t_{\rm cool} / t_{\rm ff}$ ratio relates to the classic condensation criterion $t_{\rm cool} / t_{\rm ff} \approx 1$ in the presence of dynamical disturbances.  Each of the figure's distribution functions has a median value $t_{\rm cool} / t_{\rm ff} = 10$ and is lognormal above $t_{\rm cool} / t_{\rm ff} = 1$.  The standard deviation $\sigma_{\rm \ln K}$ of entropy fluctuations determines the width of each distribution, and $\Delta \ln K$ is the entropy difference corresponding to the interval between $t_{\rm cool} / t_{\rm ff} = 10$ and $t_{\rm cool} / t_{\rm ff} = 1$.

Entropy fluctuations that are small compared to $\Delta \ln K$ oscillate as internal gravity waves and do not induce precipitation.  But disturbances within a galactic atmosphere can drive internal gravity waves to larger amplitudes resulting in correspondingly large entropy fluctuations.  If the dispersion $\sigma_{\ln K}$ of the driven fluctuations becomes comparable to $\Delta \ln K$, then the tail of the distribution function extends below $t_{\rm cool} / t_{\rm ff} = 1$, resulting in a subpopulation of perturbations that buoyancy cannot damp.  Those perturbations therefore condense, producing precipitation.  Consequently, the median $t_{\rm cool} / t_{\rm ff}$ ratio of a marginally precipitating atmosphere depends on the dispersion $\sigma_{\ln K}$ of entropy perturbations within it.

This section has described several kinds of disturbances capable of producing such perturbations in a stratified galactic atmosphere.  They include bulk uplift of low-entropy gas, turbulent levitation of low-entropy gas parcels, cosmological infall of low-entropy gas, and stripping of low-entropy gas from satellite galaxies.  The first two of these sources are natural consequences of AGN feedback, meaning that the median $t_{\rm cool} / t_{\rm ff}$ ratio of atmospheres that AGN feedback suspends in a marginally precipitating state will be significantly greater than unity.  

Many idealized simulations of kinetic AGN feedback in massive galaxies \citep[e.g.,][]{Gaspari+2012ApJ...746...94G,Li_2015ApJ...811...73L,Prasad_2015ApJ...811..108P,Yang_2016b_ApJ...818..181Y,Meece_2017ApJ...841..133M,Prasad_2020ApJ...905...50P} have demonstrated that self-regulation results in a marginally precipitating atmosphere with $5 \lesssim \min (t_{\rm cool} / t_{\rm ff}) \lesssim 20$ in its ambient phase.  The result shown in the lower panels of Figure \ref{figure:Convection} is representative. Generalizing from those results, we infer that self-regulating feedback mediated by precipitation imposes a lower limit of $t_{\rm cool} / t_{\rm ff} \gtrsim 10$ on the median timescale ratio in a massive galaxy's atmosphere, which we will call the \textit{precipitation limit}.  It is not a precise limit, derivable from first principles, but rather a phenomenological one, depending on the disturbances that a galactic atmosphere is exposed to, including uplift, turbulence, and cosmological infall.\footnote{Like the blackbody limit, the precipitation limit appears to be an emergent property of a complex system, rather than the characteristic signature of a particular physical mechanism.}  An atmosphere's precipitation limit may also depend on how strongly magnetic fields \citep{Ji_2018MNRAS.476..852J} and angular momentum \citep{Sobacchi_2019MNRAS.486..205S} interfere with buoyancy damping.  As a result, future models of precipitation will need to explore secondary dependences of the critical value of $t_{\rm cool} / t_{\rm ff}$ on the phenomena that inhibit buoyancy and excite atmospheric perturbations, such as cosmological infall, motions of satellite halos, and feedback from the satellite galaxies embedded within those halos.

\section{Evidence for a Precipitation Limit \label{sec:Evidence}}

Many complementary observations now indicate that the ambient atmospheres of present-day massive galaxies rarely have a median $t_{\rm cool} / t_{\rm ff}$ ratio below $\sim 10$, in alignment with the precipitation hypothesis.  Apparently, a galactic atmosphere's ambient pressure and density have natural upper limits depending on the circular velocity of the confining gravitational potential. Furthermore, those limits are similar in magnitude to the pressures and densities that emerge from numerical simulations of precipitation-regulated feedback.  This section presents some of that evidence, starting with observations of galaxy cluster atmospheres and proceeding to observations of galactic atmospheres in lower mass halos.

\subsection{Profiles of $t_{\rm cool} / t_{\rm ff}$}
\label{sec:tctff_profiles}

X-ray observations of a sufficiently massive galaxy can be deprojected to obtain radial profiles of electron density, gas temperature, and heavy-element abundance (\S \ref{sec:RadialProfiles}).  One can compute a radial cooling-time profile from that information (\S \ref{sec:CentralCoolingTime}).  One can also estimate a radial freefall time profile from the same X-ray data by assuming the atmosphere to be in hydrostatic equilibrium (\S \ref{sec:ClusterMass}).  However, hydrostatic equilibrium may not always be a good assumption, and the X-ray data quality frequently does not allow a high-precision measurement of $t_{\rm ff} (r)$ with adequate radial resolution.  Therefore, observers sometimes rely on parametric models of the underlying mass distribution, jointly constrained by the X-ray observations and optical observations of the central galaxy's starlight \citep[e.g,][]{Hogan_2017_mass_ApJ...837...51H}.

\begin{figure}[!t]
\centering
\includegraphics[width=5.0in]{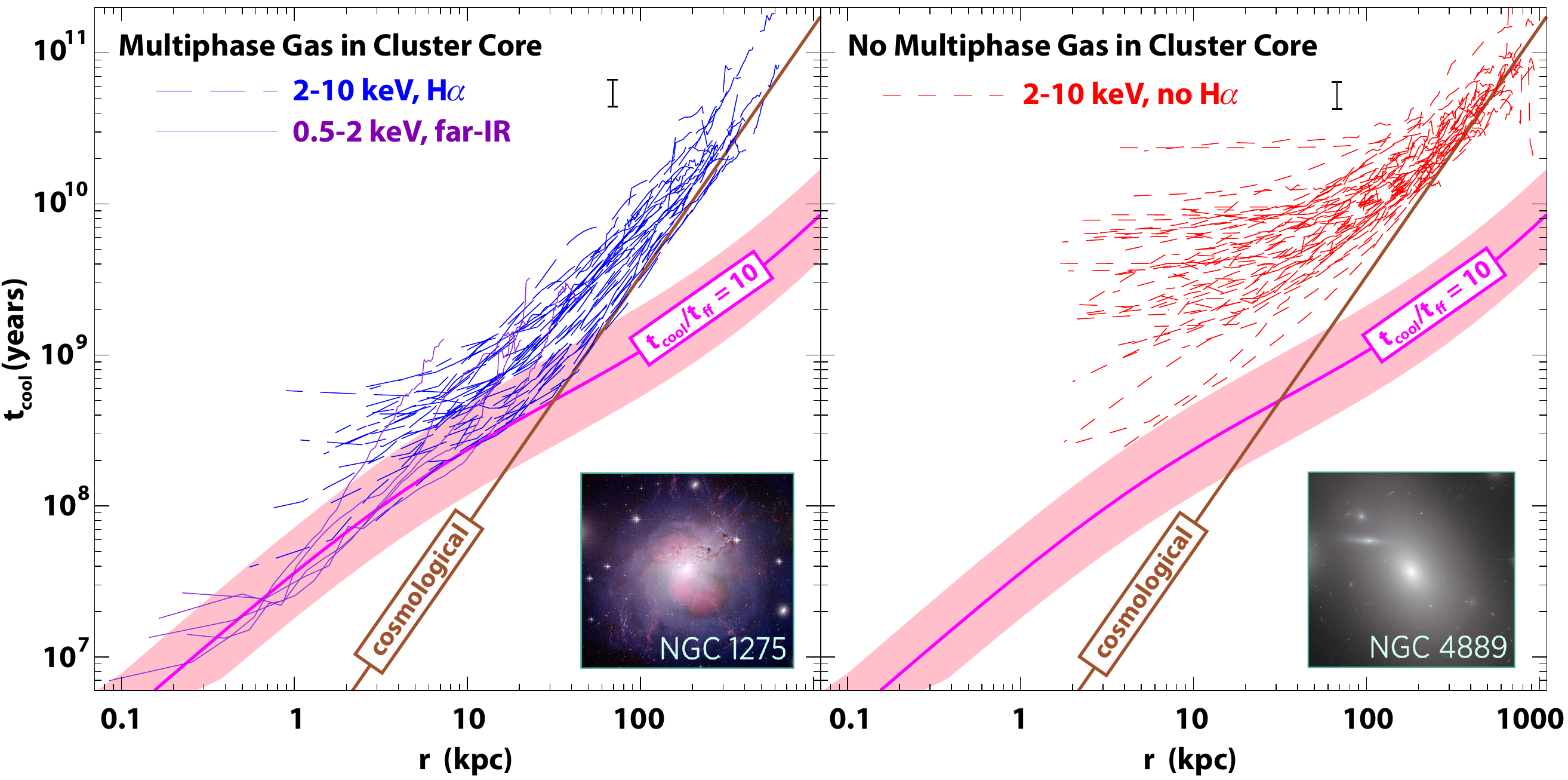}
\caption{Cooling-time profiles of multiphase (left) and single phase (right) galaxy clusters from the ACCEPT sample (adapted from \citep{Voit_2015Natur.519..203V}).  Long-dashed blue lines represent galaxy cluster atmospheres with temperatures in the 2--10~keV range that also have detectable H$\alpha$ emission in their cores, signifying the presence of BCG nebulae (like NGC 1275 in the Perseus Cluster).  Solid purple lines represent atmospheres in lower-mass halos, with temperatures of 0.5--2~keV, that have detectable far-IR emission signifying the presence of cool, dusty gas.  Short-dashed red lines represent galaxy cluster atmospheres with temperatures of 2--10~keV and no cooler gas detectable through H$\alpha$ emission (like NGC 4889 in the Coma Cluster).  Both sets of galaxy cluster profiles converge at large radii to the cosmological baseline profile (\S \ref{sec:Kbaseline}) shown by the solid brown line, but they are distinctly different at small radii.  The cluster cores with multiphase gas tend to have $10 \lesssim \min (t_{\rm cool} / t_{\rm ff}) \lesssim 30$, with a lower envelope tracing the solid magenta line representing $t_{\rm cool} / t_{\rm ff} = 10$, while the central cooling times of clusters without multiphase gas are typically an order of magnitude greater.
\label{figure:VoitNature2015}}
\end{figure}

\subsubsection{Galaxy Cluster Cores}
\label{sec:tctff_Clusters}

Our review of the evidence begins with the central regions of galaxy clusters.  The two panels of Figure \ref{figure:VoitNature2015} show two sets of observed galaxy cluster cooling-time profiles and compare them with a parametric model profile having $t_{\rm cool} / t_{\rm ff} = 10$ (magenta line).  A shaded region around the model line corresponds to the range $5 \lesssim \min (t_{\rm cool} / t_{\rm ff}) \lesssim 20$ that typically emerges from numerical simulations of precipitation-regulated galaxy cluster cores.  The figure vividly illustrates how strongly the presence of multiphase gas in a cluster core depends on the ambient atmosphere's cooling time (see also Figure \ref{figure:tc_K0_threshold} and \S \ref{sec:multiphase_tcool}).  It also reveals that the cooling-time profiles of cluster cores have a distinct lower limit that depends on radius.  At small radii, that lower limit corresponds to $\min (t_{\rm cool} / t_{\rm ff}) \approx 10$.

\citet{McCourt+2012MNRAS.419.3319M} drew attention to the significance of $\min (t_{\rm cool} / t_{\rm ff}) \approx 10$ in multiphase cluster cores and conjectured that those atmospheres became multiphase in places where $\min (t_{\rm cool} / t_{\rm ff}) < 10$.  Their conjecture was based on simulations by \citet{Sharma_2012MNRAS.420.3174S} suggesting that $\min (t_{\rm cool} / t_{\rm ff}) < 10$ might be a local criterion for multiphase condensation in spherical potential wells containing undisturbed atmospheres.  However, followup work has demonstrated that the criteria for multiphase condensation do not depend strongly on potential well geometry but rather on other global features of the atmosphere \citep{Meece_2015ApJ...808...43M,ChoudhurySharma_2016MNRAS.457.2554C,Voit_2017_BigPaper} (see also \S \ref{sec:CondensationDrivers}).  Meanwhile, numerical simulations of kinetic AGN feedback in multiphase galaxy cluster cores have demonstrated that $\min (t_{\rm cool} / t_{\rm ff})$ in the ambient gas can fluctuate over time and that uplift is often a precursor to multiphase condensation \citep{Gaspari+2012ApJ...746...94G,LiBryan2014ApJ...789...54L,Prasad_2015ApJ...811..108P}.

\begin{figure}[!t]
\centering
\includegraphics[width=5.3in]{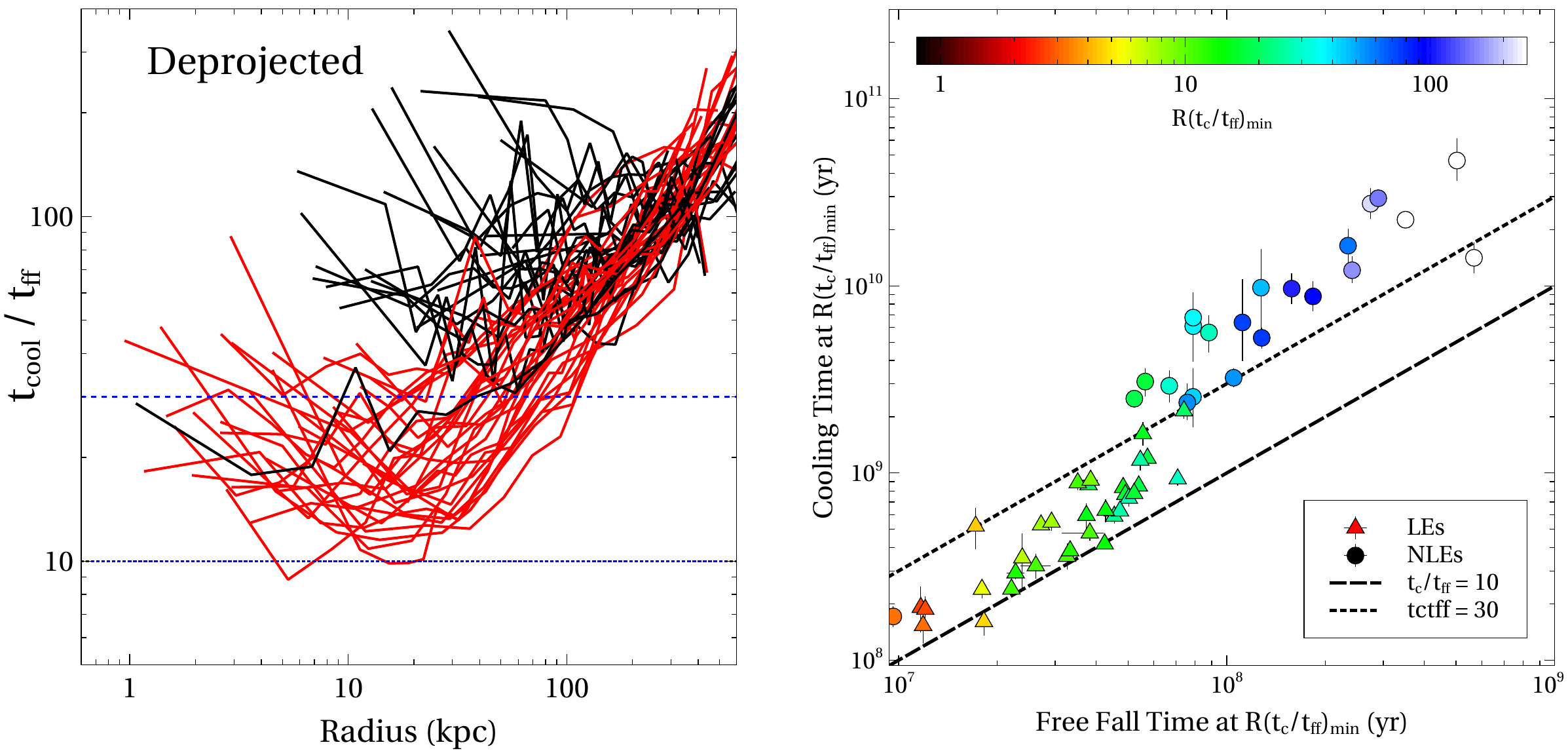}
\caption{Radial profiles of $t_{\rm cool} / t_{\rm ff}$ and the relationships between $t_{\rm cool}$, $t_{\rm ff}$, and multiphase gas (from \citet{Hogan_2017_tctff}).  The left panel shows a galaxy cluster data set similar to that in Figure \ref{figure:VoitNature2015}, except that $t_{\rm cool} (r)$ is now divided by $t_{\rm ff} (r)$.  Red lines show multiphase clusters with optical line emission (LE) and black lines show single phase clusters with no line emission (NLE).  The right panel shows the relationship between $t_{\rm cool}$ and $t_{\rm ff}$ at the radius $R(t_{\rm c} / t_{\rm ff})_{\rm min}$ where $t_{\rm cool} / t_{\rm ff}$ reaches its minimum value.  Triangles representing multiphase clusters occupy the range between the dashed line representing $t_{\rm cool} / t_{\rm ff} = 10$ and the dotted line representing $t_{\rm cool} / t_{\rm ff} = 30$.  Nearly all of the circles representing single phase galaxy clusters lie above the $t_{\rm cool} / t_{\rm ff} = 30$ line.
\label{figure:Hogan_tctff}}
\end{figure}

More recent observational investigations of $t_{\rm cool} / t_{\rm ff}$ profiles in galaxy cluster cores, based on fully deprojected X-ray data and more detailed gravitational potential modeling \citep{Hogan_2017_tctff}, have confirmed that the population as a whole generally has $\min (t_{\rm cool} / t_{\rm ff}) \approx 10$, with a few exceptions \citep{McDonald_Phoenix_2019ApJ...885...63M}. They also show that the $t_{\rm cool} / t_{\rm ff}$ profiles of individual galaxy clusters tend to flatten in the vicinity of $10$~kpc, with differing values of $\min (t_{\rm cool} / t_{\rm ff})$.  Two complementary features of the galaxy-cluster population can be seen in the data:
\begin{enumerate}
    \item Cluster cores with more than $10^8 \, M_\odot$ of detectable multiphase gas have $t_{\rm cool} \lesssim 1$~Gyr at 10~kpc, equivalent to core entropy levels $K_0 \lesssim 30 \, {\rm keV \, cm^2}$ (\S \ref{sec:multiphase_tcool}).
    \item Cluster cores with multiphase gas also have $10 \lesssim \min( t_{\rm cool} / t_{\rm ff} )  \lesssim 30$ (see Figure \ref{figure:Hogan_tctff}).
\end{enumerate}
Feature 1 implies that $\min (t_{\rm cool} / t_{\rm ff}) \lesssim 30$ in multiphase galaxy-cluster atmospheres, because $t_{\rm ff} (10 \, {\rm kpc}) \approx 33 \, \sigma_{300}^{-1} $~Myr, where $\sigma_{300}$ is the central galaxy's stellar velocity dispersion in units of 300 km s$^{-1}$.  Feature 2 implies that AGN feedback prevents the ambient $t_{\rm cool} / t_{\rm ff}$ ratio from dropping much below 10.

While there is considerable tension between Feature 2 and the early conjecture that $t_{\rm cool} / t_{\rm ff} \approx 10$ might be a threshold for multiphase condensation of initially small perturbations in a spherical potential well, there is much less tension between Feature 2 and numerical simulations of precipitation-regulated kinetic AGN feedback.  Those simulations are consistent with a picture in which precipitation-regulated feedback imposes a lower limit on ambient cooling time at small radii, by enforcing $\min (t_{\rm cool} / t_{\rm ff}) \gtrsim 10$, and cosmological structure formation imposes a lower limit on ambient cooling time at large radii, corresponding to the cosmological baseline entropy profile (\S \ref{sec:Kbaseline}).  Those limits are evident in the left panel of Figure~\ref{figure:VoitNature2015}.\footnote{The same limits are also evident in Figure~\ref{figure:UniversalEntropy}, in which the $K \propto r^{2/3}$ profile found at small radii is consistent with $10 \lesssim t_{\rm cool} / t_{\rm ff} \lesssim 20$ in an isothermal potential well \citep[e.g.,][]{Voit+2015ApJ...803L..21V,Hogan_2017_tctff}.} 

\begin{figure}[!t]
\centering
\includegraphics[width=5.3in]{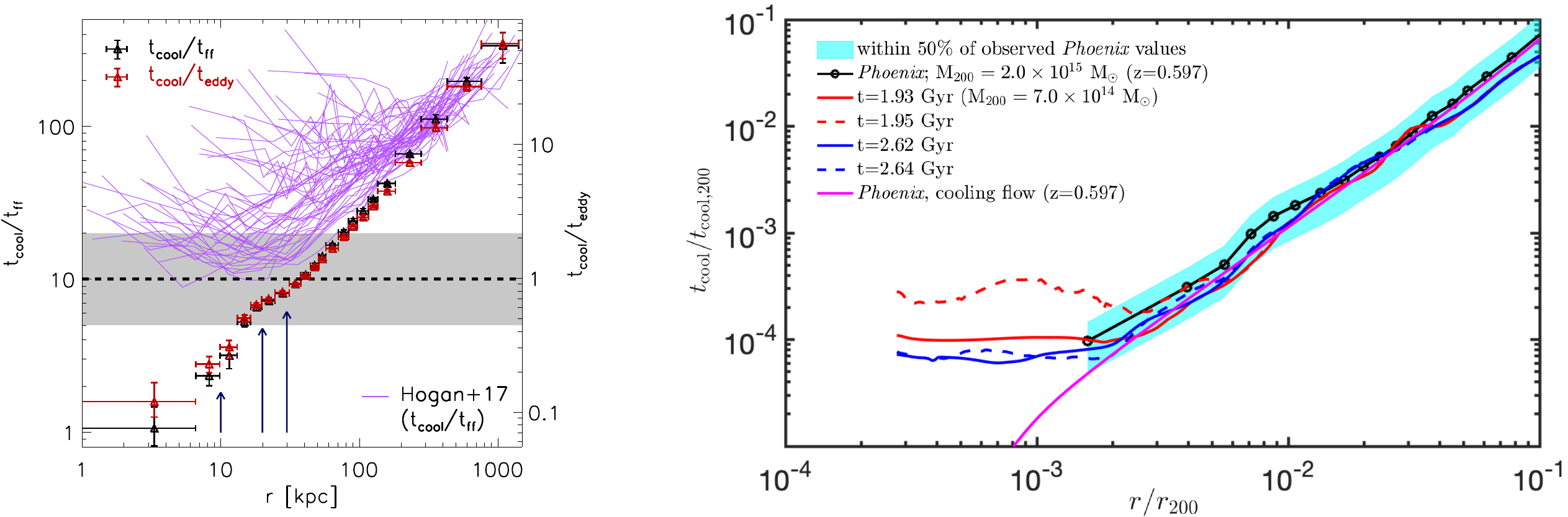}
\caption{The Phoenix Cluster's unusual cooling time profile.  It reaches $t_{\rm cool} / t_{\rm ff} \lesssim 3$ (left panel, from \citep{McDonald_Phoenix_2019ApJ...885...63M}) and is similar to the brief strong-cooling events that occasionally happen in AGN feedback simulations (right panel, from \citep{Prasad_Phoenix_2020MNRAS.495..594P}).  Black triangles in the left panel represent $t_{\rm cool} / t_{\rm ff}$ ratios observed in the Phoenix Cluster, and purple lines show the same $t_{\rm cool} / t_{\rm ff}$ profiles as in Figure~\ref{figure:Hogan_tctff}.  Black circles in the right panel show $t_{\rm cool} (r)$ for Phoenix, scaled by both $r_{200 {\rm c}}$ and $t_{\rm cool}(r_{200 {\rm c}})$, with cyan shading indicating a 50\% range around the observations.  The red and blue lines represent brief intervals during the simulation (at 1.93--1.95~Gyr and 2.62--2.64~Gyr), in which the scaled cooling profiles match those observed in Phoenix and have $\min (t_{\rm cool} / t_{\rm ff}) < 3$ as small radii.  The magenta line represents a pure cooling flow model.
\label{figure:tctff_Phoenix}}
\end{figure}

Whether or not the rarity of cluster cores observed to have $\min(t_{\rm cool} / t_{\rm ff}) \ll 10$ is consistent with numerical simulations of precipitation-regulated feedback remains an unsettled question \citep{McNamara_2016ApJ...830...79M,Hogan_2017_tctff,Pulido_2018ApJ...853..177P}.  The Phoenix Cluster is a notable exception to the usual rule (see Figure~\ref{figure:tctff_Phoenix}).  Inside of 10~kpc, the $t_{\rm cool} / t_{\rm ff}$ ratio of its ambient gas drops to less than 3, and the structure of its atmosphere is consistent with that of a pure cooling flow.  This extreme state is likely to be short-lived, because otherwise the currently enormous central star-formation rate ($\sim 800 \, M_\odot \, {\rm yr}^{-1}$) would have produced a central galaxy with many more stars. Also, the central galaxy's AGN is generating a strong feedback response that may still be in its early stages \citep{Prasad_Phoenix_2020MNRAS.495..594P}.

The right panel of Figure~\ref{figure:tctff_Phoenix} compares the extreme state of the Phoenix Cluster with two short-lived ($\sim 20$~Gyr) episodes in a numerical simulation of precipitation-regulated feedback.  Occasionally, the ambient $t_{\rm cool} / t_{\rm ff}$ ratio of the simulated galactic atmosphere drops below 5.  That happens when the supply of cold gas to the central AGN is too small to fuel sufficient feedback.  Cooling then corrects that situation on a timescale of a few tens of Myr, reigniting feedback and restoring a state with $\min (t_{\rm cool} / t_{\rm ff}) \approx 10$.  However, the incidence of those excursions to low $t_{\rm cool} / t_{\rm ff}$ in some (but not all) simulations of precipitation-regulated feedback appears to be too frequent to account for the comparative rarity of examples like the Phoenix Cluster \citep{Hogan_2017_tctff,Pulido_2018ApJ...853..177P}. 

Settling this question will require improvements in both observations and simulations.  On the observational side, we need to make sure that our current galaxy-cluster samples are not missing other Phoenix-like clusters in which the brightness of a central AGN could be masking emission from a strong central cooling flow \cite[e.g.,][]{Somboonpanyakul_2018,Somboonpanyakul_2021ApJ...910...60S,Donahue_2020ApJ...889..121D}.  On the simulation side, we need to understand the circumstances that result in occasional strong-cooling episodes.  In particular, do they result from unrealistic idealizations of the simulation conditions?  Or are they an inevitable feature of precipitation-regulated feedback itself?  And if they are inevitable, then what determines the expected fraction of cluster cores in a strong-cooling state like the one in the Phoenix Cluster?

\subsubsection{Massive Galaxies}
\label{sec:MassiveGalaxiesEvidence}

\begin{figure}[t]
\centering
\includegraphics[width=5.3in]{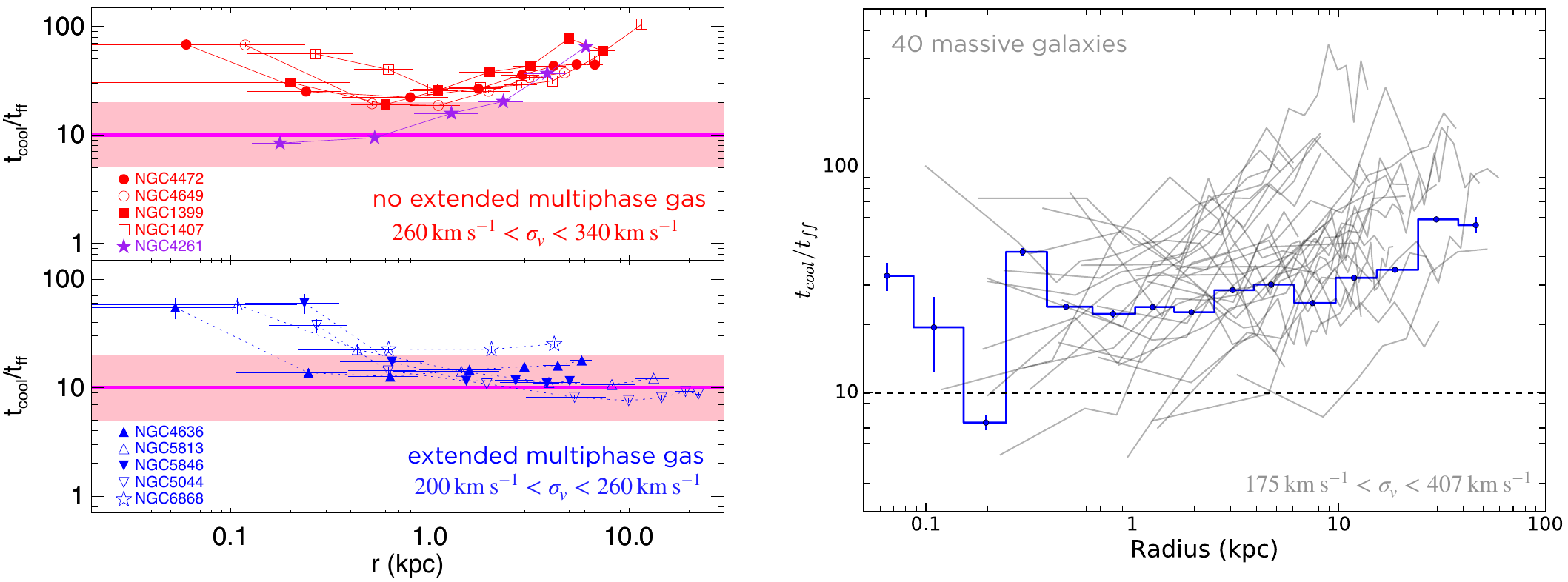}
\caption{Profiles of $t_{\rm cool}/t_{\rm ff}$ in and around massive galaxies. The left panel (from \citep{Voit+2015ApJ...803L..21V}) shows the profiles of ten massive elliptical galaxies observed by \citet{Werner+2012MNRAS.425.2731W,Werner+2014MNRAS.439.2291W}.  On top are five elliptical galaxies without multiphase gas beyond the central kiloparsec.  On the bottom are five elliptical galaxies with multiphase gas that does extend beyond the central region. Collectively, the lower envelope of this sample of ten X-ray bright galaxies tracks $t_{\rm cool} / t_{\rm ff} \sim 10$ over two orders of magnitude in radius.  The right panel (from \citep{Babyk2018ApJ...862...39B}) shows $t_{\rm cool}/t_{\rm ff}$ profiles for a sample of 40 massive galaxies that includes both elliptical and spiral galaxies. A thick blue line shows the sample's mean profile.  The lower envelope of this larger sample also tracks $t_{\rm cool} / t_{\rm ff} \approx 10$. 
\label{figure:Werner_tctff}}
\end{figure}

\begin{figure}[t]
\centering
\includegraphics[width=5.3in]{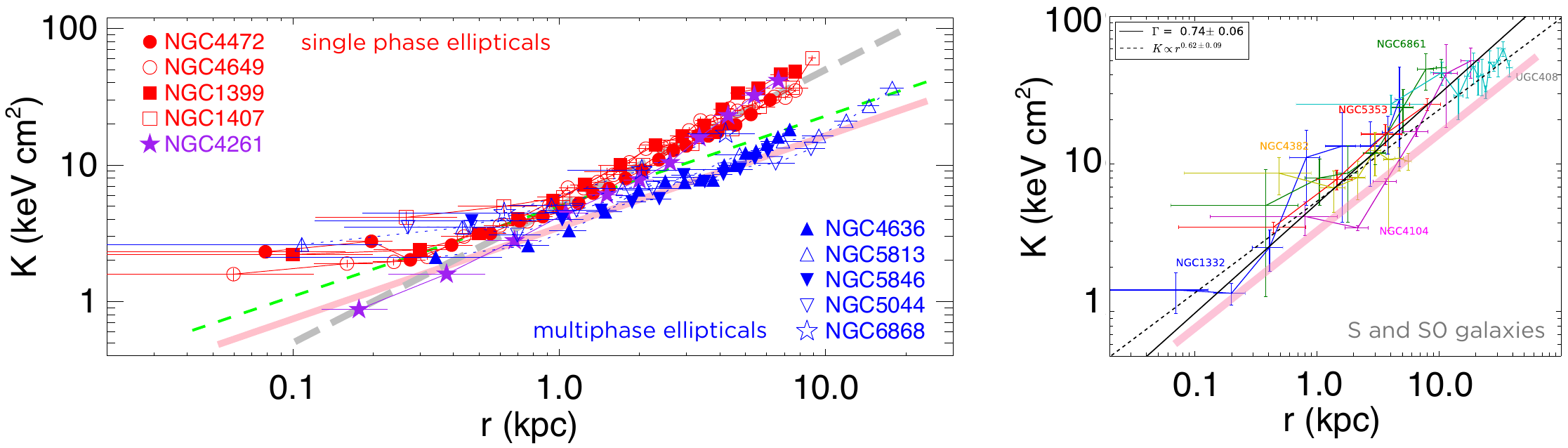}
\caption{Atmospheric entropy profiles of massive galaxies.  The left panel (from \citep{Voit+2015ApJ...803L..21V}) shows the elliptical galaxy sample from the left panel of Figure \ref{figure:Werner_tctff}. A solid pink line illustrates the entropy profile $K_{\rm pre}(r) = (3.5 \, {\rm keV \, cm^2}) \, r_{\rm kpc}^{2/3}$ corresponding to the precipitation limit ($t_{\rm cool} / t_{\rm ff} \approx 10$) in $\sim 1$~keV gas of solar composition.  A long-dashed gray line illustrates the profile $K(r) = (5 \, {\rm keV \, cm^2}) \, r_{\rm kpc}$. A short-dashed green line shows the entropy locus $K_{\rm eq}(r) = (5 \, {\rm keV \, cm^2}) \, r_{\rm kpc}^{2/3}$ above which SNIa heating supplied by an old stellar population exceeds radiative cooling.  The right panel (from \citep{Babyk2018ApJ...862...39B}) shows just the spiral (S) and lenticular (S0) galaxies among the 40 in the right panel of Figure \ref{figure:Werner_tctff}, with a pink line indicating $K_{\rm pre}(r)$.
\label{figure:K-r_galaxies}}
\end{figure}

Moving down an order of magnitude or more in halo mass, one continues to find $\min (t_{\rm cool} / t_{\rm ff}) \approx 10$ in the atmospheres of individual galaxies (see Figure \ref{figure:Werner_tctff}).  However, the relationship between atmospheric structure and the presence of extended multiphase gas appears somewhat different from the relationship that prevails in the cores of galaxy clusters, which can have much greater gas pressures.  \citet{Werner+2014MNRAS.439.2291W} noticed a distinct difference between the entropy profiles of massive elliptical galaxies with extended multiphase gas and those without it (Figure \ref{figure:K-r_galaxies}).  The ones that contain extended multiphase gas tend to have $K \propto r^{2/3}$ at 1--10~kpc, resembling the entropy profiles observed at $\sim 10$~kpc in multiphase galaxy cluster cores.  Figure \ref{figure:Werner_tctff} shows that those galaxies have $10 \lesssim t_{\rm cool} / t_{\rm ff} \lesssim 20$ in that radial range, in alignment with simulations of precipitation-regulated feedback.   But the ones without extended multiphase gas have steeper entropy profiles, with $K \propto r$ at 1--10~kpc, unlike the nearly isentropic cores of single-phase galaxy clusters. 

The entropy-profile slope observed among the single phase ellipticals in Figure \ref{figure:K-r_galaxies} ($K \propto r$) is consistent with the slope predicted by a model of steady-state outflow driven by SNIa heating of gas coming from the galaxy's old stellar population  \citep{Voit+2015ApJ...803L..21V,Voit_2020ApJ...899...70V}.  A short-dashed green line in that figure shows the entropy locus $K_{\rm eq} (r)$ along which SNIa heating would equal radiative cooling in $\sim 1$~keV gas of solar composition.  Above that line, a combination of SNIa heating and mass loss from the old stars can drive a slow outflow that has an entropy slope
\begin{equation}
    \alpha_K \: \approx \: \frac {d \ln K} {d \ln r} 
            \: = \:  
        \frac {2} {3} 
        \left[  1.1 \left( \frac {\sigma_v} {240 \, {\rm km \, s^{-1}}} \right)^{-2} - \, 0.1 \right]^{-1}
    \label{eq:alphaK_sigmav}
\end{equation}
depending primarily on the depth of the galaxy's potential well, as reflected by the galaxy's central stellar velocity dispersion ($\sigma_v$), and depending only weakly on the age of its old stellar population \citep{Voit_2020ApJ...899...70V}.  Given the typical stellar velocity dispersion of the single phase elliptical galaxies in this small sample ($\sim 300 \, {\rm km \, s^{-1}}$), one expects $\alpha_K \approx 1$, similar to what is observed.\footnote{We have already described two other mechanisms that can yield $\alpha_K \approx 1$. One is cosmological accretion (\S \ref{section:CosAtmo}). The other is a pure cooling flow in a nearly isothermal potential well (\S \ref{sec:CoolingFlows}). A cosmological origin for $\alpha_K \approx 1$ is plausible only in atmospheric regions where $t_{\rm cool}$ exceeds the age of the universe. A cooling-flow origin is plausible only where cooling substantially exceeds heat input. Neither of those conditions apply to the gas at 1--10~kpc from the centers of the massive elliptical galaxies in Figure~\ref{figure:K-r_galaxies}.}  

Interestingly, an atmospheric entropy slope steeper than $K \propto r^{2/3}$ focuses cooling and multiphase condensation onto the central black hole because the ambient $t_{\rm cool} / t_{\rm ff}$ ratio then increases with radius.   According to \citet{Voit+2015ApJ...803L..21V}, the entropy profile corresponding to $t_{\rm cool} / t_{\rm ff} = 10$ is
\begin{equation}
    K_{\rm pre} (r) \approx (3.5 \, {\rm keV \, cm^{-2}}) 
        \, T_{\rm keV}^{1/3} \, \Lambda_{3e-23}^{2/3} \, \sigma_{250}^{-2/3} \, r_{\rm kpc}^{2/3}
\end{equation}
in an isothermal potential well, given $T_{\rm keV} \equiv kT / ({\rm 1 \, keV})$, $\Lambda_{3e-23} \equiv \Lambda(T) / (3 \times 10^{-23} \, {\rm erg \, cm^3 \, s^{-1}})$, $\sigma_{250} \equiv \sigma_v / (250 \, {\rm km \, s^{-1}})$, and $r_{\rm kpc} \equiv r / (1 \, {\rm kpc})$.  Entropy profiles steeper than $K_{\rm pre} \propto r^{2/3}$ must intersect that critical locus somewhere, and the entropy profile of NGC~4261 (purple stars) does so at $r \approx 1 \, {\rm kpc}$.  Inside of that radius, the atmosphere of NGC~4261 is multiphase, with a dusty disk of condensed gas at  $\sim 0.1$~kpc \citep{Jaffe_N4261_1996ApJ...460..214J}, presumably the source of cold fuel for the powerful ($\sim 10^{44} \, {\rm erg \,s^{-1}}$) bipolar outflow emanating from the central engine (see Figure \ref{fig:narrow_jets}).  All of these features are consistent with the precipitation hypothesis.

The entropy profiles of the other four single phase ellipticals in Figure~\ref{figure:K-r_galaxies} change slope near the pink line indicating $K_{\rm pre}(r)$, reaching $\min(t_{\rm cool} / t_{\rm ff}) \approx 20$ as they bend (see Figure \ref{figure:Werner_tctff}).  Apparently, central heating has boosted their central entropy levels, relative to NGC~4261, along with their $t_{\rm cool} / t_{\rm ff}$ ratios at $< 1$~kpc. Their inner atmospheres are therefore less prone to precipitation, but perhaps only temporarily. The entropy profiles of the multiphase ellipticals likewise flatten at $< 1$~kpc, with $t_{\rm cool} / t_{\rm ff} \gg 10$ at small radii, perhaps explaining why the AGNs in these galaxies are also temporarily much less powerful than the one in NCG~4261.

The bottom line of these analyses
is that nearly all of the massive galaxies observed to date, including both the BCGs in massive galaxy clusters and individual massive elliptical galaxies, adhere to the precipitation limit corresponding to $\min(t_{\rm cool} / t_{\rm ff}) \approx 10$.  X-ray observations of massive spiral galaxies are less plentiful but reach the same conclusion (see the right panel of Figure \ref{figure:K-r_galaxies}).  The next question to address is, how far down in halo mass does this trend persist?

\subsection{Precipitation-Limited Luminosity}

Observations of radially resolved $t_{\rm cool} / t_{\rm ff}$ profiles within lower mass halos become more difficult because the X-ray surface brightness of a galactic atmosphere diminishes as halo mass declines (\S \ref{sec:MassiveGalaxies}).  However, easier measurements of total X-ray luminosity $L_X (R)$ performed within a circular aperture of radius $R$ enable a cruder test of the precipitation hypothesis, along lines first proposed by \citet{Sharma+2012MNRAS.427.1219S}.  The test stems from the condition
\begin{equation}
    n_e (r) \lesssim \frac {3kT} {10 \, t_{\rm ff} \, \Lambda(T,Z)}
    \label{eq:nelim}
\end{equation}
implied by the precipitation limit ($t_{\rm cool} / t_{\rm ff} \gtrsim 10$).  Inserting this condition into the integral giving the X-ray luminosity of a spherical atmosphere of radius $R$ leads to
\begin{equation}
    L_X(R) \: = \: \int_0^R 4 \pi r^2 \Lambda(T) \, n_e^2(r) \, dr 
           \: \lesssim \: \int_0^R 4 \pi r^2 \Lambda \left( \frac{3kT}{10 \, t_{\rm ff} \, \Lambda} \right)^2 dr 
    \; \; .
\end{equation}
The integral can be simplified by assuming the atmosphere is isothermal in a potential well that is also isothermal, with a constant circular velocity $v_{\rm c}^2 \approx 2 \sigma_v^2$.  Making those approximations yields the luminosity limit
\begin{equation}
    L_X(R) \: \lesssim \: \frac {9 \pi} {25} (kT)^2 \Lambda^{-1} \sigma_v^2 R
    \label{eq:LxLim}
\end{equation}
in which the numerical factor $9 \pi / 25 \approx 1$ can be ignored \citep{Voit2018_LX-T-R}. 

\begin{figure}[t]
\centering
\includegraphics[scale=0.5]{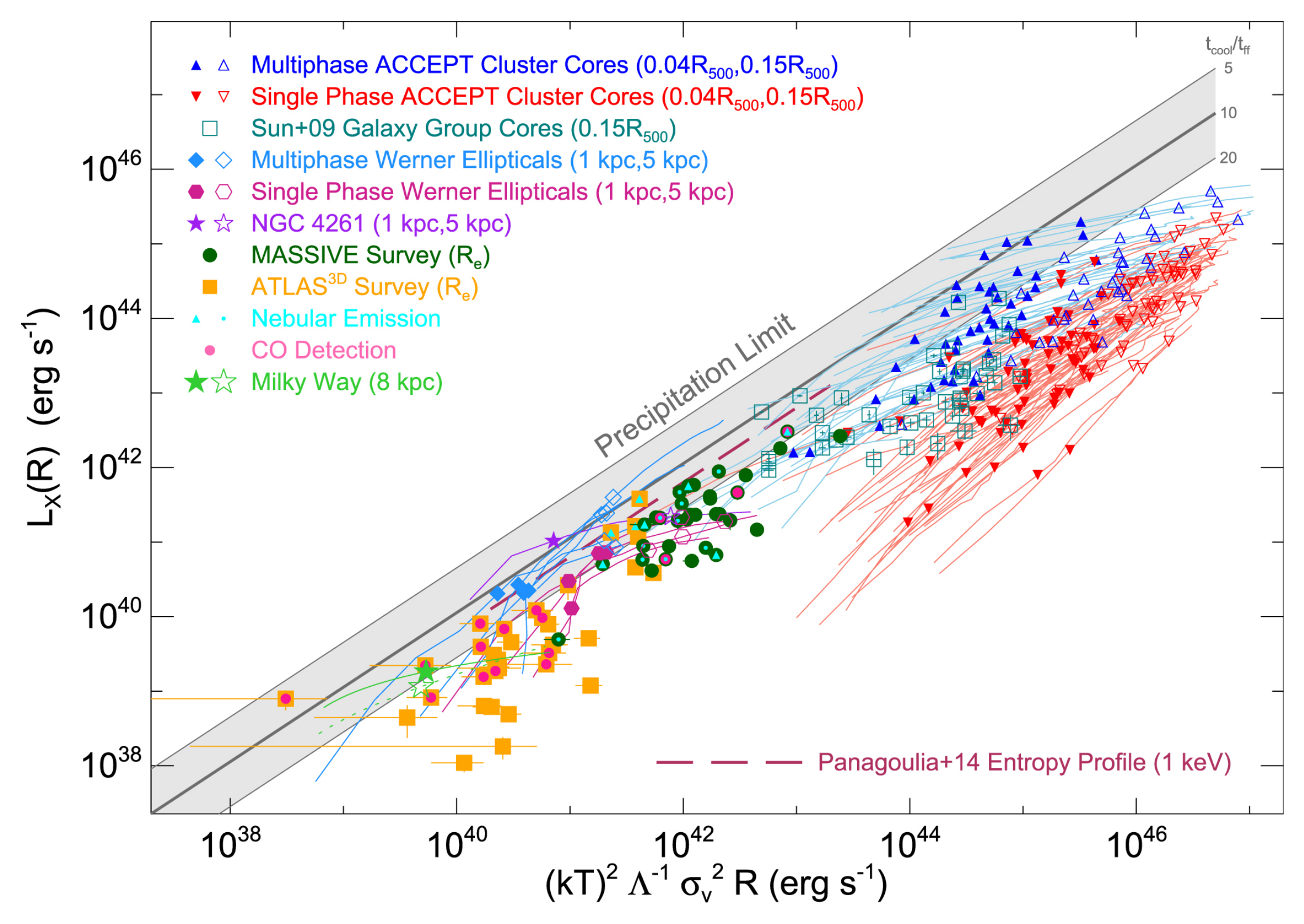}
\caption{X-ray luminosity $L_X(R)$ from within an circular aperture corresponding to a physical radius $R$ plotted as a function of the approximate precipitation-limited luminosity $(kT)^2 \Lambda^{-1} \sigma_v^2 R$ given by equation (\ref{eq:LxLim}). The figure is from \citep{Voit2018_LX-T-R} and encompasses the atmospheres of galaxies ranging from the BCGs of the most massive galaxy clusters down to the Milky Way (see the original paper for details).  Collectively, the observed upper envelope of $L_X(R)$ values tracks the thick gray line corresponding to $t_{\rm cool} / t_{\rm ff} \approx 10$ over more than seven orders of magnitude in X-ray luminosity and three orders of magnitude in halo mass.  Also, galactic atmospheres with detectable emission lines from multiphase gas tend to lie closer to the shaded region marking $5 < t_{\rm cool} / t_{\rm ff} < 20$.
\label{figure:LxLim}} 
\end{figure}

Figure \ref{figure:LxLim} shows that the atmospheric X-ray luminosities of massive galaxies respect that limit, from galaxy clusters all the way down to the Milky Way, implying that the condition $\min(t_{\rm cool} / t_{\rm ff}) \gtrsim 10$ places a strong upper limit on the ambient gas densities of galactic atmospheres in halos from $10^{12} \, M_\odot$ through $10^{15} \, M_\odot$.  The fact that the upper envelope of the observations follows a straight line in the plane of the figure is non-trivial, because the cooling function $\Lambda (T)$ has a complicated dependence on $T$ over the relevant temperature range (see Figure \ref{Figure:CoolingCurve}). Also, atmospheres closer to the precipitation limit in this figure are more likely to exhibit optical, infrared, and radio emission lines indicating the presence of cooler multiphase gas, consistent with greater susceptibility to multiphase condensation.

Two interesting features of the limit expressed in equation (\ref{eq:LxLim}) are worth noting.  One of them---the presence of the radiative cooling function $\Lambda$ in the denominator---may seem counterintuitive at first.  Ordinarily, an increase in $\Lambda$ would result in a proportional increase in luminosity, but an increase in $\Lambda$ also lowers the cooling time of gas at a given density.  In a precipitation-limited atmosphere, multiphase cooling and precipitation should then fuel feedback that expands the atmosphere until $\min (t _{\rm cool} / t_{\rm ff}) \gtrsim 10$.  The gas density associated with that limit is proportional to $\Lambda^{-1}$ (see equation \ref{eq:nelim}), and so the change in emissivity of atmospheric gas ($n_e^2 \Lambda$) results in a precipitation-limited luminosity also proportional to $\Lambda^{-1}$.  Consequently, when heavy-element enrichment of a precipitation-limited atmosphere raises $\Lambda$, feedback must then expand the atmosphere, driving it to a lower density at which it generates a smaller X-ray luminosity.

The other feature of equation (\ref{eq:LxLim}) worth noting is the linear dependence of $L_X(R)$ on $R$.  At sufficiently large radii, the cosmological baseline profile becomes more restrictive than the precipitation limit.  In galaxy clusters, that happens around 30~kpc from the center (see Figure \ref{figure:VoitNature2015}), and the crossover happens farther out in less massive halos.  Beyond the crossover radius, the decline in density of the atmosphere is closer to $n_e \propto r^{-2}$, and so its total X-ray luminosity converges.  Therefore, the total X-ray luminosity of a precipitation-limited galactic atmosphere depends on the radius at which its cosmological density profile intersects the limit in equation (\ref{eq:nelim}).


\subsection{Joint X-ray and SZE Stacking}

Almost all of the individual galaxies shown in Figure \ref{figure:LxLim} inhabit halos of mass $> 10^{13} \, M_\odot$ because the X-ray luminosity of a galactic atmosphere in a lower-mass halo is exceedingly faint.  However, stacking of many such X-ray observations places useful constraints on the population as a whole.  For example, each of the purple squares depicted at $kT < 2 \, {\rm keV}$ in the $L_X$--$T$ relation of Figure \ref{Figure:LX_T} represents a stack of several thousand galaxies done by \citet{Anderson_2015MNRAS.449.3806A}.  They started with a catalog of $\sim 250,000$ massive galaxies from the Sloan Digital Sky Survey, binned them according to stellar mass, and stacked observations from the ROSAT all-sky X-ray survey using an aperture corresponding to an estimate of $r_{500 {\rm c}}$ for each galaxy.  The resulting $L_X$--$T$ relation obviously overlaps with observations of individual atmospheres at $kT > 1 \, {\rm keV}$ and extends down in halo mass nearly to the scale of the Milky Way.  

Similar stacks of CMB observations have been made to probe how the strength of the Sunyaev-Zeldovich Effect (SZE, \S \ref{sec:SZE}) depends on halo mass \citep[e.g.,][]{Planck_LRGstacks_2013A&A...557A..52P}.  Those observations constrain the product of gas mass and temperature within a given aperture, whereas the X-ray stacks are more sensitive to the distribution of that gas as a function of radius.  Jointly fitting both data sets with a particular parametric model therefore provides constraints on the electron density distribution in the atmospheres of massive galaxies \citep{Singh_stacks_2018MNRAS.478.2909S}. 

\begin{figure}[t]
\centering
\includegraphics[width=5.4in]{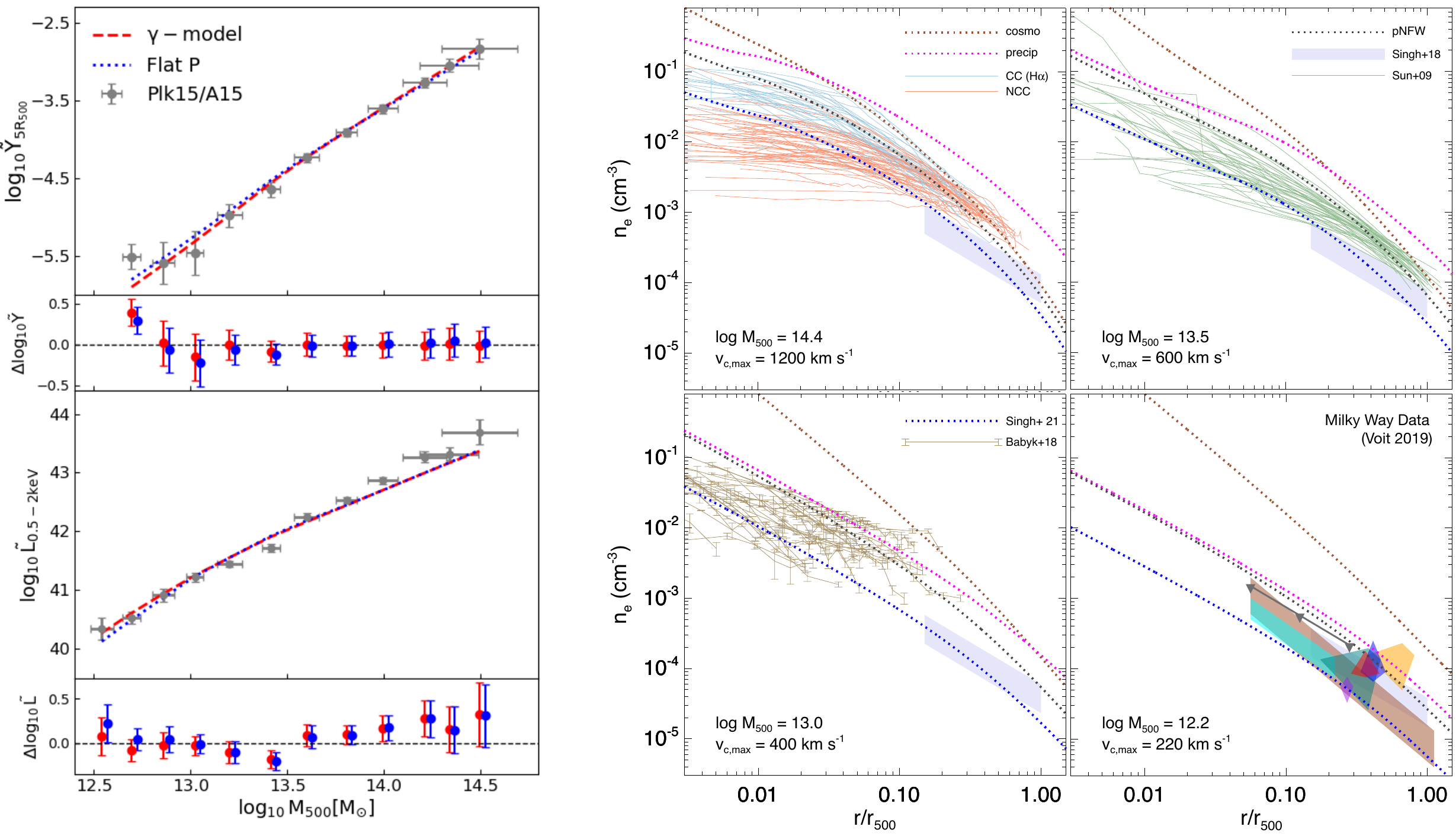}
\caption{Joint fitting of a parametric precipitation-limited model to stacked X-ray and SZE data (adapted from \citep{SinghVoitNath_2021MNRAS.501.2467S}).  Panels on the left show fits to stacks of SZE signals (top) and X-ray luminosity (bottom) along with the residuals.  Two versions of the model are shown, one with a flat pressure profile at $r > r_{200 {\rm c}}$ (blue dotted line) and one with $P \propto r^{-3.3}$ at $r > r_{200 {\rm c}}$ (red dashed line).  The group of panels on the right compare electron density profiles from the best-fitting flat-$P$ model (blue dotted line) with observations of individual sources, including galaxy clusters (upper left), galaxy groups (upper right), individual massive galaxies (lower left), and the Milky Way (lower right).  In each panel, a brown dotted line shows the cosmological baseline profile, a magenta dotted line shows the precipitation limit corresponding to $t_{\rm cool} / t_{\rm ff} \approx 10$, the gray dotted line shows a hybrid ``pNFW" profile that combines those two limits \citep{Voit_pNFW_2019ApJ...880..139V}, and a lavender parallelogram shows constraints that \citep{Singh_stacks_2018MNRAS.478.2909S} obtained from fitting just the mass range $10^{12.6}$--$10^{13} \, M_\odot$.
\label{figure:Singh2021}} 
\end{figure}

Figure \ref{figure:Singh2021} shows the results of a recent attempt by \citet{SinghVoitNath_2021MNRAS.501.2467S} to perform such a joint fit.  They explored a parametric precipitation-limited model (based on the pNFW models from \citep{Voit_pNFW_2019ApJ...880..139V}) with three free parameters: a limiting value of $t_{\rm cool} / t_{\rm ff}$, a temperature boundary condition at $r_{200 {\rm c}}$, and a power-law dependence of that boundary condition on halo mass.  Models with a single value of the $t_{\rm cool} / t_{\rm ff}$ parameter resulted in adequate fits to both the X-ray and SZE stacks across the mass range $10^{12.5}$--$10^{14.5} \, M_\odot$ (see left panels of Figure \ref{figure:Singh2021}).  However, the resulting electron density profiles differ from those measured from direct X-ray observations of individual galactic atmospheres (see right panels of Figure \ref{figure:Singh2021}).

The mismatches illustrated in Figure \ref{figure:Singh2021} probably result from a combination of systematic biases and observational selection effects.  On the observational side, X-ray surveys of galactic atmospheres tend to be biased toward the more X-ray luminous examples within each mass range, and so the objects selected for followup and further analysis tend to have electron densities exceeding the population mean.  However, there may also be biases in the statistical modeling, because halo masses are not directly measured.  Instead, they are inferred from each galaxy's stellar mass before the galaxies are placed in halo mass bins.  Scatter in that relationship, along with potential covariances in the $L_X$--$M_*$ relation at fixed halo mass, can therefore lead to systematic underestimates in $L_X$ at a given halo mass in the stacked data, because lower-mass halos are much more numerous than higher-mass halos \citep{Evrard_2014MNRAS.441.3562E,Farahi_2019NatCo..10.2504F}.

Perhaps more importantly, the data in the right-hand panels clearly show a large spread in electron density at small radii within each mass bin, indicating a large spread in $\min ( t_{\rm cool} / t_{\rm ff})$ from object to object.  The origin of this dispersion in $\min ( t_{\rm cool} / t_{\rm ff})$ is still unknown and needs to be included in the next generation of precipitation-limited models.  Also, the implementation of $\min ( t_{\rm cool} / t_{\rm ff})$ as a parameter in this particular model leads to underestimates of $n_e(r)$ at large radii, where the galaxy groups and clusters are clearly converging to cosmological baseline profile.  

Nevertheless, the observations suggest that in each halo-mass range the collective upper envelope of the electron density profiles do indeed track the precipitation-limited model atmospheres with $\min (t_{\rm cool} / t_{\rm ff}) \approx 10$.  With each step down in halo mass, the precipitation limit illustrated by the magenta line becomes progressively more stringent, compared to the cosmological baseline profile illustrated by the brown line.  In galaxy clusters, those two limiting profiles intersect near $0.03 r_{500 {\rm c}}$, implying that the precipitation limit affects only the inner $\sim 3$\% of a massive galaxy cluster.  In galaxy groups, the intersection has shifted to $0.2 r_{500 {\rm c}}$.  Among the individual massive galaxies it is approaching $r_{\rm 500}$.  Around the Milky Way, the entire atmosphere is plausibly precipitation-limited.  And in each case, galactic atmospheres appear to adhere to that limit.

\subsection{The Milky Way's Atmosphere}
\label{sec:MilkyWayCGM}

\begin{figure}[!t]
\centering
\includegraphics[width=5.3in]{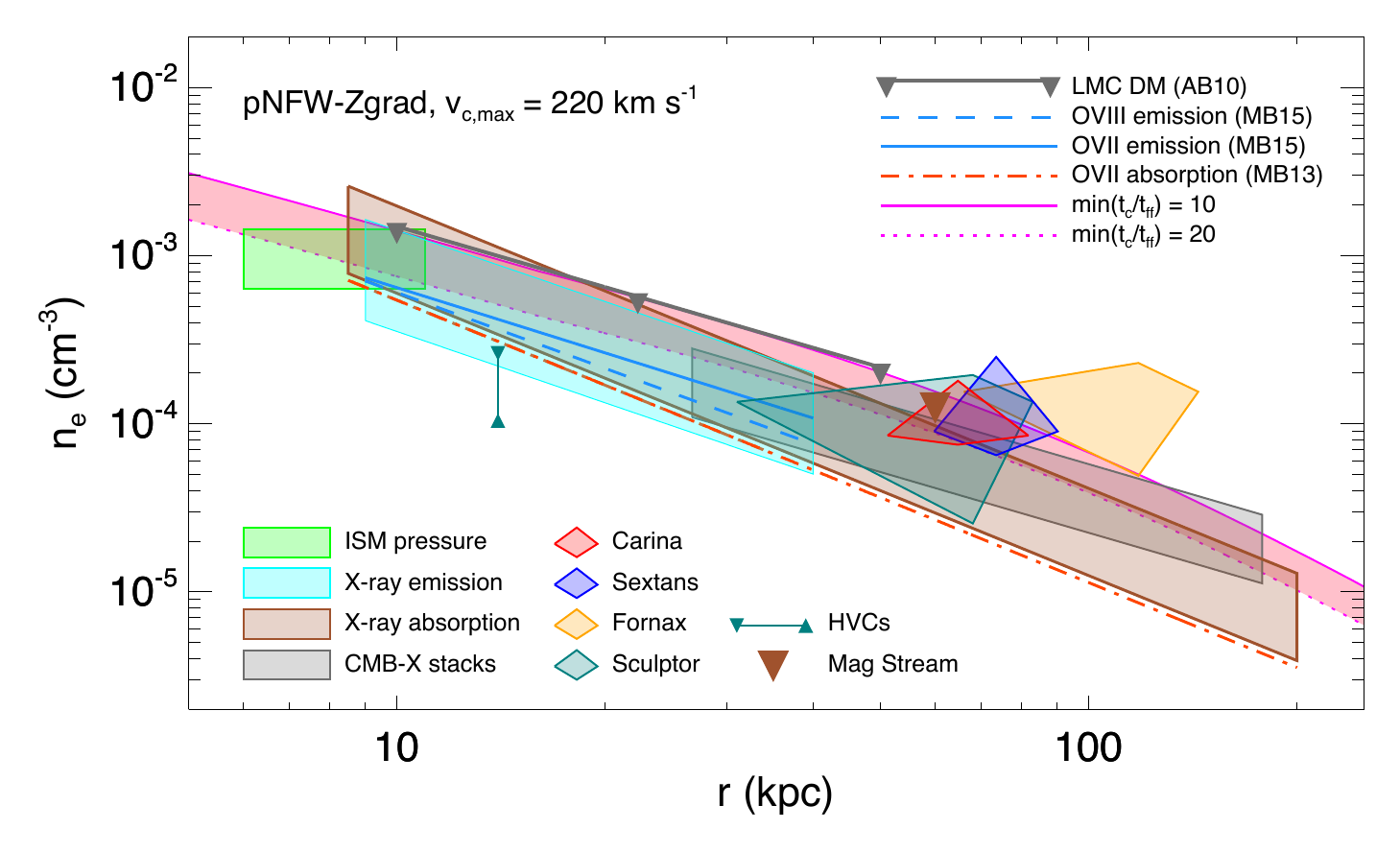}
\caption{Comparison of observational constraints on the Milky Way's ambient atmosphere with precipitation-limited models (from \citep{Voit_pNFW_2019ApJ...880..139V}).  This particular comparison assumes that heavy-element abundances in the atmosphere gradually decline from solar proportions at small radii to 30\% solar near the virial radius.  A model with $\min (t_{\rm cool} / t_{\rm ff}) \approx 20$ appears to be consistent with many different observational constraints.
\label{figure:MW_Zgrad}}
\end{figure}

Our exploration of galactic atmospheres began with the Milky Way, and now we return to it.  Figure \ref{figure:MW_Zgrad} shows the same collection of data sets as shown in Figure \ref{Figure:MW_density} and compares them with a pink band representing precipitation-limited models with $10 \leq \min ( t_{\rm cool} / t_{\rm ff}) \leq 20$.  The details of the comparison depend somewhat on the assumed elemental abundances, because of how they affect both $t_{\rm cool}$ and interpretation of the X-ray observations.  Here, an abundance gradient declining from solar proportions of heavy elements at small radii down to 30\% solar near the virial radius has been assumed.

Some of the most important constraints come from X-ray observations of O~VII absorption lines along lines of sight to X-ray bright quasars in many different directions \citep{MillerBregman_2013ApJ...770..118M,Gupta_2012ApJ...756L...8G,Fang_2015ApJS..217...21F}.  Taking advantage of our solar system's 8~kpc offset from the galactic center and assuming a spherically symmetric atmosphere centered on $r = 0$, \citet{MillerBregman_2013ApJ...770..118M} were able to constrain the power-law slope of the atmosphere's electron density profile.  Figure \ref{Figure:MW_density} shows that the best-fitting power-law slope ($n_e \propto r^{-1.7}$) is similar to that of a precipitation-limited electron density profile.\footnote{The original model of \citet{MillerBregman_2013ApJ...770..118M} was isothermal.  Correcting for the temperature gradient implied by a precipitation-limited model boosts the normalization of their profile from the orange dot-dashed line in Figure \ref{figure:MW_Zgrad} into the brown parallelogram \citep{Voit_pNFW_2019ApJ...880..139V}.}  \citet{MillerBregman_2015ApJ...800...14M} later found that X-ray observations of O~VII and O~VIII emission lines led to an atmospheric density profile for the Milky Way with a similar power-law slope and normalization. Constraints based on ram-pressure stripping analyses of dwarf galaxies (diamond-shaped polygons in Figure \ref{figure:MW_Zgrad}) tend to indicate somewhat greater gas densities at $\sim 100$~kpc but also have greater systematic uncertainties.

Taken as a whole, the constraints in Figure \ref{Figure:MW_density} are broadly consistent with a precipitation-limited model in which $\min(t_{\rm cool} / t_{\rm ff}) \approx 20$. Certain other models are consistent with those constraints, including the isentropic CGM model of \citet{Faerman_2020ApJ...893...82F} and the cooling-flow models of \citet{Qu_2018ApJ...856....5Q} and \citet{Stern_2019MNRAS.488.2549S}. However, the CGM entropy levels and cooling-flow rates in those other models are free parameters chosen to match the data. In contrast, the CGM entropy level and cooling-flow rate in a precipitation-limited model are linked to an astrophysical limiting mechanism.

\subsection{UV Probes of CGM Pressure}
\label{sec:CGM_Pressure}

Atmospheres around galaxies of Milky Way mass and smaller emit too few photons to be detectable with current X-ray telescopes.  That will change with the launch of ESA's ATHENA\footnote{https://www.the-athena-x-ray-observatory.eu/} X-ray mission, and perhaps \textit{Lynx}\footnote{https://www.lynxobservatory.com/} after that, if NASA eventually selects it.  X-ray absorption line detections may be possible sooner with the high-resolution spectrograph to be launched on XRISM,\footnote{https://xrism.isas.jaxa.jp/en/} and prospects will improve when the proposed ARCUS\footnote{http://www.arcusxray.org/} mission or something similar is launched.  In the meantime, we need to rely on less direct probes of those galactic atmospheres.

One method for indirectly measuring ambient atmospheric conditions around lower-mass galaxies relies on observations from the Hubble Space Telescope's COS spectrograph, which has revolutionized our understanding of the photoionized component of galactic atmospheres during the last decade \citep{2017ARA&A..55..389T} (see also \S \ref{sec:UV_COS}).  Cool ($10^4$~K) clouds along lines of sight to bright background quasars produce absorption lines that reveal the ionization states of common elements.  Those ionization states depend on both the intensity of intergalactic UV background radiation and the number density of free electrons. Photoelectric heating keeps the temperature of that gas close to $10^4$~K.  Photoionization modeling of the observations therefore enables estimates of cloud pressure \citep{Werk_2014ApJ...792....8W,Keeney_2017ApJS..230....6K,Zahedy_2019MNRAS.484.2257Z}.

\begin{figure}[!t]
\centering
\includegraphics[width=5.1in]{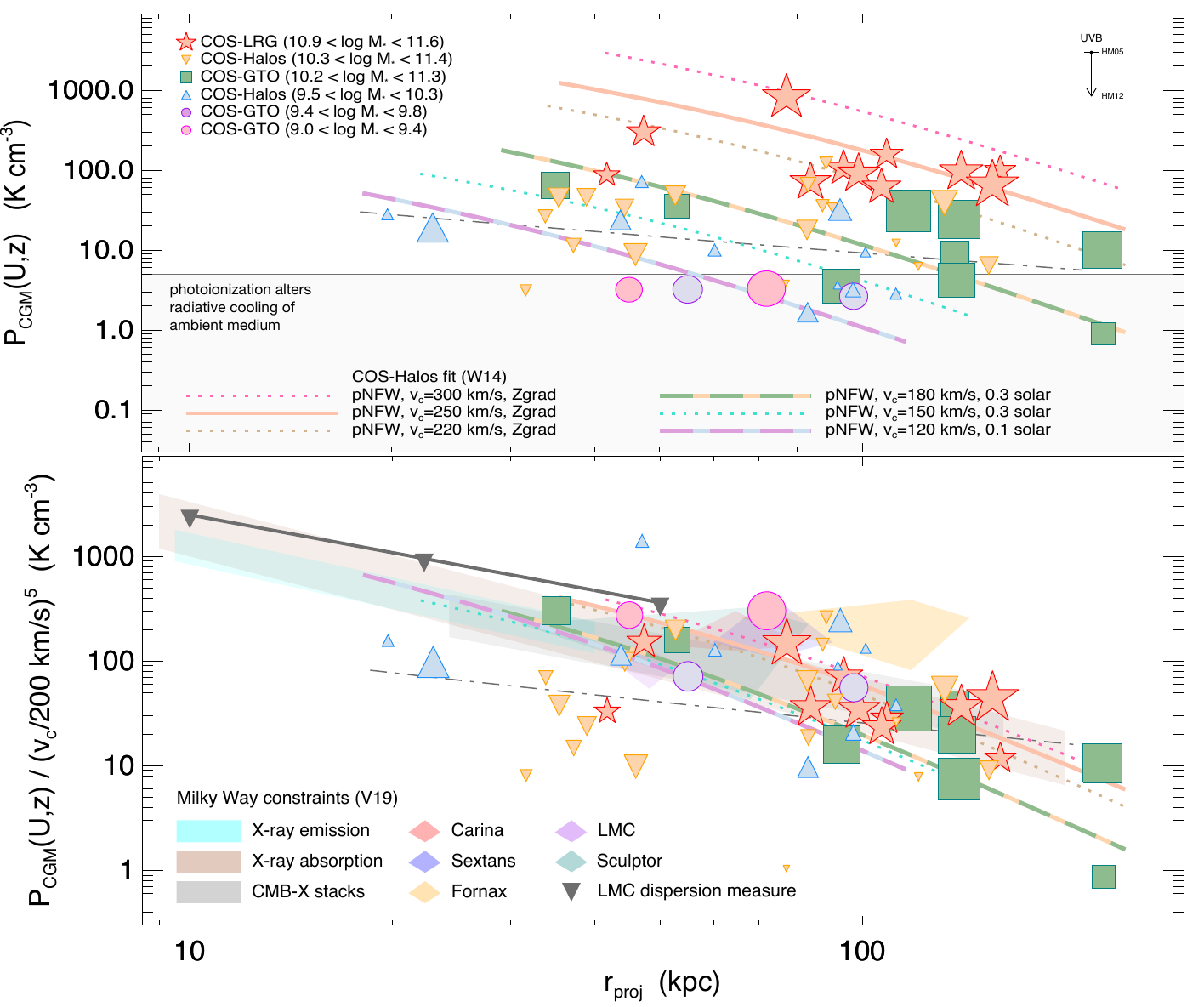}
\caption{Atmospheric pressure constraints from photoionization modeling of UV absorption lines (adapted from \citep{Voit_2019ApJ...879L...1V}).  The top panel shows atmospheric pressures ($P_{\rm CGM} = n_{\rm H} T$) inferred at various projected radii ($r_{\rm proj}$) around a large set of galaxies probed with \textit{Hubble's} Cosmic Origins Spectrograph.  Symbol types indicate galactic stellar mass and data source as given in the legend.  Symbol sizes represent relative statistical significance, with larger symbols indicating greater certainty.  Higher mass galaxies clearly have greater pressures.  Lines show precipitation-limited models that depend on halo circular velocity $v_{\rm c}$, as given in the legend.  The bottom panel shows how the data points converge when the mass trend is removed by fitting a power-law dependence on $v_{\rm c}$ (see \citep{Voit_2019ApJ...879L...1V} for details).  Shaded regions represent the Milky Way data from Figure \ref{figure:MW_Zgrad}.
\label{figure:PCGM}}
\end{figure}

Figure \ref{figure:PCGM} shows a compilation of CGM pressure estimates collected in \citep{Voit_2019ApJ...879L...1V} and representing galaxies spanning the stellar mass range $10^{9.0} \, M_\odot < M_* < 10^{11.6} \, M_\odot$.  They are plotted as a function of projected radius $r_{\rm proj}$ around the galaxy presumed to be at the center of the detected atmosphere.  The symbol colors and shapes indicate data sources and ranges in $M_*$. Their relative sizes reflect the statistical significance of each pressure measurement.  An arrow in the upper right corner indicates how systematic uncertainty in the UV background intensity affects all of the pressure measurements (see \citep{Voit_2019ApJ...879L...1V} for details). Additional uncertainty arises from assuming a constant-density cloud in the photoionization models from which pressures are determined \citep{Stern_2016ApJ...830...87S}.

Pressure estimates in the top panel of Figure \ref{figure:PCGM} exhibit a clear trend with stellar mass and therefore with halo mass.  Clouds around the most massive galaxies (red stars) tend to have pressures two orders of magnitude greater than clouds around the least massive galaxies in the sample (magenta and purple circles).  Most of the lines in the upper panel (labeled with pNFW in the legend) show the expected pressures of precipitation-limited atmospheres in halos spanning the same mass range.  The primary parameter governing those models is $v_{\rm c}$, the maximum circular velocity of the halo. Stellar mass ($M_*$) can be mapped onto $v_{\rm c}$ using scaling relationships derived from galaxy observations \citep[e.g.][]{McGaugh2005ApJ...632..859M,McGaugh+2010ApJ...708L..14M}.

The bottom panel of Figure \ref{figure:PCGM} shows how the CGM pressures measured around galaxies of different masses converge when the trend with $v_{\rm c}$ is removed.  Fitting of power-law relationships to the data gives an approximate best fit of $P_{\rm CGM} \propto v_{\rm c}^{5.1} r_{\rm proj}^{-1.7}$, with an uncertainty of roughly $\pm 1$ in the exponent of $v_{\rm c}$ \citep{Voit_2019ApJ...879L...1V}.  Dividing both the pressure measurements and the models by $(v_{\rm c} / \,  200 \, {\rm km \, s^{-1}})^5$ therefore brings them all into better alignment.  Furthermore, both the pressure measurements and the pNFW models align with the Milky Way constraints from Figure \ref{figure:MW_Zgrad}, which have been translated into pressure constraints and are depicted as shaded regions in the bottom panel of Figure \ref{figure:PCGM}.

Overall, the alignment of most of the UV data points with the pNFW models and the Milky Way constraints is consistent with pressure confinement of the photoionized $10^4$~K atmospheric phase by a hotter precipitation-limited atmospheric phase (see also \citep{Zahedy_2019MNRAS.484.2257Z}).  If that interpretation is correct, then the precipitation hypothesis might account for not only the Milky Way's CGM characteristics but also the CGM characteristics of lower-mass galaxies.  However, this finding contrasts with earlier interpretations of COS data, which appeared to indicate considerably lower CGM pressures \citep{Werk_2014ApJ...792....8W}.  

Projection effects can explain at least some of that apparent pressure discrepancy, and maybe even all of it.  The CGM absorption lines measured with COS often have significant wavelength-dependent structure indicating that multiple clouds, moving at different speeds and located at different physical distances from the central galaxy, are responsible for the absorption along a particular sightline.  The pressure gradient predicted by the models in Figure \ref{figure:PCGM} implies that $10^4$~K clouds at $r \gg r_{\rm proj}$ should have significantly lower densities and consequently greater ionization levels that those at $r \approx r_{\rm proj}$.  Wherever possible, the pressures estimated in \citep{Voit_2019ApJ...879L...1V} correspond to the lowest ionization absorption-line components along each sightline, which are the most likely to be representative of pressures at $r \approx r_{\rm proj}$. Those measurements correspond to the stars, squares, and circles in Figure \ref{figure:PCGM}.  However, the triangles represent measurements not similarly corrected for projection.  Those sightlines are the ones most likely to end up as outliers in the bottom panel of Figure \ref{figure:PCGM}, and they indicate pressures at $r_{\rm proj} < 50$~kpc an order of magnitude smaller than the model profiles and the Milky Way data.  Pressures along those sightlines are more consistent with those found at $r_{\rm proj} \gtrsim 100$~kpc in the other data sets, suggesting that the UV-absorbing gas observed along those discrepant sightlines may actually be at $r \gg r_{\rm proj}$.

\subsection{Future Tests}

Many more observational tests of the precipitation hypothesis are now becoming possible or will become possible in the coming decade.  They include:
\begin{itemize}

    \item \textbf{Sunyaev-Zeldovich Effect (SZE).}  Spatially resolved SZE observations of galactic halos are starting to accumulate.  \textit{Planck} observations of halos that are nearby and therefore large on the sky can deliver spatially resolved CGM pressure profiles, when they are stacked \citep[e.g.,][]{Pratt_SZ_2021arXiv210501123P,Bregman_2021arXiv210714281B}.  More distant halos require SZE telescopes with greater spatial resolution, and those telescopes are also beginning to deliver useful constraints on CGM pressure profiles out to and beyond the virial radius \citep[e.g.,][]{Schaan_2021PhRvD.103f3513S,Amodeo_2021PhRvD.103f3514A,Meinke_2021ApJ...913...88M}.  On the mass scales of individual galaxies up to galaxy groups, these SZE detections are consistent with suppressed electron density at $\lesssim r_{500 {\rm c}}$ as well as a compensating excess of gas pressure beyond the virial radius.

    \item \textbf{Fast Radio Bursts (FRBs).}  Thousands of FRBs have now been detected and their dispersion measures will soon be used to constrain the electron density profiles of galactic halos \citep[e.g.,][]{McQuinn2014ApJ...780L..33M,ProchaskaZheng_2019MNRAS.485..648P}.  As with the resolved SZE observations, constraints on electron density as a function of halo mass and projected radius can be compared with precipitation-limited models of the CGM.

    \item \textbf{eROSITA Stacks.}  The eROSITA\footnote{https://www.mpe.mpg.de/eROSITA} instrument is currently performing an all-sky X-ray survey that will greatly surpass the ROSAT All-Sky Survey in sensitivity.  Stacking of eROSITA pointings, binned according to galaxy mass, will therefore provide unprecedented constraints on $L_X(R)$ as a function of halo mass \citep{Oppenheimer_eROSITA_2020ApJ...893L..24O}, for comparison with the precipitation limit in Figure \ref{figure:LxLim}.

    \item \textbf{Correlations with Gas Dynamics.}  Early versions of the precipitation hypothesis considered only the role of the median $t_{\rm cool} / t_{\rm ff}$ parameter, which determines an atmosphere's susceptibility to precipitation.  But more recent models have demonstrated that a second parameter, one that reflects the amplitudes of atmospheric disturbances, is also important.  One option is $\sigma_{\ln K}$, the log-normal dispersion of entropy perturbations (see \S \ref{sec:PrecipitationLimit}).  In an atmosphere near pressure equilibrium, $\sigma_{\ln K}$ is related to the log-normal dispersion $\sigma_{\ln \rho}$ of density perturbations through $\sigma_{\ln \rho} \approx (3/5) \sigma_{\ln K}$.  And in a stratified atmosphere, it is related to the typical fractional amplitude $\sigma_{\ln r}$ of internal gravity waves via $\sigma_{\ln K} \approx \alpha_K \sigma_{\ln r} \approx \alpha_K^{1/2} (\sigma_{\rm t} / \sigma_v )$, where $\sigma_{\rm t}$ is the velocity dispersion of those disturbances and $\sigma_v \approx v_{\rm c} / \sqrt{2}$ \citep{Voit_2018ApJ...868..102V}.
    
    If the distribution function of atmospheric perturbations is similar to the schematic distribution in Figure \ref{figure:tctff}, then the atmosphere's precipitation threshold should depend jointly on both $\sigma_{\ln K}$ and the median $t_{\rm cool} / t_{\rm ff}$ ratio. The modeling that has been done so far indicates that an atmosphere with $\sigma_{\ln K} \sim 0.5$ and $t_{\rm cool} / t_{\rm ff} \approx 10$ should be in a marginally precipitating state \citep[e.g.,][]{Voit_2018ApJ...868..102V}.  More theoretical modeling is needed to map out the joint dependence of precipitation on those two parameters. Such models should be designed to support observational tests comparing correlations of precipitation with both the velocity dispersion $\sigma_{\rm t}$ of precipitating gas and the log-normal atmospheric gas density dispersion $\sigma_{\ln \rho}$ inferred from surface-brightness fluctuations.
    
\end{itemize}
It will also remain important to improve existing tests of the precipitation hypothesis with more accurate measurements of $t_{\rm ff}$ at small radii in galactic halos, based on observations of stellar dynamics, and the atmospheric abundances that enter into calculations of $t_{\rm cool}$.

\section{Implications for Galaxy Evolution \label{sec:Implications}}

This article's exploration of baryon cycles in the biggest galaxies has largely focused on what can be inferred from the current states of massive galaxies and their atmospheres.  Observations show that the most massive galaxies formed almost all of their stars within the first few billion years after the Big Bang \citep[e.g.,][]{Behroozi2019MNRAS.488.3143B} and now have atmospheres with ambient densities and pressures in which $\min ( t_{\rm cool} / t_{\rm ff} ) \gtrsim 10$ (\S \ref{sec:Evidence}).  Ambient atmospheric gas with $t_{\rm cool}$ less than the current age of the universe is susceptible to forming a cooling flow that can remain homogeneous as long as $t_{\rm cool} \gg t_{\rm ff}$, and gravitational compression in such a flow keeps the gas temperature close to the halo's gravitational temperature (\S \ref{sec:Cooling}). However, the $t_{\rm cool} / t_{\rm ff}$ ratio of a homogeneous cooling flow tends to decline as the gas moves inward, making the inflow increasingly prone to runaway thermal instability and formation of an inhomogeneous multiphase atmosphere in which stars can form (\S \ref{sec:ThermalInstability}).

The precipitation hypothesis proposes that this transition to a multiphase state fuels feedback that places upper limits on the pressure, density, and cooling flow rate of the ambient medium. It plausibly explains how self-regulating feedback can maintain the resulting limits on atmospheric density, pressure, and cooling time (\S \ref{sec:Balance}) but does not explain how galaxy evolution drives massive galaxies toward such a self-regulating state.  Nor does it provide much insight into the time required for a massive galaxy to transition away from an actively star-forming state into a nearly quiescent one.  However, the precipitation hypothesis may be helpful for interpreting several aspects of that transition.

We therefore conclude our review with a look at the following aspects of galaxy evolution, exploring how the limits that precipitation places on galactic atmospheres inform the broader story of star formation within galaxies: 
\begin{itemize}

\item \textbf{Precipitation-regulated star formation.}  Most of the gas supply fueling early star formation in cosmological numerical simulations of galaxy evolution arrives in a halo's central galaxy via streams of cold gas associated with cosmological accretion rather than through precipitation \citep{Keres_2005MNRAS.363....2K}.  But as time progresses, sufficiently massive galaxies build up extended hot atmospheres dense enough to disrupt those cold streams \citep{FaucherGiguere_2011MNRAS.417.2982F,Nelson_2013MNRAS.429.3353N,Mandelker_2016MNRAS.463.3921M}.  Also, the cold streams thought to feed galaxy growth become wider than the galaxy itself \citep{Dekel_2009Natur.457..451D}. Eventually, most of the gas entering the galaxy's disk has previously been part of the CGM's hot component, having arrived in the CGM through either a cosmological accretion shock or a hot galactic outflow \citep{Angles-Alcazar_2017MNRAS.470.4698A}.  When a galaxy's atmosphere reaches that state, the precipitation limit may then constrain the gas supply rate from the CGM into the central galaxy, and therefore the galaxy's long-term star formation rate.  Section \ref{sec:Regulation} shows that the precipitation-limited gas supply rate turns out to be similar to the observed star formation rates in present-day galaxies like the Milky Way and discusses why this quasi-equilibrium state might account for some of the star-formation scaling relations observed among lower-mass galaxies \citep{Voit_PrecipReg_2015ApJ...808L..30V}.  The most important takeaway is that a precipitation-limited gas supply allows halos with greater $v_{\rm c}$ to convert a larger proportion of their baryons into stars, until AGN feedback becomes capable of shutting down star formation at halo masses $\gtrsim 10^{12} \, M_\odot$. 

\item \textbf{Quenching of star formation in high-mass galaxies.} Section \ref{sec:Quenching} focuses on a massive galaxy's transition into star-formation quiescence, a phenomenon known as ``quenching" \citep{ManBelli_2018NatAs...2..695M}.  Observations show that star-formation quiescence is more closely related to a galaxy's central stellar velocity dispersion ($\sigma_v$) than to any other galactic property.  According to models of precipitation-limited atmospheres, long-term suppression of star formation depends on a link between $\sigma_v$ and a massive galaxy's atmospheric structure (\S \ref{sec:MassiveGalaxiesEvidence}).  The link is established when feedback lowers CGM pressure enough for supernova heating to exceed radiative cooling within a halo's central galaxy.  A galaxy with a deep central potential well (as reflected by large $\sigma_v$) then develops an atmosphere that focuses cooling, condensation, and precipitation onto the galaxy's central black hole (because large $\sigma_v$ results in $\alpha_K > 2/3$, as described in \S \ref{sec:MassiveGalaxiesEvidence}).  According to that model, outbursts of strong AGN feedback should then prevent both atmospheric cooling and star formation.  Consequently, those models specify a critical value of $\sigma_v$ above which galactic star formation should permanently cease \citep{Voit_2020ApJ...899...70V}.  

\item \textbf{Star formation and atmospheric structure.} Section \ref{sec:ConvectiveCGM} looks more generally at how atmospheric structure and multiphase CGM gas may be related to both $\sigma_v$ and halo mass. The atmosphere of a low-mass galaxy is inevitably convective, because cosmological entropy generation cannot prevent inhomogeneous cooling and star formation in halos of mass $\lesssim 10^{12} \, M_\odot$ (\S \ref{sec:Cooling}).  Centralized supernova feedback therefore heats the galaxy's inner atmosphere, possibly with assistance from an AGN, causing entropy inversions that make the atmosphere convectively unstable and promote multiphase condensation.  However, the situation changes across the halo mass range $\sim 10^{12-13} \, M_\odot$.  In that mass range, cosmological accretion produces enough entropy to limit inhomogeneous cooling (\S \ref{sec:Cooling}), and supernovae no longer produce enough energy to lift CGM gas out of the halo (see Figure \ref{figure:ECGM_EBH}). AGN feedback must supply the bulk of the energy \citep{Bower_2017MNRAS.465...32B} and appears to maintain the inner atmosphere in a precipitation-limited state. 

\item \textbf{Star formation and central black hole mass.}  Section \ref{sec:BlackHoleMass} considers the relationships observed among $\sigma_v$, central black hole mass ($M_{\rm BH}$), and a galaxy's specific star-formation rate (sSFR) and interprets them in the context of atmospheric structure. The observed correlations between $\sigma_v$, $M_{\rm BH}$, and sSFR suggest that $150 \, {\rm km \, s}^{-1} \lesssim \sigma_v \lesssim 240 \, {\rm km \, s}^{-1}$ corresponds to a transitional interval for galactic star formation.  At the low end of that interval, the atmosphere is likely to be convective, with a steady inflow of multiphase gas supplying fuel for star formation.  At the high end of the interval, the atmosphere is likely to be stratified, with radial gradients of pressure, density, and entropy that focus cooling and multiphase condensation on the central black hole.

\item \textbf{Evolution of the criteria for quenching.}  The models outlined in \S \ref{sec:Quenching} predict that the upper end of the transitional interval in $\sigma_v$ should evolve with time.  Section \ref{sec:QuenchingCriteria} summarizes those predictions, which depend on how the energy supplied by supernova heating evolves.  Early in time, the specific energy of a supernova-heated outflow is expected to be greater, which enables star formation to proceed in galaxies with greater $\sigma_v$.  As time passes, the critical value of $\sigma_v$ should decline. However, observations indicate that the critical value of $\sigma_v$ also correlates with a galaxy's total stellar mass, suggesting that the specific energy supplied by an aging galaxy's stellar population may also correlate with total stellar mass.

\item \textbf{Central black hole mass and stellar velocity dispersion.} Section \ref{sec:CauseOrEffect} summarizes matters with a discussion of the galactic properties most critical for quenching of star formation in massive galaxies ($M_{\rm halo}$, $M_{\rm BH}$, and $\sigma_v$) and some thoughts about identifying the root cause.

\end{itemize}

\subsection{Precipitation-Regulated Star Formation}
\label{sec:Regulation}

\begin{figure}[!t]
\centering
\includegraphics[width=5.0in]{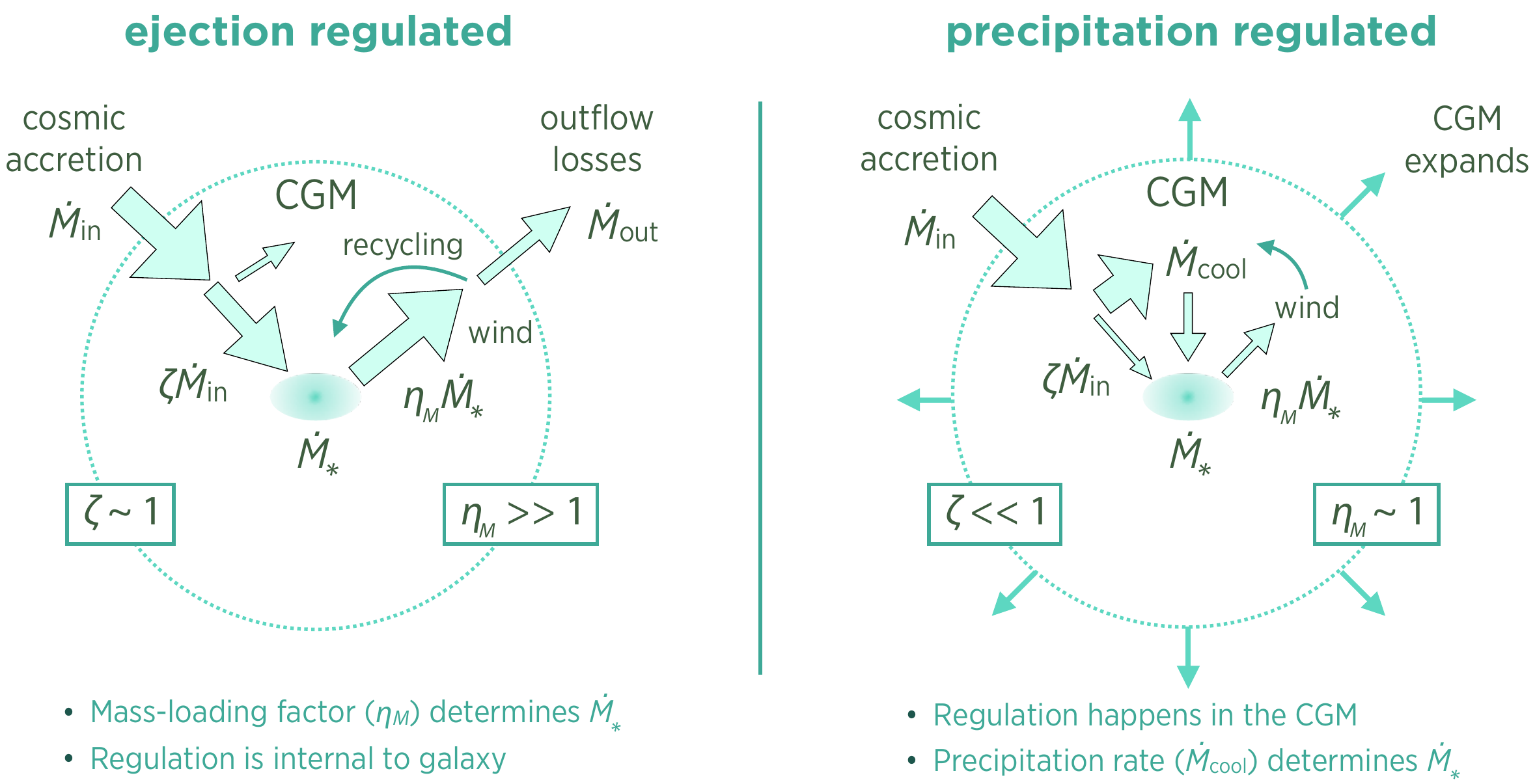}
\caption{Two schematic scenarios for supernova feedback. \textit{Left:}  In the classic ejection-regulated scenario, supernova energy needs to eject baryonic gas from a galaxy and perhaps also its halo \citep{Larson_1974MNRAS.169..229L,DekelSilk1986ApJ...303...39D,WhiteFrenk1991ApJ...379...52W}. Preventative feedback is negligible (i.e. $\zeta \sim 1$, see \S \ref{sec:SN_heating}), meaning that most of the gas entering a halo ($\dot{M}_{\rm in}$) reaches its central galaxy, fueling star formation at the rate $\dot{M}_*$.  Supernova feedback then drives an outflow at the rate $\eta_M \dot{M}_*$, with a mass loading factor $\eta_M$ depending on the gravitational potential's circular velocity ($v_{\rm c}$).  Some recycling of outflowing gas back into the galaxy is possible, but $\eta_M$ is the main parameter governing the galaxy's star-formation efficiency.  \textit{Right:} The precipitation limit enables an alternative scenario \citep{Voit_PrecipReg_2015ApJ...808L..30V}. It presumes that preventative feedback and gravitational heating are significant (i.e. $\zeta \ll 1$), meaning that most of the gas entering a halo is incorporated into the CGM and heated to the halo's gravitational temperature.  That gas must then shed entropy before entering the galaxy and cooling enough to form stars. The limiting cooling rate is therefore determined by the ambient medium's precipitation limit.  Feedback does not explosively eject CGM gas but rather expands it until $\min(t_{\rm cool} / t_{\rm ff}) \gtrsim 10$ in the ambient medium.  Precipitation then supplies star forming gas to the galaxy at a rate ($\dot{M}_{\rm cool}$) determined by $v_{\rm c}$.  Both scenarios result in a similar dependence of star-formation rate on halo mass (see text).
\label{figure:FeedbackScenarios}}
\end{figure}

Figure \ref{figure:FeedbackScenarios} schematically depicts two contrasting scenarios for supernova feedback.  The classic scenario is on the left and relies on supernova energy to limit star formation by ejecting most of the galaxy's gas.  Most of the gas falling into the galaxy's halo is assumed to enter the central galaxy, sustaining a star formation rate $\dot{M}_*$.  Supernova energy then drives an outflow at the rate $\eta_M \dot{M}_*$, and the mass-loading parameter $\eta_M$ determines the proportion of baryons that form stars \citep{SomervilleDave_2015ARA&A..53...51S}. 

On the right is a scenario in which the precipitation limit determines the galaxy's star formation rate. If a galaxy's primary long-term source of star-forming gas depends on the cooling-flow rate of its CGM, via either hot accretion or precipitation, then feedback limits that cooling-flow rate through its effects on CGM pressure and density. This scenario does not apply to galaxies fed by cold streams rooted in cosmological accretion but may apply to galaxies with atmospheres dense enough to disrupt those streams.

In a precipitation-regulated environment, the limiting star-formation rate is imposed by the upper limit on ambient density expressed in equation (\ref{eq:nelim}).  It is closely related to the precipitation-limited luminosity $L_X (R)$ from equation (\ref{eq:LxLim}).  Setting $R$ in that equation equal to the radius $r_{\rm cool}$ at which $t_{\rm cool} \sim H^{-1}(z)$ gives
\begin{equation}
    \dot{M}_{\rm cool} \: \lesssim \; \frac {2 \mu m_p} {3 kT} \, L_X (r_{\rm cool})   
                       \; \lesssim \; \frac {3 \pi \mu m_p} {25} \, \frac {kT} {\Lambda} \, v_{\rm c}^2 r_{\rm cool}
        \; \; .
    \label{eq:Mcool_precip}
\end{equation}
This expression further simplifies if the galaxy's atmosphere is precipitation-limited all the way out to the cosmological scale radius $r_{500 {\rm c}}$ at which $t_{\rm ff} \approx 0.1 \, H^{-1}(z)$ (see equation \ref{eq:r_Delta}). Then the condition $t_{\rm cool} / t_{\rm ff} \approx 10$ implies $r_{\rm cool} \approx r_{500 {\rm c}}$ and yields
\begin{equation}
    \dot{M}_{\rm cool} \: \lesssim \: \frac {3 \pi G \mu m_p} {25} \, \frac {kT} {\Lambda} \, M_{500 {\rm c}}
        \; \; .
    \label{eq:Mcool_precip_r500}
\end{equation}
As with the analogous limits on $n_e(r)$ and $L_X(R)$, the precipitation-limited gas supply into a galaxy depends inversely on $\Lambda$, meaning that heavy-element enrichment of the CGM \textit{lowers} the long-term rate at which gas can cool out of it, if feedback keeps the ambient gas in a marginally precipitating state.  We therefore need to consider how heavy-element enrichment couples with precipitation before evaluating the resulting limit on star formation.

\subsubsection{Abundance Saturation}

The inverse dependence of $\dot{M}_{\rm cool}$ on $\Lambda$ in a precipitation-limited atmosphere potentially couples a galaxy's star formation rate with chemical enrichment of its atmosphere \citep{Voit_PrecipReg_2015ApJ...808L..30V}.  For example, suppose the mass $M_Z$ of heavy elements in the atmosphere accumulates according to $\dot{M}_{Z,{\rm gas}} = (Y - Z_{\rm gas}) \dot{M}_*$, where $Y$ is the fractional heavy-element yield from a single generation of star formation, $Z_{\rm gas} = M_{Z,{\rm gas}} / M_{\rm gas}$ is the heavy-element mass fraction, and $M_{\rm gas} = f_{\rm b} M_{500 {\rm c}} - M_*$ is the atmospheric mass associated with the halo mass $M_{500 {\rm c}}$.  The equation governing the rate of change in $Z_{\rm gas}$ then has the structure 
\begin{equation}
    \dot{Z}_{\rm gas} = \left( Y \dot{M}_* - Z_{\rm gas} f_{\rm b} \dot{M}_{500 {\rm c}} \right) M_{\rm gas}^{-1}
    \label{eq:Zdot}
\end{equation}
in this highly simplified model of chemical enrichment.  The first term in parentheses is a source term for enrichment, and the second one accounts for dilution of chemical enrichment as cosmological accretion introduces low-enrichment baryonic matter into the galaxy's halo at the rate $f_{\rm b} \dot{M}_{500 {\rm c}}$.\footnote{In that respect this model resembles ``bathtub" models for galactic chemical evolution \citep[e.g.,][]{Bouche_2010ApJ...718.1001B,DaveFinlatorOppenheimer_2012MNRAS.421...98D,Lilly+2013ApJ...772..119L}.}

Equation (\ref{eq:Zdot}) is interesting because it describes a galactic atmosphere in which chemical enrichment can self-regulate, as long as CGM recycling is the main fuel supply for star formation (i.e. $\dot{M}_* \sim \dot{M}_{\rm cool}$).  In that case, $Z_{\rm gas}$ asymptotically approaches the saturation value
\begin{equation}
    Z_{\rm sat} 
      \: \equiv \: 
    \frac {Y \dot{M}_*} {f_{\rm b} \dot{M}_{500 {\rm c}}}
\end{equation}
at which $\dot{Z}_{\rm gas} = 0$.  Overenrichment makes $\dot{Z}_{\rm gas}$ negative, because it raises $\Lambda$ and lowers $\dot{M}_* \sim \dot{M}_{\rm cool}$. Underenrichment has the opposite effect.  Furthermore, combining the definition of $Z_{\rm sat}$ with equation (\ref{eq:Mcool_precip_r500}) yields an expression that relates $Z_{\rm sat}$ to the ambient gas temperature $T$: 
\begin{equation}
    \frac {Z_{\rm sat} \, \Lambda (T, Z_{\rm sat})} {kT}
      \; \approx \:
    \frac {3 \pi G \mu m_p} {25} \frac {Y} {f_{\rm b} H(z)} 
       \left[ \frac {M_{500 {\rm c}} H(z)} {\dot{M}_{500 {\rm c}}} \right]
    \; \; ,
    \label{eq:Zsat_kT}
\end{equation}
in which approximate equality stems from assuming $t_{\rm cool}/t_{\rm ff} \approx 10$ in the ambient medium.
The factor in square brackets is of order unity, depends on cosmological structure formation, and changes slowly with time.  Most of the relationship's time dependence therefore comes from the $H(z)$ factor in the denominator.  All of the other factors on the right hand side are constants.

An approximate power-law dependence of $Z_{\rm sat}$ on halo mass emerges when a power-law approximation for $\Lambda(T,Z)$ is inserted into equation (\ref{eq:Zsat_kT}).  For example, the approximation $\Lambda \approx 10^{-22} \, {\rm erg \, cm^3 \, s^{-1}} \, (Z / Z_\odot)^{4/5} (T / 10^6 K)^{-4/5}$, which is appropriate for the CGM temperature range $10^{5.5}$--$10^{6.7} \, {\rm K}$, leads to
\begin{equation}
    \frac {Z_{\rm sat}} {Z_\odot} 
      \: \approx \: 0.5 \, \left[ \frac {Y} {Z_\odot}        
          \frac {M_{500 {\rm c}} H_0} {\dot{M}_{500 {\rm c}}} \right]^{5/9}
          \left( \frac {T} {10^6 \, {\rm K}} \right)
      \: \propto \:  v_{\rm c}^2
    \label{eq:Zsat}
\end{equation}
and implies that $\Lambda(T,Z_{\rm sat})$ is approximately constant in this temperature range.  While the numerical coefficient in this relation depends on many uncertain approximations, the predicted power-law dependence of $Z$ on halo mass (via $v_{\rm c}^2$) is more robust. 

This very simple model makes some testable predictions.  For example, it implies that $M_* \propto M_{\rm halo}^{5/3}$ for galaxies with $Z \approx Z_{\rm sat}$ (because $Z \propto f_* \propto M_* / M_{500 {\rm c}}$). That prediction is similar to the $M_*$--$M_{\rm halo}$ relation observed among galaxies less massive than the Milky Way (e.g., \citep{Moster_2010ApJ...710..903M}, see also Figure \ref{fig:SF_Efficiency}). The model also implies that lower gas-phase abundances permit star formation to proceed more quickly in galaxies of a given mass.  For example, setting a galaxy's star-formation rate $\dot{M}_*$ equal to the limiting rate on the right side of equation (\ref{eq:Mcool_precip_r500}) and assuming $Z \propto M_* / M_{\rm halo}$ yields
\begin{equation}
    Z \: \propto \: M_*^{11/15} \dot{M}_*^{-1/3} 
      \: \propto \: M_*^{2/5} (\dot{M}_*/M_*)^{-1/3}
      \label{eq:MZR}
\end{equation}
for $\Lambda \propto (Z/T)^{4/5}$ and $T \propto M_{\rm halo}^{2/3}$.  

\begin{figure}[!t]
 \centering
 \includegraphics[width=5.0in]{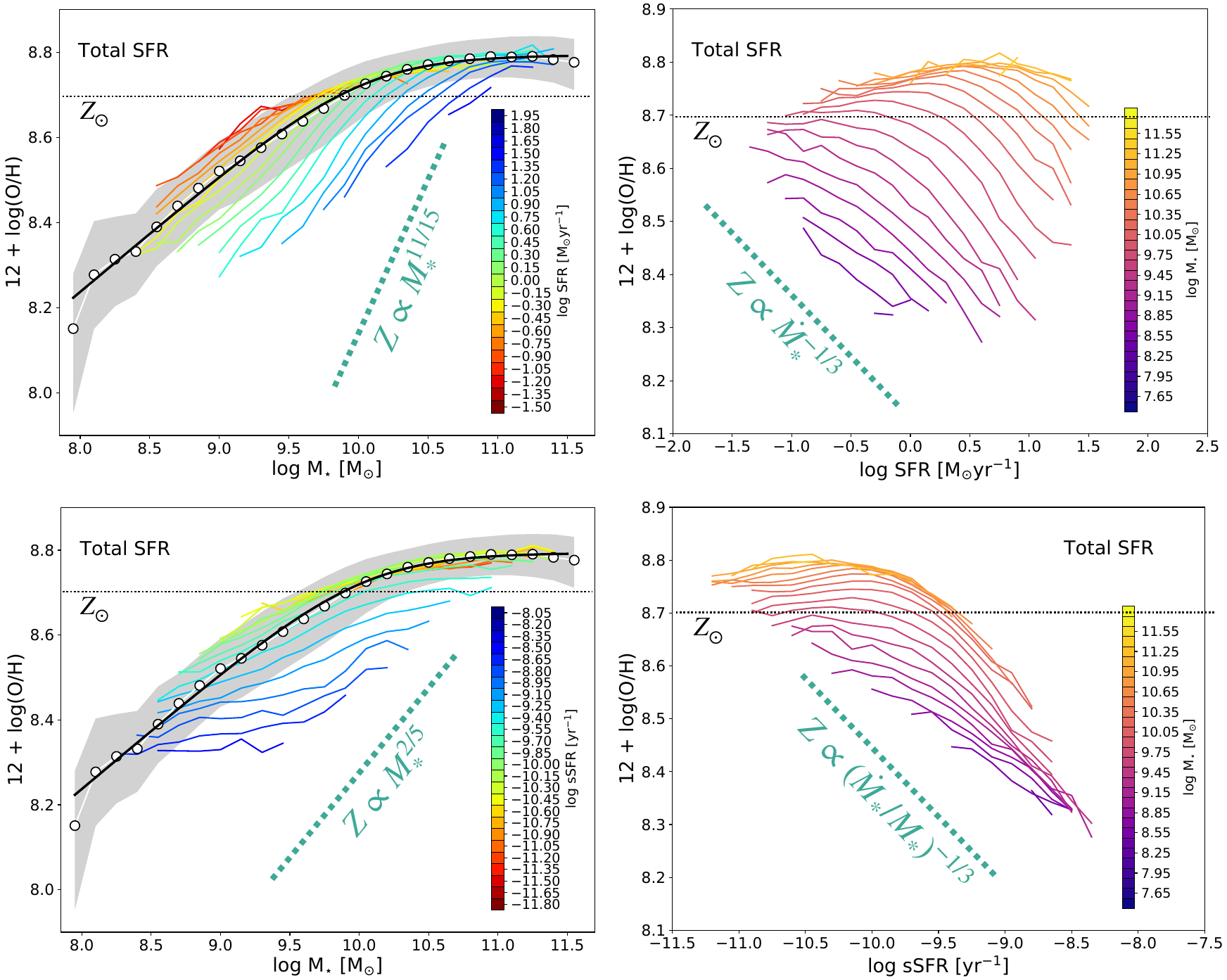}
  \caption{Comparisons of observed relationships between galactic stellar mass ($M_*$), gas-phase abundance ($Z$), star formation rate ($\dot{M}_*$), and specific star formation rate (sSFR $\equiv \dot{M}_* / M_*$) with the scaling laws predicted by abundance saturation.  The original figure is from \citet{Curti_2020MNRAS.491..944C} and represents $Z$ in terms of the quantity $12 - \log({\rm O/H})$, where O/H is the abundance ratio of oxygen to hydrogen and black dotted lines indicate $Z_\odot$. It has been adapted through the addition of green dotted lines showing the scaling predictions.  Colored solid lines in upper left panel show the $Z$--$M_*$ relation at constant $\dot{M}$.  In the lower left panel they show the $Z$--$M_*$ relation at constant sSFR. In the upper right panel they show the $Z$--$\dot{M}_*$ relation at constant $M_*$.  In the lower right panel they show the $Z$--sSFR relation at constant $M_*$. The points and solid black lines in the left-hand panels show the $Z$--$M_*$ relation for the whole galaxy sample.
  \label{figure:AbundanceSaturation}}
\end{figure}

Figure \ref{figure:AbundanceSaturation} shows that the scaling prediction in equation (\ref{eq:MZR}) resembles the relationships between halo mass, heavy-element abundance, and galactic star formation rate observed among galaxies less massive than the Milky Way \citep{Mannucci+2010MNRAS.408.2115M,Curti_2020MNRAS.491..944C}.  Its right-hand panels show that star formation rates in galaxies of similar stellar mass anticorrelate with $Z$ if their abundances are subsolar.\footnote{Solar abundance is equivalent to $12 - \log({\rm O/H}) \approx 8.7$ in the vernacular of the figure.}  Furthermore, the power-law scalings of the observations are similar to the model predictions.  In the left panels are the complementary scalings of $Z$ with $M_*$.  The observed scalings at fixed $\dot{M}_*$ are slightly shallower than the predicted slope, but the observed scalings at fixed specific star-formation rate (sSFR) are considerably shallower, perhaps signaling that the abundances observed among rapidly star forming galaxies are not as closely related to $M_* / M_{\rm 500c}$ as the model assumes.  All of those scaling relations flatten in the vicinity of the solar abundance ($Z_\odot$), as they approach the total heavy-element yield ($Y$) from a single generation of star formation.

The scaling properties shown in Figure \ref{figure:AbundanceSaturation} are usually interpreted in terms of supernova-driven galactic winds \citep{SomervilleDave_2015ARA&A..53...51S}, according to the left panel of Figure \ref{figure:FeedbackScenarios}, rather than abundance saturation, according to the right panel of Figure \ref{figure:FeedbackScenarios}.  We will return to the implicit connection between those seemingly disjoint interpretations in \S \ref{sec:ImplicitLifting}, after a look at the star formation rates predicted for galaxies that have precipitation-limited atmospheres.

\subsubsection{Star Formation Limits}
\label{sec:StarFormationLimits}

As a galaxy's atmospheric abundances converge toward $Z_{\rm sat}$, the model predicts that its limiting precipitation rate converges toward a value that depends only on the halo's gravitational properties. Combining equations (\ref{eq:Mcool_precip_r500}) and (\ref{eq:Zsat}) for a galaxy centered in a halo of mass $\sim 10^{12} \, M_\odot$ gives a precipitation-limited CGM cooling rate
\begin{equation}
    \dot{M}_{\rm cool} \: \lesssim \: 2 \, M_\odot \, {\rm yr^{-1}} \, 
            \left[ \frac {Y} {Z_\odot}        
                    \frac {M_{500 {\rm c}} H_0} {\dot{M}_{500 {\rm c}}} \right]^{-4/9}
                           \left( \frac {T} {10^6 \, {\rm K}} \right)  
            \left( \frac {M_{500 {\rm c}}} {10^{12} \, M_\odot} \right)
        \; \; .
    \label{eq:Mcool_precip_MW}
\end{equation}
This limit on CGM cooling is essentially the same as the observed star-formation rate in the present-day Milky Way \citep{Chomiuk_2011AJ....142..197C}, because all of the factors in parentheses and square brackets are of order unity.  

The result captured in equation (\ref{eq:Mcool_precip_MW}) is consistent with other analyses of atmospheric cooling rates. It aligns with the CGM model of \citet{Qu_2018ApJ...856....5Q}, in which the normalization of the CGM's gas density profile is determined by setting $\dot{M}_{\rm cool}$ of the CGM equal to the central galaxy's star-formation rate, a normalization that successfully accounts for the Milky Way's CGM characteristics.  It also aligns with cooling-flow models for the CGM developed by \citet{Stern_2019MNRAS.488.2549S}, which agree well with the both the Milky Way's atmosphere and the atmospheres of similarly massive galaxies simulated by the FIRE collaboration \citep{Hafen_2019MNRAS.tmp.1734H,Esmerian_2021MNRAS.tmp.1268E}. However, $\dot{M}_{\rm cool}$ in equation (\ref{eq:Mcool_precip_MW}) is a \textit{consequence} of the precipitation limit rather than an input parameter, as it is in the other two models.\footnote{Note that the precipitation limit determines the \textit{maximum} accretion rate of a cooling flow that manages to remain homogeneous, because a lower-entropy atmosphere with a greater cooling-flow rate \textit{by definition} would be unstable to inhomogeneous condensation and precipitation.}

An important feature of the cooling limit in equation (\ref{eq:Mcool_precip_MW}) is its temperature dependence.  It implies that star formation happens more quickly in halos of greater circular velocity.  The enhancement of star formation in this case results from the greater atmospheric density allowed by the precipitation limit in higher-mass halos.  Consequently, it predicts that galaxies with greater $v_{\rm c}$ should form most of their stars earlier than galaxies with smaller $v_{\rm c}$, even though halos with smaller $v_{\rm c}$ form earlier.  This phenomenon, sometimes called ``downsizing" \citep{Cowie+1996AJ....112..839C}, is one of the most distinctive features of galaxy evolution, but it is usually explained in terms of feedback, not precipitation and abundance saturation.  The next section discusses how those two explanations are implicitly connected.

\subsubsection{Implicit CGM Lifting Energy}
\label{sec:ImplicitLifting}

The precipitation-model ansatz specifying $t_{\rm cool} / t_{\rm ff} \gtrsim 10$ implies the presence of an energy source capable of reducing the CGM's ambient density by lifting much of the ambient gas to greater altitudes.  Section \S \ref{sec:EnergyRequirements} assessed those energy requirements, showing that supernova energy can lift the CGM in halos of mass $\lesssim 10^{12} \, M_\odot$ but not in halos of mass $\gtrsim 10^{13} \, M_\odot$.  Continual lifting of CGM gas entering a halo through cosmological accretion requires $\eta_E \epsilon_{\rm SN} \dot{M}_* \gtrsim f_{\rm b} \dot{M}_{500 {\rm c}} (kT / \mu m_p)$, which reduces to
\begin{equation}
    \dot{M}_* \: \gtrsim \: 2 \, M_\odot \, {\rm yr}^{-1} \, 
            \left[ \frac {M_{500 {\rm c}} H_0} {\dot{M}_{500 {\rm c}}} \right]^{-1}
                           \left( \frac {\eta_E} {0.2} \right)^{-1}  
                           \left( \frac {T} {10^6 \, {\rm K}} \right)  
            \left( \frac {M_{500 {\rm c}}} {10^{12} \, M_\odot} \right)
    \label{eq:SFR_lifting_MW}
\end{equation}
if core-collapse supernovae are the only energy source.

Comparing this constraint with equation (\ref{eq:Mcool_precip_MW}) reveals interesting complementarity.  In a galaxy like the Milky Way, supernova heating of the CGM needs to have an efficiency of at least $\eta_E \approx 0.2$ in order to drive the CGM into a precipitation-limited state.  The galaxy's star formation rate in a precipitation-limited state is therefore doubly constrained, by a lower limit on the supernova heating requirement and by an upper limit on the atmospheric cooling rate, as long as AGN heating is unimportant.  Both of those constraints depend similarly on halo mass and CGM temperature. Therefore, the scenarios illustrated in Figure \ref{figure:FeedbackScenarios} are not independent of each other.  Both constraints may apply to mature galaxies in which cooling of the CGM is the primary gas supply for star formation.

\subsection{Quenching of Star Formation}
\label{sec:Quenching}

Our Milky Way galaxy currently sits near a pivotal transition in galaxy evolution.  Most galaxies of lower mass have specific star formation rates $\sim 10^{-10} \, {\rm yr}^{-1}$ (similar to $H_0$). Most galaxies of greater mass have specific star formation rates $\lesssim 10^{-11} \, {\rm yr}^{-1}$ (much less than $H_0$) \citep{Baldry_2006MNRAS.373..469B,Peng_2010ApJ...721..193P,RenziniPeng_2015ApJ...801L..29R}.  

It is also near the peak of the $f_*$--$M_{\rm halo}$ relation (see green stars in Figure \ref{fig:SF_Efficiency}).  According to analyses of large galaxy surveys, this peak has remained near a halo mass of $\sim 10^{12} \, M_\odot$ during most of cosmic time \citep{Behroozi+2013ApJ...762L..31B}.  They show that star formation in a cosmological halo's central galaxy proceeds vigorously until the halo mass exceeds $\sim 10^{12} \, M_\odot$.  Star formation then subsides as halo growth raises its mass into the regime where AGN feedback power starts to exceed supernova power (see Figure \ref{figure:ECGM_EBH}).  This ``quenching" of star formation may initially happen because of a particular event, such as a particularly large AGN outburst \citep{SpringelDiMatteoHernquist_2005ApJ...620L..79S,Hopkins_2008ApJS..175..356H}, but it cannot persist unless a permanent change in the galaxy's atmospheric properties cuts off the galaxy's supply of star-forming gas \citep[e.g.,][]{Sparre_2017MNRAS.470.3946S}.  Once established, a quenched state must be maintained.

\subsubsection{The Significance of $\sigma_v$}

\begin{figure}[!t]
 \centering
 \includegraphics[width=5.3in]{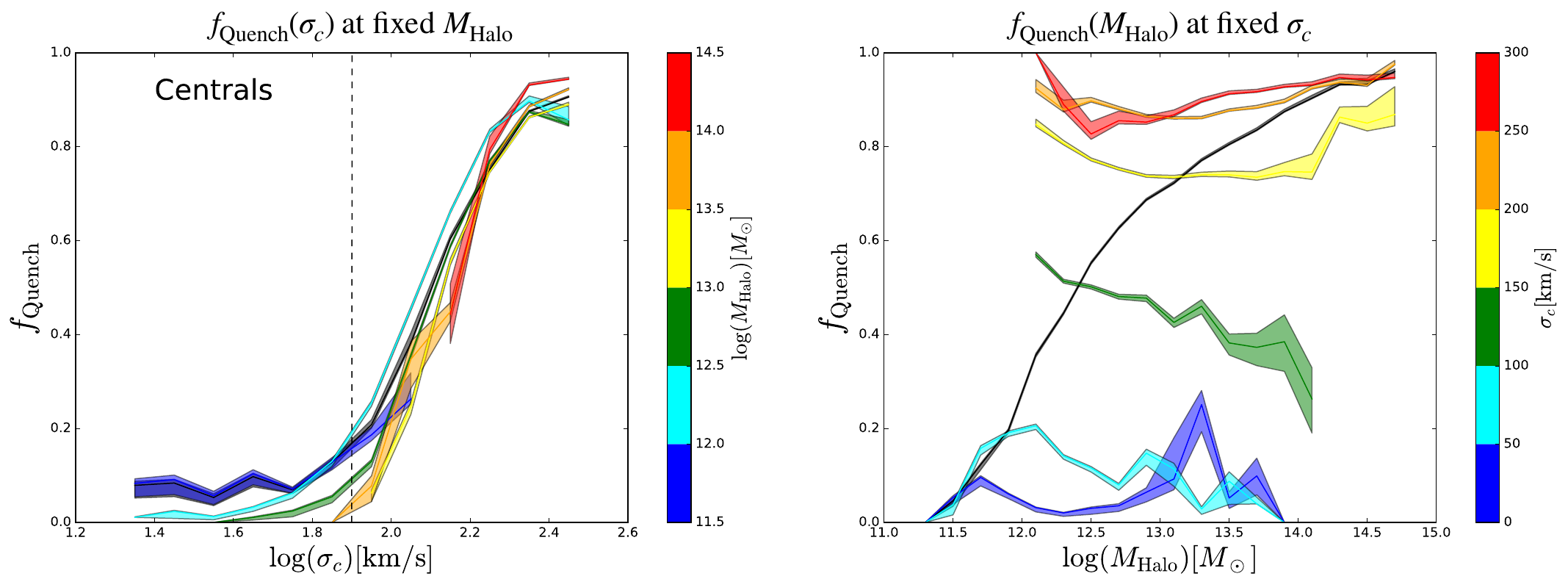}
  \caption{Observed dependence of the fraction of Sloan Digital Sky Survey (SDSS) galaxies with quenched star formation ($f_{\rm Quench}$) on central velocity dispersion (represented by $\sigma_{\rm c}$ in this figure) and halo mass (from \citep{Bluck_2016MNRAS.462.2559B}). \textit{Left:} Relationship between $f_{\rm Quench}$ and stellar velocity dispersion (solid black line), with colored regions showing the dependence of $f_{\rm Quench}$ on $\sigma_{\rm c}$ within bins of halo mass.  \textit{Right:} Relationship between $f_{\rm Quench}$ and halo mass (solid black line), with colored regions showing the dependence of $f_{\rm Quench}$ on halo mass within bins of $\sigma_{\rm c}$.  Once the dependence of $f_{\rm Quench}$ on $\sigma_{\rm c}$ is accounted for, there is essentially no residual dependence on halo mass.
  \label{figure:fQuench}}
\end{figure}

Observations show that long-term quenching of star formation in a halo's central galaxy depends even more strongly on the galaxy's central velocity dispersion ($\sigma_v$) than on its halo mass (see Figure \ref{figure:fQuench}).  In fact, star-formation quiescence depends more strongly on $\sigma_v$ than on any other galactic property \citep{Wake_2012ApJ...751L..44W,Bell_2012ApJ...753..167B,Woo_2015MNRAS.448..237W,Teimoorinia_2016MNRAS.457.2086T,Bluck_2016MNRAS.462.2559B,Bluck_2020MNRAS.492...96B}. Therefore, the atmospheric reconfiguration that shuts off star formation must depend more directly on $\sigma_v$ than on halo mass.

Several other measures of galactic structure correlate almost as strongly with star-formation quiescence in central galaxies.  One of the most commonly used alternative measures is $\Sigma_1$, the surface mass density of stars within 1~kpc of a galaxy's center, which can be determined from photometric images of galaxies, without spectroscopy.  According to \citet{Fang_2013ApJ...776...63F}, the expression
\begin{equation}
    \Sigma_1 = 6 \times 10^9 \, M_\odot \, {\rm kpc}^{-2} 
               \left( \frac {\sigma_v} {200 \, {\rm km \, s^{-1}}} \right)^2
\end{equation}
relates the two quantities. \citet{Kauffmann_2003MNRAS.341...54K} first noted the strong connection between stellar surface density and quenching in galaxies from the Sloan Digital Sky Survey.  Quenching also correlated closely with the prominence of a galaxy's stellar bulge \citep{Bell_2008ApJ...682..355B,Bluck_2014MNRAS.441..599B}.  All of these structural features correlate closely with quenching because they correlate closely with $\sigma_v$.

\subsubsection{A Black Hole Feedback Valve}
\label{sec:Valve}

The atmospheric properties of single phase and multiphase elliptical galaxies, illustrated in Figures \ref{figure:Werner_tctff} and \ref{figure:K-r_galaxies}, provide a potential clue to the significance of $\sigma_v$.  In that small sample, the galaxies with greater $\sigma_v$ tend to have steeper radial profiles of atmospheric pressure, density, and entropy (with $\alpha_K \approx 1$), and also lack extended multiphase gas \citep{Werner+2014MNRAS.439.2291W,Voit+2015ApJ...803L..21V,Voit_2020ApJ...899...70V}.  As discussed in \S \ref{sec:MassiveGalaxiesEvidence}, the median $t_{\rm cool} / t_{\rm ff}$ ratio in their atmospheres rises with radius outside of the central kiloparsec, implying that they are most susceptible to multiphase condensation at small radii, where the supermassive black hole resides.  Furthermore, SNIa heating appears to exceed radiative cooling within each of the galaxies that have greater velocities ($\sigma_v$). This heating can drive outflows with power-law entropy slopes similar to the ones observed.

\begin{figure}[!t]
\centering
\includegraphics[width=5.3in]{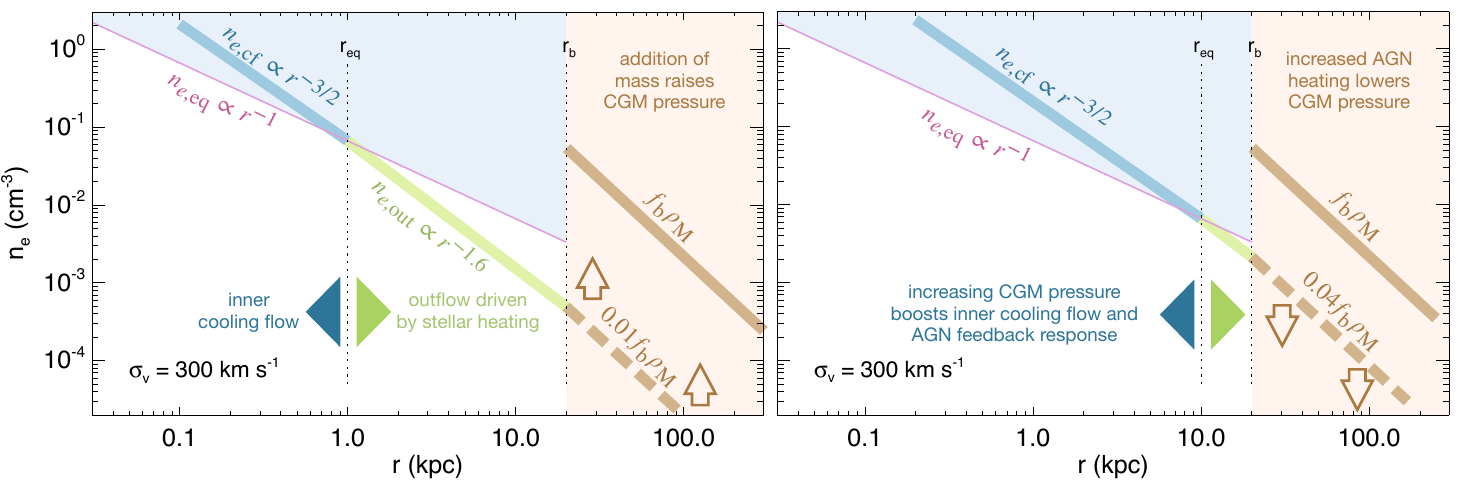}
\caption{Schematic illustration (from \citep{Voit_2020ApJ...899...70V}) of the black hole feedback valve mechanism in a galaxy with $\sigma_v = 300 \, {\rm km \, s^{-1}}$.  A thin red line in each panel represents the electron density profile ($n_{e,{\rm eq}}$) along which SNIa heating would equal radiative cooling.  Its slope ($n_{e,{\rm eq}} \propto r^{-1}$) corresponds to a stellar mass density distribution $\propto r^{-2}$.  Cooling exceeds SNIa heating in the blue region above $n_{e,{\rm eq}} (r)$.  Within $r_{\rm eq}$ the atmospheric density profile (thick blue line) is that of a steady cooling flow ($n_{e,{\rm cf}} \propto r^{-3/2}$).  Outside of $r_{\rm eq}$ there is a steady heated outflow with a density profile $(n_{e,{\rm out}}\propto r^{-1.6})$ determined by equation (\ref{eq:alphaK_sigmav}) and the depth of the isothermal potential.  The density normalization of both profiles is set by CGM confinement (dashed brown line), which applies a pressure 0.01 times the cosmological baseline pressure at the boundary radius ($r_{\rm b}$) in the left panel and 0.04 times the baseline pressure in the right panel.  This configuration focuses atmospheric cooling onto the central black hole and directly links AGN fueling to CGM pressure, thereby enabling AGN feedback to be self-tuning.
\label{figure:valve_schematic}}
\end{figure}

According to the black hole feedback valve model of \citet{Voit_2020ApJ...899...70V}, the conditions observed in single phase elliptical galaxies directly link AGN fueling to CGM pressure.  Figure \ref{figure:valve_schematic} schematically illustrates the fundamental ideas.  In a present-day galaxy with $\sigma_v \gtrsim 240 \, {\rm km \, s^{-1}}$, an outflow driven by SNIa heating can surround an inner cooling flow.  The border between those flows lies near the radius $r_{\rm eq}$ at which SNIa heating equals radiative cooling.  Both flows are subsonic, and so their densities depend on the confining CGM pressure.  Consequently, both $r_{\rm eq}$ and the mass flow rate of the inner cooling flow also depend on CGM pressure.  Under these conditions, the inner atmosphere (represented by blue and green lines in Figure \ref{figure:valve_schematic}) can act like a valve that controls AGN fueling, based on the CGM pressure level.  

The valve becomes self-tuning if AGN feedback responds to increased central cooling by driving strong bipolar jets that drill through the SNIa heated outflow, depositing their energy beyond $\sim 10$~kpc, as observed in NGC 4261 and IC 4296 (see Figure \ref{fig:narrow_jets}). These AGN outbursts cause the CGM to expand as they deposit energy into the CGM, thereby lowering the CGM pressure, reducing the central cooling-flow rate, and diminishing accretion onto the central black hole. After the AGN reverts to a low-power state, the galaxy's stars continue to drive a gaseous outflow into the CGM, where it accumulates, steadily raising the CGM pressure as long as AGN feedback remains minimal.  This buildup of CGM pressure ultimately causes the inner cooling flow rate to increase until the AGN flips back into a high-power state and once again reduces CGM pressure. 

Two conditions are necessary for this valve mechanism to operate:
\begin{enumerate}
    \item  The power-law slope of the heated outflow's electron density profile ($n_{e,{\rm out}}$) must be steeper than the locus of heating-cooling equality ($n_{e,{\rm eq}}$). 
    That slope is determined by how the specific energy $\epsilon_*$ of gas coming from the galaxy's stellar population compares with the circular velocity ($v_{\rm c}$) of its potential well.  If the $\epsilon_*/v_{\rm c}^2$ ratio is large (corresponding to small $\sigma_v$), then not much work is required to push the outflow, meaning that the atmosphere's pressure, density, and entropy profiles will have shallow slopes.  In a deeper potential well, a greater proportion of the flow's specific energy goes into the work required to lift the flow, and so its atmospheric profiles are correspondingly steeper. Consequently, there is a minimum value of $\sigma_v$ above which the feedback valve can operate as illustrated in Figure \ref{figure:valve_schematic}.  Equation (\ref{eq:alphaK_sigmav}) implies that this critical value is $\sigma_v \approx 240 \, {\rm km \, s^{-1}}$ for a stellar population age of $\sim 10$~Gyr, because $\alpha_K = 2/3$ implies $n_e \propto r^{-1}$ in an isothermal potential.  
    \item The CGM's pressure must be low enough for SNIa heating to exceed radiative cooling within the galaxy.  Figure \ref{figure:K-r_galaxies} shows that the single phase ellipticals from \citet{Werner+2014MNRAS.439.2291W} satisfy this condition, but many BCGs in multiphase cluster cores do not, even if $\sigma_v$ in those galaxies significantly exceeds $240 \, {\rm km \, s^{-1}}$.  This circumstance presumably arises because the AGNs in single phase elliptical galaxies can supply enough energy to lift the galaxy-group atmosphere filling a halo of mass $\sim 10^{13.5} \, M_\odot$, but the AGNs in multiphase cluster cores do not supply enough energy to lift the galaxy-cluster atmosphere within a halo of mass $> 10^{14} \, M_\odot$.
\end{enumerate}

Numerical simulations have explored many aspects of AGN feedback in idealized elliptical galaxies \citep[e.g.,][]{CiottiOstriker_1997ApJ...487L.105C,MathewsBrighenti2003ARAA..41..191M,Gaspari+2011MNRAS.411..349G,Gaspari+2011MNRAS.415.1549G,Gaspari+2012ApJ...746...94G}.
In high-mass ellipticals those simulations tend to exhibit cycles of inflow and outflow punctuated by strong AGN outbursts \citep{CiottiOstriker_2001ApJ...551..131C,Ciotti_2010ApJ...717..708C,Shin_2012ApJ...745...13S,Ciotti_2017ApJ...835...15C}.  The cycles qualitatively resemble the behavior outlined in Figure \ref{figure:valve_schematic}.  Furthermore, cosmological simulations of galaxies with similar halo mass demonstrate that kinetic AGN feedback results in self-regulated atmospheres in better agreement with observations than the ones regulated by purely thermal feedback \citep{Choi_2015MNRAS.449.4105C,Weinberger_2017MNRAS.465.3291W}.   However, those simulation analyses did not explicitly explore how $\sigma_v$  determines the nature of self-regulation.

\begin{figure}[!t]
\centering
\includegraphics[width=5.3in]{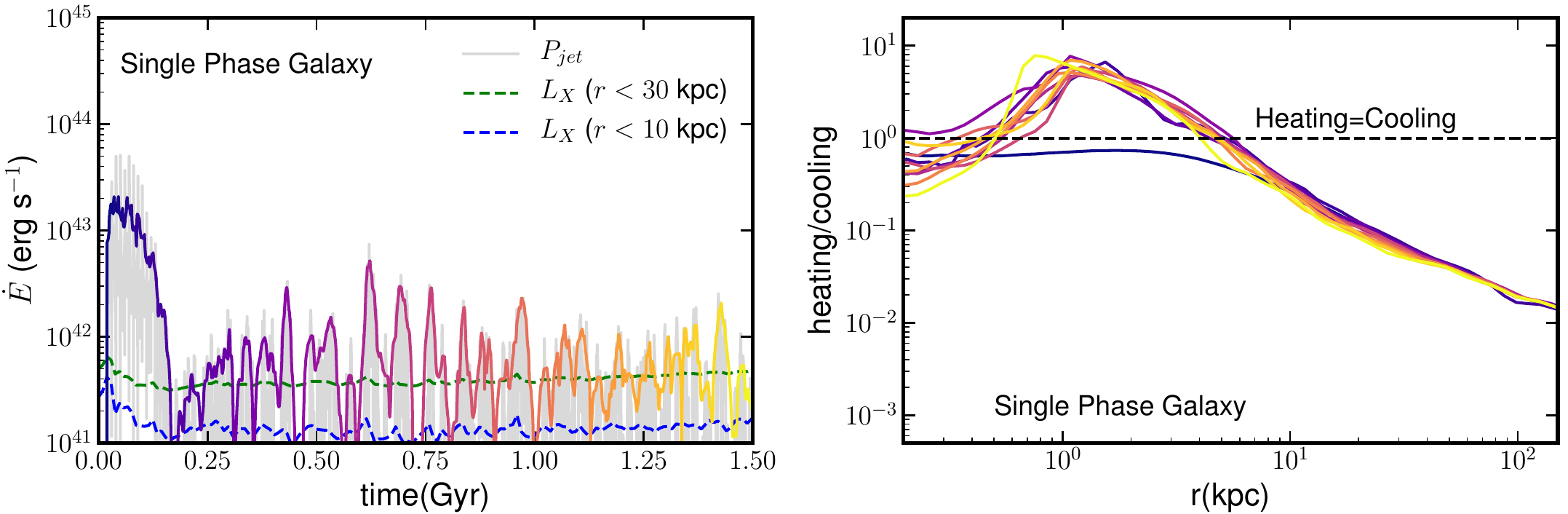}
\caption{Numerical simulation of kinetic AGN feedback in an idealized elliptical galaxy with $\sigma_v \approx 300 \, {\rm km \, s^{-1}}$ and $\alpha_K \approx 1$ (from \citep{Prasad_2020ApJ...905...50P}).  \textit{Left:} Evolution of jet power ($P_{\rm jet}$, solid gray line) and X-ray luminosity ($L_X$) from within 30~kpc (dashed green line) and 10~kpc (dashed blue line).  A thick line representing smoothed jet power is color coded according to time.  \textit{Right:} Local ratio of SNIa heating to radiative cooling as a function of radius and time.  Color coding of the lines represents time through correspondence with the colors of the thick line in the left panel.  An initial transient burst of AGN feedback reconfigures the atmosphere so that SNIa heating exceeds radiative cooling within the galaxy.  After that, AGN feedback closely self-regulates.
\label{fig:SPG_simulation}}
\end{figure}

Two recent simulation efforts have focused more directly on the significance of $\sigma_v$ and the resulting power-law slopes of pressure, density, and entropy.  \citet{WangLiRuszkowski_2019MNRAS.482.3576W} simulated idealized examples of both single phase and multiphase galaxies from \citet{Werner+2014MNRAS.439.2291W}, finding that the simulated single phase galaxy with $\sigma_v \approx 300 \, {\rm km \, s^{-1}}$ and $\alpha_K \approx 1$ successfully self-regulated without producing extended multiphase gas.  \citet{Prasad_2020ApJ...905...50P} performed similar simulations, investigating more closely the interplay between SNIa heating and radiative cooling. Figure \ref{fig:SPG_simulation} shows how feedback in their single phase galaxy with $\sigma_v \approx 300 \, {\rm km \, s^{-1}}$ and $\alpha_K \approx 1$ evolves with time.  After a transient burst of strong AGN feedback, which expands the atmosphere until SNIa heating exceeds radiative cooling from $\sim 1$~kpc to $\sim 5$~kpc, the galaxy settles into a well-regulated state in which time-averaged AGN power is comparable to radiative losses from within the central $\sim 30$~kpc.

Cosmological numerical simulations of galaxy evolution have not yet implemented bipolar kinetic AGN feedback with numerical resolution comparable to that achieved in these numerical simulations of idealized galaxies.  High resolution is important for enabling the black hole feedback valve mechanism, because the distances to which bipolar jets propagate depend on their momentum flux.  Poorly resolved jets do not propagate as far as well resolved jets and may thermalize their energy at radii that are too small because of their artificial bluntness.  Too much energy deposition at small radii will fail to replicate the valve mechanism if it inverts the inner entropy profile, resulting in convection and multiphase condensation over an extended region.  Careful investigations of how numerical resolution affects the ability of AGN feedback to drill through an elliptical galaxy's inner atmosphere without disturbing it (as in NCG 4261 and IC 4296, see Figure \ref{fig:narrow_jets}) will therefore be needed.

\subsection{Entropy, Convection, and Star Formation}
\label{sec:ConvectiveCGM}

The AGN feedback mechanism that regulates the atmospheres of single phase elliptical galaxies is clean and efficient, in the sense that their multiphase gas is restricted to the central kiloparsec.  Precipitation can then fuel the AGN without making many stars.  However, feedback is messier in multiphase elliptical galaxies and BCGs, in which extended multiphase gas is an apparent byproduct (\S \ref{sec:Weather}).  As discussed in \S\ref{sec:Evidence}, those atmospheres are plausibly precipitation-limited because they are observed to have $\alpha_K \approx 2/3$ and $10 \lesssim \min (t_{\rm cool} / t_{\rm ff}) \lesssim 30$ and are dense enough for radiative cooling to exceed SNIa heating within $\sim 10$~kpc.

Whether or not AGN feedback produces extended multiphase gas as a byproduct seems to depend on atmospheric structure.  Among halos of mass $\lesssim 10^{14} \, M_\odot$, central black holes appear to be capable of releasing enough feedback energy to significantly lift the halo's entire atmosphere (see Figure \ref{figure:ECGM_EBH}), driving some of it beyond the virial radius (see Figure \ref{figure:Singh2021}).  That capability is essential for the black hole feedback valve mechanism to come into play.  Once AGN feedback can lift the galaxy's CGM, a galaxy's ability to continually form stars depends on the entropy slope $\alpha_K$ of its atmosphere, because of how it affects both buoyancy and the radial $t_{\rm cool} /t_{\rm ff}$ profile (\S \ref{sec:BuoyancyDamping}).  After lifting of the CGM reduces its pressure enough for SNIa heating to exceed radiative cooling within $\sim 10$~kpc, $\alpha_K$ should then depend directly on $\sigma_v$ as expressed in equation ($\ref{eq:alphaK_sigmav}$).

According to this analysis, the atmospheric structure of massive galaxies should separate into three qualitatively distinct regimes, schematically represented in Figure \ref{figure:sigmav_schematics}:  

\begin{figure}[!t]
\centering
\includegraphics[width=5.3in]{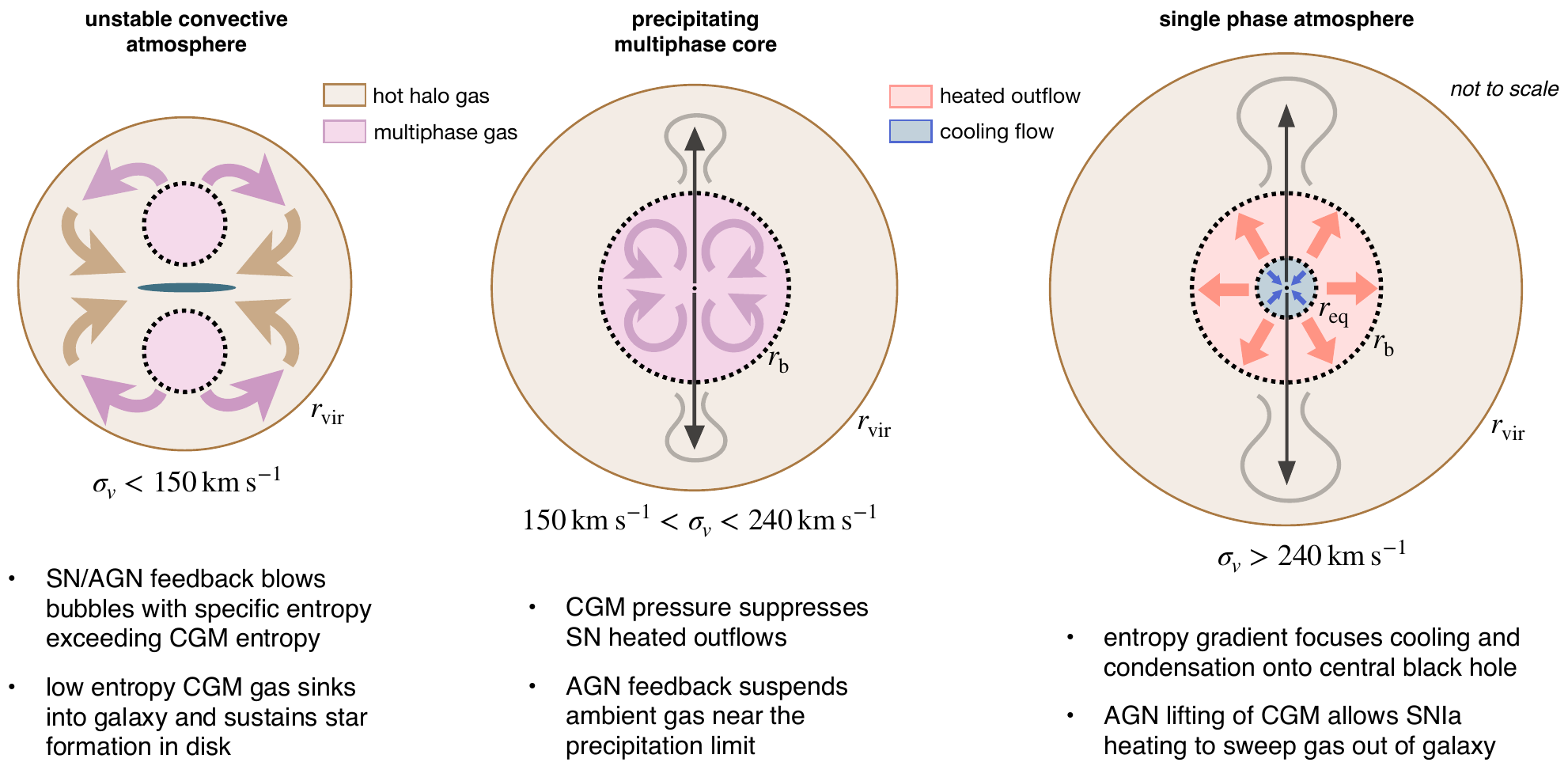}
\caption{Schematic representation of three qualitatively different regimes of coupling between AGN feedback and a massive galaxy's atmosphere.  
\label{figure:sigmav_schematics}}
\end{figure}

\begin{itemize}
    
    \item \textbf{Unstable convective atmosphere.} Accretion shocks around central galaxies in lower mass halos (approximately corresponding to $\sigma_v < 150 \, {\rm km \, s^{-1}}$) do not generate enough entropy to prevent multiphase condensation of the CGM \citep{Voit_2020ApJ...899...70V}.  Also, feedback from both supernovae and AGNs can heat the central gas to entropy levels exceeding what cosmological accretion produces.  Such halos experience entropy inversions that are inevitably convective and therefore prone to precipitation at all radii.  In this regime, AGN feedback may have difficulty suppressing star formation, because thermal energy introduced into a highly multiphase CGM tends to escape along the paths of least resistance, pushing away the hot component without preventing infall of cooler gas.  In fact, AGN feedback in this regime may  promote additional multiphase condensation and star formation by generating CGM entropy inversions.
 
    \item \textbf{Precipitating multiphase core.} Supernova heating in galaxies with intermediate central potential wells (corresponding to $150 \, {\rm km \, s^{-1}} < \sigma_v < 240 \, {\rm km \, s^{-1}}$), cannot drive gas out of the galaxy without initiating multiphase circulation, as long as the CGM pressure is great enough to prevent supersonic escape from the halo.  AGN feedback can maintain the central region in a precipitation-limited multiphase state, and cosmological shock heating outside of that multiphase region establishes a long-lasting entropy gradient that suppresses condensation of hot halo gas at larger radii. AGN feedback can suppress star formation among massive galaxies in this regime of $\sigma_v$ by lifting the CGM until precipitation diminishes, but star formation may still continue at a diminished rate.
 
    \item \textbf{Single phase atmosphere.} In galaxies with a deep central potential well ($\sigma_v > 240 \, {\rm km \, s^{-1}}$), lifting of the galaxy's entire atmosphere until supernova heating exceeds radiative cooling at 1--10~kpc can enable the black hole feedback valve mechanism, which focuses multiphase condensation onto the central black hole. The asymptotic steady-state structure of the galaxy's atmosphere should then have three layers: an inner cooling flow at small radii, an outflow heated by SNIa at intermediate radii, and an outer region of low-density hot halo gas.  The heated outflow corresponds to the galaxy's X-ray corona (\S \ref{sec:Coronae}), and the extent of that corona may be limited by ram-pressure stripping as the galaxy moves through the halo gas (see Figure \ref{Figure:NCG1265_corona}).  Star formation is fully quenched in this population because there is very little extended multiphase gas.
    
\end{itemize}

Cosmological numerical simulations support several aspects of this broad-brush picture. For example, \citet{Bower_2017MNRAS.465...32B} show that AGN feedback becomes strong in the EAGLE simulations when the hot bubbles blown in the CGM by supernova feedback no longer have an entropy exceeding the median CGM entropy.  When that happens, buoyancy can no longer lift those bubbles out of the central region of the halo, which results in a central buildup of gas.  Fueling of AGN feedback is the inevitable result, and it sets in at halo masses of $\sim 10^{12-12.5} \, M_\odot$ (corresponding to $\sigma_v \sim 100$--$150 \, {\rm km \, s^{-1}}$).  Davies et al. \citep{Davies_2019MNRAS.485.3783D,Davies_2020MNRAS.491.4462D,Davies_2021MNRAS.501..236D} and \citet{Oppenheimer_2020MNRAS.491.2939O} have demonstrated that star formation in a halo's central galaxy then becomes quenched as AGN feedback lifts the CGM, thereby reducing the density of the inner atmosphere and lowering its cooling rate.

\begin{figure}[!t]
\centering
\includegraphics[width=5.1in]{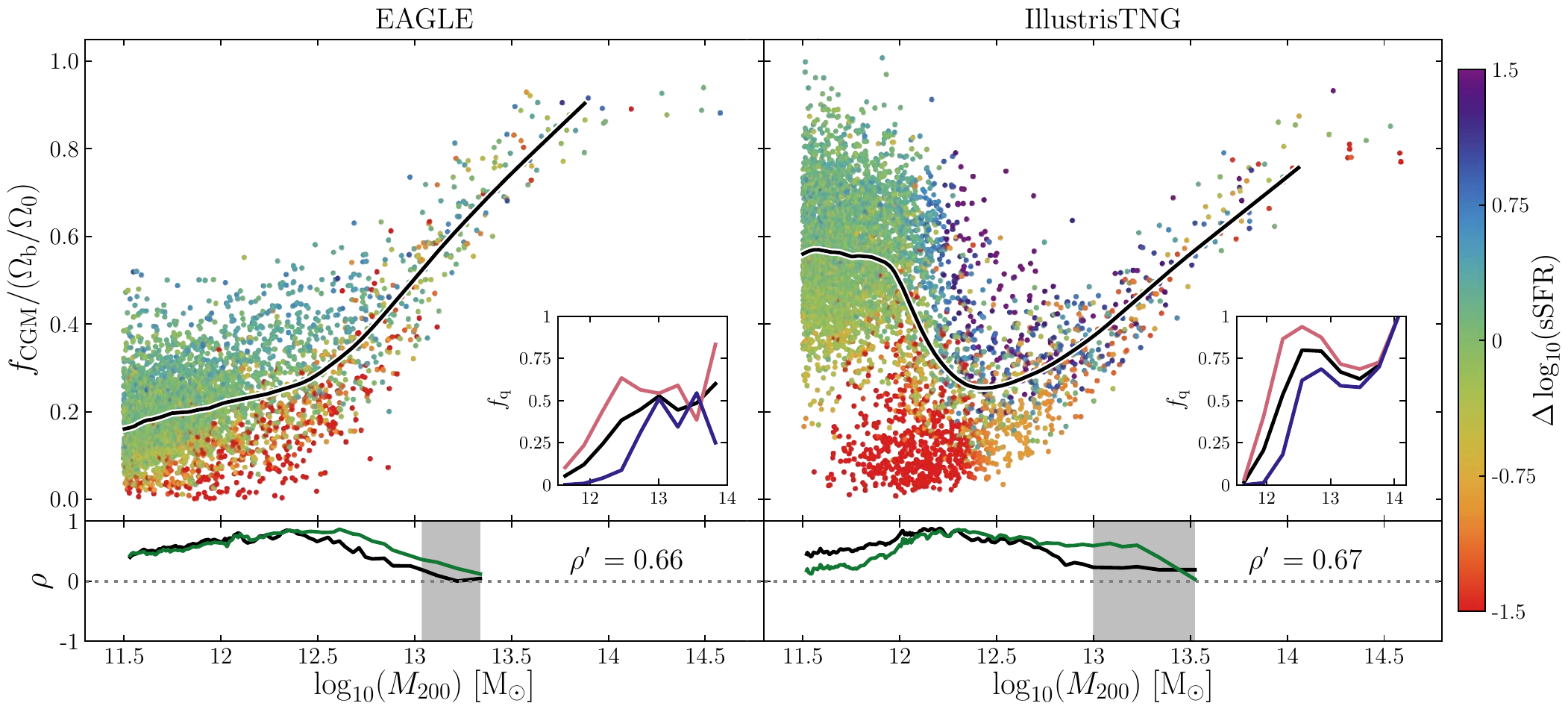}
\caption{Evidence for a causal relationship between star-formation quenching and CGM lifting, found by \citet{Davies_2020MNRAS.491.4462D} in two numerical simulations of cosmological galaxy evolution, EAGLE (left) and IllustrisTNG (right).  Each panel shows the dependence on halo mass ($M_{200}$) of the fraction of a halo's baryons remaining in the CGM, $f_{\rm CGM}/(\Omega_{\rm b}/\Omega_0)$. Black lines show the median relationship.  Dot colors indicate the deviation of a particular galaxy's specific star-formation rate (sSFR) from the median sSFR of the central galaxies in halos of similar mass.  Inset panels show the fraction of quenched galaxies as a function of halo mass, with a black line representing all central galaxies, a red line representing galaxies below the median CGM mass fraction at $M_{200}$, and a blue line representing galaxies above the median CGM mass fraction at $M_{200}$.  Sub-panels below each main panel show the Spearman rank coefficient ($\rho$) of the correlation between $\Delta \log_{10} sSFR$ and the CGM gas deficit, with a black line representing the CGM out to $r_{200c}$, a green line representing the CGM out to $0.3 r_{200c}$, and shading showing where those correlations have lower statistical significance.
\label{figure:sSFR_fCGM}}
\end{figure}

Figure \ref{figure:sSFR_fCGM} (from \citet{Davies_2020MNRAS.491.4462D}) shows the relationship between star-formation quenching and CGM lifting in the EAGLE and IllustrisTNG simulations.  Those two cosmological simulation efforts implement completely different numerical methods and also different algorithms for AGN feedback.\footnote{Note the large differences between the simulations in the median relationship between halo mass and CGM baryon fraction, which demonstrate that implementation of feedback in numerical simulations of galaxy evolution is still very much a work in progress (see \citet{Oppenheimer_2021arXiv210613257O} for a recent review).}  But in both of them, star formation in the central galaxies of halos in the mass range $\sim 10^{12-13} \, M_\odot$ is highly correlated with the fraction of the halo's baryons within $r_{200c}$.  In the galaxies with the lowest star-formation rates, the fraction of baryons remaining in the CGM is $\lesssim 0.2$. The CGM baryon fraction is approximately twice as large in the actively star forming galaxies at each halo mass.  Furthermore, \citet{Oppenheimer_2020MNRAS.491.2939O} have shown that the transition to a low star-formation rate in EAGLE coincides in time with a strong AGN feedback outburst that pushes the CGM outward.

\subsection{Central Black Hole Mass and Star Formation}
\label{sec:BlackHoleMass}

If CGM lifting is indeed essential for long-term quenching of star formation within massive galaxies, then the mass of the central black hole supplying the energy necessary for lifting the CGM should depend on the mass of the galaxy's halo as predicted by \citet{BoothSchaye_2010MNRAS.405L...1B} (see also \S \ref{sec:EnergyRequirements}).  Observations (see Figure \ref{figure:ECGM_EBH}) and simulations (see Figure \ref{figure:sSFR_fCGM}) both indicate that AGN feedback is capable of reducing the CGM density and pressure in halos of mass $\sim 10^{12-14} \, M_\odot$ (see also \citet{Truong_2020MNRAS.494..549T,Truong_2021MNRAS.501.2210T}).  And in galaxy clusters (halos of mass $\gtrsim 10^{14} \, M_\odot$), AGN feedback apparently tunes itself to balance radiative cooling of the gas with $t_{\rm cool} \lesssim H_0^{-1}$ (see Figure \ref{figure:Sun2009}).  However, the link between black hole mass and quenching of star formation must be more subtle, because quenching of star formation is more closely related to $\sigma_v$ than to halo mass (see Figure \ref{figure:fQuench}).

According to the schema outlined in Figure \ref{figure:sigmav_schematics}, quenching of star formation is closely connected to $\sigma_v$ because AGN feedback becomes ever more tightly coupled to the CGM as $\sigma_v$ increases.  In other words, growth in $\sigma_v$ as a galaxy evolves enables AGN feedback to shut off long-term star formation when $\sigma_v$ reaches a critical value depending on $\epsilon_*$.  Focusing of cooling gas onto the galaxy's central black hole mass causes the black hole to grow until it has released enough energy to significantly lift the CGM, thereby alleviating central cooling.  An amount of feedback energy comparable to the CGM binding energy is required.

\begin{figure}[!t]
\centering
\includegraphics[width=4.6in]{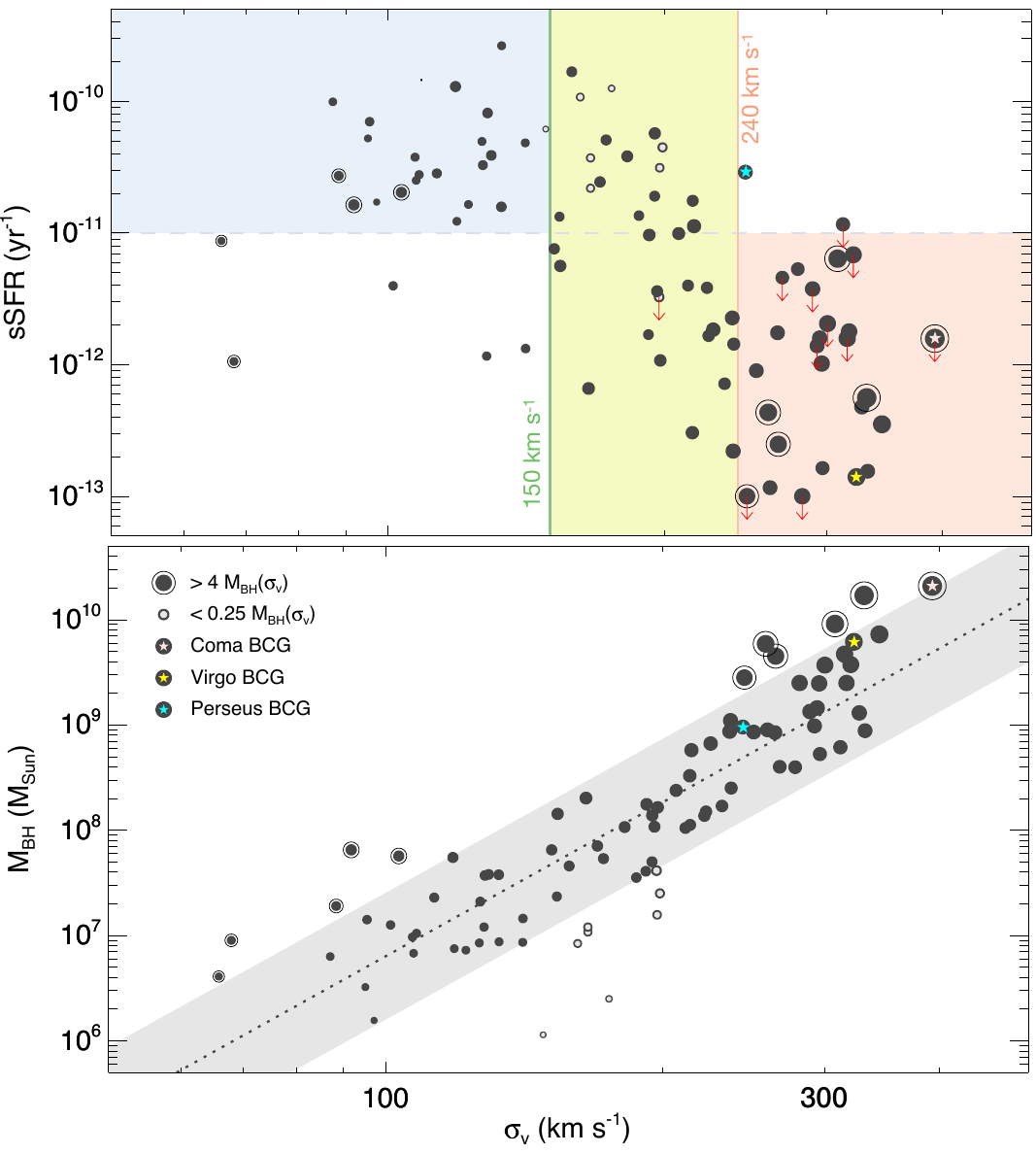}
\caption{Dependences of a galaxy's specific star formation rate (sSFR, top) and central black hole mass ($M_{\rm BH}$, bottom) on its central stellar velocity dispersion ($\sigma_v$), based on data compiled by \citet{Terrazas_2017ApJ...844..170T}.  Colors in the upper panel correspond to the intervals of $\sigma_v$ given in Figure \ref{figure:sigmav_schematics}.  Below the gray dashed line at sSFR $= 10^{-11} \, {\rm yr^{-1}}$, star formation is considered quenched.  Galaxies above that line are considered actively star forming.  Symbol sizes in both panels indicate relative black hole mass.  A dotted line in the bottom panel shows the best fitting $M_{\rm BH}$--$\sigma_v$ relation. gray shading around that line shows where $M_{\rm BH}$ is within a factor of four of that best fit.  Points corresponding to black holes with greater masses are circled, while points corresponding to black holes with lower masses are marked with gray dots.  Three famous BCGs in the sample are marked with colored stars, as listed in the legend.  Downward pointing red arrows in the upper panel indicate upper limits on star formation rates. 
\label{figure:Terrazas_composite}}
\end{figure}

Figure \ref{figure:Terrazas_composite} shows some observed relationships between $\sigma_v$, $M_{\rm BH}$, and specific star-formation rate that support this general picture.  The data were compiled by Terrazas et al. \citep{Terrazas_2016ApJ...830L..12T,Terrazas_2017ApJ...844..170T}, who showed that galactic star formation at fixed stellar mass anticorrelates with $M_{\rm BH}$, implicating the central black hole in the shutdown of star formation (see also \citet{Martin-Navarro_2016ApJ...832L..11M,Martin-Navarro_2018Natur.553..307M}).  The figure presents those data as functions of $\sigma_v$, as listed in the Hyperleda catalog.\footnote{http://leda.univ-lyon1.fr/}

The green shaded region in the figure's upper panel, marking $150 \, {\rm km \, s^{-1}} < \sigma_v < 240 \, {\rm km \, s^{-1}}$ is clearly a transitional one for star formation.  Most of the galaxies with lower $\sigma_v$ fall within the blue shaded region, in which star formation is active. Most of the galaxies with greater $\sigma_v$ fall within the red shaded region, where star formation has been quenched.  The only galaxy in the sample that has both a specific star formation rate $> 10^{-11} \, {\rm yr}^{-1}$ and $\sigma_v > 240 \, {\rm km \, s^{-1}}$ is NGC~1275, the Perseus Cluster's BCG, in which star formation may be a consequence of precipitation induced by AGN feedback.

Within the green shaded region, a galaxy's specific star formation rate anticorrelates with $\sigma_v$, but the other two regions show essentially no such anticorrelation.  Also, the specific star formation rate at a given $\sigma_v$ anticorrelates with $M_{\rm BH}$.  All but one of the galaxies falling unusually far below the mean $M_{\rm BH}$--$\sigma_v$ relation (marked by gray dots) are actively star forming, while most of their counterparts at greater $M_{\rm BH}$ reside in galaxies with quenched star formation.

\begin{figure}[!t]
\centering
\includegraphics[width=5.3in]{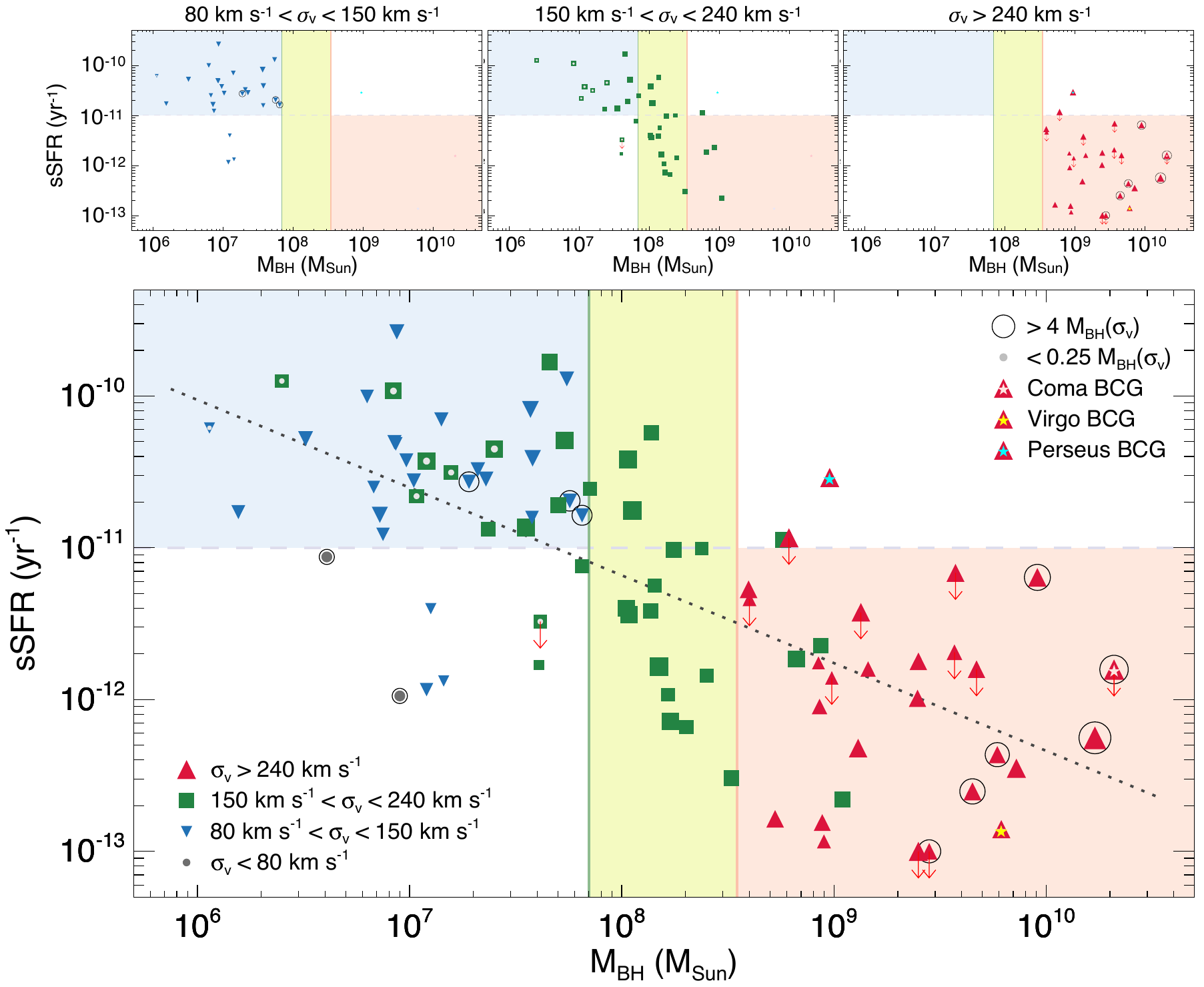}
\caption{Dependence of a galaxy's specific star formation rate (sSFR) on its central black hole mass ($M_{\rm BH}$).  The main panel shows the same data as in Figure \ref{figure:Terrazas_composite}, while the three panels across the top show subsets of it, sorted according to $\sigma_v$.  Filled gray circles show galaxies with $\sigma_v < 80 \, {\rm km \, s^{-1}}$. Inverted blue triangles show galaxies with $80 \, {\rm km \, s^{-1}} < \sigma_v < 150 \, {\rm km \, s^{-1}}$. Green squares show galaxies with $150 \, {\rm km \, s^{-1}} < \sigma_v < 240 \, {\rm km \, s^{-1}}$. Red triangles show galaxies with $\sigma_v > 240 \, {\rm km \, s^{-1}}$.  As in Figure \ref{figure:Terrazas_composite}, galaxies with black holes far above the mean $M_{\rm BH}(\sigma_v)$ relation are circled, filled gray circles mark black holes far below the mean $M_{\rm BH}(\sigma_v)$ relation, and colored stars mark the famous BCGs. 
\label{figure:Terrazas_MBH_sSFR}}
\end{figure}

Figure \ref{figure:Terrazas_MBH_sSFR} shows the same data, plotted as a function of central black hole mass.  A dotted line in the main panel shows the anticorrelation between $M_{\rm BH}$ and sSFR spotlighted by \citet{Terrazas_2016ApJ...830L..12T}.  Three subpanels across the top of the figure show the same relationship in subsets of those data, sorted according to $\sigma_v$.  Specific star formation rates clearly anticorrelate with $M_{\rm BH}$ among the galaxies with $150 \, {\rm km \, s^{-1}} < \sigma_v < 240 \, {\rm km \, s^{-1}}$.  No such anticorrelation is evident among either the galaxies with $\sigma_v < 150 \, {\rm km \, s^{-1}}$ or the galaxies with $\sigma_v > 240 \, {\rm km \, s^{-1}}$.  Apparently, star formation is independent of central black hole mass among the population expected to have highly convective atmospheres and also among the population in which the feedback valve mechanism can operate. 

\subsection{Evolution of Quenching Criteria}
\label{sec:QuenchingCriteria}

The critical threshold value of $\sigma_v$ predicted by the black hole feedback valve model is not universal.  It depends on the specific energy $\epsilon_*$ of the gas shed by a galaxy's aging stellar population, which itself depends on both the specific SNIa rate and the specific stellar mass-loss rate. Both of those rates evolve with time.  Therefore, the model predicts that the critical value of $\sigma_v$ should also evolve with time.  

\citet{Voit_2020ApJ...899...70V} estimated the evolution rate assuming a specific SNIa rate $\propto t^{-1.3}$ (following \citep{FriedmannMaoz_2018MNRAS.479.3563F}) and a specific stellar mass-loss rate $\propto t^{-1}$ (following \citep{LeitnerKravtsov_2011ApJ...734...48L}), where $t$ is the age of the stellar population.  Combining those dependences gives $\epsilon_* \propto t^{-0.3}$ and therefore a time dependence $\propto t^{-0.15}$ for the critical $\sigma_v$ threshold.  The predicted critical value for maintaining long-term quenching consequently drops from $340 \, {\rm km \, s^{-1}}$ at a stellar population age of 1~Gyr to $240 \, {\rm km \, s^{-1}}$ at a stellar population age of 10~Gyr.
Comparisons of how quenching correlates with both $\sigma_v$ and $\Sigma_1$ in \textit{Hubble} galaxy surveys at $z \sim 2$ and large galaxy surveys at lower redshifts indicate a similar decline with time \citep{Franx_2008ApJ...688..770F,vanderWel_2009ApJ...698.1232V,vanDokkum_2014ApJ...791...45V,vanDokkum_2015ApJ...813...23V,Tacchella_2015Sci...348..314T,Barro_2017ApJ...840...47B,ChenFaber_2020ApJ...897..102C}.  

\begin{figure}[!t]
 \centering
 \includegraphics[width=5.3in]{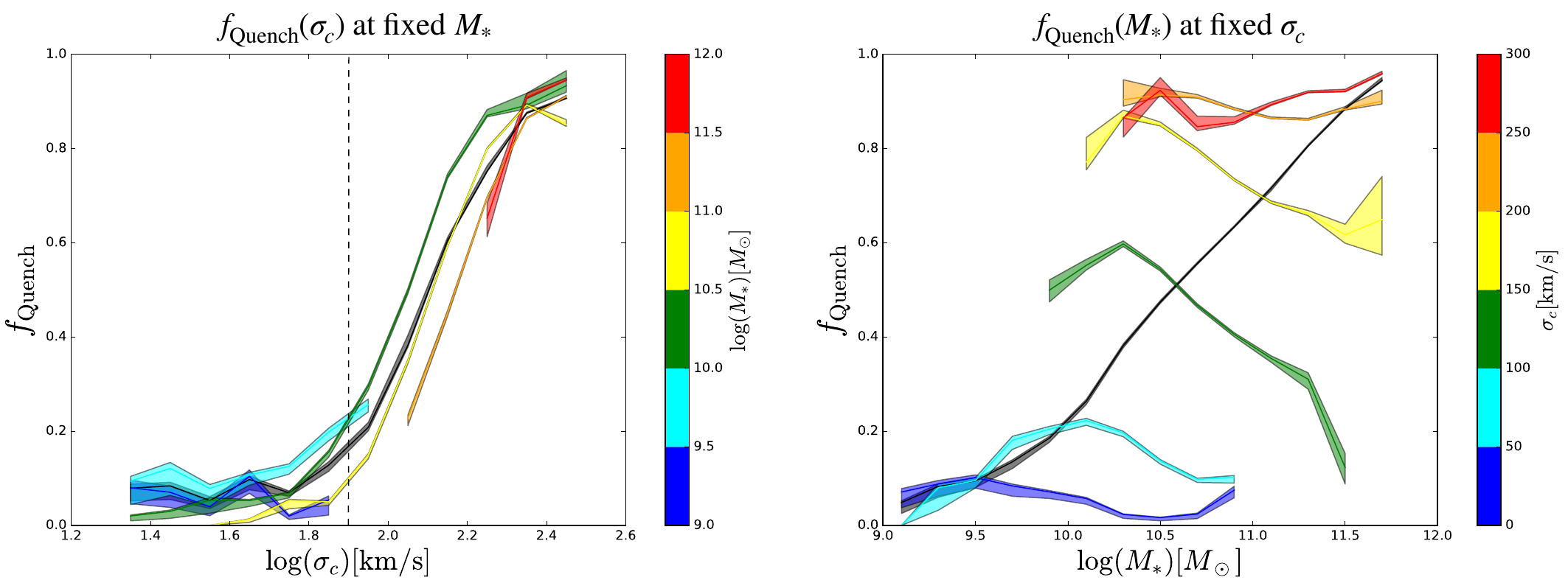}
  \caption{Observed dependence of the fraction $f_{\rm Quench}$ of SDSS galaxies with quenched star formation on central velocity dispersion (represented by $\sigma_{\rm c}$ in this figure) and the galaxy's total stellar mass (from \citep{Bluck_2016MNRAS.462.2559B}). \textit{Left:} Relationship between $f_{\rm Quench}$ and stellar velocity dispersion (solid black line), with colored regions showing the dependence of $f_{\rm Quench}$ on $\sigma_{\rm c}$ within bins of stellar mass.  \textit{Right:} Relationship between $f_{\rm Quench}$ and stellar mass (solid black line), with colored regions showing the dependence of $f_{\rm Quench}$ on stellar mass within bins of $\sigma_{\rm c}$.  After the dependence of $f_{\rm Quench}$ on $\sigma_{\rm c}$ is accounted for, quenching anticorrelates with total stellar mass in the 100--$200 \, {\rm km \, s^{-1}}$ range of stellar velocity dispersion, in contrast to the lack of residual dependence on halo mass in Figure \ref{figure:fQuench}.
  \label{figure:fQuench_Mstar}}
\end{figure}

Both the black hole feedback valve model and its dependence on $\epsilon_*$ need to be validated with numerical simulations before more quantitative comparisons with quenching observations can be made.  However, another feature of the quenching threshold suggests an independent test of the model.  The critical value of $\sigma_v$ for quenching appears to depend on a galaxy's total stellar mass as well as on time \citep{Fang_2013ApJ...776...63F,Barro_2017ApJ...840...47B,ChenFaber_2020ApJ...897..102C}, producing a ``tilt" of the quenching threshold in the $M_*$-$\sigma_v$ plane approximately following $\sigma_v \propto M_*^{1/3}$.  Figure \ref{figure:fQuench_Mstar} (from \citep{Bluck_2016MNRAS.462.2559B}) shows the residual dependence of quenching on $M_*$, once the dependence on $\sigma_v$ is removed.   Quenching clearly anticorrelates with $M_*$ for galaxies with central velocity dispersion in the range 100--200~km~s$^{-1}$.  The galaxy's stellar population apparently plays a role in quenching that is independent of halo mass, which shows no such residual correlation (see Figure \ref{figure:fQuench}).

If a galaxy's $\epsilon_*/\sigma_v^2$ ratio is indeed the primary factor determining the quenching threshold, as envisioned by the black hole feedback valve model, then the anticorrelation in Figure \ref{figure:fQuench_Mstar} implies that $\epsilon_*$ should correlate with $M_*$ at fixed $t$.  One possible origin for a dependence of $\epsilon_*$ on $M_*$ is a dependence of a galaxy's specific SNIa rate on stellar mass.  Efforts to measure the dependence of a stellar population's specific SNIa rate on time suggest that this rate depends on more than just the age of the stellar population. For example, \citet{FriedmannMaoz_2018MNRAS.479.3563F} find that the specific SNIa rate is $\sim 3 \times 10^{-14} \, {\rm SNIa \, yr^{-1}} (t / 10 \, {\rm Gyr})^{-1.3}$ for elliptical galaxies belonging to galaxy clusters, while the rate among elliptical galaxies not belonging to clusters is $\sim 2 \times 10^{-14} \, {\rm SNIa \, yr^{-1}} (t / 10 \, {\rm Gyr})^{-1.1}$ \citep{MaozGraur_2017ApJ...848...25M}.  The origin of this difference is unknown, but it may correlate with $M_*$ at fixed $t$, because the galaxies belonging to clusters are generally more massive that those that are not in clusters.\footnote{Observations indicate that the specific SNIa rate among all galaxies anticorrelates with $M_*$, but that anticorrelation arises primarily from the dependence of SNIa rate on stellar population age, because the ages of the stars correlate with $M_*$ \citep{Graur_2015MNRAS.450..905G}. More SNIa observations are needed to constrain the dependence of the specific SNIa rate on $M_*$ at fixed population age.}

\subsection{Causes and Effects}
\label{sec:CauseOrEffect}

A consensus supported by several complementary lines of evidence is now growing around the idea that AGN feedback quenches star formation in massive galaxies by pumping an amount of energy equivalent to the CGM binding energy into a galaxy's atmosphere \citep[e.g.,][]{Bower_2017MNRAS.465...32B,Davies_2020MNRAS.491.4462D,ChenFaber_2020ApJ...897..102C,Oppenheimer_2020MNRAS.491.2939O,Voit_2020ApJ...899...70V,Zinger_2020MNRAS.499..768Z}.  The AGN energy supply appears necessary for explaining the observed atmospheric configurations in halos of mass $\sim 10^{12-14} \, M_\odot$ (\S \ref{sec:GalaxyGroups}, \S \ref{sec:MassiveGalaxies}, \S \ref{sec:MassiveGalaxiesEvidence}).  But what is the root cause of the transition from active star formation to quiescence?

\subsubsection{Black Hole Growth}
\label{sec:BH_growth}

One commonly expressed viewpoint is that quenching simply results from an episode of rapid black hole growth.  The root cause is taken to be a sudden increase in the black hole accretion rate \citep[e.g.,][]{Bower_2017MNRAS.465...32B,ChenFaber_2020ApJ...897..102C}.  From that viewpoint, the strong observed correlation between $\sigma_v$ and star-formation quiescence arises from a more fundamental link between black hole mass and quenching, paired with a less fundamental link between $\sigma_v$ and $M_{\rm BH}$.  

\citet{Bower_2017MNRAS.465...32B} base their interpretation of the black hole's role on what they observe in the EAGLE simulations and connect it to growth in halo mass.  As mentioned in \S \ref{sec:ConvectiveCGM}, supernova feedback is effective as long as it generates hot bubbles containing gas with specific entropy exceeding the surrounding CGM entropy (see also \citet{Keller_2016MNRAS.463.1431K,Keller_2020MNRAS.493.2149K}).  Buoyancy is then able to lift the heated gas out of the halo.  But that mechanism falters in the halo mass range $\sim 10^{12-12.5} \, M_\odot$, because of the larger CGM entropy in higher-mass halos.  In simulated EAGLE galaxies, the resulting central buildup of gas then fuels rapid black hole growth and strong AGN feedback, which lifts the CGM and quenches star formation \citep{Davies_2020MNRAS.491.4462D,Oppenheimer_2020MNRAS.491.2939O}.

\citet{ChenFaber_2020ApJ...897..102C} interpret the black hole's role differently.  Their phenomenological model for quenching is inspired by the patterns observed in large galaxy surveys. Black hole growth is assumed to be tied to galaxy growth during the time when star formation is active, resulting in a well-defined $M_{\rm BH}$--$\sigma_v$ relation.  Quenching then begins when the black hole has pumped enough energy into the CGM to significantly lift it.  What follows is a period of black hole growth during which $M_{\rm BH}$ rises without much change in $\sigma_v$ or $\Sigma_1$.  After that period of black hole growth ends, star-formation quenching is complete, and the galaxy resides on a new $M_{\rm BH}$--$\sigma_v$ relation having the same power-law slope as the original one but translated upward in $M_{\rm BH}$ by an order of magnitude.

\subsubsection{Growth of $\sigma_v$}

Quenching of star formation follows from black hole growth in the scenarios that \S \ref{sec:BH_growth} describes.  However, the evidence presented in \S \ref{sec:Quenching} suggests that black hole growth follows from something more fundamental: a change in the galaxy's central structure.  That is, growth of $\sigma_v$ may be what triggers growth of $M_{\rm BH}$ and the quenching that ensues.  First, observations indicate that quenching depends directly on $\sigma_v$, with halo mass playing little or no role once the correlation of halo mass with $\sigma_v$ is factored out.  Second, a massive galaxy's atmospheric structure within the central few kiloparsecs appears to depend more critically on $\sigma_v$ than on either halo mass or total stellar mass.  Therefore, a galaxy's central stellar mass density, as reflected by either $\sigma_v$ or $\Sigma_1$, may be the key galactic property that ensures rapid black hole growth, rather than being just a proxy for $M_{\rm BH}$.

In the scenario that \S \ref{sec:Valve} outlines, the period of black hole growth that brings about quenching depends on the relationship between $\sigma_v$ and supernova energy input.  Quenching via AGN feedback happens when supernova feedback starts to fail, but it entails more than just a central pileup of gas, as envisioned by \citet{Bower_2017MNRAS.465...32B}.  Its other important feature is focusing of atmospheric cooling and condensation onto the central black hole, as the galaxy's $\epsilon_* / \sigma_v^2$ ratio declines.  This second feature is what enables AGN feedback to lift the entire CGM, rather than just driving multiphase circulation. Quenching of star formation then follows from CGM lifting.  

A recent numerical experiment by \citet{Davies_2021MNRAS.501..236D} supports these conjectures.  The experiment compares three simulations of star formation using the EAGLE galaxy formation model within a halo that reaches $M_{200c} = 10^{12.53} \, M_\odot$ at $z = 0$.  Each simulation begins with slightly different initial conditions.  An ``organic" simulation is the standard to which two other simulations are compared.  An ``early" simulation begins with slightly denser initial conditions, resulting in earlier mass assembly.  And a ``late" simulation begins with slightly less dense initial conditions, resulting in later mass assembly.  Star formation in the halo's central galaxy becomes quenched only in the ``early" simulation, as AGN feedback lifts the CGM between $z \approx 2$ and $z \approx 1$.  The ``early" halo is also the one with the most centrally concentrated mass distribution.  In the ``organic" simulation, star formation is somewhat suppressed but not fully quenched at $z=0$.  AGN feedback has begun to lift the CGM by $z \sim 0$ in that simulation, but the CGM gas mass is still twice that in the ``early" simulation.  And in the ``late" simulation, with the lowest central mass concentration of the three, AGN feedback has neither suppressed star formation nor lifted the CGM by $z = 0$.  Yet all three simulated galaxies have entered the halo mass range in which stellar feedback is thought to be insufficient for regulating star formation. 

According to the qualitative picture outlined in Figure \ref{figure:sigmav_schematics}, crossing the halo mass threshold above which supernova feedback is insufficient shifts a galaxy from the left panel to the central panel.  However, the galaxy's central potential well is not yet deep enough to focus atmospheric cooling onto the central black hole, allowing multiphase gas and some star formation to persist.  In order to reach the panel on the right, a galaxy must cross the critical $\sigma_v$ threshold for focusing of atmospheric cooling, establishing a direct link between black-hole fueling and CGM pressure. 

A galaxy may cross the $\sigma_v$ threshold for quenching because of either growth in $\sigma_v$ or a decline in the threshold value resulting from a decline in specific stellar heat input.  Once the galaxy crosses that threshold, black hole accretion proceeds until AGN feedback has lifted the CGM.  The energy input required for lifting the CGM then brings about the strong correlation between black hole mass and CGM temperature shown in Figure \ref{figure:ECGM_EBH}.\footnote{Almost all of the black holes in Figure \ref{figure:ECGM_EBH} have $M_{\rm BH} \gtrsim 10^8 \, M_\odot$, suggesting that the correlation might be strongest among black holes that have already lifted the CGM.  Note that the spread in $\sigma_v$ at fixed $M_{\rm BH}$ in Figure \ref{figure:Terrazas_composite} is considerably greater below $10^8 \, M_\odot$ than above it, suggesting that the black holes with $M_{\rm BH} < 10^8 \, M_\odot$ are less closely connected with the larger galactic environment.}  But what causes $\sigma_v$ to grow? 

\subsubsection{Mergers and Compaction}

At long last, we arrive at a role for galaxy mergers.  Mergers are not necessary for quenching of star formation, but they can result in long-term quenching if they cause central star formation that increases the central stellar mass density and $\sigma_v$.  In simulations of galaxy evolution, mergers of gas-rich galaxies do tend to cause a central buildup of gas that initially leads to a burst of star formation \citep[e.g.,][]{BarnesHernquist_1991ApJ...370L..65B,MihosHernquist_1996ApJ...464..641M,DiMatteo_2005Natur.433..604D}.  During this starburst stage, the galaxy's central mass density increases as dissipative cooling allows the gas to sink toward the center.  After the centrally concentrated gas forms stars, the remaining stellar population has an increased central velocity dispersion.  If the central stellar mass density and post-merger velocity dispersion are large enough, then the black hole feedback valve mechanism can ensure long-term quenching.

More generally, the increase in $\sigma_v$ that ensures long-term quenching can arise from a phenomenon that has been called ``compaction" \citep[e.g.][]{Zolotov_2015MNRAS.450.2327Z}.  Gas-rich mergers are one route to a compact central stellar population with large $\sigma_v$, but there are others.  Dynamical instabilities of a galactic disk can also channel gas toward the center and are inevitable in gas disks that are being rapidly fed by cold streams of accreting gas \citep[e.g.,][]{Shlosman_1990Natur.345..679S,Gammie_2001ApJ...553..174G,DekelBurkert_2014MNRAS.438.1870D}.  

In the context of the black hole feedback valve model, the route to compaction is less important than the outcome.  A starburst or a strong AGN feedback event may accompany compaction and expel much of the galaxy's gas, temporarily halting further star formation.  But according to the model, long-term quenching results from the outcome of compaction: a dense central collection of stars that focuses atmospheric cooling onto the central black hole. 

\subsubsection{Future Investigations}

This concluding section of our review has tried to interpret galaxy evolution, and particularly quenching of star formation, by applying the physical principles governing galactic atmospheres, as outlined in Sections \ref{sec:AmbientCGM} through \ref{sec:Evidence}.  There are still many loose ends.  The proposed black hole feedback valve mechanism, while plausible, is far from proven.  Many other aspects of atmospheric physics, such as the implications of rotation, have yet to be thoroughly explored in the context of that model.  We will therefore finish the review by pointing out some potentially fruitful avenues for further research: 

\begin{itemize}

\item The black hole feedback valve model needs to be tested for robustness with high-resolution numerical simulations.  An authentic test of the mechanism requires AGN jets narrow enough to drill through the inner several kiloparsecs of an elliptical galaxy's atmosphere (i.e. the X-ray corona) without depositing much energy there, yet strong enough to lift the galaxy's CGM.  Under those conditions, the mechanism is expected to self-tune, remaining near a state with SNIa heating slightly greater than radiative cooling in the $\sim 1$--10~kpc region.  

\item An algorithm that successfully self-tunes in idealized simulations then needs to be implemented in cosmological simulations and tested to see whether it establishes a quasi-steady state similar to observations.  Simulations that succeed can then be tested to see if they correctly predict the redshifts at which quenching of star formation happens and how they are related to the observed correlation of $\sigma_v$ and $\Sigma_1$ with quenching in large galaxy surveys.

\item The predicted relationship between $\sigma_v$, star-formation quenching, and atmospheric structure at 1--10~kpc needs to be assessed with X-ray observations.  High spatial resolution will be essential for constraining the gradient of $t_{\rm cool} / t_{\rm ff}$ at small radii.  Approximately 50 galaxies are both bright enough and close enough for \textit{Chandra} observations to be adequate \citep[e.g.,][]{Babyk2018ApJ...862...39B,Lakhchaura_2018MNRAS.481.4472L}, but more examples are needed.  The proposed \textit{Lynx} mission or something similar will be critical for expanding the sample of massive galaxies with well-resolved atmospheric properties.

\item In the meantime, it will be interesting to examine the dependence of $L_X(R)$ on $\sigma_v$ and its relationship to star-formation quenching among large samples of eROSITA observations.  The model predicts that the atmospheric properties of elliptical galaxies in halos less massive than $\sim 10^{14} \, M_\odot$ should depend on $\sigma_v$.  Among halos of mass $\sim 10^{13.5} \, M_\odot$, the halos with central galaxies having $\sigma_v \approx 300 \, {\rm km \, s^{-1}}$ are predicted to have lower X-ray luminosity than those with $\sigma_v \approx 240 \, {\rm km \, s^{-1}}$, because their atmospheres are predicted have steeper density profiles.  Also, star-formation quenching should be inversely correlated with $L_X(R)$ at fixed $\sigma_v$, if lifting of the CGM is indeed what ensures permanent quenching of star formation.

\item The apparent ``tilt" of the quenching threshold in the $M_*$--$\sigma_v$ plane merits more exploration.  The black hole feedback valve model predicts that permanent quenching of star formation should depend closely on a $\sigma_v$ threshold that decreases with time, but the observed tilt of that threshold requires a second parameter related to $M_*$.  One potential explanation is a supernova heating rate that correlates with $M_*$.  However, the tilt may depend on other aspects of how stars interact with galactic atmospheres, including the mass loading factor ($\eta_M$) of supernova-driven outflows.

\item Rotation is clearly an important factor in precipitation because it can eliminate buoyancy, enabling CGM gas to cool without descending to smaller radii, as long as  rotational support dominates pressure support (see \citep[e.g.,][]{Oppenheimer_2018MNRAS.474.4740O,Stern_2020MNRAS.492.6042S}).  Rotating atmospheres should therefore be more prone to precipitation, resulting in a precipitation limit with a greater $t_{\rm cool}/t_{\rm ff}$ ratio, perhaps explaining why massive spiral galaxies are observed to have lower X-ray luminosities than similarly massive elliptical galaxies \citep[e.g.,][]{Anderson_2016MNRAS.455..227A,Bogdan_2017ApJ...850...98B,Li_2018ApJ...855L..24L}.  Also, observations indicate that ellipticals with greater rotation rates are more likely to contain extended multiphase gas \citep[e.g.,][]{Crocker_2011MNRAS.410.1197C,Young_2011MNRAS.414..940Y,Davis_2011MNRAS.417..882D,Juranova_2020MNRAS.499.5163J}.  These relationships need to be compared with models of precipitation-limited atmospheres that incorporate the dependence of precipitation on angular momentum.

\item How AGN feedback energy propagates to radii beyond where X-ray cavities are commonly observed ($\sim 10-20$~kpc) and thermalizes at those larger radii remains an unanswered question. More precise measurements of sound waves, internal gravity waves, turbulence, cosmic rays, and bulk flows are needed to inform the many different theoretical models currently under investigation. X-ray spectroscopy of galaxy clusters and groups during the next two decades, first with XRISM and later with Athena, will be essential to progress toward an answer.

\end{itemize}

\section*{Acknowledgments} 
\noindent This article made extensive use of NASA's Astrophysics Data System.  It benefited from the KITP program on \textit{Fundamentals of Gaseous Halos}, which was supported in part by the National Science Foundation under Grant No.~NSF PHY-1748958. MD acknowledges the support of the Caroline Herschel visiting scholar program at the Space Telescope Science Institute, where the first drafts were written. MD has been partially supported by NASA, Chandra X-ray Observatory/Smithsonian Astrophysical Observatory, and the Space Telescope Science Institute (SAO-AR1-22011X, SAO-NASA-GO0-21114B, STScI-HST-GO-15661.002-A, NASA-SAO-GO0-21116X, SAO-AR7-18008X, STSI HST-GO-13367.001-A, and NASA NNG05GD82G). We are grateful to the many colleagues and collaborators who have shared their thoughts and shaped our thinking in conversations over the last three decades, particularly Arif Babul, Nick Battaglia, Stefano Borgani, Joel Bregman, Richard Bower, Greg Bryan, Hsiao-Wen Chen, Eugene Churazov, Romeel Dave, August Evrard, Drummond Fielding, Kevin Fogarty, Massimo Gaspari, Tim Heckman, Andrey Kravtsov, Miao Li, Yuan Li, Greg Meece, Ian McCarthy, Mike McCourt, Michael McDonald, Brian McNamara, Richard Mushotsky, Paul Nulsen, Peng Oh, Benjamin Oppenheimer, Brian O'Shea, Annalisa Pillepich, Marc Postman, Deovrat Prasad, Eliot Quataert, Chris Reynolds, Mateusz Ruszkowski, Prateek Sharma, J. Michael Shull, Aurora Simonescu, Noam Soker, Rachel Somerville, John Stocke, Ming Sun, Grant Tremblay, Jess Werk, Norbert Werner, and Yong Zheng. MD is grateful to her Michigan State University students and alumni Kenneth Cavagnolo, Seth Bruch, Aaron Hoffer, Thomas Connor, Rachel Frisbie, and Dana Koeppe, for their patience, encouragement, enthusiasm, and scientific insights along this journey.

\bibliographystyle{elsarticle-num-names}


\bibliography{precipitation}

\end{document}